\documentclass{tADP2e}
\usepackage{graphicx}
\usepackage{dcolumn}
\usepackage{bm}
\usepackage{epsfig}
\usepackage{epstopdf}
\usepackage{color}
\usepackage{pdflscape}

\newcommand{\red}{\textcolor{red}}

\newcommand{\onlinecite}{\cite}

\begin{document}

\articletype{REVIEW ARTICLE}

\title{Intermediate Coupling Model of the Cuprates}

\author{Tanmoy Das$^{\rm a,b}$, R.S. Markiewicz$^{\rm a}$ and A. Bansil$^{\rm a,*}$\\\vspace{6pt}  $^{\rm a}${\em{Physics Department, Northeastern University, Boston, MA, 87545, USA}; $^{\rm b}$Theoretical Division, Los Alamos National Laboratory, Los Alamos, NM, 87545, USA.}
}

%
\date{\today}
\maketitle

\begin{abstract}
We review the intermediate coupling model for treating electronic correlations in the cuprates. Spectral signatures of the intermediate coupling scenario are identified and used to adduce that the cuprates fall in the intermediate rather than the weak or the strong coupling limits. A robust, `beyond LDA' framework for obtaining wide-ranging properties of the cuprates via a GW-approximation based self-consistent self-energy correction for incorporating correlation effects is delineated. In this way, doping and temperature dependent spectra, from the undoped insulator to the overdoped metal, in the normal as well as the superconducting state, with features of both weak and strong coupling can be modeled in a material-specific manner with very few parameters. Efficacy of the model is shown by considering available spectroscopic data on electron and hole doped cuprates from angle-resolved photoemission (ARPES), scanning tunneling microscopy/spectroscopy (STM/STS), neutron scattering, inelastic light scattering, optical and other experiments. Generalizations to treat systems with multiple correlated bands such as the heavy-fermions, the ruthenates, and the actinides are discussed. 
\end{abstract}


\vspace{1cm}
* To whom correspondence should be sent at {\it Ar.Bansil@neu.edu}.
\newline\\
\begin{center}
{\bf CONTENTS}\\
\end{center}
1. Introduction \hspace{10.8cm} 3 \\
2. Correlation strength and fingerprints of intermediate coupling \hspace{2.5cm} 6 \\
$~~~~$2.1 What is intermediate coupling? \hspace{7.0cm} 6 \\
$~~~~$2.2 Quantifying the degree of correlation \hspace{6.0cm} 7 \\
3. Survey of intermediate coupling models \hspace{6.1cm} 9 \\
$~~~~$3.1 DMFT \hspace{11.3cm} 9 \\
$~~~~$3.2 Extensions of DMFT \hspace{8.8cm} 10 \\
$~~$3.2.1 Cluster extensions: CPT, DCA, CDMFT \hspace{5.1cm} 10 \\
$~~~~~~~$3.2.2 Diagrammatic extensions \hspace{7.2cm} 10 \\
$~~~~$3.3 QMC \hspace{11.4cm} 11 \\
$~~~~$3.4 GW theory and the QP-GW model \hspace{6.4cm} 11 \\
$~~~~~~~~$3.4.1 Introduction \hspace{9.6cm} 11 \\
$~~~~~~~$3.4.2 Other `QP-GW' type models \hspace{6.7cm} 12 \\
4. Theory of the intermediate coupling model \hspace{5.7cm} 13 \\
$~~~~$4.1 Self-consistency loop\hspace{8.9cm} 13 \\
$~~~~$4.2 Cuprates including mean-field pseudogap and superconducting gap \hspace{0.8cm} 14 \\
$~~~~$4.3 SDW order in RPA \hspace{8.9cm} 17 \\
$~~~~$4.4 Hartree-Fock self-energy with SDW \hspace{6.1cm} 18 \\
$~~~~$4.5 Self-energy \hspace{10.2cm} 19 \\
$~~~~$4.6 Off-diagonal self-energy \hspace{8.3cm} 21 \\
$~~~~$4.7 Comparison with strong coupling \hspace{6.6cm} 21 \\
$~~~~$4.8 G$_Z$ vs G$_{exp}$ \hspace{10.3cm} 22 \\
5. Bosonic features \hspace{10.0cm} 23 \\
$~~~~$5.1 Spin-wave spectrum in undoped cuprates \hspace{5.3cm} 23 \\
$~~~~$5.2 Spin-waves in the SDW state at finite doping \hspace{4.8cm} 24 \\
6. Single-particle spectra \hspace{9.0cm} 26 \\
$~~~~$6.1 Electron-doped cuprates \hspace{8.1cm} 26 \\
$~~~~$6.2 Hole-doped cuprates \hspace{8.9cm} 27 \\
$~~~~$6.3 ARPES: renormalization and HEK \hspace{6.3cm} 29 \\
$~~~~~~~~$6.3.1 Renormalized quasiparticle spectra and the high energy kink \hspace{1.6cm} 30 \\
$~~~~~~~~$6.3.2 ARPES matrix element effects \hspace{6.4cm} 32 \\
$~~~~~~~~$6.3.3 Low energy kink \hspace{8.9cm} 33 \\
$~~~~~~~~$6.3.4 Lower-energy kinks \hspace{8.4cm} 35 \\
$~~~~$6.4 STM \hspace{11.3cm} 35 \\
$~~~~~~~~$6.4.1 Matrix element effects in STM/STS \hspace{5.7cm} 35 \\
$~~~~~~~~$6.4.2 Doping dependent dispersion in Bi2212 \hspace{5.1cm} 36 \\
$~~~~~~~~$6.4.3 Gap collapse: two topological transitions \hspace{4.6cm} 37 \\
$~~~~$6.5 X-ray absorption \hspace{9.3cm} 39 \\
$~~~~$6.6 Momentum density and Compton scattering \hspace{4.8cm} 40 \\
7. Two-particle spectroscopies \hspace{8.1cm} 42 \\
$~~~~$7.1 Optical absorption spectroscopy\hspace{6.8cm} 42 \\
$~~~~~~~$7.1.1 Optical spectra in electron and hole doped cuprates \hspace{3.0cm} 42 \\
$~~~~~~~~$7.1.2 Optical sum rule \hspace{9.0cm} 45 \\
$~~~~$7.2 RIXS \hspace{11.3cm} 46 \\
$~~~~$7.3 Neutron scattering \hspace{9.2cm} 49 \\
8. Doping dependent gaps \hspace{8.7cm} 52 \\
9. Non-Fermi liquid physics \hspace{8.5cm} 54 \\
$~~~~$9.1 Anomalous spectral weight transfer \hspace{6.2cm} 54 \\
$~~~~$9.2 Non-Fermi-liquid effects due to broken symmetry phase \hspace{2.8cm} 56 \\
$~~~~$9.3 Role of VHS \hspace{10.0cm} 59 \\
10. Superconducting state \hspace{8.9cm} 60 \\
$~~~~$10.1 Separating the SC gap from competing orders \hspace{4.3cm} 60 \\
$~~~~$10.2 Nodeless d-wave gap from competition with SDW \hspace{3.6cm} 61 \\
$~~~~$10.3 Glue functions \hspace{9.6cm} 61 \\
$~~~~~~~~$10.3.1 Optical extraction techniques \hspace{6.4cm} 62 \\
$~~~~~~~~$10.3.2 Application to Bi2212 \hspace{7.8cm} 63 \\
$~~~~$10.4 Calculation of T$_c$ \hspace{9.3cm} 66 \\
$~~~~~~~~$10.4.1 Formalism \hspace{9.8cm} 66 \\
$~~~~~~~~$10.4.2 Pure d-wave solution \hspace{8.0cm} 67 \\
$~~~~~~~~$10.4.3 Low vs high energy pairing glue \hspace{6.0cm} 70 \\
$~~~~~~~~$10.4.4 Competing SC gap symmetries \hspace{6.2cm} 71 \\
$~~~~~~~~$10.4.5 Comparison with other calculations including DCA and CDMFT \hspace{0.1cm} 71 \\
$~~~~~~~~$10.4.6 Universal superconducting dome \hspace{6.0cm} 73 \\
11. Competing phases \hspace{9.5cm} 74 \\
$~~~~$11.1. ($\pi,~\pi$)-order \hspace{10.0cm} 74 \\
$~~~~~~~~$11.1.1 Coherent-incoherent crossover \hspace{6.4cm} 74 \\
$~~~~~~$11.1.2 Other calculations; effects of finite $q$-resolution \hspace{3.6cm} 75 \\
$~~~~$11.2 Incommensurate order \hspace{8.5cm} 79 \\
$~~~~~~~~$11.2.1 SDWs \hspace{10.5cm} 80 \\
$~~~~~~~~$11.2.2 CDWs \hspace{10.4cm} 81 \\
12. Extensions of present model calculations \hspace{5.7cm} 83 \\
$~~~$12.1 Mermin-Wagner physics and quantum critical phenomena\hspace{2.5cm} 83 \\
$~~~~$12.2 Disorder effects \hspace{9.6cm} 83 \\
$~~~~$12.3 Stronger correlations \hspace{8.6cm} 84 \\
$~~~~~~~~$12.3.1 Suppression of double occupancy \hspace{6.1cm} 84 \\
$~~~~~~~~$12.3.2 Spin wave dispersion \hspace{8.1cm} 84 \\
$~~~~~~~~$12.3.3 Mott vs Slater physics \hspace{7.5cm} 84 \\
$~~~~~~~~$12.3.4 Quasiparticle dispersion \hspace{7.5cm} 85 \\
$~~~~~~$12.3.5 Transition temperature at strong coupling \hspace{4.2cm} 85 \\
$~~~~~~~~$12.3.6 Anomalous spectral weight transfer (ASWT) \hspace{3.7cm} 86 \\
$~~~~~~~~$12.3.7 Zhang-Rice physics \hspace{8.0cm} 86 \\
$~~~~~~~~$12.3.8. One-dimensional Hubbard model \hspace{5.9cm} 86 \\
$~~~~$12.4 3- and 4-band models \hspace{8.7cm} 86 \\
13. Other materials and multiband systems \hspace{5.9cm} 88 \\
$~~~~$13.1. Sr$_2$RuO$_4$ \hspace{10.6cm} 88 \\
$~~~~$13.2 UCoGa$_5$ \hspace{10.8cm} 89 \\
$~~~~$13.3. Pnictides \hspace{10.6cm} 89 \\
14. Conclusions and outlook \hspace{8.3cm} 90 \\
Appendix A. Coexisting antiferromagnetic and superconducting orders \hspace{1.5cm} 91 \\
$~~~~$A1. Phase coexistence \hspace{9.0cm} 91 \\
$~~~~$A2. Tensor Green's function \hspace{8.3cm} 92 \\
$~~~~$A3. Transverse spin susceptibility \hspace{7.4cm} 94 \\
$~~~~$A4. Writing a 4$\times$ 4 matrix for susceptibility \hspace{5.5cm} 96 \\
$~~~~$A5. Charge and longitudinal spin susceptibility: \hspace{4.6cm} 100 \\
$~~~~$A6. RPA susceptibility: \hspace{8.8cm} 102 \\
Appendix B. Antiferromagnetic order only \hspace{6.0cm} 102 \\
Appendix C. Superconducting order only \hspace{6.3cm} 103\\
Appendix D. Calculating the self-energy \hspace{6.3cm} 104 \\
Appendix E. Optical conductivity \hspace{7.5cm} 105 \\
Appendix F. Parameters \hspace{9.0cm} 106 \\
Appendix G. Acronyms \hspace{9.0cm} 108 \\
References \hspace{11.2cm} 110

\section{Introduction} \label{S:Intro}

Ever since their discovery, modeling the electronic structure of the cuprate high-T$_c$ superconductors has remained a fundamental theoretical challenge. Although the density functional theory (DFT) provides an accurate theory of predictive value for weakly correlated materials$-$the topological insulators being the most spectacular recent example where first-principles band theory predictions have often led to the discovery of new materials classes\cite{TI1,TI2}, the DFT fails just as spectacularly for the cuprates in that it produces a metallic state and not the insulating state assumed by the parent half-filled compounds from which superconductivity arises via electron or hole doping. The DFT appears to reasonably describe the overdoped metallic phase of the cuprates, where correlations have presumably weakened sufficiently, but since it fails in the undoped limit, DFT cannot be expected to provide a meaningful theory of the doping dependencies of  electronic spectra in the cuprates. 

While cuprates have been treated traditionally via strong coupling
formalisms, recent work has revealed that intermediate coupling
models can capture many salient features of cuprate physics, including the doping dependencies of dispersion, and neutron and optical properties\cite{markiewater,comanac,comanac2,DMFT2,tanmoyop}. Generally speaking, in weak coupling models a sharp dispersion can be defined, while in intermediate coupling a self-energy correction is invoked, which can describe coupling to electronic and phononic bosons, leading to a significant broadening of the spectrum, including a splitting of the spectrum into low-energy `coherent' and higher energy `incoherent' features.  This is typically incorporated via a GW, quantum Monte Carlo (QMC), or dynamic mean-field theory (DMFT) scheme or the related cluster extensions. Interestingly, a recent variational calculation finds a smooth evolution from a spin-density wave (SDW) to a Mott gap with increasing $U$, without an intervening phase transition or spin-liquid phase in the cuprate parameter range.\cite{TBPS}

Our purpose in this review is to discuss a `beyond DFT', GW-approximation based comprehensive modeling scheme, which we have developed, for treating the electronic spectra of correlated materials, including their doping and temperature dependencies. We describe the methodology underlying this quasiparticle-GW (QP-GW) scheme, and discuss its implications for various spectroscopies, casting this discussion in the broader context of current models of the cuprates. Our QP-GW self-energy reasonably captures in the cuprates the dressing of low-energy quasiparticles by spin and charge fluctuations,\cite{markiewater} the high-energy kink (HEK) seen in ARPES,\cite{markiewater} the residual high-energy Mott features in the optical spectra,\cite{tanmoyop} gossamer features,\cite{tanmoygoss} anomalous spectral weight transfer (ASWT) with doping\cite{ASWT}, and the magnetic resonance in neutron scattering.\cite{tanmoymagres,ArunSNS} Our
model also captures a number of characteristic signatures of strong coupling physics of the Hubbard model, including suppression of double-occupancy, the $t-J$-model-like dispersions, spin-wave dispersion, and the $1/U$ scaling of the magnetic order. 

Our focus on the cuprates is a natural one. The reason is that substantial insight into the physics of the cuprates can be gained within the framework of a minimal, single-band model, without the need to address interband contributions to the susceptibility and the associated complexities resulting from the increased degrees of freedom. Nevertheless, extension to a three-band model is quite practical, especially if important correlation effects are assumed to be limited to the band nearest to the Fermi level $E_F$, allowing exploration of relative roles of copper and oxygen electrons in, for example, the evolution of Zhang-Rice physics with doping. 
Moreover, a tremendous amount of systematic experimental data is now available on the cuprates as a function of doping, energy, momentum, and other external controls such as pressure and magnetic field. Therefore, model predictions can be directly tested in some depth against the corresponding experimental results, and one is in a position thus to assess the robustness of the models and their limitations.  

Finally, we emphasize that robust first-principles methodologies for treating the electronic spectra of strongly correlated materials at the level of predictive capabilities comparable to those available currently for weakly correlated systems are likely to remain elusive for some time to come. We hope that the beyond-DFT modeling schemes, such as the present comprehensive QP-GW scheme, will be viewed in this larger context as a pathway for making progress toward realistic  modeling of wide-ranging properties of complex quantum matter by helping to isolate spectral features that require more sophisticated approaches for analysis and interpretation.

This review is organized as follows.  Section~\ref{S:Str} analyzes the strength of correlations, and shows how intermediate coupling models can capture many salient features of the physics of the cuprates.  In Section~\ref{S:IntCII} we give an overview of intermediate coupling models, while in Section~\ref{S:QP-GW} we discuss our QP-GW scheme, with further details taken up in Appendices A-F. The spectral functions associated with electronic bosons, and the resulting electronic susceptibilities are described in Section~\ref{S:Boson}. An emphasis is placed on the spin waves and spin fluctuations, and the resulting self-energies are presented in Section~\ref{S:QP-GW}.5 and Appendix~\ref{S:AF}.   Sections~\ref{S:ElSpec} and~\ref{S:ElSpec2} turn to the comparison of theoretical spectra with experiments for a large variety of one-body (Section~\ref{S:ElSpec}) and two-body (Section~\ref{S:ElSpec2}) spectroscopies.  After briefly summarizing electron (\ref{S:ElSpec}.1) and hole-doped cuprates (\ref{S:ElSpec}.2), we discuss ARPES (\ref{S:ElSpec}.3), STM/STS (\ref{S:ElSpec}.4), x-ray absorption (XAS) (\ref{S:ElSpec}.5), momentum density and Compton scattering (\ref{S:ElSpec}.6), optical (\ref{S:ElSpec2}.1), resonant inelastic x-ray scattering (RIXS) (\ref{S:ElSpec2}.2), and neutron scattering (\ref{S:ElSpec2}.3) results.  In Section~\ref{S:Gaps} we consider general features of the doping dependence of the cuprates derived from various analyses, while aspects of non-Fermi liquid physics are discussed in Section~\ref{S:NFL}.  Section~\ref{S:SC} discusses superconductivity, both the extraction of possible `glue' functions (\ref{S:SC}.3) and the calculation of the gaps and $T_c$s assuming an Eliashberg formalism (\ref{S:SC}.4).  Possible competing phases, charge and spin density waves (C/SDWs), are described in Section~\ref{S:DWs}.  Section~\ref{S:Ext} summarizes several extensions of the model for the cuprates, while Section~\ref{S:Other} considers applications to other materials, and Section~\ref{S:Disc} presents our conclusions and suggestions for future work.  Comparisons with strong coupling models are made in Sections~\ref{S:QP-GW}.7 and~\ref{S:Ext}.3.  Acronyms are summarized in Appendix~\ref{S:AI}.

\section{Correlation Strength and Fingerprints of Intermediate Coupling}\label{S:Str}

\subsection{What is Intermediate Coupling?}

\begin{figure}
\centering
\rotatebox{0}{\scalebox{0.6}{\includegraphics{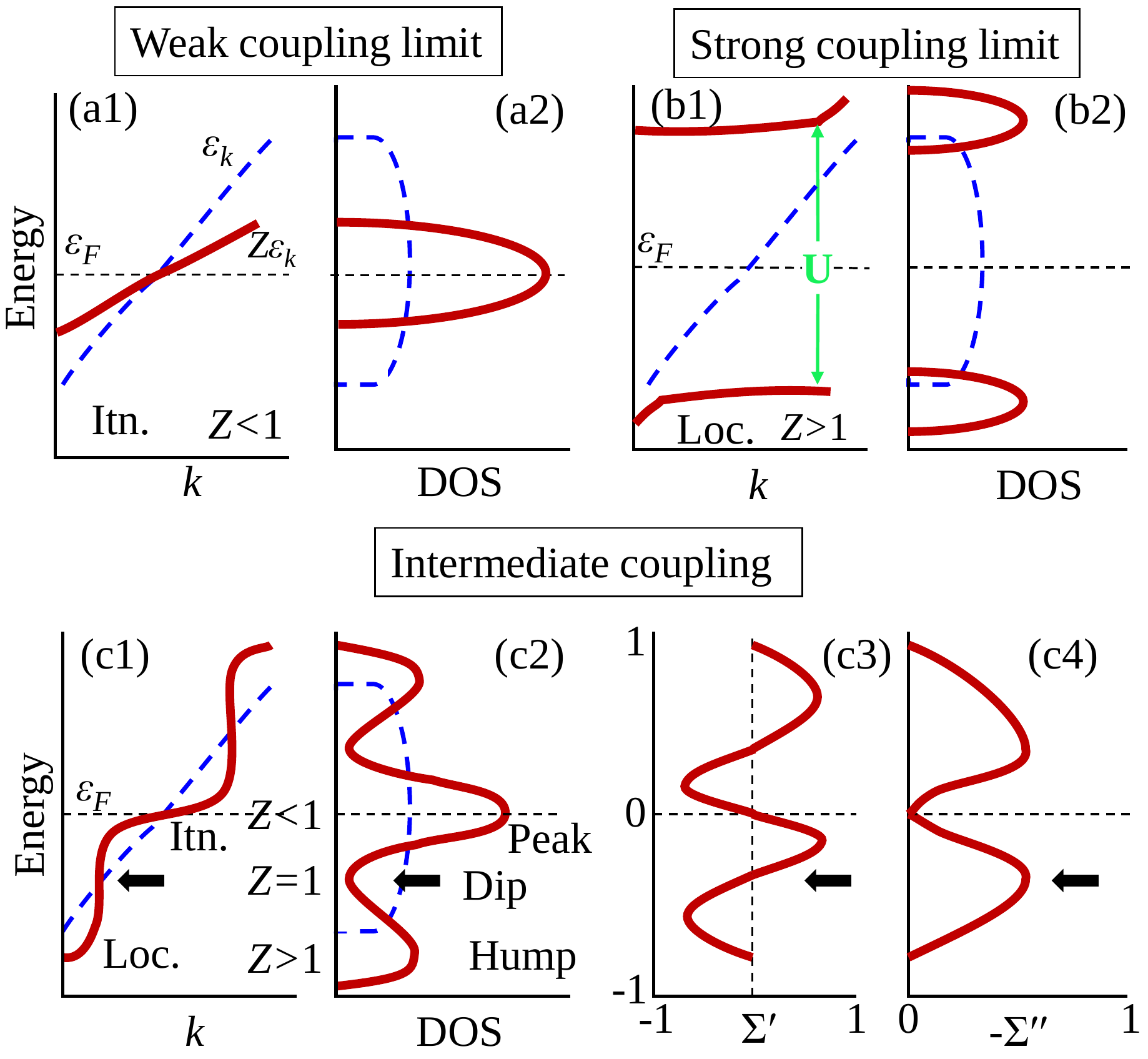}}}
\caption{Schematic illustration of characteristics of the electronic spectrum in the weak, strong and intermediate coupling cases. Bare (dashed blue lines) and correlated (solid red lines) band dispersions are compared in panels a1, b1, and c1 with itinerant (Itn.) and localized (Loc.) regions of the spectrum indicated. Related densities of states (DOS) are shown in panels a2, b2 and c2. Real and imaginary parts of the self-energy, $\Sigma$, are shown in panels c3 and c4 for intermediate coupling. $Z$ is the dispersion renormalization factor. [Figure adapted from Ref~\cite{DasOtherMat}.]}
\label{Intermediate}
\end{figure}
Correlated materials can be sorted into three categories based on the relative strength of the Coulomb interaction ($U$) compared to the DFT bandwidth ($W$). We have weak coupling when $U \ll W$, strong coupling when $U\gg W$, and an intermediate coupling scenario applies when $U$ lies between these two limiting cases. [More specifically, we suggest in Section~\ref{S:Str}.2 below that intermediate coupling for cuprates corresponds to $W/2<U<2W$.]  Fig.~\ref{Intermediate}\cite{DasOtherMat} illustrates characteristic features of the electronic spectrum in the three cases. For weak coupling (e.g. Landau's Fermi liquid theory), the quasiparticle band retains nearly the same momentum information as the corresponding non-interacting system, the renormalization of the dispersion given by the factor $Z<1$ is not drastic, and the spectral function is dominated by sharp quasiparticle peaks.
Moreover, the renormalization factor for the spectral weight, $Z_{\omega}$ (see Section~\ref{S:NFL}.2), is the same as the dispersion renormalization factor $Z$. In sharp contrast, in the strong coupling limit, the non-interacting band splits into two distinct subbands separated by the insulating gap $U$, so that at half-filling the quasiparticle weight at $E_F$ is nearly zero.  Each subband has very small dispersion, representing nearly localized electrons. 
Since the subbands are separated by $U$, the spectral weight is spread over an energy range $>W$, so that the bandwidth effectively becomes larger than the bare bandwidth $W$. If we now think of the renormalization factor as the ratio of $U$ to bandwidth, we obtain an effective $Z>$1 for the high-energy band. We will see in Section~\ref{S:NFL}.2 that the effective $Z$ values can display quite complex behavior with doping and energy. In the QP-GW scheme, which is the focus of this review, we will mainly be concerned with the near-Fermi level value of $Z$.


Properties of electrons in the intermediate coupling limit, $U\sim W$, interpolate smoothly between those for the weak and strong coupling limits as depicted in Figs.~\ref{Intermediate}(c1) and \ref{Intermediate}(c2). In particular, states near $E_F$ resemble the itinerant states of the weak coupling limit, but with a larger renormalization, $Z<1$.  At the same time, there are incoherent states at higher energies, so that the total bandwidth remains large ($\sim U$), yielding an effective $Z>1$ at high energies as in the strong coupling case. There is a crossover energy scale between the coherent and incoherent parts of the spectrum where the real part of the self-energy changes sign. This causes the band near $E_F$, where $\Sigma^{\prime}(-\omega)>0$, to become renormalized toward $E_F$, while the band at higher energy, where $\Sigma^{\prime}(-\omega)<0$, shifts to even higher energy. These two parts are connected by a crossover energy where the band is effectively unrenormalized.  The interpolative nature of the resulting spectrum is clear: the coherent bands near $E_F$ retain many properties of weakly correlated Fermi liquids, albeit with narrower bands, while the incoherent features far from $E_F$ appear to be nearly localized precursors of strong coupling effects.  Corresponding to the crossover energy scale, there is a temperature scale above which 
coherence at $E_F$ is destroyed.  Such coherence temperatures are well known in heavy-fermion materials, where the localized ($f$) and itinerant states begin to hybridize.\cite{DavisHF}

Anomalous energy dependence of $\Sigma^{\prime}$ yields via Kramer's-Kronig relation a peak in its imaginary part $\Sigma^{\prime\prime}$ where $\Sigma^{\prime}$ changes sign. As a result, spectral weight is transferred away from the crossover energy toward both the low-energy quasiparticle peak and the high-energy incoherent hump features. This is the mechanism for forming {\it kinks} in the dispersion, also referred to as `S'-shapes or `waterfalls', and the corresponding peak-dip-hump features in the density-of-states (DOS) found in the one-particle spectra, Fig.~\ref{Intermediate}(c2). This anomalous electronic structure induces anomalies in correlation functions, first explored by Moriya in his classic book.\cite{Moriya}   

The peak in $\Sigma^{\prime\prime}$ also distinguishes intermediate coupling case from the weak coupling limit.  Whereas in weak coupling the dispersion is always sharp and quasiparticles are well defined, the peak in $\Sigma^{\prime\prime}$ splits the dispersion into two almost distinct branches lying below and above $E_F$. The peak in $\Sigma^{\prime\prime}$ represents a large concentration of electronic bosons [electron-hole pairs], which act to split and dress electrons into coherent electronic states near $E_F$, and incoherent states across the kink energy where electrons are nearly localized.  Correlations thus play an increasing role as their strength increases. Note that the bosonic strength is given by the imaginary part of the susceptibility, but when electrons are renormalized by the bosons, the associated dispersions will change and shift the peaks in the susceptibility. Therefore, calculations must be carried out self-consistently so that the susceptibility peaks line up properly with the peaks in self-energy.  Stated differently, at low energies, electrons
can be considered to behave either as quasiparticles or as components of bosons -- electron-hole or electron-electron pairs.  The renormalization $Z$ must be chosen self-consistently so that a fraction of the electronic spectral weight contributes to the Fermi quasiparticles, and a fraction $1-Z$ to the bosons, which shows up at higher energies as the incoherent contribution to the electronic spectral weight.

\subsection{Quantifying the Degree of Correlation}

In order to quantify how strong the correlations are in the cuprates, we must first get a handle on the boundaries between weak, intermediate, and strong coupling regimes.  Since mean-field theories such as the random-phase approximation (RPA) provide a reasonable description of the weak coupling case, we may locate the crossover between weak and intermediate regime via the interaction strength at which RPA still predicts the correct phases, but overestimates the transition temperature by say up to $\sim$20\%. In a recent Gutzwiller approximation (GA) + RPA study of the $t-t'-U$ Hubbard model\cite{MGu} (where $t$ and $t'$ are nearest and next-nearest neighbor tight-binding hopping parameters), this criterion yielded the weak/intermediate coupling boundary at $\approx U=W/2$, where $U$ is the Hubbard on-site interaction and $W\sim 8t$ is the electronic bandwidth.

\begin{figure}
\centering
\includegraphics[width=9cm,clip=true]{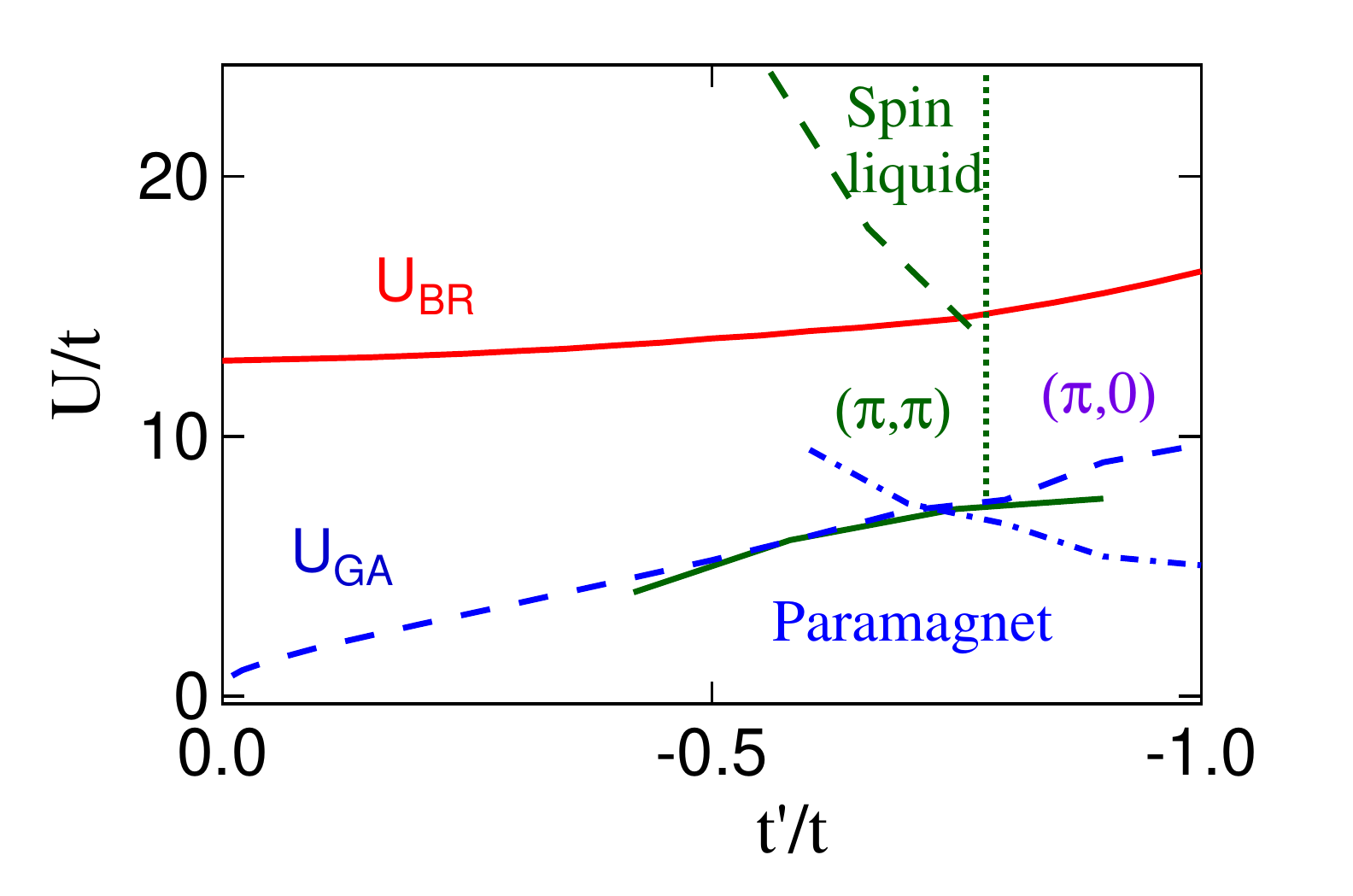}
\caption{
Phase diagram of the $t-t'-U$ Hubbard model discussed in the text. [From Ref.~\onlinecite{MGu}.]
}
\label{fig:1d}
\end{figure}
Turning to the strongly correlated case, one typically associates this limit with a very small probability of double-occupancy, i.e. when the renormalization factor $Z\rightarrow 0$, or equivalently, the effective mass $m^*\rightarrow\infty$.  Within the GA, this occurs at the Brinkman-Rice (BR) transition when $U=U_{BR}\simeq 2W$\cite{BR}. The exact value of $U_{BR}$ for the $t-t'-U$ model is shown in Fig.~\ref{fig:1d} (red solid line). Also shown are results of the $t'/t - U$ phase diagram of Tocchio et al.\cite{TBPS} (green lines), along with results for the Gutzwiller transition\cite{MGu} 
$\tilde U_{GA}$ obtained by considering only the $(\pi ,\pi )$ [$(\pi ,0)$] magnetic order (dashed [dotted] blue line).  A spin-liquid phase is found only above $U_{BR}$ in the presence of frustration, close to the $(\pi,\pi)-(\pi,0)$ crossover.  In contrast, RPA-like magnetic phases arise for much lower values of $U$. For $|t'/t|<0.5$ there is no obvious change in the structure on crossing the BR line.  The commensurate GA transitions are in excellent agreement with the results of Ref.~\onlinecite{TBPS}.  We thus adduce that the intermediate coupling regime corresponds approximately to $W/2<U<2W$.\cite{footSh,Shast}  For the cuprates, $U$ is usually estimated to be $\sim 8t=W$, while $-0.5<t'/t<0$, placing the cuprates far from the strong correlation regime or any spin-liquid phase.  For this reason, we will generally speak of upper and lower magnetic bands (UMB/LMBs) rather than upper and lower Hubbard bands (UHB/LHBs).

\begin{figure}
\centering
\includegraphics[width=12cm,clip=true]{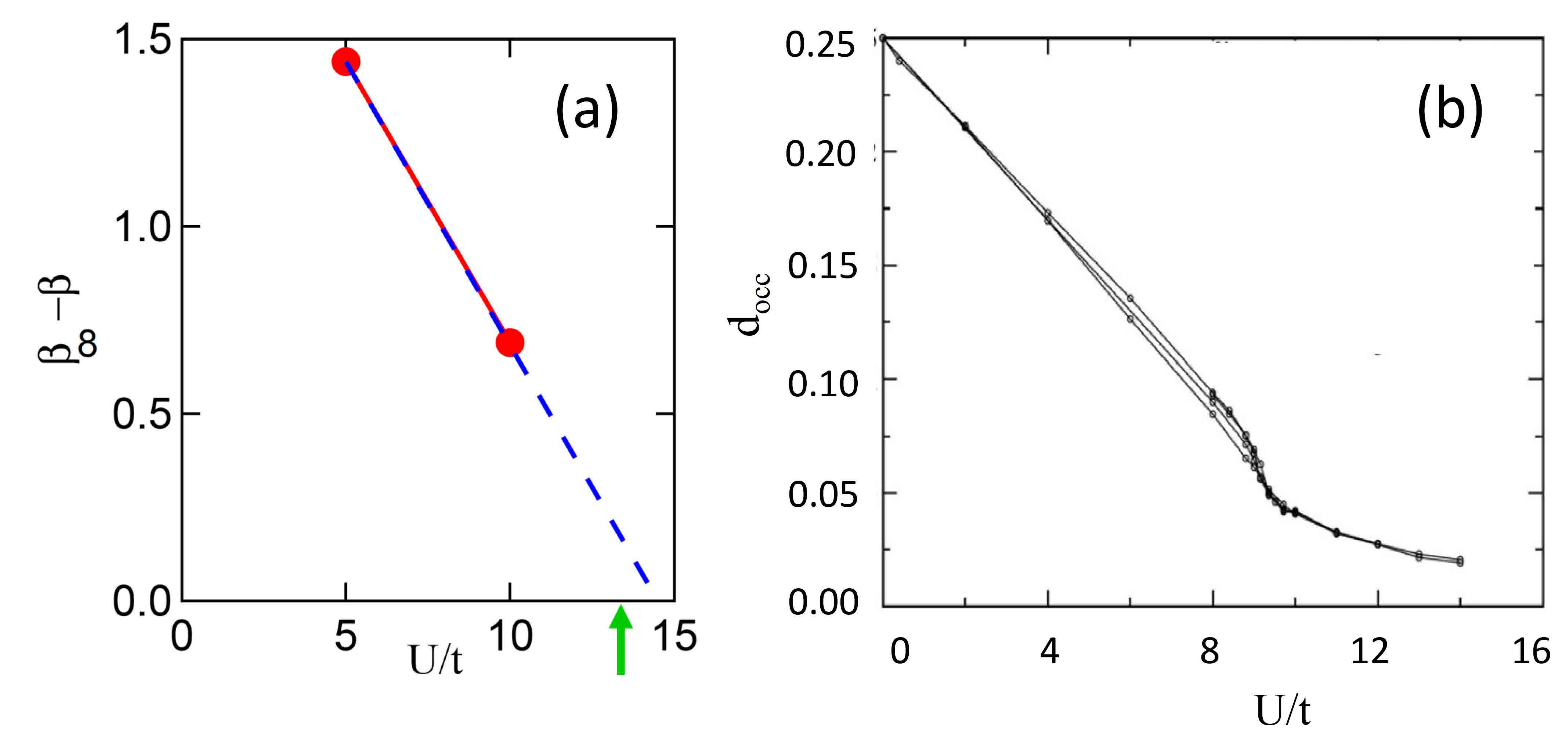}
\caption{
(a) Rate of anomalous spectral weight transfer, $\beta$, at half-filling, as a function of $U/t$.\cite{EMS} Blue dashed line shows linear extrapolation to $\beta_{\infty}$ at $U=14.5t$, close to the calculated $U_{BR}=13.5t$  (green arrow).  
(b) Fractional double occupancy, $d_{occ}$, obtained via cellular DMFT computations for the $t-t'-U$ Hubbard model using $t'=t$ at various temperatures. 
[Frame (b) from Ref.~\onlinecite{DMFT5}.]. }
\label{fig:1bc}
\end{figure}

One property that depends sensitively on double-occupancy is the anomalous spectral weight transfer (ASWT)\cite{EMS,ASWT}, see also Section~\ref{S:NFL}.1 below.  We expect that the rate of spectral weight transfer from the upper to the lower magnetic band, $\beta$, defined by the slope of the spectral weight transfer versus doping, will vary linearly with the degree of double-occupancy, and hence linearly with $U-U_{BR}$.  Based on the current exact diagonalization calculations\cite{EMS}, which provide only two data points, $\beta$ extrapolates to the infinite-$U$ limit $\beta_{\infty}=-1$ at $U=14.5t$, close to the expected $U_{BR}=13.5$, Fig.~\ref{fig:1bc}(a).  Further insight comes from a cellular-DMFT calculation by Parcollet {\it et al.}\cite{DMFT5}, who frustrate the system with a large anisotropic $t'=t$, forcing it to be in the paramagnetic phase, where double-occupancy was found to decrease linearly with $U$ for small $U$, extrapolating to zero at $U=13t$, Fig.~\ref{fig:1bc}(b).  However, Ref. \onlinecite{DMFT5} also found a metal-insulator transition (MIT) near $U=9t$, a signature of the true Mott transition, even though double-occupancy retained $\sim 20$\% of its uncorrelated value at the transition, falling to only half that value at their largest $U=14t$. This MIT has been further analyzed in Refs.~\onlinecite{MIT1,MIT2}. 
Hence, it appears that the Mott transition is controlled by a Brinkman-Rice transition near $U_{BR}\sim 2W$, superceded usually by a more conventional AFM transition at smaller $U$. A Mott-type MIT can be found at smaller values of $U$, but to realize this transition the AFM order must usually be suppressed.

\section{Survey of Intermediate Coupling Models}\label{S:IntCII}
Here we provide a broader overview of the intermediate coupling models, even though our focus in this review is on a particular intermediate coupling model, the QP-GW model. Extensive reviews of the DMFT formalism and its cluster extensions are available in the literature\cite{DMFT,DMFT3,LDADMFT,LDADMFT2}. We note that an essential feature of intermediate coupling is the introduction of a self-energy to describe the interaction of electrons with bosonic modes.  In this spirit, one may consider local-density approximation (LDA)+U to be a weak-coupling model since this scheme describes gap opening in different orbitals via a Hartree-Fock description, but fails to provide a self-energy. 

\subsection{DMFT}

The dynamical mean field theory (DMFT) generalizes earlier mean field theories (MFTs) such as the coherent-potential-approximation (CPA, KKR-CPA)\cite{ABD1,ABD2,ABD3,ABD4} by computing the dynamical interaction between a quantum impurity and an electronic bath to obtain an optimal local ($k$-independent) self-consistent self-energy.\cite{DMFT} DMFT can treat the coexistence of localized or atom-like and low-energy band-like features in the electronic spectrum, and the renormalized mass of the quasiparticles.\cite{DMFTPu}  DMFT calculations have been combined with first-principles band structure theories giving rise to the LDA+DMFT schemes.\cite{LDADMFT,LDADMFT2}. The DMFT was one of the first to show how a Mott gap at half-filling can develop a narrow coherent in-gap band at $E_F$ with doping\cite{DMFT2}. 

Since the interaction is solved on a single site, DMFT lacks momentum resolution and 
a number of approaches have been developed to extend the DMFT to an impurity cluster\cite{DMFT3}. These extensions attempt to solve the impurity problem on a cluster exactly, embedding the cluster into a lattice bath to self-consistently obtain the self-energy.  An obvious limitation of any cluster model [including DMFT as a 1$\times$1 cluster] is that it can only accurately describe fluctuations up to the scale of the cluster size.  For example, DMFT includes only on-site fluctuations, and therefore, it cannot describe either the AFM or $d$-wave SC order, although it can describe a Mott transition since the AFM order is frustrated.  While a $2\times 2$ cluster can describe both the AFM and $d$-SC order, longer range fluctuations are underestimated, resulting in mean-field phase transitions.  Finally, cluster models can only determine correlation lengths smaller than the cluster size, limiting the use of cluster approaches for investigating critical dynamics.  Diagrammatic extensions of DMFT, which calculate correlation functions beyond the self-energy, promise overcoming these limitations.

The preceding considerations also underlie a number of limitations of the DMFT, which have been discussed in the literature as follows. Since DMFT cannot describe the AFM order, it does not account for $\sim Nln 2$ spin-entropy, where $N\rightarrow\infty$ is the number of sites, which distorts the doping-dependent phase diagram\cite{Trem1}. 
The mean-field phase transitions predicted by DMFT are particularly problematic in lower-dimensional systems, but even in the 3D Hubbard model significant fluctuation effects are missed below $T\sim 5T_N$, where $T_N$ is the Neel temperature.\cite{Therm3D}
Millis' group has explored the underpinnings of DMFT to find that vertex corrections appear to be as significant in the DMFT as in GW calculations.\cite{Millis1,Millis2,Millis3} The vertex corrections are important not only for the self-energy, but also for calculating two-body spectroscopies, and for capturing the correct orbital ordering physics in multi-orbital systems such as the pnictides.  
It has been suggested that DMFT could be improved by including a self-consistent GW calculation\cite{YKK} (see also Refs.~\cite{KotGW1,KotGW2}).

\subsection{Extensions of DMFT}
\subsubsection{Cluster Extensions: CPT, DCA, Cluster DMFT}	
As we already pointed out above, the premise of various cluster extensions of the DMFT is straightforward: an interacting Hamiltonian can be exactly diagonalized on a small cluster, even when it is coupled to a self-consistent bath.  The results, however, are often sensitive to how the cluster impurity problem is solved, and there is now a growing emphasis on developing improved `impurity solvers'.  To date, three main approaches have been followed. Perhaps the simplest is the {\it cluster perturbation theory} (CPT)\cite{refcpt}, in which the lattice is treated as a lattice of clusters, the self-energy is found exactly for an isolated cluster, and the lattice Green's function is found by treating the inter-cluster hopping as a perturbation.

The other two approaches also solve for the self-energy exactly on a cluster, but then embed the cluster self-consistently into an infinite lattice.  They differ in the nature of the embedding.  The {\it cellular dynamical mean-field theory} (CDMFT)\cite{refcdmft} is a direct cluster generalization of DMFT.  The embedding is done at a local cluster in real space, so that the translational symmetry is violated.  In contrast, the {\it dynamical cluster approximation} (DCA)\cite{refdca} employs a periodic cluster to preserve translational symmetry.  Although we will return to comment on applications of various DMFT models to the cuprates below, we note that the DCA procedure amounts to coarse-graining in momentum space, so that the size of the cluster dictates the number of $k$-points where the self-energy and susceptibility can be calculated.  The CDMFT encounters a similar problem. A DCA calculation can only determine the  temperature at which a correlation length becomes larger than the cluster size (see Fig.~\ref{fig:64C}).  This presents a problem in dealing with quantum phase transitions and competing phases, especially for 2D systems, where the Mermin-Wagner theorem shows that fluctuations can drive the phase transition to $T=0$.\cite{MW} For further discussion of these and related issues bearing on the stability of competing phases, see  Refs.~\onlinecite{DMFT4,DMFT5,DMFT6,DMFT7,DMFT8,DMFT9,DMFT10,DMFT11,DMFT12,DMFT13,DMFT14} and Section~\ref{S:DWs}.1.2 below. 

\subsubsection{Diagrammatic Extensions}
While cluster extensions are suitable for {\it short-range} correlations, another approach is needed when {\it long-range} correlations become important as, for example, for treating quantum phase transitions, Luttinger liquid physics, and Van Hove singularities.
Instead of going from a single impurity site to a cluster of impurities, in either real- or $k$-space, DMFT can be extended diagrammatically by including the local two-particle vertex of the Anderson impurity model in the computations. The momentum dependence of the self-energy can now be computed using this two-particle vertex.  Various approaches for approximating the vertex include the dynamical vertex approximation (D$\Gamma$A) [full two-particle irreducible local vertex]\cite{DGA1,DGA3,DGA2}, dual fermion (DF) [one- and two-particle reducible local vertex]\cite{DF1,DF2}, and the one-particle irreducible approach (1PI).\cite{OPI}  Since the D$\Gamma$A approach includes both GW and DMFT contributions, {\it ab initio} calculations with D$\Gamma$A should be superior to GW+DMFT.\cite{DGA2}  In this way, Mermin-Wagner physics has been accessed down to low temperatures\cite{DF2,DGA3}, where the spin-wave spectrum is found.\cite{DF2} Fluctuations also cause large reductions of the Neel temperature in the 3D Hubbard model\cite{DGA4} (see Refs.~\cite{Therm3D,DCA3D} for related cluster results).

\subsection{QMC}
A variety of quantum Monte Carlo (QMC) techniques have been applied to study the Hubbard model to determine its phase diagram and the possibility of superconductivity.\cite{jarrellb,maier,macridin} Also, the DCA calculations typically use QMC as an `impurity solver' to treat the cluster Hamiltonian.\cite{refdca,grober}  At various points in the text below we will compare our results with QMC and other approaches as appropriate.

\subsection{GW theory and the QP-GW model}

\subsubsection{Introduction}
The one-particle Green's function $G({\bf k},\omega)$ can be written in terms of the bare  Green's function $G_0({\bf k},\omega)$ via Dyson's equation as $G^{-1}=G_0^{-1}-\Sigma$, where $\Sigma({\bf k},\omega)$ is the self-energy.  The perturbation series for $\Sigma$ can be solved exactly to give 
\begin{eqnarray}\label{selfen0}
&&\Sigma({\bf k},\omega)= \sum_{{\bf q}}
\int_{0}^{\infty}\frac{d\omega'}{2\pi}\Gamma({\bf k},{\bf q},\omega,\omega') G({\bf k}+{\bf q},\omega+\omega')
{\rm Im}[W({\bf q},\omega')],
\end{eqnarray}
where $\Gamma$ is a vertex correction and $W\sim U^2\chi$, with $U$ denoting an appropriate interaction parameter, and $\chi$ is an electronic susceptibility whose imaginary part yields the spectrum of electronic bosons in the model.  Hedin proposed a simpler alternative in which vertex corrections are ignored [$\Gamma\rightarrow 1$], yielding the so-called GW model of self-energy.\cite{Hedin,GWRev}

While the exact Eq.~\ref{selfen0} involves the fully dressed Green's function $G$, the simpler approximation of using the bare $G_0$ is often made when the interactions are not too strong.  Since $\chi$, and hence $W$, is built from the Green's functions, one can also use $W$ or $W_0$.  Interestingly, a fully dressed $GW$ calculation often performs worse than the bare $G_0W_0$ computation.  This is because when correlations modify the electronic degrees of freedom, the electronic bosons can form preformed pairs, excitons for neutral pairs or Cooper pairs for charged bosons.  Therefore the vertex correction $\Gamma$, which describes the interactions between the electron and hole or between two electrons that make up the boson, must be included. Vertex corrections seem to be important whenever the imaginary part of the self-energy, $\Sigma^{\prime\prime}$, displays a significant frequency dependence.  



Correlations in the cuprates are sufficiently strong that the $G_0W_0$ approach fails, and must be replaced by a self-consistent calculation.\cite{markiewater}  
This may be viewed as a form of spin-charge separation in higher dimensions.  
More precisely, bare electrons contribute to both the fermionic [quasiparticle] and bosonic [electron-hole pair] excitations, and these two contributions must be reasonably well aligned in energy and momentum, i.e., peaks in $\chi^{\prime\prime}$ [corresponding to maxima in bosonic spectral weight] must line up with peaks in $\Sigma^{\prime\prime}$ [maxima in scattering]; see Ref.~\cite{Trem2} for an example with important corrections due to superconductivity.  In other words, peaks in the susceptibility must fall approximately midway between $E_F$ and the band edges, so that the peak in $\Sigma^{\prime\prime}$ will compress the coherent, dressed quasiparticles toward $E_F$, while shifting to higher energies the incoherent weight resulting from the residues of the electrons used to make electron-hole [and electron-electron] pairs.  This is where $G_0W_0$ fails: when the bandwidth is renormalized by a factor of order $\ge$2, peaks in $\chi_0^{\prime\prime}$ lie outside the dressed bandwidth, yielding only coherent states, which are unphysically compressed into a very narrow bandwidth, and no incoherent spectral weight is left to make up the bosons.  

It seems, however, that correlations in cuprates are not so strong that the full apparatus of the $GW\Gamma$ approach is needed.  We have therefore introduced an `intermediate' model, which we call the quasiparticle (QP)-GW, or $G_ZW_Z$,  method, where we attempt to retain the simplicity of the $G_0W_0$ scheme while overcoming its key shortcomings.  Here the Green's function, $G_Z$, which enters the convolution integral, is the bare Green's function renormalized by an energy independent renormalization constant $Z$, and $W_Z=U^2\chi_Z$, with $\chi_Z$ calculated from $G_Z$.  Equivalently,
\begin{equation}\label{eq:gz}
G_Z^{-1}=G_0^{-1}-\Sigma_Z,
\end{equation}
\begin{equation}\label{eq:sigmaz}
\Sigma_Z(\omega)=(1-Z^{-1})\omega.
\end{equation}
The parameter $Z$ is then found self-consistently such that the spectral weight of $G_Z$ matches the coherent or the low-energy part of the spectral weight of $G$.  Equation~\ref{eq:sigmaz} is the key to the QP-GW model: $\Sigma_Z$ is the most complicated self-energy one can introduce with $\Sigma^{\prime\prime}_Z=0$, so that the vertex correction is unimportant. There is some ambiguity in the choice of $\Gamma$ for the model.  For a strict GW model, $\Gamma$ should be taken as 1.  However, for the present $\Sigma_Z$, the corresponding $\Gamma=1/Z$, and we generally use this value, although results are not very sensitive to its precise value.
We further discuss this point in the following section and in Appendix~\ref{S:AC}. Typical values of $Z$ are listed in Table F1 in Appendix~\ref{S:AH}.  When only the low-energy physics is of interest, one may work just with the $G_Z$ Green's function as a quasi-Landau Fermi liquid model [`quasi', since $Z<1$.]

\subsubsection{Other `QP-GW' Type Models}

Other QP-GW type methods in the literature with a similar underlying conceptual framework include FLEX and quasiparticle self-consistent GW (QS-GW)\cite{QPGWJapan1,QPGWJapan2}, although the way self-consistency is obtained in each method makes a substantial difference.  In Refs.~\onlinecite{QPGWJapan1,QPGWJapan2}, self-consistency involves evaluating the band renormalization $Z$ only at $E_F$. But bands are renormalized by a variety of different processes spread over multiple energy scales (high-energy kink, low-energy kink, and phonons), and these effects are not captured in the value of $Z$ at $E_F$, leading to a greatly reduced bandwidth and inconsistencies with experiments. Our QP-GW method, in contrast, invokes an effective renormalization factor, $Z$, which is based on self-energy corrections extending to the high-energy kink scale, which marks the crossover between the coherent and incoherent states. As a consequence, different scales involving bosons and kinks get aligned reasonably, resulting in a wider coherent bandwidth, and the crossover or waterfall energy scale is pushed into the experimental range (300-500meV in cuprates and around 500meV in actinides).

\section{Theory of the intermediate coupling model}\label{S:QP-GW}

\subsection{Self-consistency loop for computing momentum-dependent dynamical fluctuations}

\begin{figure}[top]
\centering
\rotatebox{0}{\scalebox{0.48}{\includegraphics{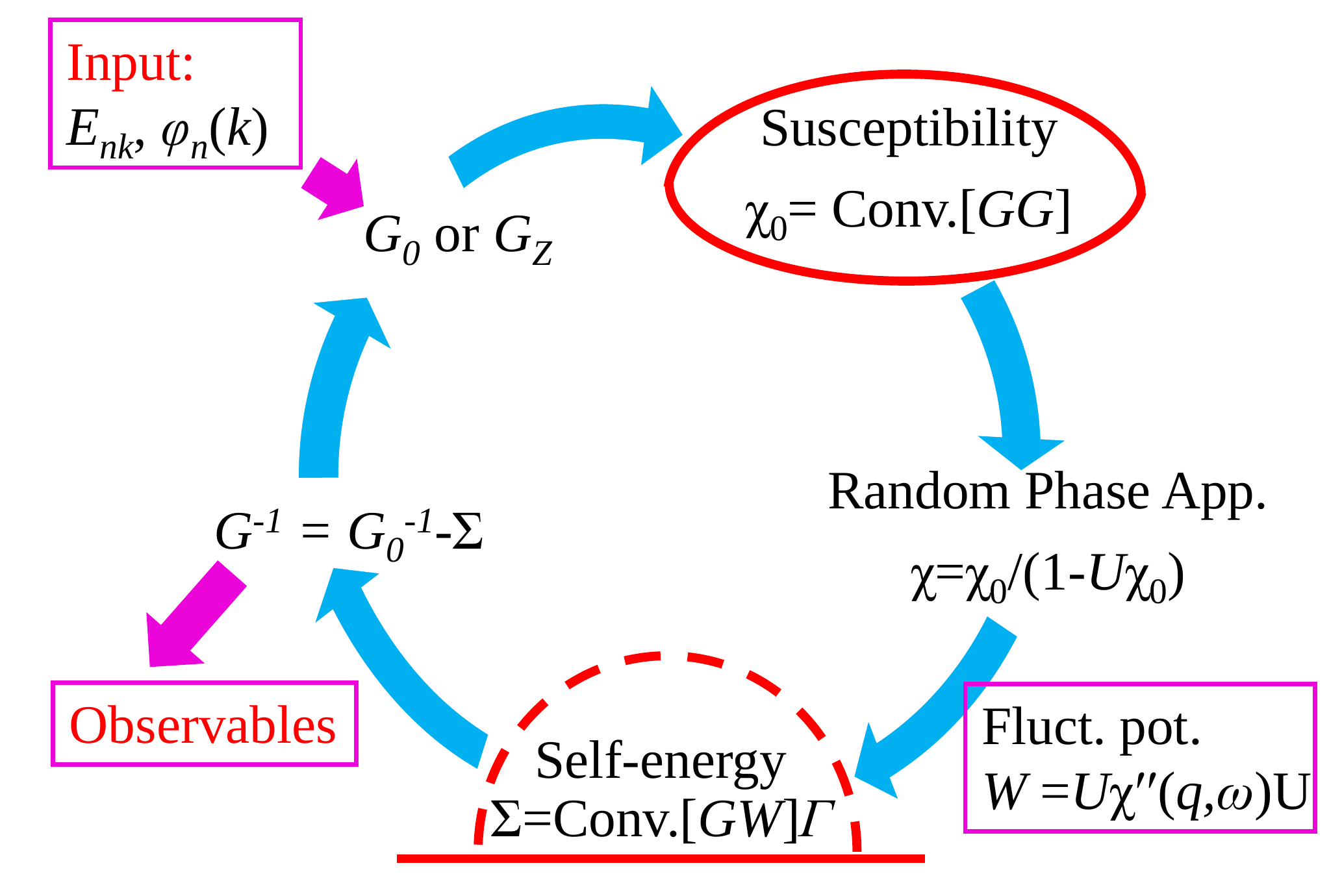}}}
\caption{Schematic of the self-consistency loop for a QP-GW computation of susceptibility and self-energy. `Conv.' denotes a convolution.  Figure taken from Ref.~\cite{DasOtherMat}.}
\label{GWLoop}
\end{figure}

Fig.~\ref{GWLoop}\cite{DasOtherMat} schematically lays out the self-consistency loop in our QP-GW scheme. The matrix of the starting Hamiltonian is generated by extracting a tight-binding representation of the LDA band structure and the associated wavefunctions. The wavefunctions are important in evaluating matrix element effects in various spectroscopies discussed throughout this review.\cite{AA1,AA2}  Using the LDA parameters, we construct both a bare Green's function $G_0$ and a $Z$-renormalized Green's function $G_Z$ as in Eq.~\ref{eq:gz}.  With the latter Green's function, we evaluate the susceptibilities, and from the susceptibilities, the self-energy $\Sigma$ and the dressed Green's function $G$ via $G^{-1}=G_0^{-1}-\Sigma$. We then impose self-consistency by requiring that the DOS calculated from $G_Z$, which is the Green's function used to calculate the bosonic spectrum $\chi^{\prime\prime}$, is the same as the coherent part of the DOS calculated from the full $G$. This ensures that the bosons and the resulting kinks line up reasonably in energy.  This can lead to inaccuracies at energies above the HEK, but these are not important for many purposes.


By including a self-energy correction, our model captures effects of the important bosons with which the electrons interact.  The resulting dispersion is naturally broadened and splits into coherent and incoherent branches separated by kinks.  The electronic bosons are weighted to reflect their importance via the spectral weight $\sim\chi^{\prime\prime}(q,\omega )$ and renormalized self-consistently, leading to the related (multiple) kinks in the dispersion. Spin- and charge-excitations are separated in the RPA (random phase approximation) susceptibility.  The computationally intensive steps are the calculation of the full RPA spin- and charge-susceptibility $\chi (q,\omega)$ and the associated self-energy $\Sigma (k,\omega)$, the latter involving  a convolution over $\chi$ and $G (k,\omega)$.  

We note that if one is interested mainly in the coherent part of the spectrum, the GW correlations play little role beyond renormalizing the bandwidth, and one can use the simpler Green's function $G_Z$ of Eq.~\ref{eq:gz} above.  On the other hand, in cuprates it is mainly the band closest to $E_F$ which is correlated, and the associated self-energy can be used in multi-band models.  Indeed, we have exported this one-band self-energy into first-principles methodologies for realistic modeling of ARPES, inelastic X-ray scattering, scanning tunneling, and other highly resolved spectroscopies including matrix element effects.\cite{AA1,AA2,waterfall,Towfiq,Jouko}  It is important here to understand the mean-field theory for the three-band model, which reproduces the spectral features of the one-band model, a point to which we return in Section~\ref{S:Ext}.4 below. 

We use a mean-field RPA to account for broken-symmetries, leading to tensor susceptibilities and Green's functions.  Typically, when both superconductivity and a density-wave order are involved one obtains $4\times 4$ tensors.  We have modeled the simplest density wave order, a commensurate $Q=(\pi,\pi)$ spin-density-wave (SDW) state,\cite{SWZ,SinT,chubukov,ZSch,Cr} although the methodology is  readily generalized to treat other orders, including incommensurate orders, at the expense of working with larger tensors.   The $Q=(\pi,\pi)$ order, however, appears to be an ideal choice for a variety of reasons: (1) It captures many salient features of the competing order in electron-doped cuprates over a wide doping range and in hole-doped cuprates for doping $x\le 0.03$; (2) It has a special place in a number of strong-coupling calculations at arbitrary hole-doping\cite{YRZ,YRZ3,YRZ1,YRZ2}; (3) As discussed in Section~\ref{S:DWs}.1.2  below, the cluster DMFT extensions will tend to find only $(\pi,\pi)$-order due to limitations of momentum resolution; (4) A number of factors, including high temperatures, high energies $\omega$, and large disorder cause the susceptibility to smear out, shifting the peak susceptibility toward $(\pi,\pi)$ (Section~\ref{S:DWs}); (5) Finally, many spectroscopies are insensitive to the exact nesting vector, and insight can be obtained by using the $(\pi, \pi)-$SDW to estimate the self-energy.\cite{tanmoy2gap}

That said, note that at higher hole-doping there is evidence for a variety of incommensurate spin- and charge-density waves, possibly in the form of stripes,\cite{vignolle,tranquada} which we have explored using only $G_Z$ to model the coherent part of the spectrum\cite{MGu2}, see Section~\ref{S:DWs} below for further discussion. Treating the pseudogap as a single ordered state competing with superconductivity also allows us to avoid the complications arising from {\it competing density-wave phases}, see Fig.~\ref{fig:8A} below.  In 3D materials, the correlation length diverges as soon as the Stoner criterion for a particular phase is satisfied, which cuts off other competing phases.  In 2D, Mermin-Wagner\cite{MW} physics (strong fluctuations) restricts the divergence of the correlation length to $T=0$, while Imry-Ma physics\cite{ImMa} (strong impurity pinning) can eliminate this divergence completely.  The issue of what would happen if two or more phases would have diverged (at slightly different temperatures) in the absence of fluctuations/disorder has to our knowledge not been addressed in the literature.  

\subsection{Cuprates including mean-field pseudogap and superconducting gap}

The Hamiltonian which includes competing pseudogap (modeled as $(\pi,\pi)$- order) and superconducting orders is:\cite{tanmoy2gap}
\begin{eqnarray}\label{Ham}
H&=&\sum_{{\bf k},\sigma}(\epsilon_{{\bf k}}-\epsilon_F)c^{\dag}_{{\bf k},\sigma}
c_{{\bf k},\sigma}+U\sum_{{\bf k},{\bf k}^{\prime}}c^{\dag}_{{\bf k}+{\bf Q},\uparrow}
c_{{\bf k},\uparrow}c^{\dag}_{{\bf k}^{\prime}-{\bf Q},\downarrow}
c_{{\bf k}^{\prime},\downarrow}\nonumber\\
&&~~+\sum_{{\bf k},{\bf k}^{\prime}}V({\bf k},{{\bf k}^{\prime}})c^{\dag}_{{\bf k},\uparrow}
c^{\dag}_{-{\bf k},\downarrow}c_{-{\bf k}^{\prime},\downarrow}c_{{\bf k}^{\prime},\uparrow}
\end{eqnarray}
where $c^{\dag}_{{\bf k},\sigma} (c_{{\bf k},\sigma})$ is the
electronic creation (destruction) operator with momentum ${\bf k}$
and spin $\sigma$, $\epsilon_{\bf k}$ is the material-specific 
dispersion taken from a tight-binding parameterization of the LDA
band structure with no adjustable parameters [listed
in Table ~F1], $U$ is the on-site Hubbard interaction energy, and $\epsilon_F$ is the chemical potential.
By expanding the quartic terms as in Hartree-Fock, the
$d-$wave superconducting gap can be calculated using the  BCS formalism as
$\Delta_{k}=Vg_k\sum_{k^{\prime}}g_{k^{\prime}} \left<\sigma
c^{\dag}_{{\bf k^{\prime}},\uparrow}c^{\prime}_{{-\bf
k^{\prime}},\downarrow}\right>$ and $V$ is the pairing interaction.  The average here is taken over the
ground state with combined superconducting (SC) and SDW (pseudogap) order.  Similarly, the SDW gap $\Delta_{SDW}=US$ is found from the self-consistent mean-field solution of the order parameter
$S=\left<\sum_{\bf k}\sigma c^{\dag}_{{\bf k+Q},\sigma}c_{{\bf
k},\sigma}\right>$ (spin index $\sigma=\pm$).  The resulting Hartree-Fock (HF) Hamiltonian is\cite{Das_nonmonotonic}
\begin{eqnarray}\label{HFHam}
H_{HF}&=&\sum_{{\bf k},\sigma}(\epsilon_{{\bf k}}-\epsilon_F)c^{\dag}_{{\bf k},\sigma}
c_{{\bf k},\sigma}+\Delta_{SDW}\sum_{{\bf k},{\bf k}^{\prime}}[c^{\dag}_{{\bf k}+{\bf Q},\uparrow}
c_{{\bf k},\uparrow}-c^{\dag}_{{\bf k}^{\prime}-{\bf Q},\downarrow}
c_{{\bf k}^{\prime},\downarrow}]\nonumber\\
&&~~+\sum_{{\bf k}}[\Delta_{\bf k}c^{\dag}_{{\bf k},\uparrow}
c^{\dag}_{-{\bf k},\downarrow}+\Delta^*_{\bf k}c_{-{\bf k},\downarrow}c_{{\bf k},\uparrow}].
\end{eqnarray}

The parameters employed for various materials in our QP-GW calculations are summarized in Tables F1 and F2 of Appendix~\ref{S:AH}.  We emphasize that the effective onsite energy $U$ in a one-band Hubbard model must be doping dependent to correctly capture the observed gap evolution in both electron- and hole-doped cuprates.\cite{kusko}  Physically, this reflects effects of changes in screening near the metal-insulator transition, and would seemingly require corrections beyond the one-band Hubbard model in the form of longer-range Coulomb interactions\cite{markiewater,tanmoyop,tanmoysw} or a three-band model\cite{Hsin,MBRIXS}.  The doping-dependent $U$-values are given in Figure~\ref{Ux}, and these and other relevant parameters are summarized in Tables F1 and F2.  A doping and temperature dependent $U$ has also been found in DCA calculations,\cite{DCA2} shown as orange stars in Fig.~\ref{Ux}.  For hole-doped cuprates, we find the Hubbard $U$s to be quite similar for different materials, and to be nearly independent of doping after a large drop near half-filling.

While we have carried out some Eliashberg equation based calculations of the superconducting gap, see Section~\ref{S:SC}.4, these computations do not incorporate effects of the pseudogap, prompting us to typically introduce momentum independent interaction strengths $V$, which reproduce experimental $d$-wave gaps.  The resulting gap values are 
given in Table F2.  As the strength of the pairing interaction $V$ is adjusted to reproduce the experimental superconducting gaps, and the bare dispersion is taken from LDA, no free parameters are invoked in the modeling of the various spectroscopies discussed below.

The Hamiltonian in Eq.~\ref{HFHam} can be diagonalized straightforwardly with quasiparticle dispersion
consisting of upper ($\nu=+$) and lower ($\nu=-$) magnetic bands (U/LMB) further split by
superconductivity:
\begin{eqnarray}\label{eigen}
E_{\bf k}^{\nu}=\pm\sqrt{\left(E^{s,\nu}_{k}\right)^2+\Delta_{k}^2}.
\end{eqnarray}
Here $E_{\bf k}^{s,\nu}=\xi_{\bf k}^++\nu
E_{0k}$ are the quasiparticle energies in the non-superconducting SDW state, $E_{0{\bf k}}=\sqrt{\left(\xi_{\bf k}^-\right)^2+(US)^2}$, and $\xi_{\bf k}^{\pm}=(\xi_{\bf k}\pm\xi_{{\bf k}+{\bf Q}})/2$. The SDW and SC coherence factors for these two bands are given by
\begin{eqnarray}\label{eigenvec}
\alpha_{\bf k}(\beta_{\bf k})&=&\sqrt{(1\pm\xi_{\bf k}^-/E_{0{\bf k}})/2},\nonumber\\
u_{\bf k}^{\nu} (v_{\bf k}^{\nu})&=&\sqrt{[1\pm(\xi^+_{\bf k}+\nu E_{0{\bf k}})/E^{\nu}_{\bf k}]/2}.
\end{eqnarray}
In term of the SDW and SC coherence factors, the coupled self-consistent gap equations become
%
\begin{eqnarray}
\Delta_{0}&=&-V\sum_{{\bf k}}g_{{\bf k}}\left[u_{{\bf k}}^{+}v_{{\bf k}}^{+}\tanh{}(\beta E_{{\bf k}}^{+}/2)+u_{{\bf k}}^{-}v_{{\bf k}}^{-}\tanh{}(\beta E_{{\bf k}}^{-}/2)\right]\nonumber\\
&=&-V\Delta_0\sum_{{\bf k}}g_{{\bf k}}^2\left[\frac{1}{2E_{{\bf k}}^{+}}\tanh{}(\beta
E_{{\bf k}}^{+}/2)+\frac{1}{2E_{{\bf k}}^{-}}\tanh{}(\beta
E_{{\bf k}}^{-}/2)\right]
\hspace{0.3in}\\
S &=&\frac{1}{N}\sum_{{\bf k}}\alpha_{{\bf k}}\beta_{{\bf k}}\big[\big((v_{{\bf k}}^{-})^2-(v_{{\bf k}}^{+})^2\big)+\big((v_{{\bf k}}^{+})^2-(u_{{\bf k}}^{+})^2\big)f(E_{{\bf k}}^{+})\nonumber\\
&&~~~~~~-\big((v_{{\bf k}}^{-})^2-(u_{{\bf k}}^{-})^2\big)f(E_{{\bf k}}^{-})\big],\nonumber\\
&=&\frac{US}{N}\sum_{{\bf k}}\frac{1}{4E_{0{\bf k}}}\left[\frac{E^{s,+}_{\bf k}}{E_{\bf k}^+}\tanh{}(\beta
E_{{\bf k}}^{+}/2)-\frac{E^{s,-}_{\bf k}}{E_{\bf k}^-}\tanh{}(\beta
E_{{\bf k}}^{-}/2)\right].
\hspace{0.3in}
\label{delselffinal}
\end{eqnarray}
%

The spin-susceptibility in the SDW state is complicated by the
associated unit cell doubling, so that the correlation functions
(Lindhard susceptibility) have off-diagonal terms in a momentum
space representation\cite{SWZ,chubukov}, which arise from Umklapp processes with respect to ${\bf Q}$.
Therefore, we define susceptibilities as the standard linear
response functions\cite{SWZ}
\begin{eqnarray}\label{Eq:ch3_chi_1}
\chi^{ij}({\bf q},{\bf q}^{\prime},\tau)&=&
\frac{1}{2N}\Big\langle T_{\tau}\Pi_{{\bf q}}^i(\tau)\Pi^j_{-{\bf q}^{\prime}}(0)\Big\rangle
\end{eqnarray}
where the response operators ($\Pi$) for the charge and spin density
correlations, respectively, are
\begin{eqnarray}\label{Eq:ch3_rho_s}
\rho_{\bf q}(\tau)&=&\sum_{{\bf k},\sigma}
c_{{\bf k}+{\bf q},\sigma}^{\dag}(\tau)c_{{\bf k},\sigma}(\tau),~~{\rm and}~\nonumber\\
S^i_{{\bf q}}(\tau)&=&\sum_{{\bf k},\sigma,\gamma}c_{{\bf k}+{\bf q},\sigma}^{\dag}
(\tau)\sigma^i_{\sigma,\gamma}c_{{\bf k},\gamma}(\tau).
\end{eqnarray}
The $\sigma^i$ are 2D Pauli matrices along the
$i^{th}$ direction. For the transverse spin response, $S^{\pm}=S_x\pm
iS_y$, the longitudinal fluctuations are along the
$z-$direction only.  In the present ($\pi,\pi$)-commensurate
state, charge and longitudinal spin-fluctuations become coupled
at finite doping\cite{chubukov}. In common practice, the
transverse, longitudinal spin- and charge-susceptibilities are
denoted as $\chi^{+-}, \chi^{zz}$ and $\chi^{\rho\rho}$, respectively. We
collect all the terms into a single notation as
$\chi^{\sigma\bar{\sigma}}$, where $\bar{\sigma}=\sigma$ gives the
charge and longitudinal components and $\bar{\sigma}=-\sigma$
stands for the transverse component. The noninteracting Lindhard susceptibility in the SDW-BCS case is a $2\times2$ matrix whose components are
\cite{chubukov,SWZ}
%
\begin{eqnarray}
\chi_{ij}^{\sigma\bar{\sigma}}({\bf q},\omega)&=&\frac{1}{N\beta}\sum_{{\bf k}}\sum_n\sum_s G_{is}({\bf k},\sigma,i\omega_n)G_{sj}({\bf k}+{\bf q},\bar{\sigma},i\omega_n+\omega)\label{chi1}\\
&=&\frac{1}{N}\sum_{{\bf k}}^{\prime}\sum_{\nu\nu^{\prime}=\pm}B^{\nu\nu^{\prime},\sigma\bar{\sigma}}_{ij}({\bf k},{\bf q})
\sum_{n=1}^3 A_{n}^{\nu\nu^{\prime}}({\bf k},{\bf q})\chi_{0n}^{\nu\nu^{\prime}}({\bf k},{\bf q},\omega).
\label{chi2}
\end{eqnarray}
Eq.~\ref{chi2} is obtained from Eq.~\ref{chi1} after performing a Matsubara summation over $n$. $G$ is the $4\times 4$ single-particle Green's function. The summation indices $\nu (\nu^{\prime}) =
\pm$ refer to the UMB and LMB, respectively, and $B$ is the coherence factor due to the SDW order in the particle-hole channel,
\begin{eqnarray}\label{chiSDW}
B_{11/22}^{\nu\nu^{\prime},\sigma\bar{\sigma}}({\bf k},{\bf q})&=&\frac{1}{2}\left(1\pm\nu\nu^{\prime}\frac{\xi_{\bf k}^-\xi_{{\bf k}+{\bf q}}^-+\sigma\bar{\sigma}(US)^2}{E_{0{\bf k}}E_{0{\bf k}+{\bf q}}}\right)\nonumber\\
B_{12/21}^{\nu\nu^{\prime},\sigma\bar{\sigma}}({\bf k},{\bf q})&=&-\nu\frac{US}{2}\left(\frac{\sigma}{E_{0{\bf k}}}+\nu\nu^{\prime}\frac{\bar{\sigma}}{E_{0{\bf k}+{\bf q}}}\right)\nonumber\\
\end{eqnarray}
The pair-scattering coherence factors are
\begin{eqnarray}\label{t11}
A_{1}^{\nu\nu^{\prime}}({\bf k},{\bf q})
&=&\frac{1}{2}\left(1+\frac{E^{s,\nu}_{\bf{k}}
E^{s,\nu^{\prime}}_{\bf{k}+\bf{q}}+\Delta_{\bf{k}}\Delta_{\bf{k}+\bf{q}}}
{E^{\nu}_{\bf{k}}E^{\nu^{\prime}}_{\bf{k}+\bf{q}}}\right)\nonumber\\
A_{2/3}^{\nu\nu^{\prime}}({\bf k},{\bf q})&=&\frac{1}{4}\left(1\pm\frac{E^{s,\nu}_{\bf{k}}}{E^{\nu}_{\bf{k}}}
\mp\frac{E^{s,\nu^{\prime}}_{\bf{k}+\bf{q}}}{E^{\nu^{\prime}}_{\bf{k}+\bf{q}}}
-\frac{E^{s,\nu}_{\bf{k}}E^{s,\nu^{\prime}}_{\bf{k}+\bf{q}}+\Delta_{\bf{k}}\Delta_{\bf{k}+\bf{q}}}
{E^{\nu}_{\bf{k}}E^{\nu^{\prime}}_{\bf{k}+\bf{q}}}\right).\nonumber\\
\end{eqnarray}
Lastly, the index $m$ represents summation over three possible quasiparticle polarization bubbles
related to quasiparticle scattering ($m=1)$, quasiparticle pair
creation ($m=2$), and pair annihilation ($m=3$), defined by
\begin{eqnarray}\label{Eq:ch3_chi_0_sc}
\chi_{01}^{\nu,\nu^{\prime}}({\bf k},{\bf q},\omega)&=&
-\frac{f(E^{\nu}_{{\bf k}})-f(E^{\nu^{\prime}}_{{\bf k}+{\bf q}})}
{\omega+i\delta+(E^{\nu}_{{\bf k}}-E^{\nu^{\prime}}_{{\bf k}+{\bf q}})},\\
\chi_{02,03}^{\nu,\nu^{\prime}}({\bf k},{\bf q},\omega)&=&\mp
\frac{1-f(E^{\nu}_{{\bf k}})-f(E^{\nu^{\prime}}_{{\bf k}+{\bf q}})}
{\omega+i\delta\mp(E^{\nu}_{{\bf k}}+E^{\nu^{\prime}}_{{\bf k}+{\bf q}})}.\label{Eq:ch3_chi_0_scB}
\end{eqnarray}
Once the mean-field Green's functions and susceptibilities are known, they can be renormalized by $Z$ and used to calculate the QP-GW self-energy, as discussed above. 

\subsection{SDW Order in RPA} 

For the case with only SDW order, the $2\times2$ transverse susceptibility within the RPA is given by the standard formula\cite{SWZ}
\begin{eqnarray}\label{chirpat}
\chi_{RPA}^{\sigma\bar{\sigma}}({\bf q},\omega)&=&\frac{\chi^{\sigma\bar{\sigma}}({\bf q},\omega)}{{\bf I}- \tilde{U}^{\sigma\bar{\sigma}}\chi^{\sigma\bar{\sigma}}({\bf q},\omega)}.
\end{eqnarray}
Away from half-filling, the charge and longitudinal parts get mixed.\cite{chubukov} Therefore, the interaction vertex becomes,
\begin{eqnarray}\label{tildeu}
\tilde U^{\sigma\bar{\sigma}}&=&\begin{pmatrix}U&0\cr
                    0&-\sigma\bar{\sigma}U\cr\end{pmatrix}.
\end{eqnarray}
We write the non-interacting charge and longitudinal susceptibility matrix explicitly as
$\chi^{z\rho}_{11}=\chi^{zz}({\bf q},\omega )$, $\chi^{z\rho}_{22}=\chi^{\rho\rho}({\bf q+Q},\omega)$ and $\chi^{z\rho}_{12}=\chi^{z\rho}({\bf q,q+Q},\omega ).$
The explicit forms of the RPA susceptibilities are instructive and bear on important physical insights. In this spirit, we expand Eq.~\ref{chirpat} to obtain
%
%
\begin{eqnarray}\label{RPASpinSucscomp}
\chi_{RPA,11}^{\sigma\bar{\sigma}}({\bf q},\omega)
 &=&\frac{\bigl[1+\sigma\bar{\sigma} U\chi_{22}^{\sigma\bar{\sigma}}({\bf q},\omega)\bigr]\chi_{11}^{\sigma\bar{\sigma}}({\bf q},\omega) +
  U\bigl[\chi_{12}^{\sigma\bar{\sigma}}({\bf q},\omega)\bigr]^2}
{\bigl[1-U\chi_{11}^{\sigma\bar{\sigma}}({\bf q},\omega)\bigr]\bigl[1+\sigma\bar{\sigma}U\chi_{22}^{\sigma\bar{\sigma}}({\bf q},\omega)\bigr]
+\sigma\bar{\sigma}\bigl[U\chi_{12}^{\sigma\bar{\sigma}}({\bf q},\omega)\bigr]^2},\nonumber\\
\chi_{RPA,22}^{\sigma\bar{\sigma}}({\bf q},\omega)
 &=&\frac{\bigl[1-U\chi_{11}^{\sigma\bar{\sigma}}({\bf q},\omega)\bigr]\chi_{22}^{\sigma\bar{\sigma}}({\bf q},\omega) + U\bigl[\chi_{12}^{\sigma\bar{\sigma}}({\bf q},\omega)\bigr]^2}
{\bigl[1-U\chi_{11}^{\sigma\bar{\sigma}}({\bf q},\omega)\bigr]\bigl[1+\sigma\bar{\sigma}U\chi_{22}^{\sigma\bar{\sigma}}({\bf q},\omega)]
+\sigma\bar{\sigma}\bigl[U\chi_{12}^{\sigma\bar{\sigma}}({\bf q},\omega)\bigr]^2},\nonumber\\
\chi_{RPA}^{\sigma\bar{\sigma}}({\bf q},\omega)
 &=&\frac{\chi_{12}^{\sigma\bar{\sigma}}({\bf q},\omega)}
{\bigl[1-U\chi_{11}^{\sigma\bar{\sigma}}({\bf q},\omega)\bigr]\bigl[1+\sigma\bar{\sigma}U\chi_{22}^{\sigma\bar{\sigma}}({\bf q},\omega)]
+\sigma\bar{\sigma}\bigl[U\chi_{12}^{\sigma\bar{\sigma}}({\bf q},\omega)\bigr]^2}.\nonumber\\
\end{eqnarray}
%
%

The gap equation for $\Delta_{SDW}$ is  given by the vanishing of the Stoner denominator 
\begin{equation}\label{RPAgap2}
U\chi_0^{zz}(Q,\omega=0)=1,
\end{equation}
where $Q$ is the momentum at which $\chi_0^{zz}(Q,\omega=0)$ has its maximum. For $Q=(\pi,\pi)$, this gives the self-consistent SDW gap equation, Eq.~\ref{delselffinal} above, which at $T=0$ becomes
\begin{equation}\label{SDWgap}
1 = U\sum_k\frac{f(E_{k}^+)-f(E_{k}^-)}{2E_{0k}}.
\end{equation}
The highest temperature at which a solution of Eq.~\ref{delselffinal} is found is the Neel temperature $T_N$.  For a second order transition, $\Delta_{SDW}\rightarrow 0$ as $T\rightarrow T_N$.  However, mean-field calculations find a weakly discontinuous, first-order transition near the quantum critical point\cite{Mkstripes}.   We will discuss nesting at various other $q$-vectors, as is appropriate for most hole-doped cuprates, in Section~\ref{S:DWs} below.

\subsection{Hartree-Fock Self-energy from SDW order}
 The Green's function for the $(\pi ,\pi )$-SDW can be written as
\begin{eqnarray}\label{SDW}
G({\bf k},\omega) = \frac{\omega-\xi_{{\bf k}+{\bf Q}}}{ (\omega-\xi_{{\bf k}})(\omega-\xi_{{\bf k}+{\bf Q}}) + \Delta_{SDW}^2},
= \frac{1}{\omega-\xi_{{\bf k}} - \Sigma_{HF}({\bf k},\omega)},
%
\end{eqnarray}
where the HF self-energy is
\begin{equation}\label{SDWSE}
\Sigma_{HF}(k,\omega) = \frac{\Delta_{SDW}^2}{\omega -\xi_{{\bf k}+{\bf Q}}}.
\end{equation}

Generalization of Eqs.~\ref{SDW} and~\ref{SDWSE} to the full tensor QP-GW formulation is presented in Appendix~\ref{S:AC}.2.
We have previously used the HF model to study AFM gap collapse\cite{tanmoyprl,tanmoy2gap,kusko,Das_nonmonotonic}. Since the coherent states in the QP-GW model are self-consistently related to $G_Z$, these results are nearly unchanged by the introduction of a GW-like self-energy
correction.\cite{foottb,ABD1,ABD4,ft_sophi1,ft_sophi3,ft_sophi4}

\subsection{Self-energy} 
The self-energy (a $4\times 4$ tensor) due to all magnetic and charge modes is
\begin{eqnarray}\label{Aselfeng}
&&\Sigma({\bf k},\sigma,i\omega_n)=\frac{1}{2}U^2Z \sum_{{\bf q},\sigma^{\prime}}^
{\prime}\eta_{\sigma,\sigma^{\prime}}
\int_{0}^{\infty}\frac{d\omega_p}{2\pi}\nonumber\\
&&~~~G({\bf k}+{\bf q},\sigma^{\prime},i\omega_n+\omega_p)
\Gamma({\bf k},{\bf q},i\omega_n,\omega_p){\rm Im}[\chi_{\rm
RPA}^{\sigma\sigma^{\prime}}({\bf q},\omega_p)].
\end{eqnarray}
Here the spin degrees of freedom $\eta_{\sigma,\sigma^{\prime}}$
take the value of 2 for the transverse and 1 for both the
longitudinal and charge modes. $G$ is the $4\times4$ single particle
Green's function including the Umklapp part from SDWs and the anomalous term coming from the SC gap. The
real part of the self-energy is linear-in-$\omega$ near $E_F$, which gives the self-consistent dispersion renormalization
$\Sigma(\omega)=(1-Z^{-1})\omega{\bf 1}$ to the input LDA band. Here {\bf 1} denotes a $4\times 4$ unit matrix.  
\begin{figure}[top]
\centering
\rotatebox{0}{\scalebox{0.55}{\includegraphics{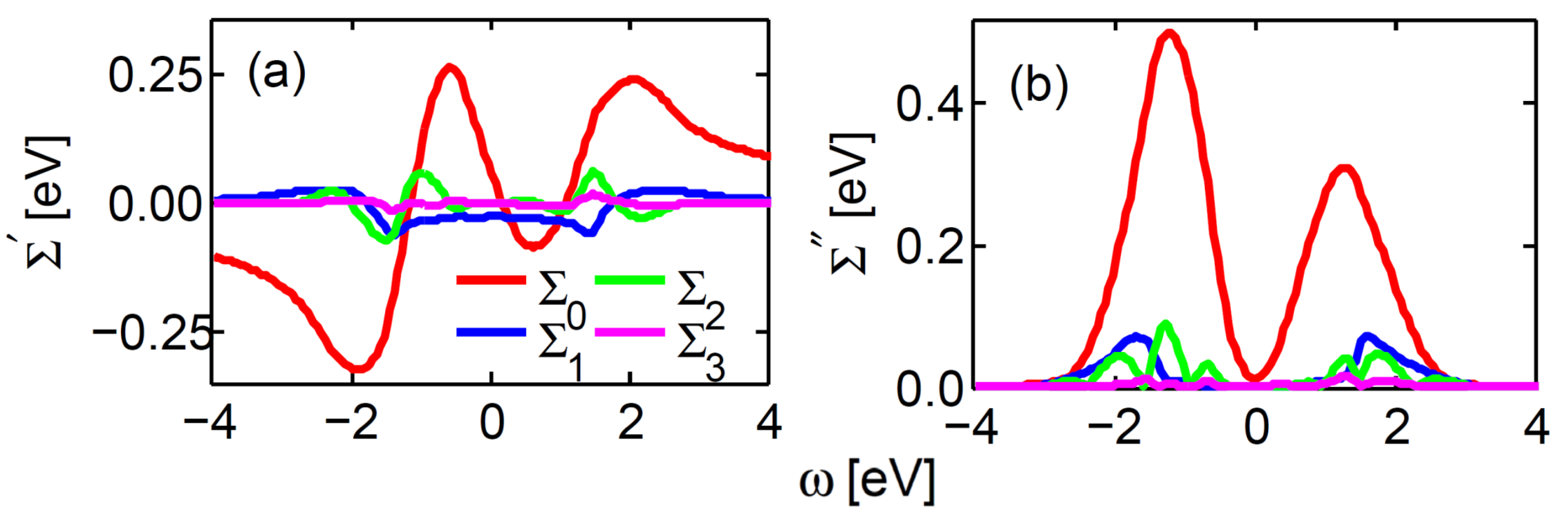}}}
\caption{ { Calculated angle-dependence of self-energy.} Various curves show energy dependence of the coefficients of the cosine expansion, Eq.~\ref{SelfenQ}. Only the constant term $\Sigma_0$ (red line) displays a large magnitude and $\omega$-dependence. [From Ref.~\cite{tanmoygoss}.]}
\label{SelfQ}
\end{figure}

While many models employ empirical hopping parameters to fit the experimental dispersions, we use first-principles band
parameters, which we self-consistently renormalize by the GW self-energy.  This in turn renormalizes the  coupling to bosonic excitations, i.e., the effective coupling $U$ is similarly renormalized,
$U_{eff}\sim ZU$, which largely offsets the enhanced bare
susceptibility $\chi_{0,eff}\sim \chi_0/Z$. This suggests that  theories which invoke dressed [experimental] dispersion as a
staring point will exaggerate coupling to bosonic modes,
and hence the strength of the associated kinks and the magnetic resonance mode.


The efficacy of our procedure is clear from the fact that we not only reproduce the low-energy dispersion found in ARPES, but also the high-energy kink or waterfall seen in ARPES, which represents an `undressing' of the electrons. Moreover, we are able to capture the doping evolution of both the mid-infrared feature and the high-energy `residual Mott gap' seen in optical studies.  Details are summarized in Sections~\ref{S:ElSpec} and~\ref{S:ElSpec2}  below.  The relevant parameters used are given in Appendix~\ref{S:AH}.

While we can calculate the self-energy at an arbitrary wave-vector $q$, we have found that its $k$-dependence is rather weak, and the use of a k-independent value is typically justified. Figure~\ref{SelfQ} shows an estimate of this angle-dependence, which is based on fitting four QP-GW calculations at different $k$-values to the tight-binding form
\begin{eqnarray}\label{SelfenQ}
\Sigma({\bf k},\omega)=\Sigma_0(\omega)+\Sigma_1(\omega)(\cos(k_xa)+\cos(k_ya))+
\nonumber\\
+\Sigma_2(\omega)\cos(k_xa)\cos(k_ya)+\Sigma_3(\omega)(\cos(2k_xa)+\cos(2k_ya)).
\end{eqnarray}
The energy dependence in Fig.~\ref{SelfQ} is seen to be dominated by the $k$-independent $\Sigma_0$ term.
Note that reference to strong $k$-dependence of $\Sigma$ in the literature usually implies inclusion of the SDW self-energy, Eq.~\ref{SDWSE}, in a scalar $\Sigma$.  In contrast, when one employs a tensor form of $\Sigma$, this contribution is automatically included, as is the case here, and the tensor components of $\Sigma$ are less sensitive to $k$.

\begin{figure}[top]
\centering
\rotatebox{0}{\scalebox{0.22}{\includegraphics{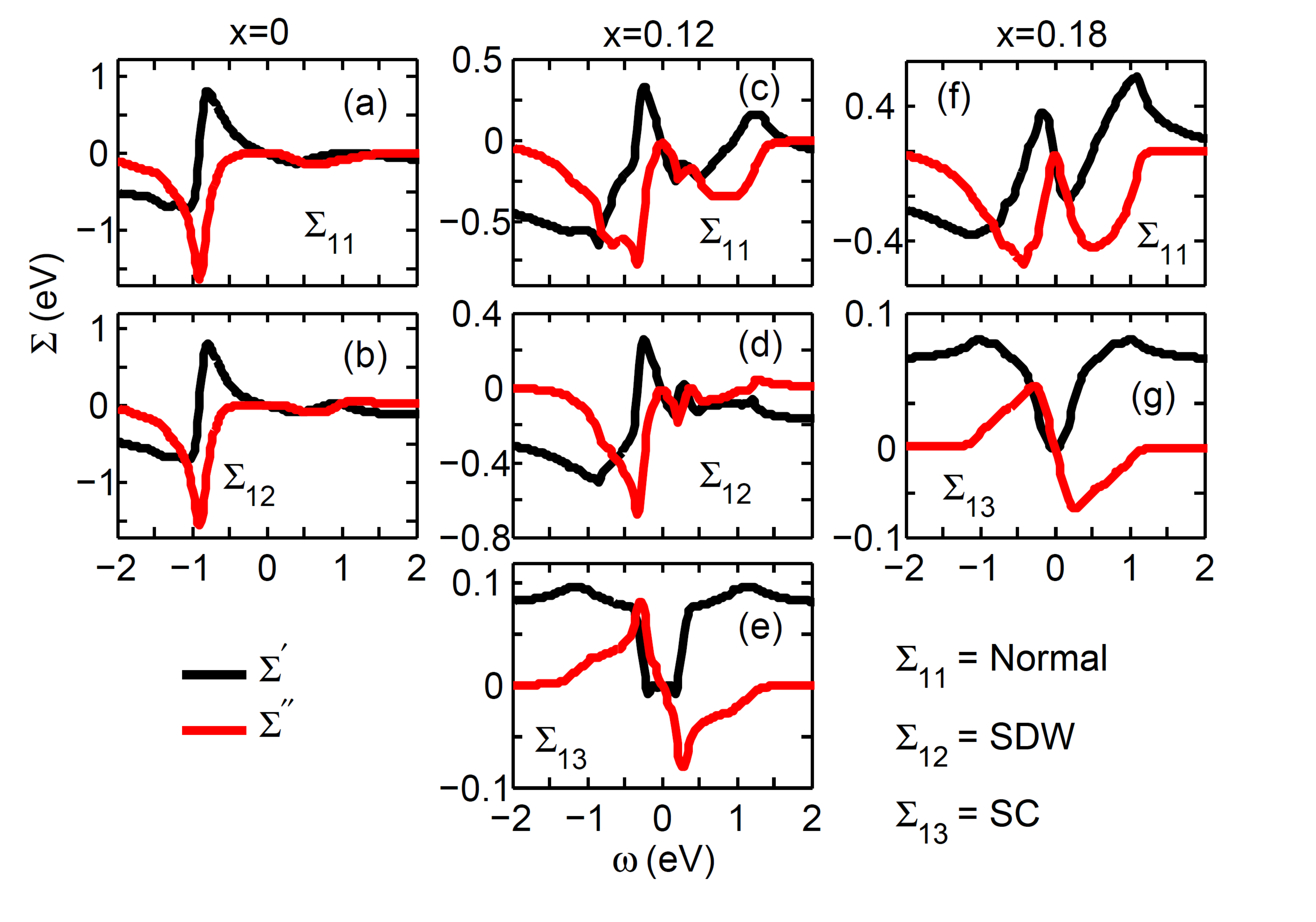}}}
\caption{Doping dependence of diagonal- and off-diagonal components of the self-energy matrix in LSCO.  (a)-(b) $x=0$ (half-filling); (c)-(e) $x=0.12$, with coexisting SDW and d-wave SC orders; (f)-(g) $x=0.18$, where the SDW induced pseudogap is absent. [From Ref.~\cite{Whiffnium}.]
}
\label{SelfRPA}
\end{figure}

Figure~\ref{SelfRPA}\cite{Whiffnium} illustrates the doping evolution of the diagonal and off-diagonal elements of the self-energy matrix with the example of La$_{2-x}$Sr$_x$CuO$_4$ (LSCO). At half-filling (x=0), both the real and imaginary part of the normal ($\Sigma_{11}$) and SDW gap ($\Sigma_{12}$) self-energy are featureless over the energy range of the insulating gap. The dramatic particle-hole asymmetry in self-energy seen in Figs~\ref{SelfRPA}(a) and (b) arises from the presence of the VHS below $E_F$. The real part of the self-energy changes sign just outside the insulating gap, splitting the LMB into coherent and incoherent portions, while the imaginary part has a peak at this energy scale, which distributes the spectral weight between these two parts of the spectrum. Therefore, a waterfall feature will be present even at half-filling, as has been observed in the ARPES spectra of Ca$_2$CuO$_2$Cl$_2$ (CCOC), see Fig.~\ref{md2d4}(f). Notably, this waterfall feature is also produced by QMC calculations [Fig.~\ref{md2d4}(i)], but not by the DMFT [Fig.~\ref{md2d4}(g)]. At finite doping ($x=0.12$), where the SDW and d-wave SC orders coexist, many features can be seen in the self-energy (middle row in Fig.~\ref{SelfRPA}). The low-energy feature stems from the total SDW+SC gap, which renormalizes the quasiparticle pseudogap. The high-energy peak in $\Sigma^{\prime\prime}$ and the corresponding change-of-sign in $\Sigma^{\prime}$ creates the UHB and LHB. In the overdoped region at $x=0.18$ where the pseudogap has disappeared, the self-energy shows a linear-in-energy dependence in the low-energy spectrum, signaling the recovery of pure d-wave character in the SC state.     

Many interesting aspects of the self-energy correction to the quasiparticle spectrum occur when the pseudogap and the SC gap are present simultaneously. For example,  Fig.~\ref{SelfRPA} shows that the slope of $\Sigma_{11}^{\prime}$ at $E_F$  gradually decreases with decreasing doping, as the size of the pseudogap increases. This implies that the dispersion renormalization $Z_d$ (defined in Eq.~\ref{Zd}) increases (towards 1) with decreasing doping, giving the surprising result that the bands are less renormalized as the system approaches half-filling. In sharp contrast, the spectral weight renormalization $Z_{\omega}$, defined in Eq.~\ref{Zw}, at $E_F$ decreases with decreasing doping. Such non-Fermi liquid or `gossamer'-like behavior is further discussed in connection with Fig.~\ref{md1d} below.

As noted earlier, the waterfall feature is present from half-filling to the overdoped regime, and disappears for $x>$0.30 (precise doping is material dependent), where the VHS crosses $E_F$ and the particle-hole fluctuations which drive the waterfall feature (see Sec.~\ref{S:ElSpec}.3) are destroyed. At the same time, the anomalous spectral weight transfer with doping from the high-energy `Hubband bands' to the quasiparticle states also disappears as seen in ARPES, optical spectra, XAS, and other spectroscopies, see Fig.~\ref{aswt}.

\subsection{Off-diagonal self-energy}

Note that the anomalous (off-diagonal) self-energies $\Sigma_{12}$ and $\Sigma_{13}$ renormalize the SDW gap and SC gap, respectively. Fig.~\ref{SelfRPA} shows that the SC gap self-energy $\Sigma_{13}$ exhibits the expected particle-hole symmetry. When both SDW and SC gaps are present, the GW calculation generates a new gap, associated with $\Sigma_{14}$, see Section~\ref{S:SC}.2.  That is, the QP-GW model naturally leads to a theory with a [broken] SO(5) symmetry\cite{SO5} or to a related extension\cite{SO8}.

\subsection{Comparison with Strong Coupling}

Insight into the QP-GW model is obtained by seeing how it evolves into the strong-coupling limit.  Here we briefly consider a recent phenomenological strong-coupling model introduced by Yang, Rice, and Zhang (YRZ)\cite{YRZ}, while further comparisons are discussed in Section~\ref{S:Ext}.3 below. YRZ invoke a no-double-occupancy picture of strong coupling in which for hole-doping $x$ the upper Hubbard band (UHB) contains $1-x$ states, 
while the lower Hubbard band (LHB) contains $1+x$ electrons, of which $1-x$ lie below $E_F$. 
Introducing the parameter $b$, the LHB is assumed to have an incoherent part, with $1-b$ electrons and a coherent part containing $x+b$ states.  The model then ignores the incoherent states and assumes that the low-energy physics is controlled by the coherent states.  Luttinger's theorem is approximately satisfied by requiring that the fractional filling of the coherent band is equal to the fractional filling of the full band, or $(b-x)/(b+x)=(1-x)/2$.  Defining $b+x\equiv 2g_t$, leads to 
\begin{equation}\label{YRZ1}
g_t=2x/(1+x),
\end{equation}
where $g_t$ is the dispersion renormalization factor of $t-J$ models.

What distinguishes the YRZ model from other $t-J$ models is the introduction of a pseudogap, which is treated at the mean field level as a phase transition described by the self-energy 
\begin{equation}\label{YRZ3}
\Sigma_k^{YRZ} = \frac{\Delta_{SDW}^2}{\omega-\xi_{0{\bf k}}}, 
\end{equation}
where $\xi_{0{\bf k}}=\epsilon_{0{\bf k}}-\mu$, and $\mu$ is the chemical potential.  So far, this would be the strong coupling version of any density-wave instability, as long as the energy $\epsilon_0=\epsilon_{k+Q}$, where $Q$ is the nesting wavevector, e.g., $(\pi,\pi)$ for the commensurate SDW order, as in Eq.~\ref{SDWSE}.  YRZ, however, introduce: 
\begin{equation}\label{YRZ2}
\epsilon_{0{\bf k}}=+2t(\cos(k_xa)+\cos(k_ya)), 
\end{equation}
so that the resulting Green's function has a zero at $E_F$ along the AFM zone boundary.

How does this model compare with the QP-GW model?  Since both YRZ and the conventional SDW model satisfy Eq.~\ref{SDWSE}, and both have the same Fermi surface up to the AFM zone boundary, both models would make the same predictions for most experiments.  Indeed, many calculations purporting to test the YRZ model\cite{YRZ1,YRZ2,YRZ3} should probably be redone to see if the results are distinguishable from the SDW model predictions.  As noted above, Eq.~\ref{YRZ2} is unique to YRZ theory, and this can be tested experimentally.  With the conventional AFM choice of $\epsilon_0$, the model would constitute a strong coupling version of the SDW phase of QP-GW, where only the coherent part, $G_ZW_Z$ is analyzed.  However, in many experiments the incoherent parts of the bands can be explored, which requires the full QP-GW Green's function.  Moreover, experiments seem to require a renormalization factor $Z>>g_t$, implying that double-occupancy is not strictly forbidden.  To describe this situation, the QP-GW model instead assumes that the occupancy of the UHB is always 1 and that the ASWT is controlled by the collapse of the SDW gap and not by strong correlations.  

\subsection{$G_Z$ vs $G_{exp}$}

Considering the paramagnetic state for simplicity, the intermediate Green's function $G_Z$ can be written as 
\begin{equation}\label{Gexp1}
G_Z=\frac{Z}{\omega-Z\epsilon_{k,LDA}+i\delta},
\end{equation}
with the associated susceptibility
\begin{equation}\label{Gexp2}
\chi_Z(\omega,q)=Z^2\sum_k\frac{f(Z\epsilon_{k,LDA})-f(Z\epsilon_{k+q,LDA})}{\omega-Z(\epsilon_{k,LDA}-\epsilon_{k+q,LDA})+i\delta}.
\end{equation}
Equation~\ref{Gexp1} is to be compared to
\begin{equation}\label{Gexp3}
G_{exp}=\frac{1}{\omega-\epsilon_{k,exp}+i\delta}.
\end{equation}
Here we have introduced $\epsilon_{LDA}$ for the LDA dispersion, and $\epsilon_{exp}$ for the experimental dispersion.  Equation~\ref{Gexp3} is a common approximation in weak coupling calculations add leads to the susceptibility
\begin{equation}\label{Gexp2}
\chi_{exp}(\omega,q)=\sum_k\frac{f(\epsilon_{k,exp})-f(\epsilon_{k+q,exp})}{\omega-(\epsilon_{k,exp}-\epsilon_{k+q,exp})+i\delta}.
\end{equation}
We have found that $\epsilon_{k,exp}\sim Z\epsilon_{k,LDA}$, i.e., the experimental bands are the coherent part of the dressed LDA bands, so that $G_Z=ZG_{exp}$, and $U\chi_Z=U_{exp}\chi_{exp}$, where $U_{exp}=Z^2U$.  Finally,
we take $W=UU_{exp}\chi_{exp}$.
This suggests that models, which are based on experimental dispersions, will exaggerate the tendency of the system toward instability [$U\chi=1$] unless $U$ is renormalized downward because one assumes that all the electronic states are included in the coherent part of the band.

While the QP-GW approach captures these features reasonably, it should be kept in mind that the intermediate function $G_Z$ is not a full Green's function, but only its coherent part in that its imaginary part integrates to $Z$ and not unity.  Similarly, while the final $G$ is a full Green's function, it underestimates the incoherent part of the spectrum. Further, the distribution of incoherent spectral weight is relatively uniform, leading to reduced electron-hole asymmetry.  

\section{Bosonic Features}\label{S:Boson}

\begin{figure}[top]
\centering
\rotatebox{0}{\scalebox{0.5}{\includegraphics{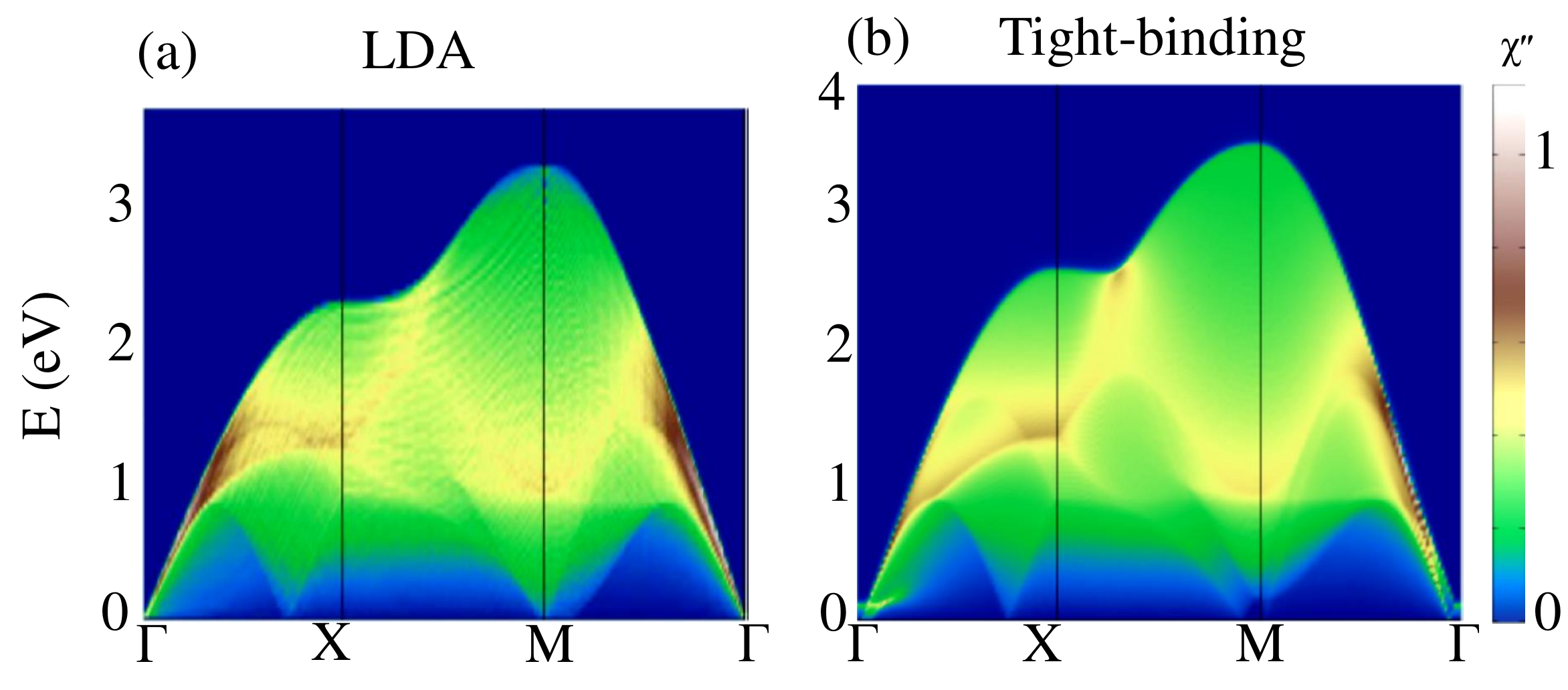}}}
\caption{Bare susceptibility $\chi^{\prime\prime}$ in NCCO at half-filling in the paramagnetic state in $(eV)^{-1}$ units, comparing LDA-based results\cite{RWang} (a) with the corresponding results based on a tight-binding model (b) along high symmetry lines $\Gamma = (0,0)$ to $X = (\pi,0)$ to $M = (\pi,\pi)$ to $\Gamma$ in the Brillouin zone (BZ).  [From Ref.~\cite{Whaffnium}.]}
\label{RayTB}
\end{figure}

In condensed matter systems, bosonic excitations such as `magnons', `plasmons', and `Cooperons' can arise from the spin, charge and pairing degrees of freedom of electrons when the corresponding scattering cross-sections are favorable. In one dimension, this includes the well-known spinon and holon excitations.  These bosons modify electronic dispersions, giving rise to a variety of dispersion and spectral weight renormalizations.  At low temperatures a mode can `soften',  
leading to various forms of  long-range density or orbital order, or superconductivity.  Hence an important aspect of the physics of correlated systems is the identification of the dominant bosonic modes. The spectral weights of various excitations are quantified by the imaginary part of the susceptibility $\chi_{RPA}^{\prime\prime}$, and can be directly probed by Raman, RIXS and inelastic neutron scattering (INS). The modes can be sorted by the susceptibility channel, and accordingly the fluctuations can be detected by different measurements. For example, magnon dispersions will show up only in the transverse channel of the spin susceptibility (observed in spin-flip INS, RIXS and Raman) and plasmons in the longitudinal + charge channel (measured by RIXS, optical probes), while the Cooperons appear along any spin- or charge channel (measured by spin-flip and non-spin-flip INS, RIXS, Raman and optical probe). Furthermore, these excitations appear in different regions of the $({\bf q},\omega)$ phase space and are strongly doping dependent, which makes it easier to observe them separately.  Notably, proper treatment of plasmons, and especially of acoustic plasmons in 2D-materials, requires extension of the QP-GW model from Hubbard interaction $U$ to include long-range Coulomb interaction.\cite{markiecharge,Mkacplas}
Figure~\ref{RayTB}\cite{Whaffnium} shows a typical map of the bare susceptibility in Nd$_{2-x}$Ce$_x$CuO$_4$ (NCCO), comparing our calculations based on tight-binding dispersions with LDA calculations\cite{RWang}.  The agreement is quite good, confirming that our dispersions accurately represent the LDA bands.  

\subsection{Spin-wave spectrum in undoped cuprates}

\begin{figure*}[top]
\centering
\rotatebox{270}{\scalebox{0.54}{\includegraphics{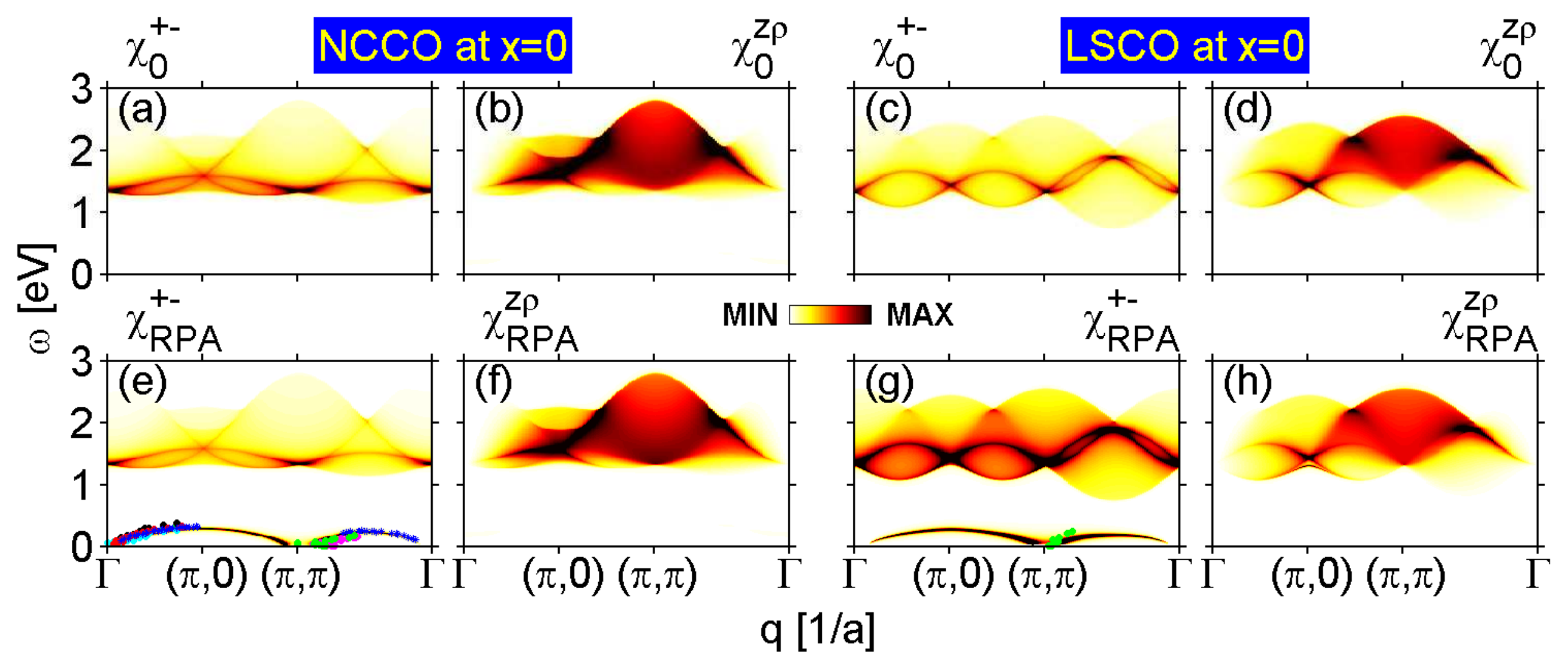}}}
\caption{{ $\chi^{\prime\prime}$ in the insulating half-filled state,} modeled via an SDW order parameter. Bare values of $\chi^{\prime\prime}$ along the high-symmetry directions in the BZ are shown for the transverse spin channel in (a), and along the longitudinal plus charge channel in (b) for electron doped Nd$_2$CuO$_4$ (NCO). (c)-(d) Corresponding results for hole-doped NCO. (e)-(h): RPA components of susceptibility in the corresponding  channels. Results are compared with RIXS data for
insulating LCO [blue symbols from Ref~\cite{braicovich}], underdoped LSCO
[black and cyan symbols from Ref.~\cite{braicovich}], and undoped
SCOC [grey symbols].\cite{guarise} Magenta and green symbols are from
neutron data for undoped and optimally doped
LSCO\cite{coldea,vignolle}, which are compared with the LBCO data
at 1/8 doping (deep green symbols)\cite{tranquada}.   [From Ref.~\cite{Whiffnium}.]}
\label{SDWx0}
\end{figure*}

Gapless spin waves are the Goldstone modes associated with any ferro- or antiferromagnet. The spin-wave spectrum of the Hubbard model was calculated by Schrieffer, Wen, and Zhang\cite{SWZ}. Figure~\ref{SDWx0}\cite{Whiffnium} shows that the spin-wave spectra are quite similar in the parent compounds of both electron-doped NCCO and hole-doped LSCO, in good agreement with experiment\cite{braicovich,guarise,coldea,vignolle,tranquada}. For the non-interacting susceptibility, Eq.~\ref{chi2}, all intraband excitations are gapped in all spin and charge channels, as shown in the upper panel of Fig.~\ref{SDWx0}. At the RPA level, the longitudinal ($\Delta S_z=0$) and charge channels remain gapped, some intensity modulation notwithstanding. But the transverse spin-flip channel ($\Delta S_z=1$) undergoes dramatic changes.

For the pure Hubbard model (nearest-neighbor hopping $t$ only) Schrieffer {\it et al.}\cite{SWZ} showed that gapless spin-wave modes occur at the AFM vector ${\bf Q}$ in the undoped SDW state, finding an analytic solution for the mean-field problem, which has been extended to a more general dispersion.\cite{SinT,chubukov,ZSch} The necessary condition for the occurrence of a Goldstone mode is that the off-diagonal term of the non-interacting susceptibility, $\chi_{12/21}$ in Eq.~\ref{chi2}, which drives the system away from the instability [at $(1-U\chi_{11}^{+-}({\bf q},\omega))(1-U\chi_{22}^{+-}({\bf q},\omega))=0$], reduces exactly to the SDW gap equation at $q=Q$ given in Eq.~\ref{delselffinal}. At ${\bf q}=(\pi,\pi)$, the spin-wave shows linear-dispersion as in AFM Cr [Ref.~\onlinecite{Cr}], which extends to zero energy. In addition, the spectra along the zone diagonal possess a symmetry between the $\Gamma-$ and $(\pi,\pi)-$points at low-energies, consistent with experiments and $t-J$ model calculations\cite{guarise}, reflecting the presence of long-range magnetic order. 


\subsection{Spin-waves in the SDW state at finite doping}

\begin{figure}[top]
\centering
\rotatebox{270}{\scalebox{0.55}{\includegraphics{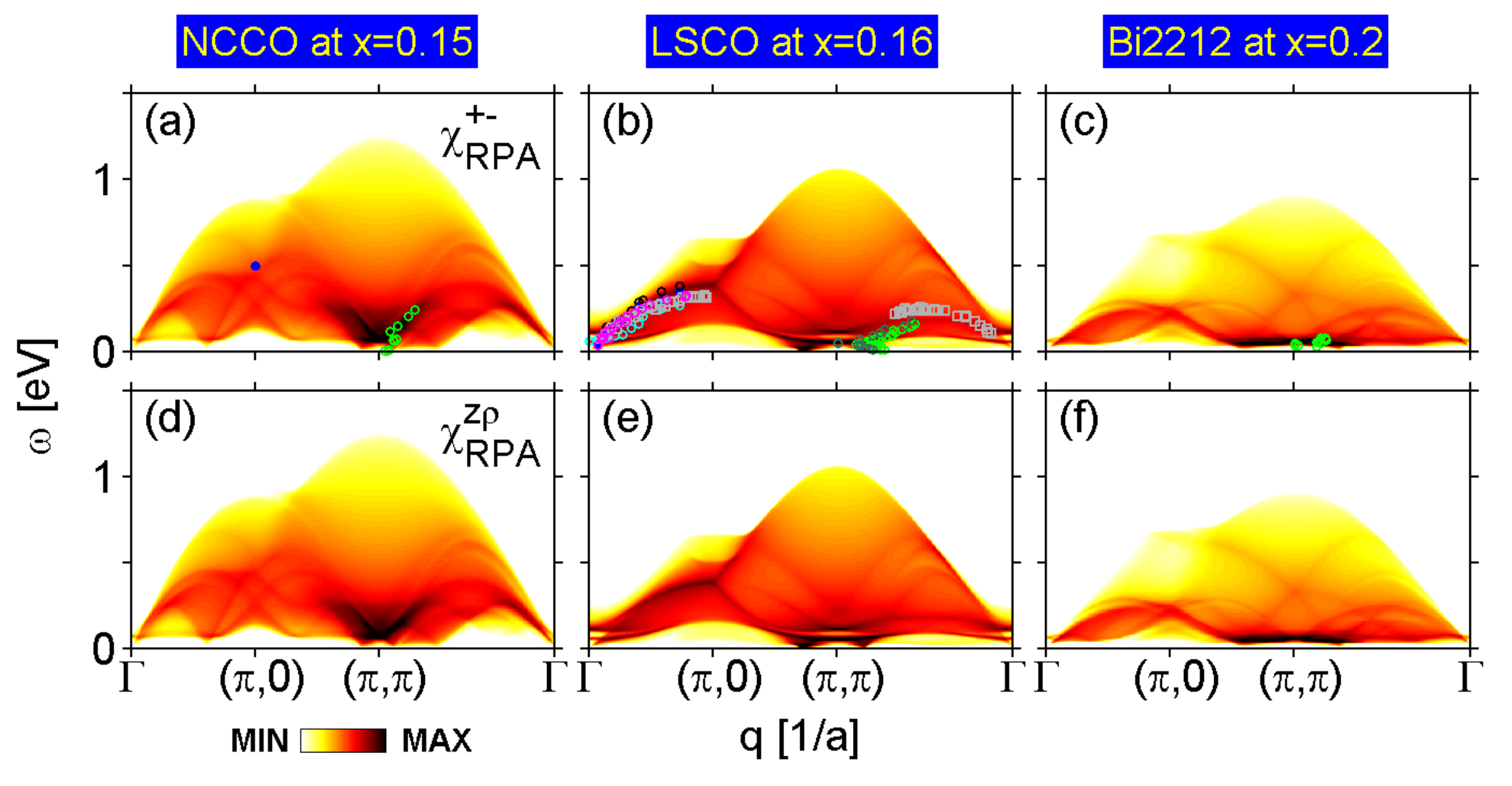}}}
\caption{{ Spin and charge spectral dispersion 
in doped LSCO, Bi2212 and
NCCO.} The imaginary part of the susceptibility is plotted along
the high-symmetry lines in the BZ for transverse
[top row] and longitudinal plus charge channels [bottom row]
for two hole-doped and one electron-doped cuprate near 
optimal doping. Experimental data are the same as in Fig.~\ref{SDWx0}.
[From Ref.~\onlinecite{tanmoyLEK}.]}
\label{chiop}
\end{figure}

Upon doping, a spin-wave dispersion persists up to the overdoped regime, but becomes gapped in the SC state to give rise to a magnetic resonance mode.  Here we discuss the example of the near-optimal doping region in the SC state, where a residual SDW dressed pseudogap is still present in the low-energy region along all directions. Computed  $\chi^{\prime\prime}$ spectra at finite doping are presented in 
Fig.~\ref{chiop}\cite{tanmoyLEK} in the transverse (top row) and longitudinal spin plus charge (bottom row) channels for three different materials. The spectra in the transverse channel can be determined directly by INS in the spin-flip mode\cite{coldea,vignolle,tranquada}, whereas transverse, longitudinal, and charge channels can all be observed by RIXS\cite{braicovich,guarise} or indirectly via optical spectroscopy\cite{hwangoplscobi2212,hwang2,schachinger}.  Also shown are the experimental single magnon RIXS results for undoped La$_2$CuO$_4$ (LCO), neutron data from the same sample, and RIXS data of undoped Sr$_2$CuO$_2$Cl$_2$ (SCOC) and NCO. The spin-waves persist throughout the BZ, are doping dependent, and become gapped in the SC state near $Q$, as discussed in Section~\ref{S:ElSpec2}.3. Along $\Gamma\rightarrow(\pi,0)$, an additional paramagnon feature appears which extends from $\omega=0$ at $\Gamma$ to $\omega\sim500$meV at ${\bf q}=(\pi,0)$ at half-filling which reduces to $\omega\sim300$meV near optimal doping, Fig.~\ref{chiop}. 

This dispersing paramagnon like spin-wave feature reflects inelastic scattering from the Van-Hove singularity (VHS) in the bands near $k=(\pi,\pi)$ in cuprates, and it is not from the low-lying SDW Goldstone mode or magnetic resonance mode around $q=(\pi,\pi)$. This feature is representative of the bi-magnon spectrum as observed in Raman spectra at the same energy scale [see, e.g, Refs.~\onlinecite{machtoub,sugai}], and it is present in both longitudinal (Figs.~\ref{chiop}(b,d,f)) as well as transverse spectra (Figs.~\ref{chiop}(a,c,e)). The extent to which these high-energy joint density-of-states (JDOS) features evolve into multi-magnon features in more sophisticated models of the self-energy\cite{singh} remains to be seen. We emphasize that the overall agreement with RIXS [blue symbols in Fig.~\ref{chiop}(a)] is remarkable, even without the inclusion of core-hole and other matrix element effects. This is the main susceptibility feature responsible for the HEK, and it is consistent with the fairly doping independent energy range of the HEK as discussed in Section~\ref{S:ElSpec}.3.1 below. Note that susceptibility in the SC state up to the gap energy ($\le 2\Delta$) also controls Friedel oscillations around impurities, and the closely related quasiparticle interference (QPI) patterns seen in STM measurements.\cite{QPI1,QPI2,QPI3,tanmoy2gap}

\section{Single-particle spectra}\label{S:ElSpec}


We begin this section with a few general remarks on electron- and hole-doped cuprates, followed by a detailed comparison of our calculations with various one-body spectroscopies, particularly ARPES and STM, and a brief discussion of  x-ray absorption (XAS), and Compton scattering results. 

\subsection{Electron-doped cuprates}
\begin{figure}[top]
\centering
\rotatebox{0}{\scalebox{0.28}{\includegraphics{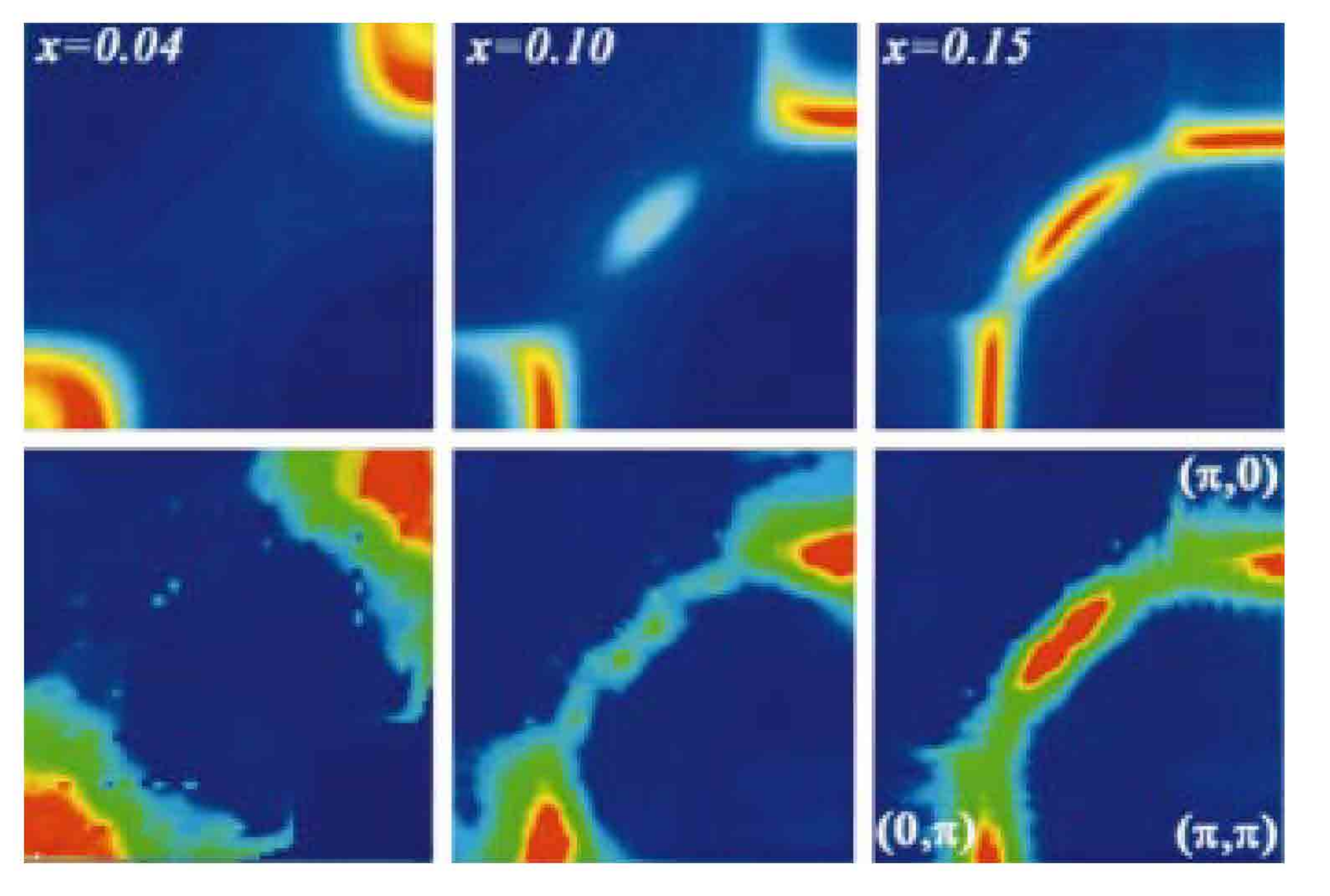}}}
\caption{{ Fermi surface (FS) maps (defined as spectral intensity at $E_F$) in NCCO:} Theoretical FS maps (top row) at three dopings, $x$ = 0.04, 0.10, and 0.15, compared to the corresponding experimental ARPES data\cite{armitage} (bottom row).
[From Ref.~\onlinecite{kusko}.]  }
\label{Kusko1}
\end{figure}
An important clue to the role of self-energy corrections was provided by the striking ARPES results on electron-doped NCCO\cite{armitage}, where early analysis\cite{markietb} indicated that the experimental dispersions are close to the LDA-bands renormalized by a constant factor $Z\sim1/2$. 

\begin{figure}[top]
\centering
\rotatebox{0}{\scalebox{0.105}{\includegraphics{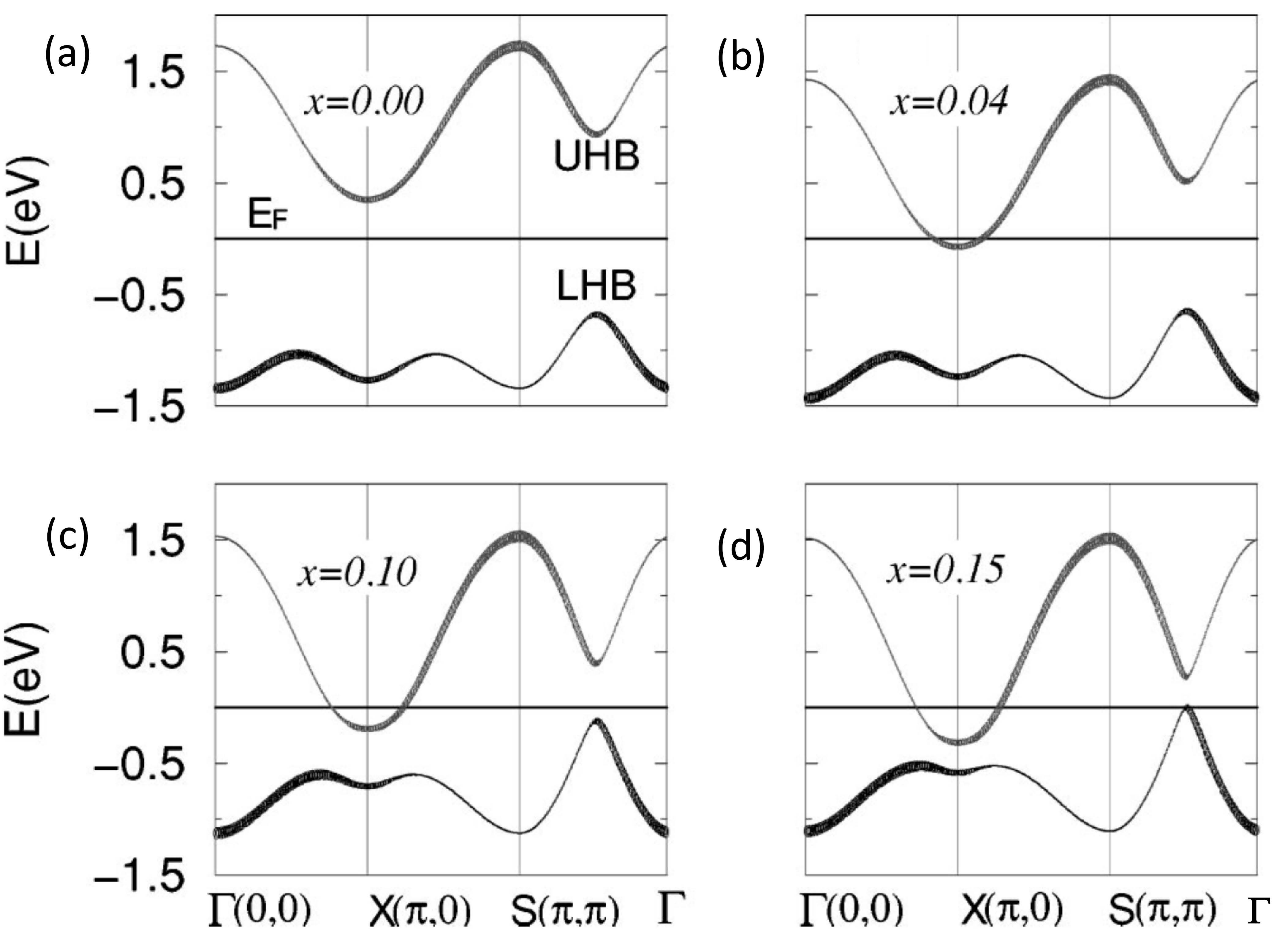}}}
\caption{ { Theoretical dispersion at four different dopings $x$ in NCCO, showing collapse of the magnetic gap.} 
[From Ref.~\onlinecite{kusko}.]}
\label{Kusko2}
\end{figure}

While proximity of the VHS and `stripe' phenomena greatly complicate the analysis of hole-doped cuprates, NCCO seems to be free of these complications, and the doping evolution of the normal state band structure can be readily captured by a model of $(\pi,\pi)$ AFM order, Fig.~\ref{Kusko1}\cite{kusko}.  At low doping, electrons enter the bottom of the UMB near $(\pi,0)$, Figs.~\ref{Kusko2}(b),~(c).  As doping increases the magnetic gap collapses, and just below 15\% doping the LMB crosses $E_F$ near $(\pi/2,\pi/2)$, Fig.~\ref{Kusko2}(d), leading to the appearance of a {\it hole-pocket} near the zone diagonal, Fig.~\ref{Kusko1}.  The electron- and hole-pockets form a necklace, separated by `hot-spots' along the zone diagonal, which represent the residual magnetic gap separating the UMB and LMB.  At a higher doping, near $x$=0.18, the magnetic gap collapses, and the pockets merge into the large FS predicted by the LDA.

\begin{figure}[top]
\centering
\rotatebox{0}{\scalebox{0.3}{\includegraphics{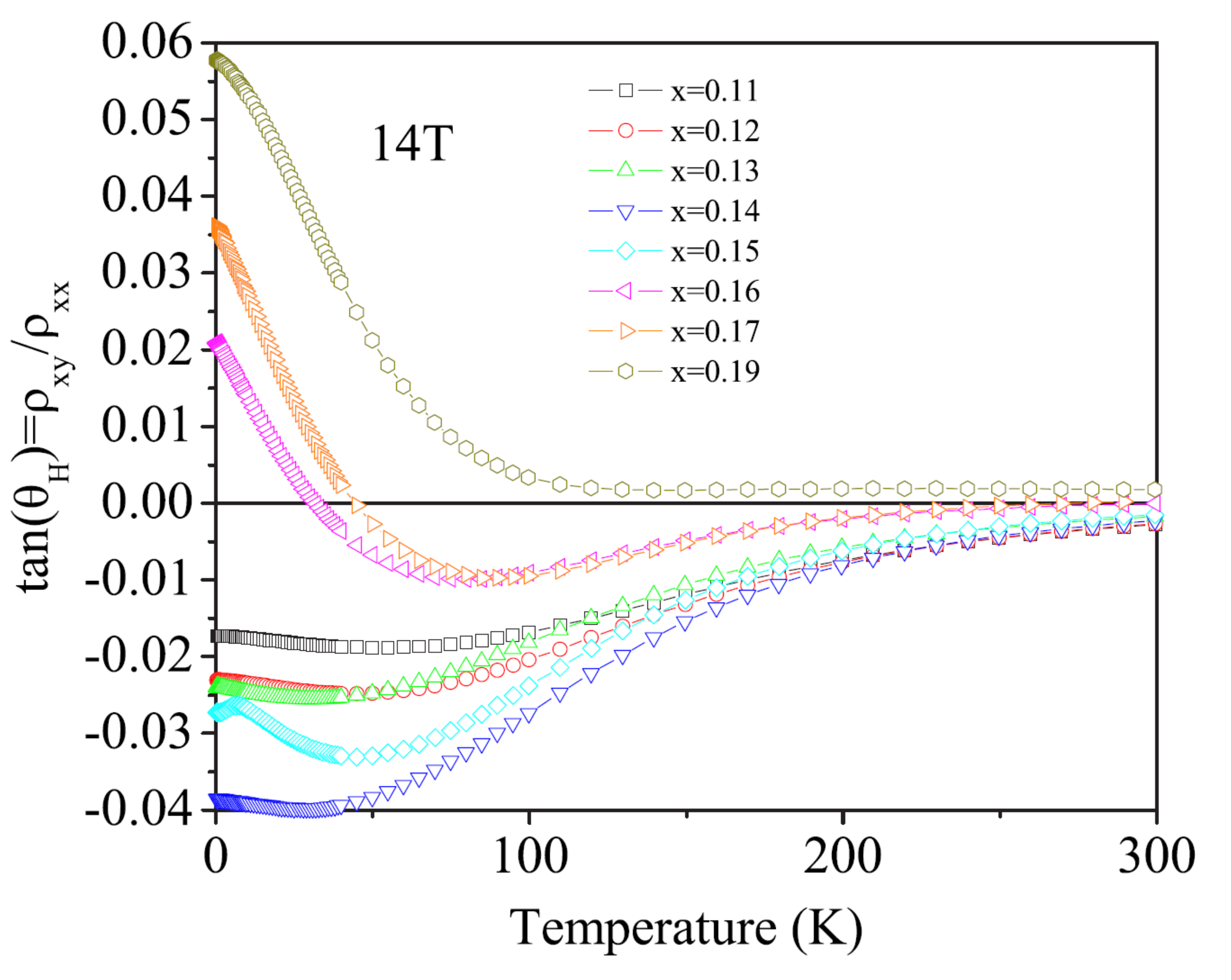}}}
\caption{Hall angle $\theta_H = tan^{-1}(\rho_{xy}/\rho_{xx})$ at $B$=14T vs temperature for a variety of dopings in PCCO. Note the low-$T$ upturn starting near $x=0.15$. [From Ref.~\onlinecite{GreeneHall}.]  }
\label{Greene1}
\end{figure}

\begin{figure}[top]
\centering
\rotatebox{0}{\scalebox{0.40}{\includegraphics{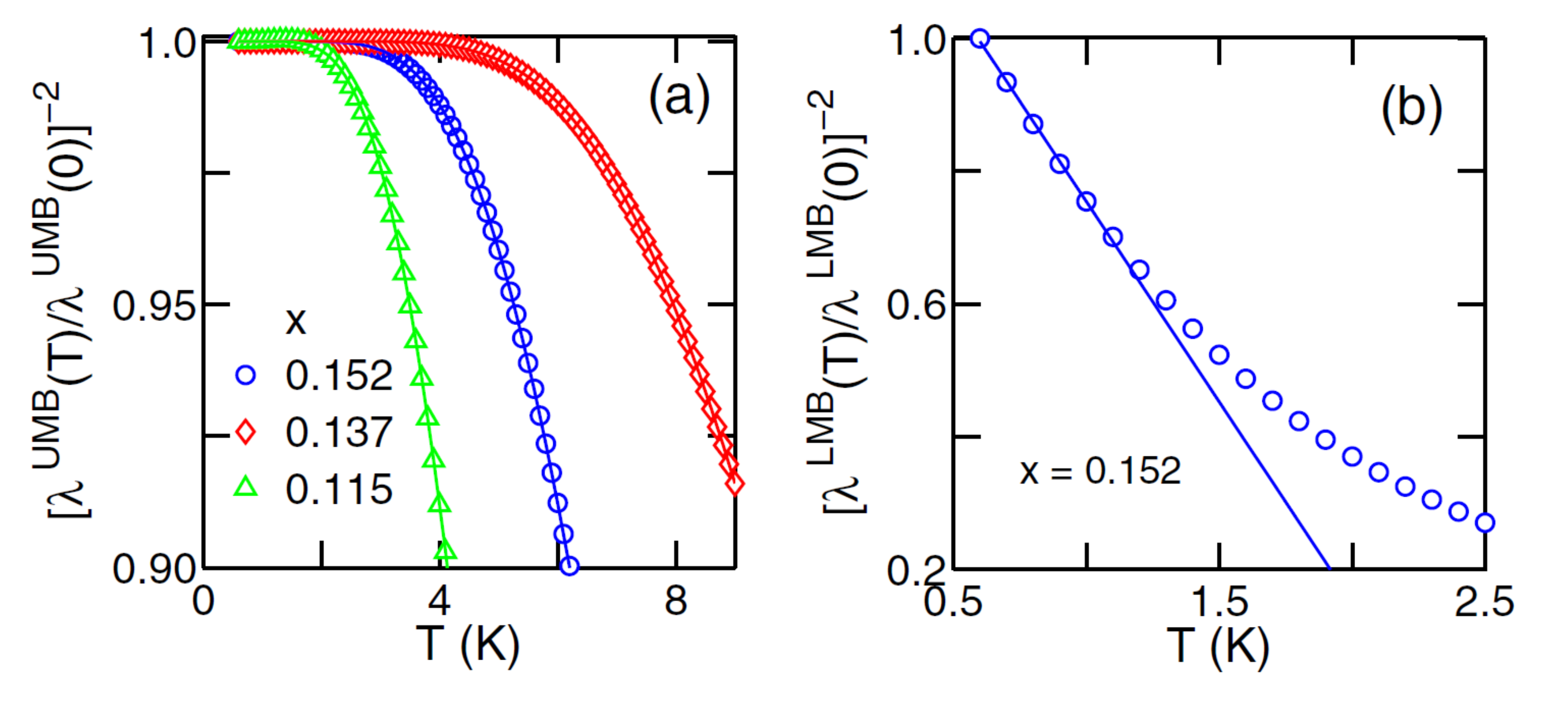}}}
\caption{Penetration depth in PCCO \cite{PCCOpenet} at several dopings, separated into a part varying exponentially in temperature (a) and a part varying linearly in $T$ (b), with the latter only present at $x$=0.152. 
[From Ref.~\onlinecite{tanmoyprl}.] 
 } 
\label{nodeless1}
\end{figure}

\begin{figure}[top]
\centering
\rotatebox{0}{\scalebox{0.34}{\includegraphics{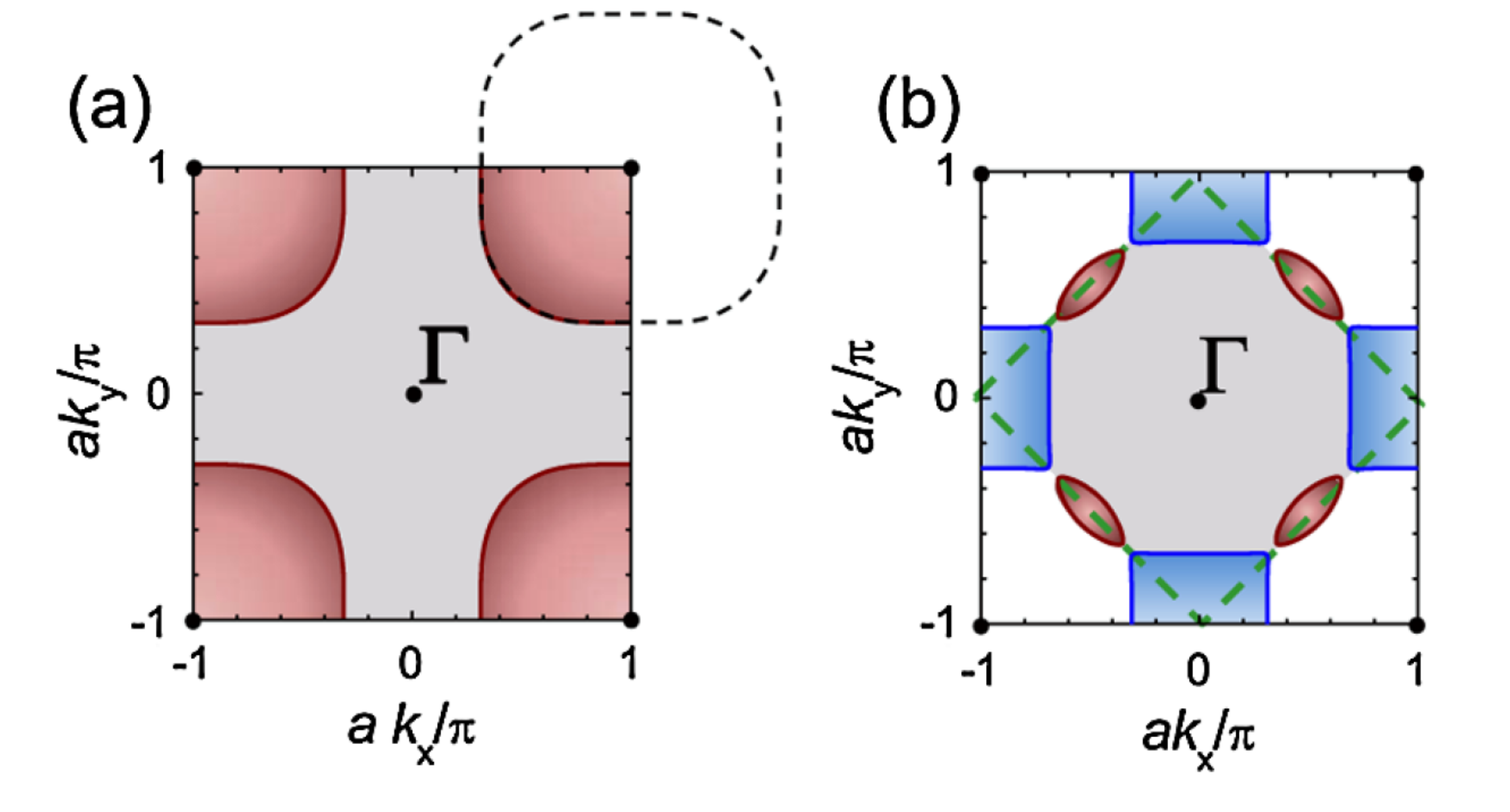}}}
\rotatebox{0}{\scalebox{0.35}{\includegraphics{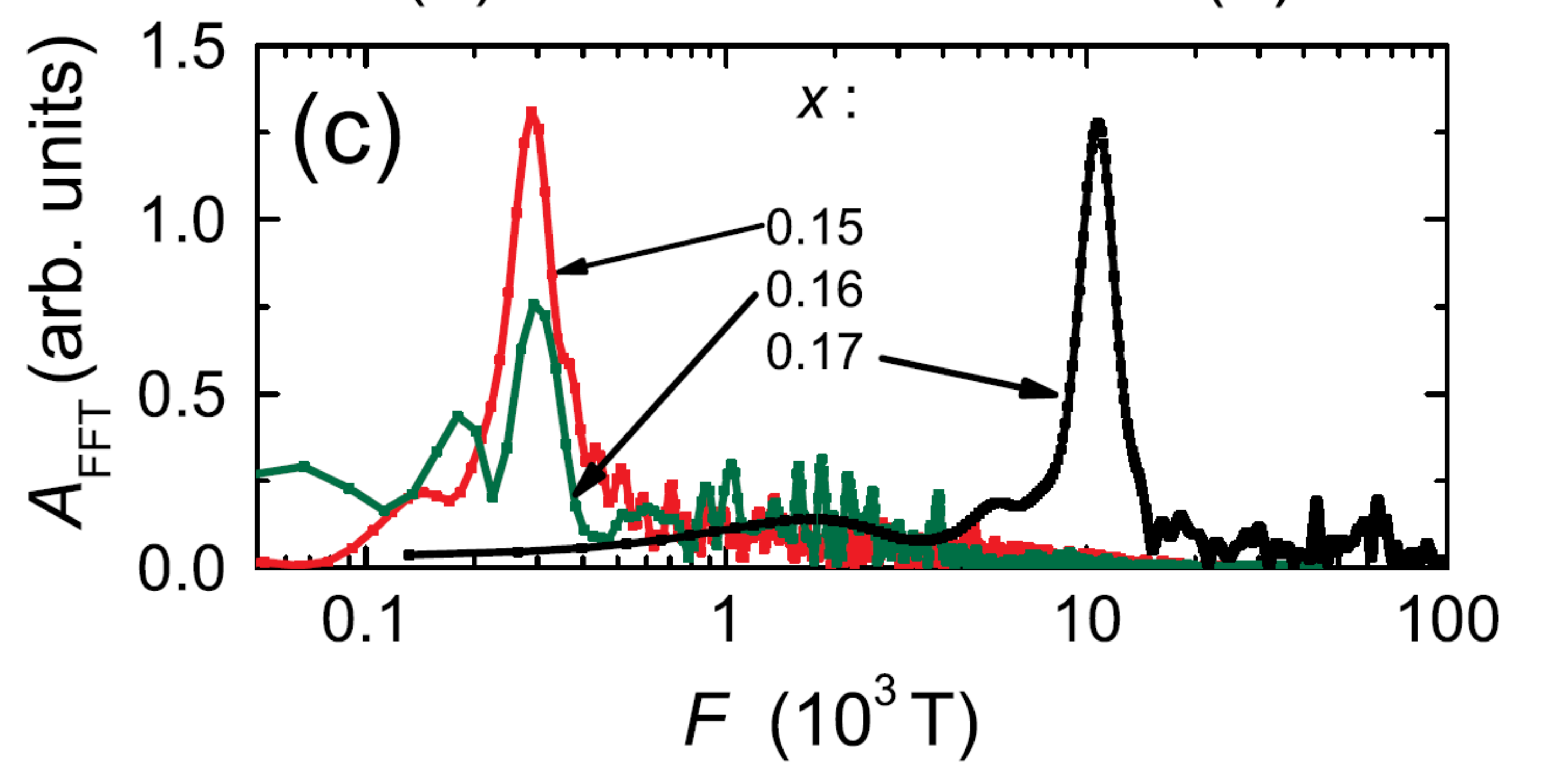}}}
\caption{{ Quantum oscillations in NCCO.} Top Row: (a) Large FS expected above the second topological transition; (b) electron (blue) and hole (red) pockets expected above the first topological transition. Bottom Row: (c) QO frequencies found at $x$ =0.15, 0.16, and 0.17. The smaller areas (lower frequencies) at the lower two dopings are consistent with the nodal hole pocket, while the larger area at $x$=0.17 is consistent with the large FS. [From Ref.~\onlinecite{qohelm}.] }
\label{NCCOQO1}
\end{figure}

A key signature of the magnetic-gap-collapse scenario is the presence of a sequence of {\it two `topological transitions'} or changes in the FS topology, the first from an initial electron-pocket to the appearance of a hole-pocket, and the second to a merging of the pockets into a single large FS.  An important datum of the model is the doping at which the hole pocket first appears.  This would affect many properties, and the consistency of its value in different experiments provides a test of the model.  In photoemission the pocket appears between 10 and 15\% doping, and from the model dispersion, Fig.~\ref{Kusko2}, probably closer to the latter.  There are three other independent indications of this transition, close to the same doping in NCCO or Pr$_{2-x}$Ce$_x$CuO$_4$ (PCCO): (1) The Hall effect starts to change sign between 14 and 15\% doping\cite{GreeneHall}, Fig.~\ref{Greene1}; (2) The penetration depth crosses over from an exponential temperature dependence to an exponential-plus-linear $T$-dependence near 14\% doping\cite{PCCOpenet,tanmoyprl}, Fig.~\ref{nodeless1}, as expected, since the nodal gap is present only on the hole-pocket;  and, (3) Quantum oscillations (QO) have been observed in NCCO\cite{qohelm}, finding a clear nodal pocket at 15\% and 16\%, and an apparent crossover to the large LDA FS (second topological transition) at 17\% doping, Fig.~\ref{NCCOQO1}.  An extension to lower doping, $x$=0.13, was unable to find QOs below 15\% doping, again consistent with the first topological transition\cite{qohelm3}.  It is surprising that QOs associated with the electron pockets have not been observed, but they are expected to be fairly weak.\cite{qohelm3} Note that the penetration depth results are particularly important as they require the superconducting electrons to see the gapped FSs produced by the AFM order. In other words, magnetism and superconductivity must coexist on the same electrons, and cannot be separated in different parts of the sample.  


The second topological transition was predicted to occur near 19\% doping, but it has not proven practical to fabricate bulk samples at such high dopings.  However, films can be made, and the Hall effect measurements\cite{GreeneHall} find a purely positive Hall effect at $x$=0.19, Fig.~\ref{Greene1}.  While the early QO measurements\cite{qohelm} reported the large FS at $x$=0.17, more recent measurements find that this is a magnetic breakdown effect\cite{qohelm2,qohelm3} with a small but finite residual gap of $\sim$ 14~meV for $x$=0.16, and 5~meV for $x$=0.17.  Note that details of theoretical predictions are sensitive to band parameters. For example, in a $t-t'$ model, the hole-pocket never crosses $E_F$, and for this transition to occur a $t''$ parameter is needed. For hole-doped LSCO, we find a complementary evolution of the pockets with doping where the electron-pocket first appears at $x=0.15$ (see the blue line in Fig.~2(a) of Ref.~\onlinecite{tanmoygoss}); see also Fig.~\ref{phasediagN}(b) for Bi$_2$Sr$_2$CaCu$_2$O$_8$ (Bi2212).


\subsection{Hole-doped cuprates}

The preceding modeling is most appropriate for electron doping, where only the $(\pi ,\pi )$ commensurate SDW order is observed, and theoretical predictions are in very good accord with experiments. Remarkably, however, the same model applied to hole-doped cuprates seems to capture many aspects of the two-gap
scenario\cite{tanmoy2gap,AndHir}, despite the fact that it does not describe the incommensurate magnetization. For example, the SDW order is observed to survive up to $\sim$10-12\% doping in several families of hole-doped cuprates, although at incommensurate $q$-vectors\cite{NMRBi12,NMRTlHg12,NMRYBCO12,NeutronYBCO10,NeutronYBCO11}. We have analyzed other candidates for the competing order including charge, flux, and $d-$density waves\cite{tanmoy2gap}, and find that results are insensitive to the detailed nature of the competing order state.  Thus, Eqs.~\ref{HFHam}-\ref{delselffinal} continue to hold for any $Q=(\pi,\pi )$ order, as long as the appropriate gap $\Delta$ is used. This is important because of the 3D computations involved first in evaluating susceptibilities, and then the self-energies, and each of these quantities is $q$- and $\omega$-dependent, as well as being a 4$\times$4 tensor when treating competing superconducting and $(\pi,\pi)$ order in QP-GW modeling. Treating incommensurate orders would require higher order tensors and many additional calculations, as the nesting $q$ vector is doping-dependent and would also need to be determined self-consistently.  Hence we have divided our computational strategy into two parts: a full QP-GW calculation assuming $(\pi,\pi)$ magnetic order, and a more comprehensive analysis in which correlations are included via the  Gutzwiller approximation + RPA (GA+RPA) and both the spin and charge orders are included to obtain $q$-dependent phase diagrams.  The QP-GW calculations are discussed below, while the  GA+RPA results are discussed in Section~\ref{S:DWs}.2.1.  We shall see that results are sensitive to the band structure, and whereas most cuprates have properties similar to Bi2212, LSCO may possess significant differences. To some extent, self-energy corrections and the nesting vectors are addressing complementary aspects of the problem, and it is not too surprising that the good agreement of this section is insensitive to the proper $q$-value.


\subsection{ARPES: Renormalization and Kinks}

\begin{figure}[h]
\rotatebox{0}{\scalebox{0.55}{\includegraphics{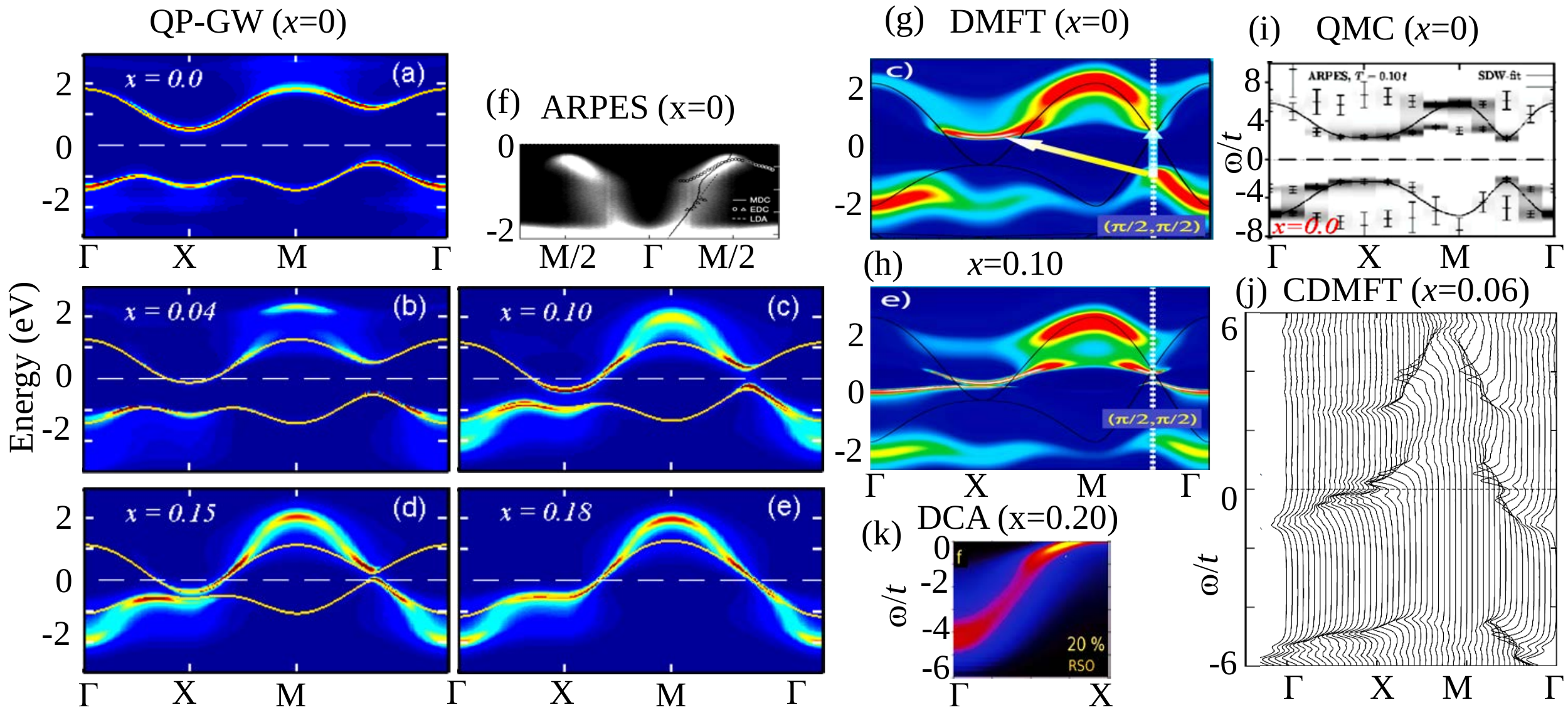}}}
\caption{ Spectral intensity as a function of energy and momentum along the high-symmetry lines in the BZ for several dopings at $T=0$. (a)-(e): QP-GW derived spectral functions for NCCO as a function of doping, ranging from half-filling to overdoping through the SDW state [after Ref.~\cite{tanmoysw}]. (f) ARPES results at half-filling in CCOC.\cite{Ronning} (g)-(h) LDA+DMFT results at half-filling and a finite doping for NCCO.\cite{DMFTcuprate} (i) QMC calculation at half-filling,\cite{grober}, (j) spectral intensity in NCCO from CDMFT computations for $U=8t$, $t'=-0.3t$\cite{DMFT13}, and (k) DCA calculation for an overdoped cuprate.\cite{macridin} At half-filling both QP-GW in (a) and QMC in (i) demonstrate the presence of an incoherent Mott band outside the AFM insulating gap, consistent with the ARPES result in (f). The waterfall shape is present at all dopings, and is consistently reproduced by various calculations. [From Ref.~\cite{Whiffnium}.]
}
 \label{md2d4}
\end{figure}

\begin{figure}[h]
\centering
\rotatebox{0}{\scalebox{0.65}{\includegraphics{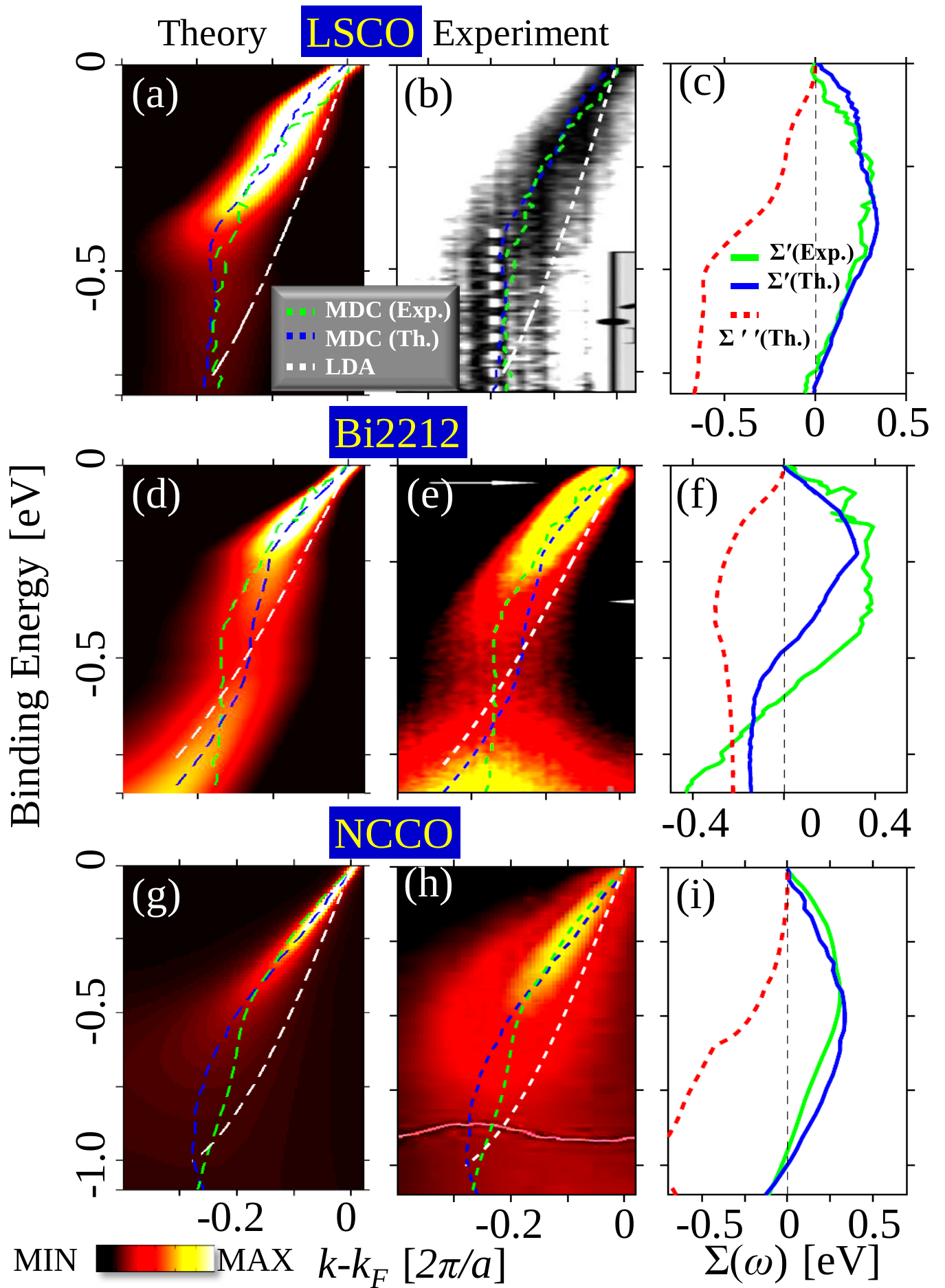}}}
\caption{HEK in LSCO, Bi2212 and NCCO at optimal dopings. (a)-(b) Theoretical and experimental spectra for LSCO. Blue and green dashed lines give MDC peak positions. The corresponding real parts of the self-energy as extracted from the MDC cuts with reference to LDA dispersion (red dashed line) are shown in (c), together with the computed real and imaginary parts of the self-energy. (d-f) Same as (a-c), except that these panels refer to Bi2212. (g-i) Same as (a-c), except that these panels refer to NCCO. Experimental spectra are from Refs.~\cite{graf2} (b),~\cite{graf} (e), and~\cite{moritz} (h).  [From Ref.~\onlinecite{Whiffnium}.]} 
\label{md2d5}
\end{figure}
The classical picture of dressing an electronic dispersion by bosons is well understood.\cite{AshcroftMermin}  For phonons, for example, the dressed bosonic spectral weight is generally given by the Eliashberg function $\alpha^2F$, where $F$ gives the bare phonon density-of-states and $\alpha$ is a measure of electron-phonon coupling.  For electronic bosons the equivalent function is $U^2\chi^{\prime\prime}$, where $\chi^{\prime\prime}$ is an appropriate susceptibility, usually evaluated within the RPA.  A peak in $\alpha^2F$ leads to a peak in $\Sigma^{\prime\prime}$, i.e., a kink or waterfall, which splits the dispersion into two parts: a dressed branch at low energies with sharp dispersion and enhanced effective mass, and a bare branch of incoherent excitations further from $E_F$.  The latter follows approximately the bare dispersion, but it is broadened by coupling with bosons.  The crossover between the two branches occurs near a peak in $\Sigma^{\prime\prime}$.  Depending on the strength of bosonic coupling, the two branches can make up a single dispersion joined by a region of more rapid dispersion, creating a kink or waterfall, or they can appear as two nearly disconnected branches.  Figs.~\ref{md2d4}(a-e)\cite{Whiffnium} shows calculations of spectral functions for NCCO at a number of dopings\cite{tanmoysw}.  While the coherent bands near $E_F$ resemble dispersions of Fig.~\ref{Kusko2}, they are separated by kinks from incoherent features at more distant energies, both below and above the $E_F$.  The remaining frames of Fig.~\ref{md2d4} show that similar kinks and incoherent subbands arise in experiment\cite{Ronning}, and in a variety of other calculations in cuprates.\cite{grober,DMFTcuprate,macridin}  

\begin{figure}[h]
\centering
\rotatebox{0}{\scalebox{0.34}{\includegraphics{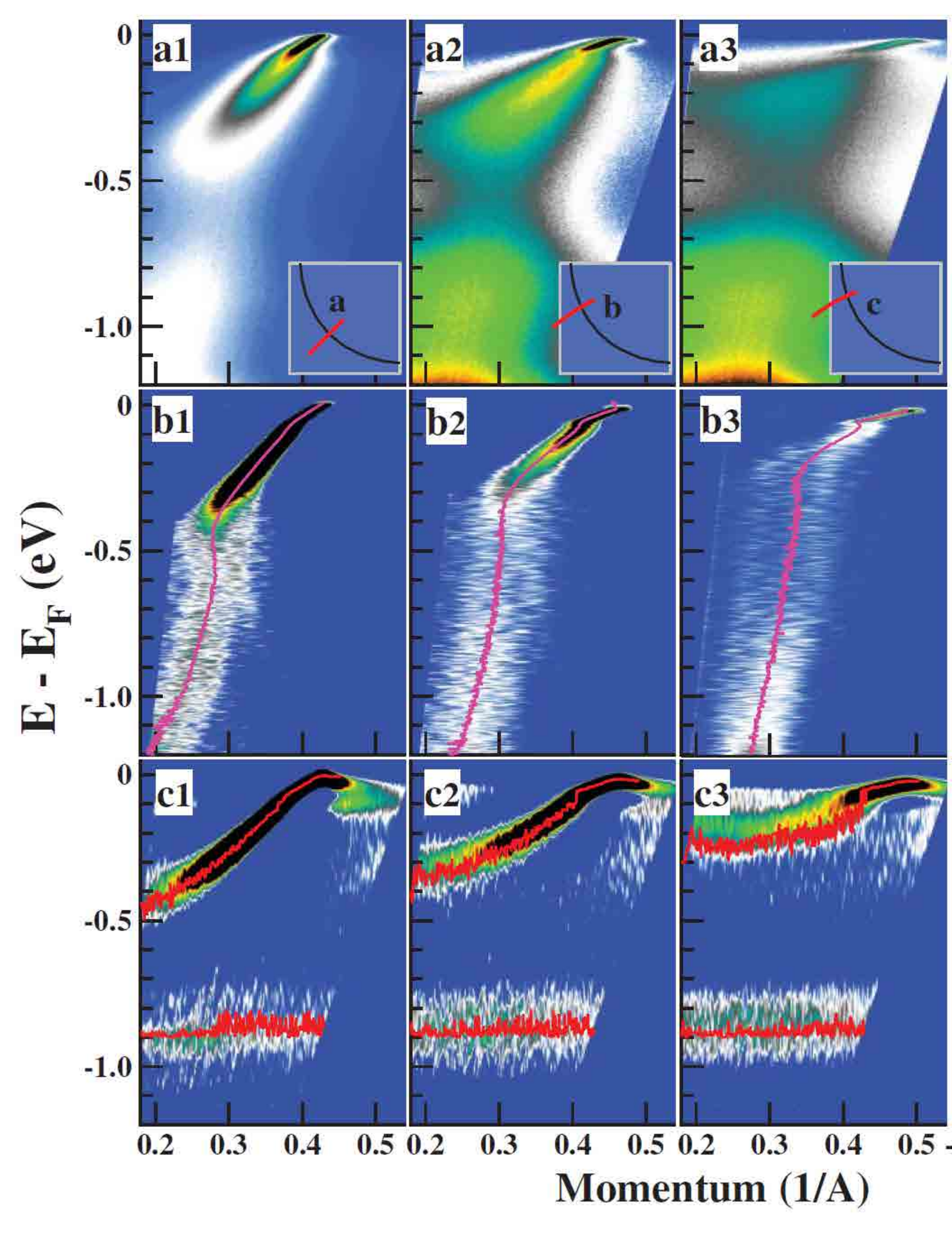}}}
\caption{HEK in Bi2212 at optimal doping along three different cuts shown in insets in the top row. Photoemission data are shown in the top row (a1-a3), and replotted as second-derivative MDC images in the second row (b1-b2), and as second-derivative EDC images in the third row (c1-c3). The third row shows the separate dispersions for the coherent and incoherent parts of the band most clearly. [From Ref.~\onlinecite{Zhou}.]    }
\label{Zhoulaser}
\end{figure}

At finite doping, our QP-GW results are in good agreement with the QMC result in terms of producing an underlying band connecting the coherent and incoherent bands, and with experiments in Fig.~\ref{md2d5}, while the DMFT results in Figs.~\ref{md2d4}(g,h) show these two bands to be disconnected. This is presumably related to an overestimation of the coupling strength and/or due to neglect of the momentum dependence of self-energy. Such disconnected splitting of the bands also persists in the CDMFT calculations for large U/t as shown in Fig.~\ref{md2d4}(j).\cite{DMFT13} However, if $U\sim 2$~eV is fixed by the optical gap, $U/t=8$ would require $t=0.25$~eV, much smaller than the expected bare LDA value. Some of the variety in the shape of the HEK is illustrated in Figs.~\ref{md2d5} and~\ref{Zhoulaser}.  In particular, Fig.~\ref{Zhoulaser} shows high-resolution laser ARPES results from optimally doped Bi2212\cite{Zhou}, with features enhanced by taking second-derivative of either the momentum distribution curves (MDCs) or energy distribution curves (EDCs), the latter clearly displaying a split dispersion.

Just as $\alpha^2F$ may have multiple peaks associated with different phonons, $\chi^{\prime\prime}$ can also display multiple peaks.  These can be either electron-hole continuum peaks associated with structure in the bare susceptibility $\chi_0$ or collective peaks associated with near zeroes of the RPA denominator, indicating proximity to a density-wave or SC instability.  In the cuprates there are two prominent peaks: the HEK and LEK, which are discussed in the following subsections.

\subsubsection{Renormalized quasiparticle spectra and the high energy kink}

The LEK, which was discovered first, falls in the low energy range of $\sim$50-70meV in the cuprates. It has been associated with either phonons or the magnetic resonance mode, there being a continuing debate as to which effect is dominant. The HEK, found above 200meV, provided the first unambiguous evidence for strong coupling to electronic bosons in the cuprates.\cite{Ronning,graf,markiewater,JChang1,JChang2} Figure~\ref{md2d5} compares theoretical predictions of the HEK (left column) to the corresponding ARPES experiments for LSCO, Bi2212, and NCCO near optimal doping.\cite{Whiffnium,graf,graf2,moritz}  How the kinks develop from the self-energy can be seen by referring to the right column of Fig.~\ref{md2d5}.  The various waterfalls occur when the spectra are broadened by peaks in $\Sigma^{\prime\prime}$ (red dashed lines).  These peaks fall around 200-400 meV, depending on the band-structure. As pointed out in connection with Fig.~\ref{Intermediate}(c) above, when the real part of the self-energy $\Sigma'$ (blue solid lines) is positive, it pushes the states toward $E_F$, increasing the low energy effective mass, while negative $\Sigma^{\prime}$ pushes weight away from $E_F$, into the incoherent spectral weight. Note that all the spin and charge components of $\Sigma^{\prime}$ are linear in the low-energy region as a result of linear dispersion of the fluctuation spectrum along $\Gamma\rightarrow(\pi,0)$ and $\Gamma\rightarrow(\pi/2,\pi/2)$ (two left columns), yielding a total dispersion renormalization of the order of 2-3, consistent with experiments.  From the excellent agreement between theory and experiment with respect to both the shape and magnitude of the HEK, we conclude that near optimal doping the HEK arises primarily 
from the paramagnon branch around $q=(\pi,0)$ of the particle-hole continuum as discussed in connection with Fig.~\ref{chiop}. 
As discussed in Section~\ref{S:Boson}, these paramagnons can be directly seen in neutron, RIXS, and Raman scattering.

DMFT has also been used to examine kinks in electronic bands\cite{DMFTcuprate}.  However, 
dispersion of the paramagnon branch noted above suggests a strong momentum transfer mechanism in dynamical fluctuations, which would not be captured by the DMFT. 


\subsubsection{ARPES Matrix Element effects on the high energy kink}

\begin{figure}[htp]
\centering
\rotatebox{270}{\scalebox{0.58}{\includegraphics{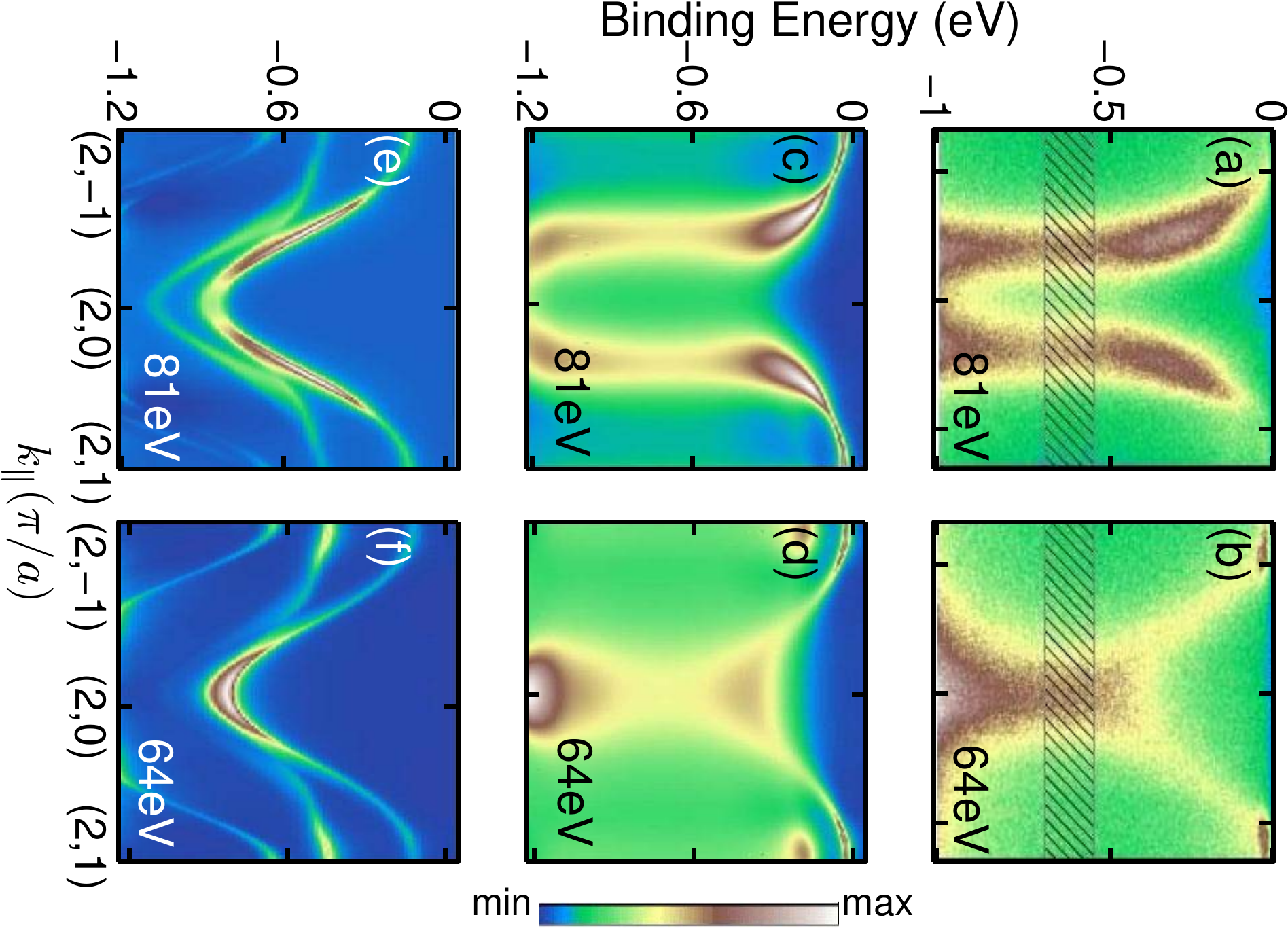}}}
\caption{
{ ARPES spectra in Bi2212.}
(a)-(b): Experimental photoemission spectra at photon energies of 81 eV and 64 eV\cite{Int2}. (c)-(d): Corresponding theoretical spectra based on first-principle one-step calculations in which the self-energy correction is {\it included}. (e)-(f): Computations where the self-energy correction is {\it excluded} to highlight the key role of self-energy corrections in capturing the experimentally observed waterfall effect. [From Ref.~\onlinecite{waterfall}.]
}\label{Bi}
\end{figure}

In Bi2212, the waterfall feature observed in ARPES changes its spectral shape substantially at different photon energies.\cite{Int,Zhou,DanD,Int2}  To understand this effect, in Ref. \onlinecite{waterfall} we incorporated our QP-GW self-energy into the first-principles one-step ARPES methodology \cite{ABC1,arpesab,LDA,me1,me2,me3,ABC2,ABC3, me4}.
Figure~\ref{Bi} compares the theoretical predictions with the corresponding experimental spectra, and shows a good accord \cite{waterfall}.

It can be useful to consider photoemission intensities within the framework of a tight-binding model in parallel with first-principles modeling. The ARPES matrix element $M$ in a tight-binding scheme can be written in terms of a structure factor.~\onlinecite{waterfall} For a bilayer system, the relevant structure factor involves the separation of the two bilayers along the c-axis:
\begin{equation}
M^{\pm}({\bf k}_f)=M^0({\bf k}_f^{\parallel})
[1\pm e^{-ik_f^\perp d}], \label{eq:6}
\end{equation}
where the + sign refers to the anti-bonding band and the - sign to the bonding
band, $M^0({\bf k}_f^{\parallel})$ is the matrix element of a single layer,  independent of $k_f^\perp$, and $d$ denotes the separation of the CuO$_2$ layers
in a bilayer\cite{foot6,dimpling}. The key feature of Eq.~\ref{eq:6} is the interference term in brackets, where $k_f^{\perp}$ depends on the photon energy\cite{waterfall}. Since the two bilayer terms in Eq.~\ref{eq:6} are out of phase [note $\pm$-sign], whenever $k_f^{\perp}d$ changes by $\pi$, the spectrum switches from the odd to the even bilayer, a change that can be induced 
via the photon frequency $\nu$. This behavior is indeed seen in panels (c)
and (d) of Fig.~\ref{Bi}, and is reproduced by the model calculations using Eq.~\ref{eq:6}. In particular, at $64$ eV in panel (d), the anti-bonding band gets highlighted resulting in a Y-shaped spectrum with a tail extending to high energies. In contrast, in panel (c) at $81$ eV, the bonding band dominates and the spectral shape reverts to that of a waterfall with a double tail. Due to the frequency dependence of $k_f^{\perp}d$, Eq.~\ref{eq:6} predicts that the spectrum would oscillate between the anti-bonding and bonding bands as a function of photon energy, consistent with experiments.\cite{waterfall}

\subsubsection{Low energy kink}\label{S:LEK}

\begin{figure}[top]
\centering
\rotatebox{0}{\scalebox{0.58}{\includegraphics{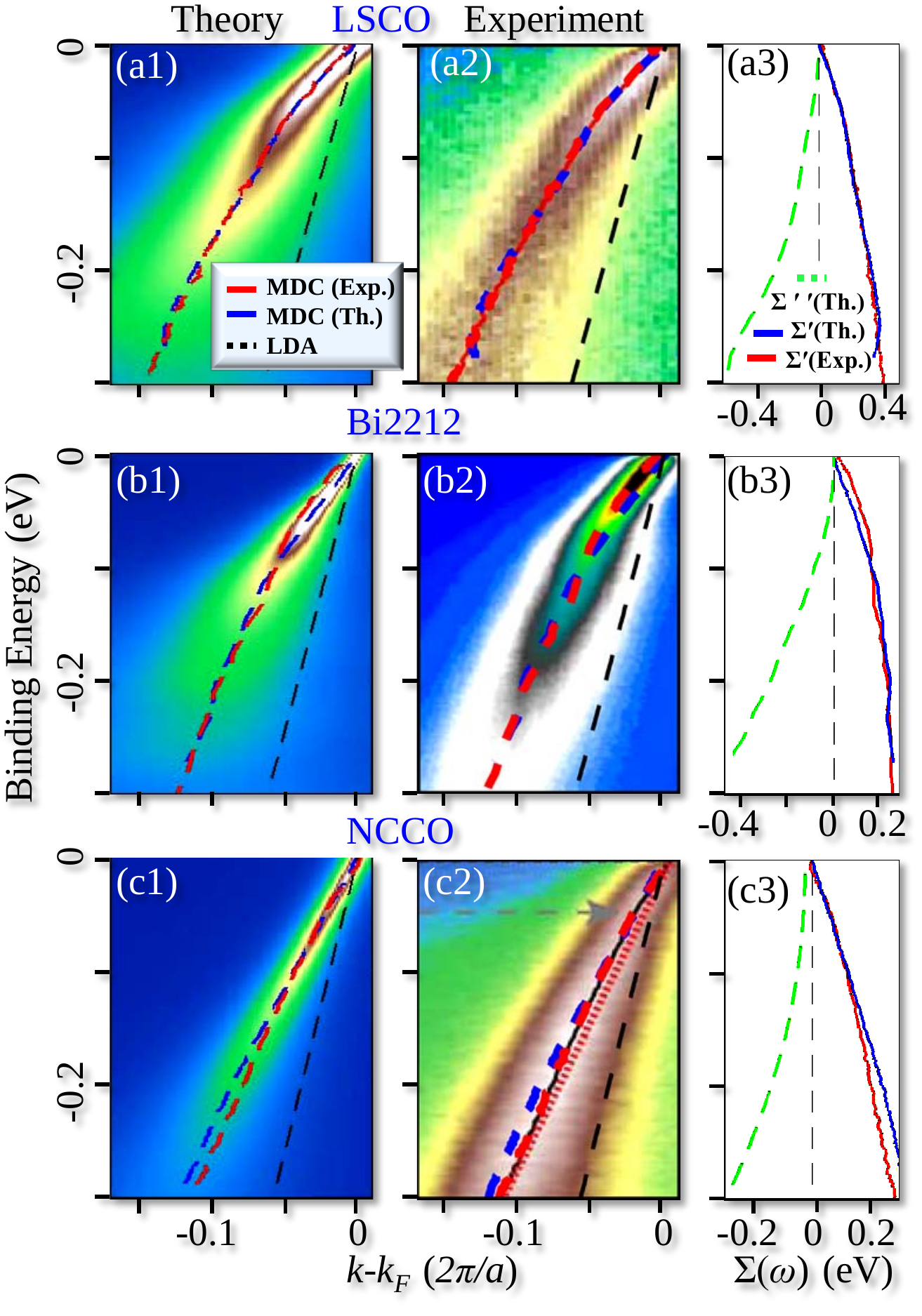}}}
\caption{{ ARPES spectra of LEK.} (a1) Theoretical and (a2) experimental\cite{kordyuk} single-particle dispersions in the energy range of the LEK for LSCO. Blue and red lines give MDC peak positions, while the non-interacting LDA dispersion is plotted as a black dashed line. (a3) Real part of the self-energy, defined as the difference between MDC peaks and the LDA dispersion. Dashed green line gives the computed imaginary part of the self-energy. Similar results are shown for Bi2212 [Ref.~\protect\onlinecite{WZhang}] in panels (b1)-(b3) and for NCCO [Ref.~\protect\onlinecite{schmitt}] in panels (c1)-(c3). [From Ref.~\onlinecite{tanmoyLEK}.]}
\label{alp2F}
\end{figure}

Experimental observation of kinks in the quasiparticle dispersion near $50-70$meV in the cuprates suggested that the bosonic excitations responsible for these kinks might also mediate electron pairing in these materials. The nature of the associated bosons (or even whether a boson model is appropriate) has been a subject of intense scrutiny, with the leading candidates being  phonons\cite{lanzara,lee} and spin fluctuations\cite{johnsonx,Niestemski}. 
An important property of the LEK is that it has a strong temperature dependence, being intense in the superconducting phase and much weaker above $T_c$, losing intensity without significant shift in energy as $T$ increases.  This $T$-dependence is suggestive of a close connection with magnetic resonance scattering as discussed in Section~\ref{S:ElSpec2}.3 below.  Here we use the theory of the magnetic resonance peak discussed there to show that the salient experimentally observed features of the LEK can be explained in terms of electronic bosons. Figure~\ref{alp2F} compares our QP-GW based spectral weights (frames a1-c1) with the corresponding experimental ARPES data (frames a2-c2) for three different cuprates near optimal doping.\cite{kordyuk,WZhang,schmitt}

Whereas for HEKs, Fig.~\ref{md2d5} in Section~\ref{S:ElSpec}.3.1, the self-energy has a peak in $\Sigma^{\prime\prime}$ accompanied with a characteristic change in sign in $\Sigma'$, the origin of the LEK is different.  At an LEK, the transverse spin $\Sigma^{\prime}$ shows a break in slope with no significant structure in $\Sigma''$. 
The reason is that in this low-energy region, the self-energy is controlled by the linear dispersion of the magnetic scattering near $(\pi,\pi)$, and the break in slope occurs at the energy of the magnetic resonance peak.

Fig.~\ref{alp2F} presents a quantitative comparison of our calculated LEKs with the corresponding experimental results\cite{kordyuk,WZhang,schmitt} in the nodal [$\Gamma\rightarrow (\pi,\pi)$] direction for three different materials near optimal doping. For ease of comparison, the dashed lines give dispersions defined as peak positions in the MDCs.  Our theory (left column) is seen to reproduce the experimental behavior (central column) reasonably well both in shape of the dispersion and in the associated spectral weight. For single layer systems, the LEK is around 70meV in LSCO, but 50meV in NCCO both in theory and experiment, while in Bi2212 our theory (neglecting bilayer splitting) finds a larger value of the kink energy around 100meV whereas the experimental data show a kink near 70 meV. 
Since the strength of this mode is weak compared to the HEK, it seems unlikely that the LEK is by itself responsible for the intermediate coupling strength of cuprates.

Experimental (red lines) and theoretical (blue lines) results for $\Sigma^{\prime}$ are compared in the rightmost column of Fig.~\ref{alp2F}. Theoretical $\Sigma^{\prime}$ here is defined as the difference between the MDC peaks and the bare LDA dispersion (black dashed lines in the left/central columns). The ARPES-derived $\Sigma'$ often shows a more pronounced peak at the LEK, rather than a break in slope.  This is due partly to the assumed form of the bare dispersion, which is taken as a straight line from $E_F$ to the dressed band at a high energy usually chosen at -300meV, rather than the correct LDA band. Ref.~\onlinecite{tanmoyLEK} further explores the temperature and doping dependence of the LEK. Overall, the calculations are in very good agreement with the ARPES data, with some discrepancies in Bi2212, which may be related to the neglect of bilayer splitting in the calculations, or to possible additional contributions associated with phonons.

\subsubsection{Lower-energy kinks}

At even lower energies, additional kinks have been reported, especially a mode near 15~meV\cite{VLEK1,VLEK2,VLEK3,VLEK4,VLEK5}.  These features are not reproduced in our calculations, suggesting that they may be signatures of electron-phonon coupling. In addition, phonon fine structure has been reported at energies of 40-46 meV and 58-63 meV, and possibly at 23-29 meV and 75-85 meV in LSCO\cite{VLEK7}, suggesting contributions from multi-phonon modes to the 70 meV nodal kink\cite{VLEK6}.

%


\subsection{STM}

\subsubsection{Matrix element effects in STM/STS}

Modeling scanning tunneling microscopy/spectroscopy (STM/STS) spectra is quite challenging. In order to trace the path of electrons from the active CuO$_2$ plane to the STM tip requires modeling of the hopping path from the cuprate to the surface layer in the presence of overlayers. For this purpose, we have developed a multiband tight-binding methodology based on the LDA band structures including effects of SDW-SC order.\cite{nieminenPRB2,Jouko,nieminenPRB,nieminenPRB3,nieminenPRB4}  The Hamiltonian is a multiband generalization of Eq.~\ref{HFHam}. The tensor (Nambu-Gorkov) Green's function ${\cal G}$ is found from Dyson's
equation:\cite{Fetter}
\begin{equation}
 {\cal G}^{-1} =  {\cal G}^{0-1} -  {\cal V},
\label{dyson1}
\end{equation}
where
\begin{displaymath}
  {\cal G} =
\left(
   \begin{array}{cccc}
G_{e \uparrow}& 0& 0&  F_{\uparrow \downarrow}\\
0&G_{e \downarrow}&  F_{\downarrow \uparrow} & 0\\
0& F^{\dagger}_{\downarrow \uparrow}& G_{h \uparrow}&0\\
F^{\dagger}_{\uparrow \downarrow}& 0 &0&  G_{h \downarrow}
   \end{array}
\right)~\textrm{with}~
{\boldmath c}_{\alpha} =
\left(
  \begin{array}{c}
c_{\alpha \uparrow}\\
c_{\alpha \downarrow}\\
c^{\dagger}_{\alpha \uparrow}\\
c^{\dagger}_{\alpha \downarrow}
  \end{array}
\right)
\end{displaymath}
$G_{e}$ and $G_{h}$ denote the tensor Green's functions for electrons and holes, respectively\cite{foot11}, $F$ is the anomalous Green's function tensor, and matrix elements of the operator ${\cal V}$ represent interaction terms in the Hamiltonian, which can include self-energy corrections to model effects of phonons and impurity scattering, in addition to those of electron-electron interactions. 

Within the Todorov-Pendry approach\cite{Todorov}, the differential conductance $\sigma$ between orbitals of the tip ($t,t'$) and the sample ($s,s'$) can be written in the form\cite{Jouko,nieminenPRB}
\begin{equation}
\sigma = \frac{dI}{dV} = \frac{2 \pi e^2 }{ \hbar} \sum_{t t' s s'}
\rho_{tt'}(E_F)t_{t's} \rho_{ss'}^{}(E_F+eV)t_{s't}^{\dagger},
\label{conductance}
\end{equation}
where the $t_{ts}$ are hopping parameters and the density matrix
\begin{eqnarray}
\rho_{s s'} = -\frac{1}{\pi}Im[G_{s s'}^{+}]
 = \frac{1}{2\pi i} \left( G^{-}_{s s'}
- G^{+}_ {s s'} \right) 
\nonumber\\
= -\frac{1}{\pi} \sum_{\alpha}
(G^{+}_{s\alpha}\Sigma{''}_{\alpha}G^{-}_{\alpha
   s'}+F^{+}_{s\alpha}\Sigma{''}_{\alpha}F^{-}_{\alpha s'})
\label{spectralfunctiona}
\end{eqnarray}
is expressed in terms of the retarded and advanced electron Green's functions or
propagators. 

STM/STS spectra are strongly influenced by the tunneling matrix element. For example, when the tip is above a Bi atom, it turns out that there is no direct signal from the $d_{x^2-y^2}$ orbital on the Cu atom lying below the Bi atom due to the symmetries of the orbitals involved. Instead, the signal comes indirectly from the neighboring Cu's.\cite{Jouko,MBZ}  In fact, tunneling matrix element from $d_{x^2-y^2}$ orbitals is much weaker than from $d_{z^2}$ orbitals, which is not symmetry-forbidden.\cite{Jouko}

\subsubsection{Doping-dependent dispersion in Bi2212}

In STM/STS spectra, cuprates appear remarkably inhomogeneous on exceedingly fine length scales: electronic density, pairing gap, and even the local FS, all differ on patches of $\sim 30\AA$ diameter.  These patches appear to be pinned by local doping\cite{McElroy}, especially by apical oxygen vacancies\cite{Jenny}. The observed concentration of impurities is consistent with the bulk average oxygen nonstoichiometry, suggesting that similar patches are responsible for the short correlation lengths in bulk Bi2212\cite{Abbamaybe}. In a recent study, we analyzed the doping dependence of STM/STS spectra of Bi2212 by assuming that the patches represent regions of local doping\cite{nieminenPRB2}.

\begin{figure}
\centering
  \includegraphics[width=0.6\textwidth,angle=0]{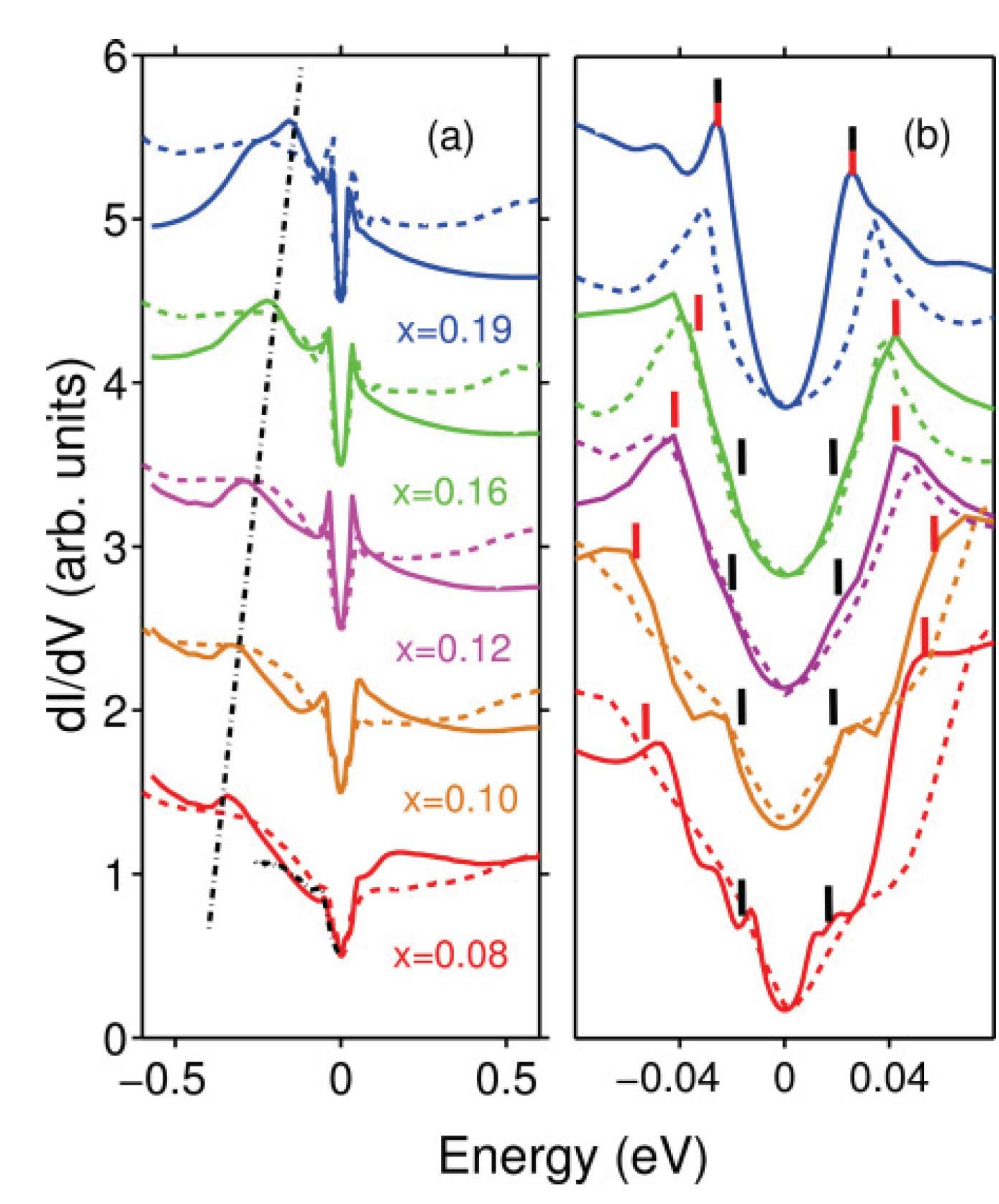}
\caption{ {Doping-dependent tunneling spectra of Bi2212.} (a) Theoretical (solid) and experimental (dashed, Ref. \onlinecite{McElroy}) SC-state spectra
over the hole doping range from $x=0.19$ at the top to $x=0.08$ at the bottom of the figure.\cite{footB1} For comparison, the experimental ARPES spectrum\cite{ARPES} at $x=0.08$ (black dashed curve) is shown. The dot-dashed line is drawn through the hump feature in the computed spectra to guide the eye. (b) Blowup of theoretical spectra near $E_F$ is compared with the
corresponding data of Ref.~\onlinecite{Lawler}. 
[From Ref.~\onlinecite{nieminenPRB2}.]}
\label{phasediag0}
\end{figure}

Figure~\ref{phasediag0} compares the calculated STS spectra with the corresponding experiments\cite{McElroy,Lawler,footB1} in the SC-state, including 
some ARPES data\cite{ARPES}. Similar comparisons have also been made for the normal state STS spectra.\cite{nieminenPRB2} At low energies, there are two gap features in the SC state spectra, Fig.~\ref{phasediag0}(b)\cite{Lawler}, which display a doping dependence consistent with two-gap physics\cite{huefner} seen in Raman scattering\cite{Tacon,Tacon2}, ARPES\cite{Lee,Tanaka}, and recent STM studies\cite{JCD,pushp}. One gap component is a nodal pairing gap $\Delta_n$ with a parabolic doping dependence reminiscent of the superconducting $T_c$, while the second component is an antinodal gap $\Delta^*$ which increases roughly linearly with underdoping, similar to the pseudogap onset temperature $T^*$. The gap features are absent in normal state spectra taken at 93K, above the highest superconducting transition temperature.\cite{Yazdani,footUD}  All these low energy features are seen to be well reproduced in the theoretical spectra of Fig.~\ref{phasediag0}.  

\begin{figure}
\centering
  \includegraphics[width=1.0\textwidth,angle=0]{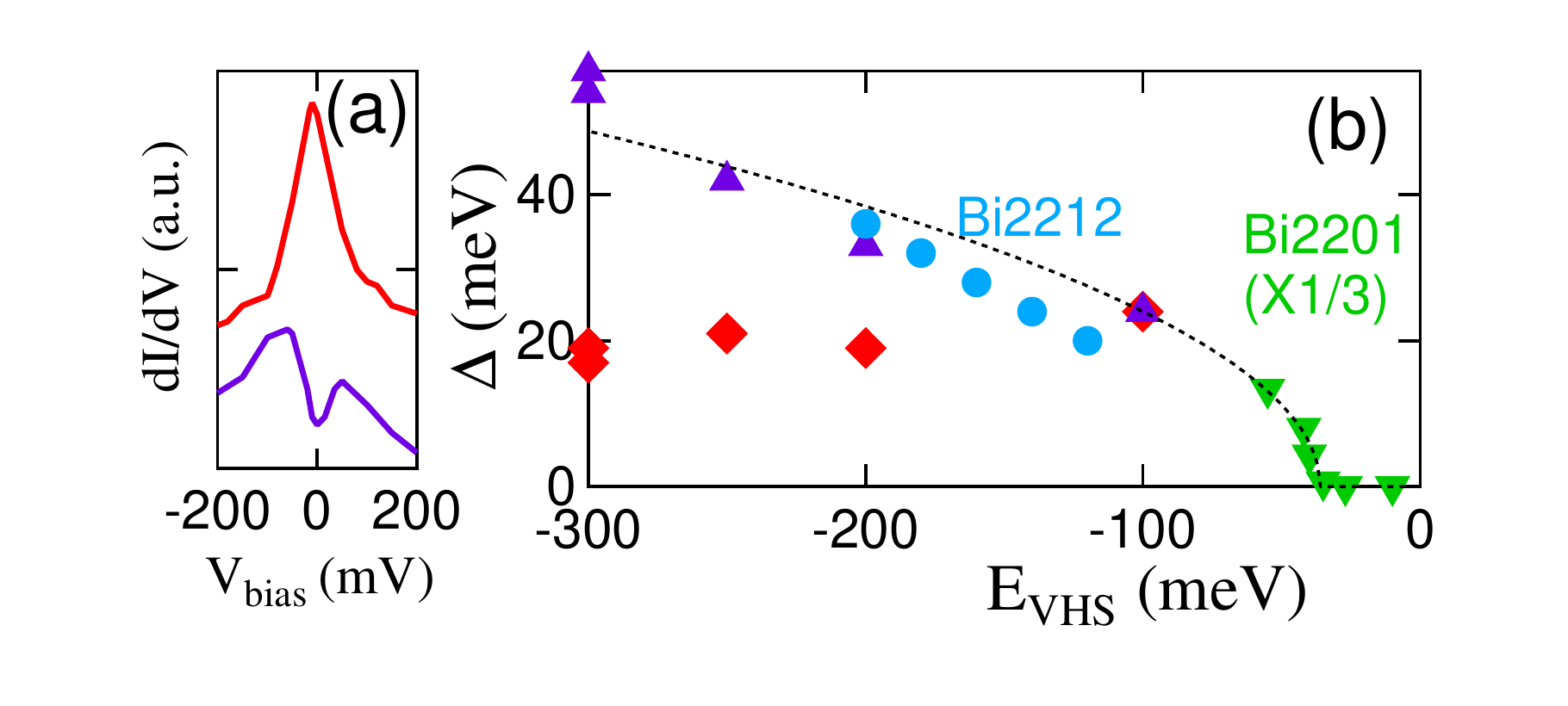}
\caption{ (a) Tunneling $dI/dV$ spectra for Bi2201 corresponding to $E_{VHS}$ = -10 meV (upper, red line) 
and -56~meV (lower, violet line, offset for clarity), after Ref.~\onlinecite{VHS3}. (b)
Plots of $\Delta$ vs $E_{VHS}$ for Bi2212 (red diamonds and violet triangles from Ref.~\onlinecite{Lawler} and blue 
circles from Ref.~\onlinecite{Yazdani}) and Bi2201 (inverted green triangles from Ref.~\onlinecite{VHS3}). Dotted line indicates the square-root dependence of $\Delta$ on $E_{VHS}+36$~meV, Eq.~\ref{eq:VHSeq}. [From Ref.~\onlinecite{nieminenPRB2}.]
}\label{Bi2201}
\end{figure}

STS spectra of Fig.~\ref{phasediag0}(a) also show a prominent doping dependent peak at negative energies, which evolves from $\sim$-500 to -100~meV as doping increases. We identify this peak as the VHS, which is enhanced by the AFM order. More accurately, it is the VHS of the lower magnetic band, which evolves into the conventional VHS as the AFM-gap collapses. A similar VHS feature is found in Bi$_2$Sr$_2$CuO$_6$ (Bi2201), Fig.~\ref{Bi2201}(a).\cite{VHS3}  This negative energy hump bears a striking relationship to superconductivity: as the hump gets closer to $E_F$, the superconducting gap decreases monotonically,  Fig.~\ref{Bi2201}(b). This is opposite to the behavior expected in the BCS theory, where a peak in the DOS would enhance $T_c$.\cite{Yazdani} Scaling of this feature (dotted curve) is captured by the relation
\begin{equation}
  \label{eq:VHSeq}
  \Delta^2+\Delta_0^2 = A|E_{VHS}|,
\end{equation}
with empirical parameters $\Delta_0\sim$18~meV and $A\sim\Delta_0/2$, suggesting a quantum critical point (QCP) associated with the VHS.\cite{JarVHS,MGu2}  
The anti-correlation of the VHS peak position with superconductivity could arise because the VHS can drive a competing ferromagnetic instability.\cite{TallStorey,markiesc}

\subsubsection{Gap Collapse: Two Topological Transitions}

\begin{figure}
\centering
 \includegraphics[width=1.0\textwidth,angle=0]{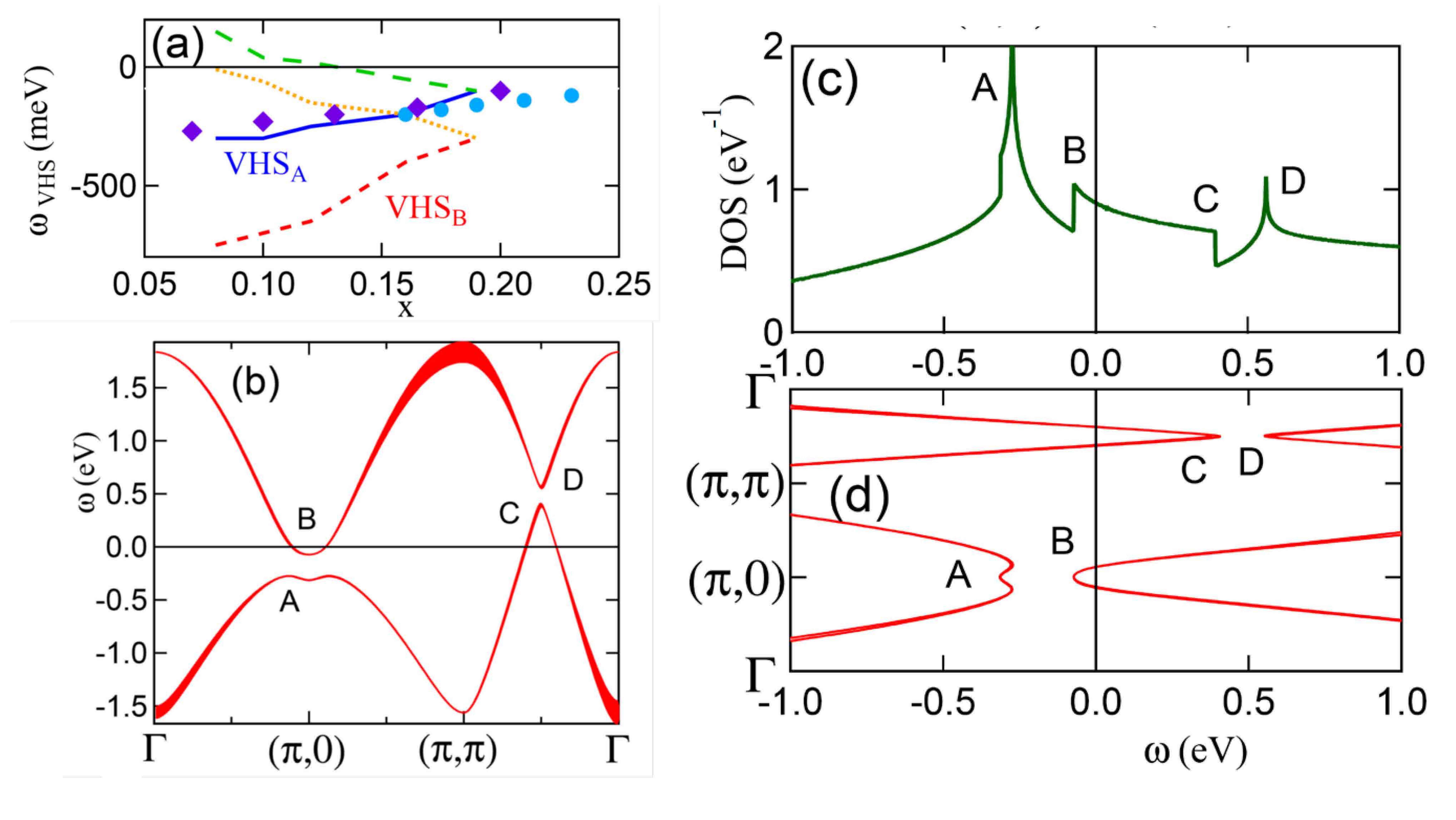}
 \caption{(a) Doping dependence of VHS-related features in the SDW phase. The filled blue circles [Ref.~\onlinecite{Yazdani}] and violet diamonds
[Ref.~\onlinecite{McElroy}] are the hump features derived from the STS spectra of Bi2212. These are compared to the calculated doping dependence of the bonding [B] and antibonding [A] VHS (red short-dashed and blue solid lines, respectively) in Bi2212, based on the analysis of Ref.~\cite{nieminenPRB2}. (b) Calculated dispersion at $x=0.15$, with feature A, the VHS of the lower magnetic band, corresponding to the blue solid line [or red short-dashed line, for the bonding band] in (a), and feature B, the bottom of the upper magnetic band, corresponding to the green long-dashed line [orange dotted line] in (a). The thickness of the lines represents the spectral weight due to the SDW structure factor. (c) Corresponding DOS, showing features derived from A, B, C, and D in (b). (d) Blow-up of the near-$E_F$ dispersion. [From Ref.~\onlinecite{nieminenPRB2}.]}

\label{phasediagN}
\end{figure}

Collapse of the SDW gap with doping is shown in Fig.~\ref{phasediagN}. Dispersions of the theoretical and experimental VHS features are compared in Fig.~\ref{phasediagN}(a). The model includes bilayer splitting. The solid blue [short-dashed red] curve represents the VHS peak of the antibonding
[bonding] band.  We see that the experimental hump agrees with the doping dependence of the antibonding VHS, both in the normal (blue filled circles)\cite{Yazdani} and  in the superconducting state (violet filled diamonds)\cite{McElroy}.  When the SDW pseudogap turns on, a second feature appears [green long-dashed and orange dotted lines for the antibonding and bonding bands, respectively].  To better understand the doping evolution of this feature, labeled $B$, frames (b) and (c) show the dispersion and the near-$E_F$ DOS of one band in the SDW state. From Fig.~\ref{phasediagN}(b), it can be seen that the SDW gap has a strong $k$-dependence, splitting the band into UMBs and LMBs.  Features $A$ and $D$ are the VHSs of
the LMB and UMB, respectively, while feature $B$ is the bottom of the
UMB, and $C$ is correlated with the top of the LMB.  Note the characteristic form of the DOS associated with each feature in Fig.~\ref{phasediagN}(c).

Figure~\ref{phasediagN}(b) should be compared with the equivalent figure for electron doping, Fig.~ \ref{md2d4}(e).  It can be seen that in both cases there are {\it two pockets} crossing $E_F$, an electron-pocket near $(\pi,0)$ and a hole-pocket near $(\pi/2,\pi/2)$.  Thus, for hole doping in Bi2212, the FS must also undergo two topological transitions (TTs), starting with just a hole-pocket near half-filling, then developing an electron-pocket when the upper magnetic band crosses $E_F$, and finally developing a single large FS as the gap collapses and the electron- and hole-pockets merge.  [This process will, of course, repeat itself for each of the bilayer split bands.]  At the second TT, as the SDW pseudogap shrinks to zero, features $A$ and $B$ as well as $C$ and $D$ merge concomitantly.  This is the reason that there is no near-$E_F$ feature in the three normal state spectra at higher dopings.\cite{nieminenPRB2} At lower dopings, a feature corresponding to $B$ appears in the low-T theoretical curves, but it seems to be absent in the 93K experimental data, suggesting a $T$-dependent pseudogap closing. Note that at lower doping, feature $B$ moves closer to $E_F$, and should persist
to higher temperatures since the pseudogap is larger.  Feature $B$ crosses $E_F$ at the first TT, near $x=0.08$ or 0.12 for the two bilayers.

By analyzing the anomalous Green's function, we find that the gaps on both pockets are superconducting. The $(\pi ,0)$ peak is the total gap $\sqrt{\Delta_{SDW}^2 + \Delta_{SC}^2}$, which predominantly represents the SDW order with a Bilbro-McMillan-like dressing by superconductivity\cite{bilbro}, while the $(\pi /2,\pi /2)$ peak is a pure SC gap. With decreasing doping, the gap between the LMB and UMB increases, leading to a monotonically increasing antinodal gap energy.  In contrast, the FS on the LMB near $(\pi /2,\pi /2)$ shrinks monotonically since its area is proportional to $x$. For a $d$-wave gap, which vanishes at $(\pi /2,\pi /2)$, its maximum value must ultimately
decrease with underdoping, thereby explaining the opposite doping dependencies of the two gaps. Notably, a similar double transition has now been found in Bi2201, although so far there is no clear evidence for two separate pockets.\cite{Jclaws} Our calculations of Fig.~\ref{phasediagN}(a) also provide insight into the transition to a large-gap insulator at half-filling. The AF pseudogap is essentially given by the separation between the VHS and the bottom of the upper magnetic band, i.e. features A and B in Figs.~\ref{phasediagN}(b) and \ref{phasediagN}(c). From Fig.~\ref{phasediagN}(a), this is seen to be approximately 0.3~eV at $x=0.08$, and the route to a 2~eV gap at half-filling is clear.


\subsection{X-ray absorption}
\begin{figure}
\centering
  \includegraphics[width=0.95\textwidth,angle=0]{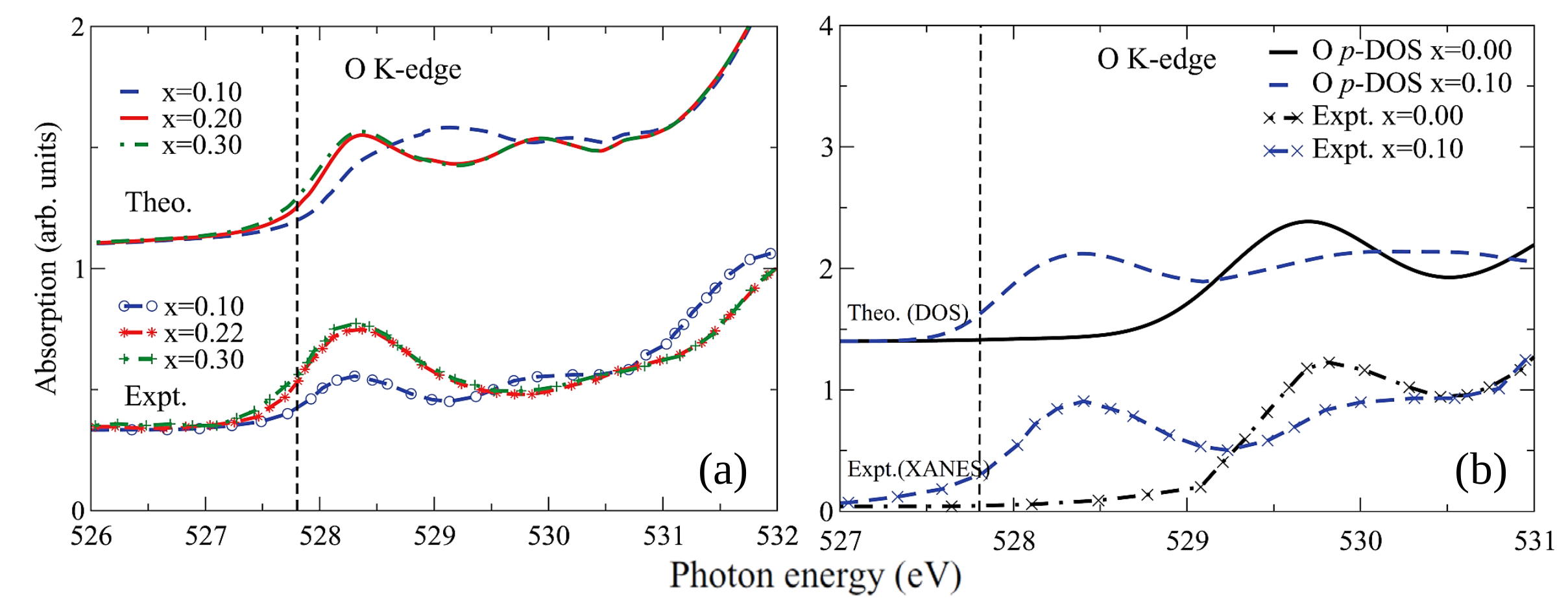}
\caption{(a) Doping-dependent theoretical (upper curves) and experimental\cite{XANESexp} (lower curves) O K-edge XANES spectra for LSCO. Computations are performed using DFT-based FEFF code in which realistic QP-GW self-energy corrections were incorporated. Results for different dopings $x$ are shown by lines of various colors (see legend). The vertical line marks the edge energy. (b) Corresponding experimental data\cite{XANESexpx0} (lower set of broken curves) are compared with the theoretical single-band p-DOS (upper curves) for O$_{pl}$ for $x$ = 0 and $x$ = 0.10. [From Ref.~\cite{Towfiq}.]}
\label{F:XANES}
\end{figure}

In order to understand how the Mott insulating gap at half-filling evolves into the SC state, we need to probe the spectral weight transfer from the Mott-Hubbard band to the in-gap states near $E_F$. In electron-doped cuprates, the Mott gap and the associated lower Hubbard band can be directly probed by photoemission spectroscopy as discussed earlier. For hole-doping, on the other hand, this gap lies above $E_F$, so that techniques sensitive to empty states within a few eV above $E_F$ must be deployed. Among the light scattering techniques, x-ray absorption near edge spectroscopy (XANES) can directly probe the single-particle DOS of empty states above $E_F$ via excitations from core levels.\cite{XANESexp,XANESexpx0,Towfiq} Accordingly, we studied the XANES spectrum of LSCO as an exemplar hole-doped cuprate in order to compare and contrast our QP-GW predictions with available experimental data.\cite{XANESexp,XANESexpx0} The analysis is based on using a real-space Green's function approach as implemented in the FEFF9 code\cite{FEFF1,FEFF2}, which was modified to include self-energy corrections to the electronic states near $E_F$.\cite{Towfiq} FEFF simulations compute the DFT based real-space Green's function by incorporating multiple scattering processes due to core-hole excitations.\cite{FEFF1,FEFF2} The QP-GW generated self-energy was included via Dyson's formula to dress the low-energy CuO$_2$ bands. Given the fact that O $p$-orbitals are strongly hybridized with Cu $d$-bands, the O K-edge spectrum is the most relevant data to access Mott physics in LSCO.

Fig.~\ref{F:XANES} compares experimental results\cite{XANESexp} with the corresponding theoretical O K-edge XANES spectra in overdoped LSCO. All spectra are calculated in the paramagnetic phase except that an insulating phase is used at half-filling. Focusing first on the overdoped regime for hole doping between $x$ = 0.20 and 0.30, both theory and experiment display a small, systematic shift of the edge to lower energies with increasing doping. The low-energy peak at 528.5 eV in Fig.~\ref{F:XANES}(a) shows the experimental and theoretical edges, both of which are seen to display a similar shift of $E_F$ with doping. Otherwise, the XANES spectra undergo relatively little change in the overdoped system. As the doping is reduced, intensity of the 528.5 eV peak decreases while a new peak appears near 530 eV and rapidly grows with underdoping until at $x$ = 0, the 528.5 eV peak disappears, see Fig.~\ref{F:XANES}(b). The remaining 530 eV peak represents the upper Hubbard band, and its shift in energy from $E_F$ is consistent with optical measurements. Turning to the x = 0.10 spectra in Fig.~\ref{F:XANES}(a), we see that now theory differs substantially from experimental results.  This is expected, since we have not included the effect of the pseudogap opening in the calculations.  Although theory correctly reproduces the reduced intensity of the 528.5 eV peak, it does not show the observed enhanced intensity of the upper peak at 530 eV. Instead, the spectral weight is shifted halfway between the lower and upper peaks. In Fig.~\ref{F:XANES}(b), experimental results indicate the opening of a gap in the spectrum which is not captured in our modeling. However, the experimental doping dependence\cite{XANESexpx0} can be reproduced in a simpler calculation in which the XANES spectrum is modeled via the empty DOS, including self-energy corrections with the magnetic gap. Fig.~\ref{F:XANES}(b) compares the experimental XANES spectrum with this approximation at $x$ = 0.10. The splitting of the spectrum into two peaks with the appropriate gap is well reproduced. We will see in Sections~\ref{S:ElSpec2}.1 and~\ref{S:ElSpec2}.2 that the same self-energy reproduces the doping dependent optical and RIXS gaps.

\subsection{Momentum density and Compton scattering }

\begin{figure}
  \includegraphics[width=1.00\textwidth,angle=0]{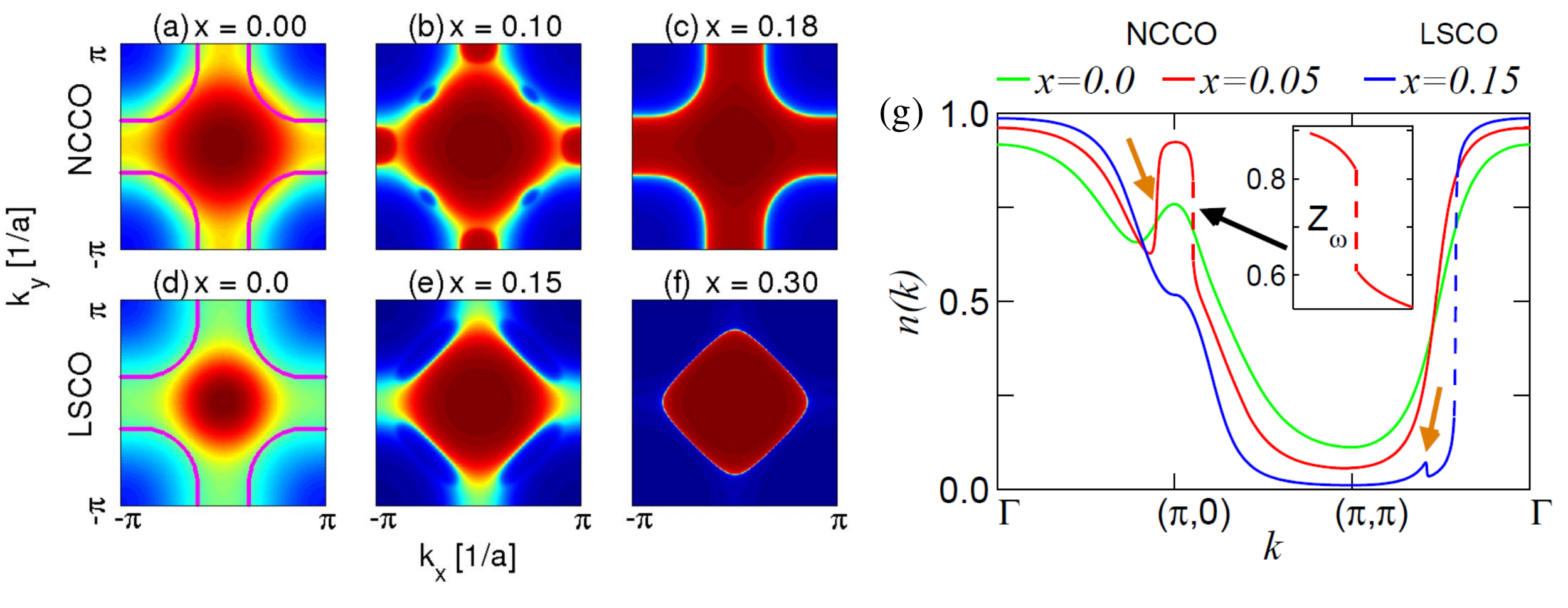}
\caption{Theoretical occupation density in momentum space, $n({\bf k})$, using the QP-GW scheme is shown for various dopings in NCCO in (a)-(c), and for LSCO in (d)-(f). (g) $n({\bf k})$ for NCCO and LSCO along the high-symmetry lines in the BZ at several representative dopings with dashed lines showing the discontinuous jump at the Fermi momenta (highlighted in inset). Gold arrows mark features from the shadow bands which cross $E_F$. [From Ref.~\cite{tanmoygoss}.]}
\label{Compton}
\end{figure}

The bulk FS of a system can be measured via the x-ray Compton scattering technique, which probes the momentum density of the correlated many-body ground state of the electronic system in the extended momentum space in ordered and disordered materials.\cite{Bansilbook, AB1,AB2}  It has been shown that by taking the difference between the momentum density $n({\bf p})$ at two different dopings, one can visualize the correlated wave function of the doped holes in momentum space.\cite{SakuraiCompton} From the 3D momentum density $n({\bf p})$, one can obtain the occupation density, $n({\bf k})$, of electronic states by folding $n({\bf p})$ at higher momenta into the first BZ.\cite{Bansilbook,AB3}  Distinction between the two densities $n({\bf k})$ and $n({\bf p})$ should be understood clearly. Here, ${\bf k}$ is the Bloch momentum which is restricted to the first BZ.  Magnitude of the FS discontinuity of $n({\bf k})$ provides a direct measure of the spectral weight renormalization factor $Z_{\omega}$ in a correlated  electronic system.\cite{AB1,AB3,PRTRVB,tanmoygoss,AB4,AB5} Note that the $l^{th}$ order moment $M_l({\bf k})$ of the spectral function is, $\int_{-\infty}^{\infty} \omega^l A({\bf k},\omega)f(\omega)$, where $A$ is the spectral function and $f$ is the Fermi function. In this section, we are considering the 0$^{th}$-order moment, which is the electron occupation density, while higher order moments are discussed in Sec.~\ref{S:NFL}.2 below. 

While $Z_{\omega}$ is controlled by the singularities at the Fermi momentum, which are characteristic of coherent gapless QP excitations, the spectral density $A$ involves both coherent (QP) and incoherent parts. In fact, the shape of $n({\bf k})$ is modified substantially by the incoherent part. Fig.~\ref{Compton} shows maps of $n({\bf k})$ as a function of doping for NCCO and LSCO. In the present QP-GW case, the combination of self-energy and SDW coherence factors leads to characteristic structures in $n({\bf k})$ at all dopings. At half-filling $n({\bf k})$ shows a maximum at the $\Gamma$-point, and away from $\Gamma$, it decreases gradually and smoothly from inside to outside the LDA-like FS [magenta solid line in Figs.~\ref{Compton}(a) and \ref{Compton}(d)]. As we dope the system with electrons, the spectral weight increases at the $\Gamma$-point and, in addition, $(\pi,0)$ and its equivalent points largely gain spectral weight due to the development of electron pockets in NCCO [Fig.~\ref{Compton}(b)]. With further increase of doping, the FS undergoes two topological transitions [see Sections~\ref{S:ElSpec}.1 and~\ref{S:ElSpec}.4 above], as reflected also in the momentum density calculations here. The first topological transition in $n({\bf k})$ occurs in NCCO when the LMB approaches $E_F$ and forms hole-like pockets at ($\pm\pi/2, \pm\pi/2$). At $x$ = 0.15, the hole pockets are fully formed and they as well as the electron pockets increase in size with further doping. At $x$ = 0.18 the electron and hole pockets merge at the hot-spot, the SDW gap collapses, and the full metal-like $n({\bf k})$ appears [second topological transition]. These results are consistent with the other observation of two topological FS transitions, discussed in Sec.~\ref{S:ElSpec}.1 above.  For hole doping, the FS topological transition is complimentary to the electron doped one in that the hole pocket appears first as shown in Fig.~\ref{Compton}(e) and above the QCP, the electron-like full FS appears in Fig.~\ref{Compton}(f). 

A more quantitative account of the effect of self-energy corrections on the residual coherent QP spectral weight is provided in Fig.~\ref{Compton}(g), which shows $n({\bf k})$ along high-symmetry lines for NCCO and LSCO. Some important effects of correlations on the insulating state should be noted here. At $x$ = 0, $n[{\bf k} = \Gamma]\approx 0.9$ and $n[{\bf k} = (\pi,\pi)]\approx 0.1$, implying that the self-energy redistributes the spectral weight from the filled states to the unfilled regions even in the insulating phase.  Starting from the resonating valence bond (RVB) limit and allowing some double occupancy, Paramekanti {\it et al},\cite{PRTRVB} generally find similar results.  At half-filling $n({\bf k})$  is a smoothly varying function throughout the BZ, due to the absence of gapless quasiparticles for both NCCO (green line) and LSCO (not shown).  As we increase electron doping, the spectral weight at $\Gamma$ [$(\pi,\pi)$] gradually increases [decreases] whereas the spectral weight increases rapidly at $(\pi,0)$ due to the development of electron pockets, while discontinuities in $n({\bf k})$  arise at the FS. In the underdoped region,  $n({\bf k})$  shows additional singularities along $\Gamma\rightarrow (\pi,0)$ and $(\pi/2, \pi/2)\rightarrow (\pi,\pi)$ due to the presence of shadow bands as marked by gold arrows. Since the shadow bands are usually weak in the cuprates, experimental data are available predominantly along the arcs - that is, along the antinodal direction for NCCO and nodal direction for LSCO. These are compared with our theory in Fig.~\ref{md1d} below.


\section{Two-particle spectroscopies}\label{S:ElSpec2}

\subsection{Optical absorption spectroscopy: Explaining the opposite doping dependences of the Mott gap and pseudogap}



\subsubsection{Optical spectra of electron and hole doped cuprates}

Most experiments on cuprates find that doping the insulating state leads to a gap collapse at a quantum critical point\cite{zxshen,armitage,Tacon}. However, the large $\sim$2~eV optical gap seen at half-filling actually shifts to higher energies as doping increases, and persists into the overdoped regime, suggesting that strong electron correlation continue to play an important role in the cuprates at all dopings.  The intensity of this high energy feature decreases systematically with doping,\cite{cooper,hwang,arima,uchida,onose,onoseprb}
as its spectral weight is transferred into the Drude peak and the mid-infrared (MIR) feature\cite{haule,chakraborty,mishchenko,vidmar} that shifts to lower energies with increasing doping. Modeling and understanding the optical spectra of the cuprates thus becomes of key importance in unraveling the routes by which the insulator becomes a superconductor.

The QP-GW model provides a viable explanation of the doping dependencies of the optical spectra. In our one-band model, the MIR gap is associated with a Slater-type gap near $E_F$ due to the presence of AFM order\cite{kusko}, which splits the CuO$_2$-band into UMB and LMB. On the other hand, the large optical gap is produced by correlation or fluctuation effects, involving transitions from the coherent states near the $E_F$ to higher energy incoherent states, which are separated by a spectral weight loss due to the HEK. Note that in a three-band model of the cuprates, the LMB has strong oxygen character and the UMB is mainly copper-like, making this an effective charge transfer (CT) gap.  At half filling the CT and magnetic gaps are indistinguishable, but they split at finite doping with the magnetic gap collapsing at a finite doping and a residual CT gap associated with incoherent spectral weight persisting at high energies.\cite{MBRIXS,foot7,ZR} [see Section~\ref{S:Ext_d}]

We compute the optical conductivity from a standard linear response theory in the presence of an AFM pseudogap including self-energy corrections.\cite{tanmoyop,tanmoysw} In the paramagnetic state, $\sigma(\omega)$ takes a simplified form for $k$-independent $\Sigma$:\cite{allen}
\begin{eqnarray}\label{sigma}
\sigma(\omega) = \frac{i\omega_p^2}{4\pi\omega}\int_{-\infty}^{\infty} d 
\omega^{\prime}\frac{f(\omega^{\prime})-f(\omega^{\prime}+\omega)}
{\omega+\Sigma^*(\omega^{\prime})-\Sigma (\omega^{\prime}+\omega)}.
\end{eqnarray}   

\begin{figure}[top]
\centering
\rotatebox{0}{\scalebox{0.15}{\includegraphics{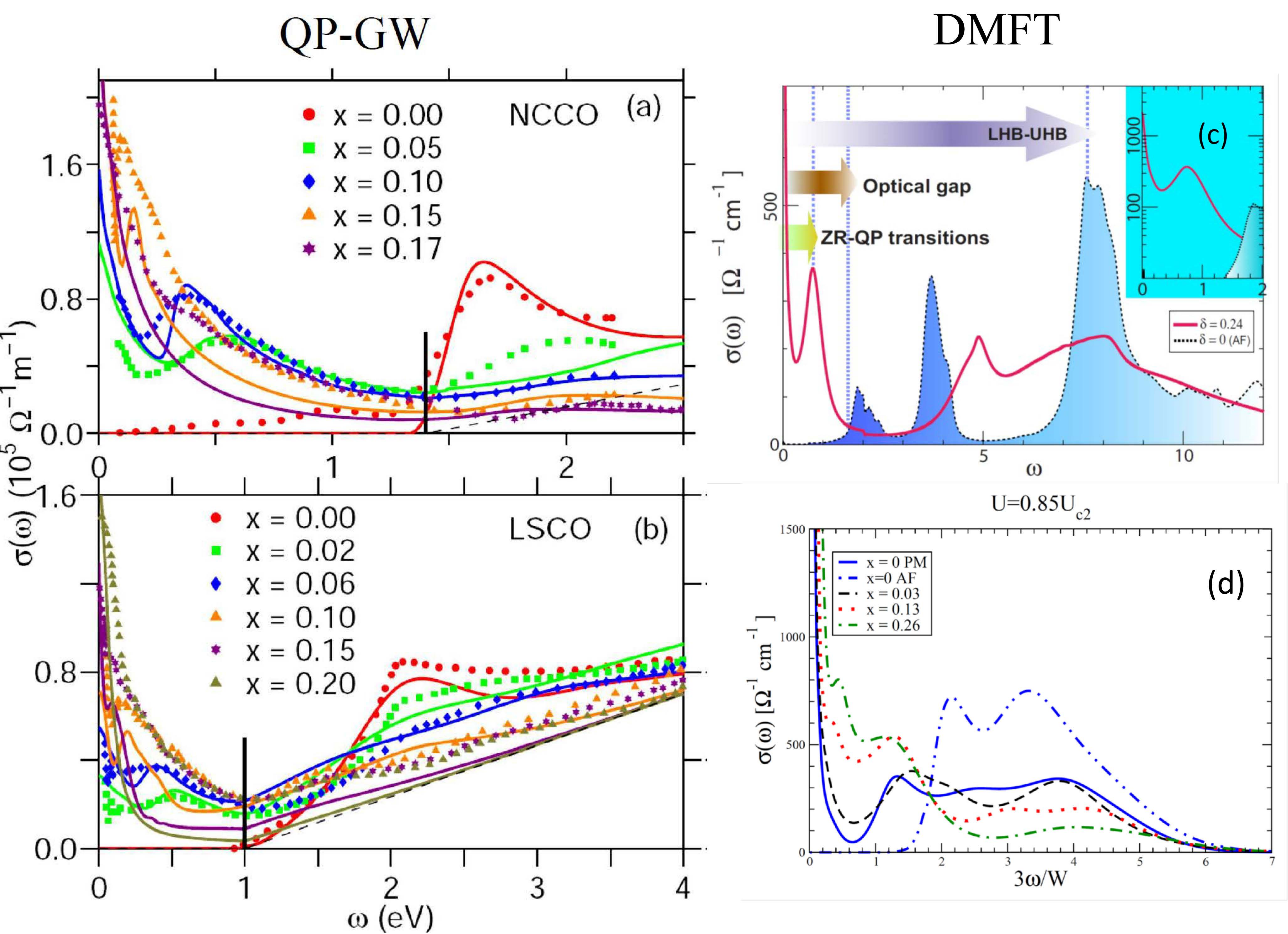}}}
\caption{ Comparison between calculated and experimental optical spectra in NCCO (a) and LSCO (b). Experimental results for NCCO are taken from
Refs.~\onlinecite{onoseprb} and ~\onlinecite{uchida}, and for LSCO from Ref.~\onlinecite{uchida}. Background subtractions [dashed line for $x=0$] are discussed in Ref.~\protect\onlinecite{tanmoyop}.  [From Ref.~\protect\onlinecite{tanmoyop}.] (c,d) DMFT based optical conductivity for LSCO from Refs.~\protect\onlinecite{WHK} and \protect\onlinecite{comanac}.  For the latter, $U_{c2}=4.4W$, bandwidth $W$ estimated as 3eV, and for $x>0$ only paramagnetic results are shown.}
\label{alp2F2}
\end{figure}
Extension to the AFM phase is discussed in Appendix~\ref{S:AG}. The computed optical conductivities $\sigma(\omega)$ for the electron-doped NCCO and hole-doped LSCO are compared with experiments\cite{onoseprb,uchida} and with DMFT results for LSCO\cite{WHK,comanac} over a wide doping range in Fig.~\ref{alp2F2}, and show that the computed evolution of $\sigma(\omega)$ is in very good accord with experiments. The theoretical spectra show an isosbetic or equal absorption point near 1.3 eV for NCCO and 1 eV for LSCO
[black vertical line], consistent with the experimental behavior. Within our model, the isosbetic point is a signature of strong magnon scattering,
closely related to the HEK seen in ARPES\cite{foot2b,moritz}. The doping evolution is completely different on the opposite sides of the isosbetic point. Above this point, the spectrum is dominated by a broad hump feature associated with the CT gap.  At half-filling, only this feature is present and the
calculated optical spectra show an insulating gap whose energy, structure, and intensity match remarkably well with measurements\cite{onoseprb,uchida}. As doping increases, the high energy peak shifts to higher energy and broadens, and its spectral weight is transferred to the Drude and MIR peaks. The MIR peak
shifts to lower energy with doping and gradually sharpens. Note that in both NCCO and LSCO at the highest doping, when the MIR peak collapses into the Drude peak, CT-gap features are still present in the spectrum. Our calculations also describe the anomalous $\sigma\sim 1/\omega$-dependence found in most cuprates and associated with magnetic scattering. A similar doping evolution is found in other cuprates\cite{cooper,hwang,arima,mcguire,uchida,onose,onoseprb,QT}

\begin{figure*}
\rotatebox{0}{\scalebox{0.25}{\includegraphics{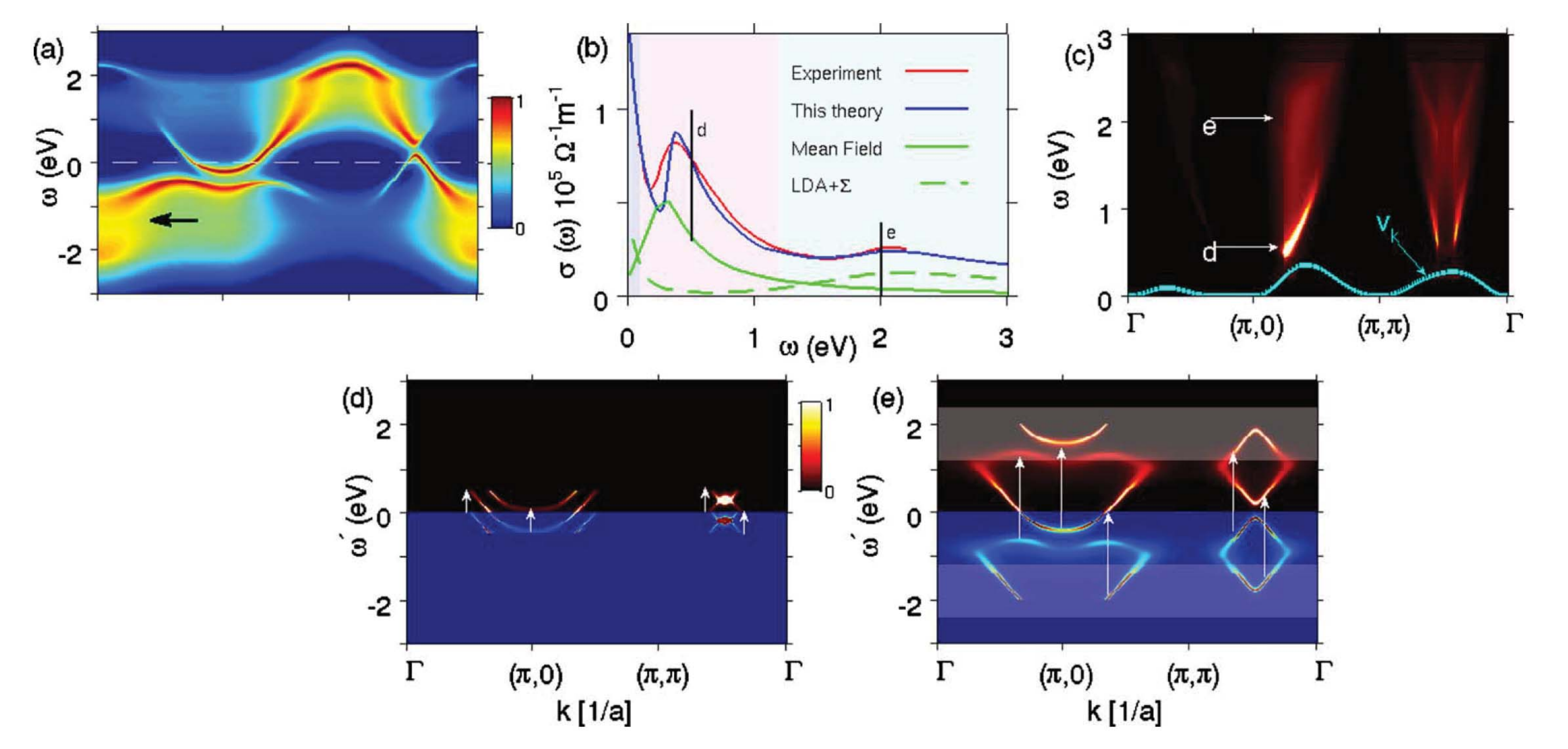}}}
\caption{ { Connection between optical spectra and spectral intensity maps in NCCO.} (a) Computed spectral intensity (log scale) as a function of $\omega$ along the high symmetry lines in the BZ at a representative doping of $x=0.10$. (b) Calculated optical spectra are compared with the corresponding experimental
results\cite{onoseprb}. Green solid line gives the spectrum calculated with the coherent bands only, i.e. without the self-energy corrections. Dashed green line is based on using LDA bands with self-energy corrections but without the pseudogap. The shaded regions of different colors (left to right) approximately mark Drude, MIR and high energy regions. Two vertical black lines indicate the photon energies at which frames (d) and (e) are calculated. (d) and (e) give
source (initial state) and sink (final state) maps corresponding to the optical transitions at fixed photon energy $\omega$ along the high symmetry directions. White vertical arrows indicate the photon energy connecting source and sink points involved in a particular transition. White shaded region in (e) highlights the incoherent part of the spectral weight, not visible in (d). (c) Optical spectrum as a function of photon energy is plotted along the high symmetry lines. Cyan line gives the band velocity as a function of $k$ (arbitrary units). The two white arrows indicate the two photon energies at which (d) and (e) are calculated. [From Ref.~\protect\onlinecite{tanmoyop}.]} 
\label{arpesop}
\end{figure*}

We can readily understand the microscopic origin of the optical spectra by considering ARPES results. Fig.~\ref{arpesop} compares the spectral intensities relevant for ARPES and optical spectra at a representative doping of $x=0.10$ for NCCO. In the computed ARPES spectrum in Fig.~\ref{arpesop}(a), the underlying LDA dispersion is clearly visible, but the spectral weight has split into four subbands, as was found originally in variational cluster
calculations\cite{grober}. The highest and lowest bands are an incoherent residue of the undressed bands, which we will refer to as the upper and lower charge transfer (CT) bands. The two inner bands are coherent in-gap states split by the AFM gap into UMB and LMB. The in-gap states and the high energy CT bands are separated by the HEKs associated predominantly with magnetic excitations, as observed universally in all cuprates by ARPES\cite{graf,moritz,markiewater}, and found in QMC\cite{macridin} and variational\cite{grober} calculations.

The optical spectrum in Fig.~\ref{arpesop}(b) consists of three main regions marked by shadings of different colors: (1) The low frequency Drude region for $\omega \lesssim 30$meV; (2) The MIR region; and, (3) the high-energy CT gap region for $\omega \gtrsim 1.5$eV. We concentrate here on the MIR and CT-gap regions, and return below to comment on the Drude region. The interband optical
absorption is proportional to the JDOS, so that at each energy we can construct a `source map' showing the filled states which make a strong contribution to the transition, and a `sink map' showing the contribution of the corresponding empty states. Figs.~\ref{arpesop}(d) and (e) show the states responsible for optical transitions along high symmetry lines in the BZ at representative photon energies of $\omega=0.5$~eV near the MIR peak
and $\omega=2$~eV around the high energy peak (vertical bars in
Fig.~\ref{arpesop}(b)). At $\omega=0.5$~eV, the transitions are confined within the in-gap [coherent] states only, whereas at $\omega=2$~eV, the optical subbands predominantly involve the incoherent region. The depletion in the optical spectral weight near the isosbetic point in Figs.~\ref{alp2F2} and \ref{arpesop}(b) is thus associated with the waterfall region marked by arrows in the ARPES spectrum in Fig.~\ref{arpesop}(a)\cite{foot2b}. We summarize the doping dependence of these competing gaps in Section~\ref{S:Gaps} below.

The total optical spectral weight obtained in Fig.~\ref{arpesop}(b) is the
integral of JDOS times the band velocity. The role of the latter factor is explicated in Fig.~\ref{arpesop}(c), where contributions to $\sigma$ are plotted as a function of photon energy $\omega$ and momentum $k$. Although the source-sink map shows a symmetry about the $(\pi,0)$ point, the quasiparticle velocity is low along the $\Gamma\rightarrow(\pi, 0)$ direction [cyan solid line in
Fig.~\ref{arpesop}(c)], leading to greatly reduced spectral intensity
associated with those regions. In contrast, the large quasiparticle velocity in the other two directions is responsible for two distinctive intense streaks labeled d and e. The MIR peak is clearly dominated by the antinodal quasiparticles, whereas the high energy hump stems from a wider $k$-range.


\subsubsection{Optical sum rule}

\begin{figure}[top]
\centering
\rotatebox{0}{\scalebox{0.35}{\includegraphics{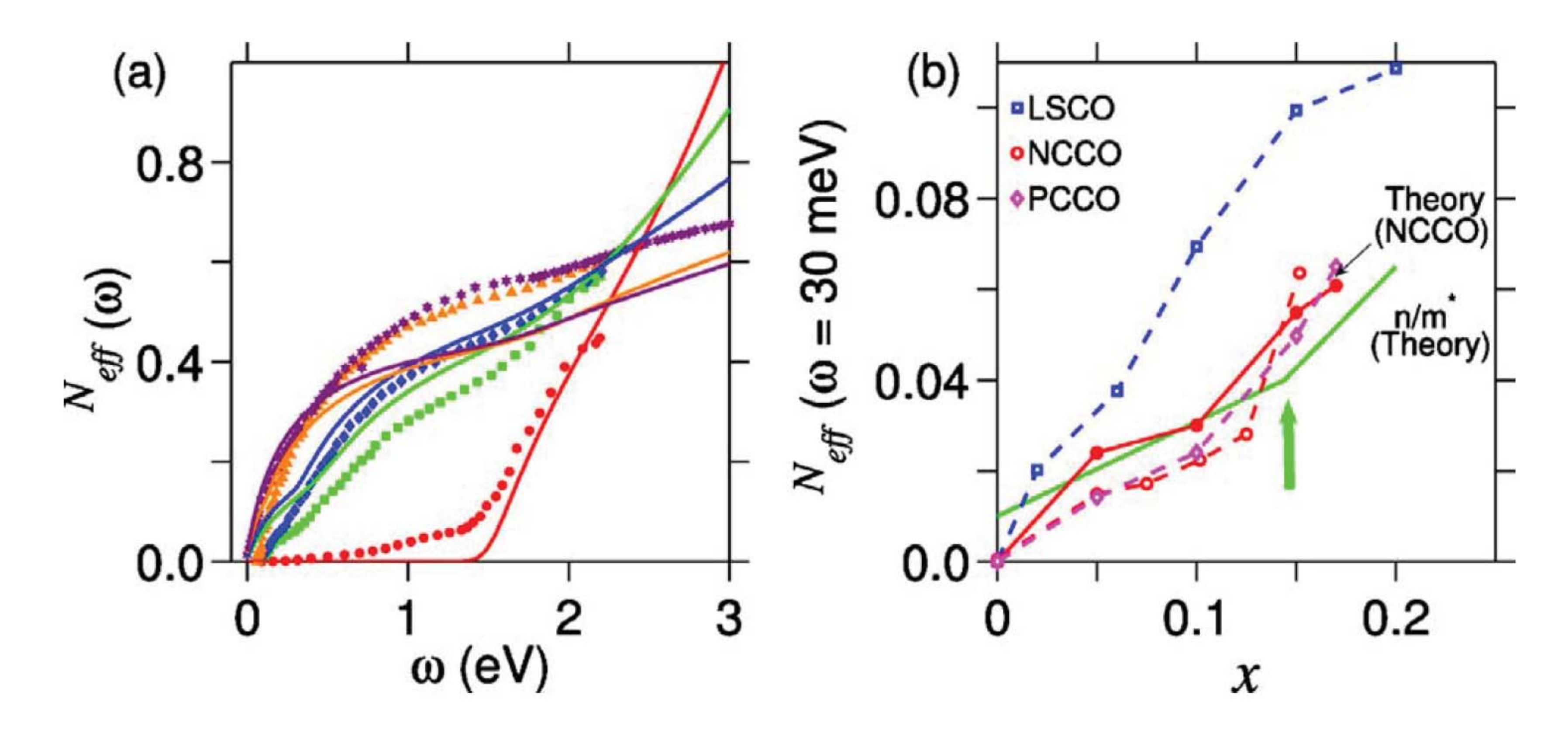}}}
\caption{ { Optical sum rule and spectral weight transfer with doping.} (a) Effective number of electrons $N_{eff}(\omega)$ obtained from the optical spectra in Fig.~\ref{alp2F2} is compared with the corresponding experimental results\cite{onose,onoseprb} on NCCO. (b) Low energy weight $N_{eff}$ at $\omega=30$ meV as an estimate of Drude weight, compared with experimental results on various cuprates\cite{arima,uchida,onose}. Green line gives a direct computation of the normalized Drude weight $n/m^*$. [From Ref.~\protect\onlinecite{tanmoyop}.]} 
\label{fopsw}
\end{figure}

It is interesting to consider the integrated optical spectral weight to illustrate the magnetic gap collapse. The effective electron number (per Cu atom), $N_{eff} (\omega)$, can be defined in terms of the optical conductivity integrated up to an energy $\omega$:
\begin{equation}\label{eq:3}
N_{eff}(\omega) = \frac{2m_0V}{\pi e^2\hbar N}
\int_0^{\omega}\sigma(\omega^{\prime})d\omega^{\prime},
\end{equation}
where $m_0$ and $e$ are the free electron mass and charge, respectively, and $N$ is the number of Cu-atoms in a cell of volume $V$. The results in Fig.~\ref{fopsw}(a) show how rapidly spectral weight shifts to low energies with doping, correctly reproducing the experimental behavior. These results, however, are incompatible with strong coupling models such as the $t-J$ or $U\rightarrow\infty$ Hubbard model, where one strictly assumes no double occupancy of Cu sites. The present intermediate coupling model of Mott gap collapse, on the other hand, properly captures these spectral weight transfers as a function of doping.

Finally, we consider the Drude weight $\propto\sum_{\bf k}n_{\bf k}/m^*_{\bf k}$, which can be compared to an experimental estimate, taken as $N_{eff}$ at a characteristic energy $\omega=30$~meV in Fig.~\ref{fopsw}(b) for NCCO\cite{onose}, and with results for PCCO\cite{arima} and LSCO\cite{uchida}. The agreement is seen to be quite good. Interestingly, at low doping, the Drude weight increases almost linearly with $x$. Within our model, this simply reflects scaling with the area of the FS
pockets in the pseudogap state.  At optimal doping, the weight shows a sharp jump (green arrow) associated with the appearance of the hole pocket. This transition in the FS topology is an intrinsic feature of the model, and it has been found in NCCO near $x\approx 0.15$ by several experimental probes, namely, 
ARPES\cite{armitage}, Hall effect\cite{Das_nonmonotonic,tanmoysns}, and superconducting penetration depth\cite{tanmoyprl}. For LSCO (blue symbols), experiments\cite{uchida} show a similar linear behavior of the Drude weight, which also corresponds to the doping dependence of the area of the FS pocket\cite{yoshida}. The peak around $x\sim0.2$ corresponds to the doping of the VHS.



%
%


\subsection{ RIXS}

In Cu K-edge resonant inelastic x-ray scattering (RIXS), an incident x-ray excites a Cu $1s\rightarrow 4p$ transition with an intermediate state shakeup involving mainly Cu $d_{x^2-y^2}$ and O $p$ states, after which the $1s$ hole and $4p$ electron recombine, emitting a second photon.\cite{KoShin,Amen} The RIXS cross section is a function of the energy ($\omega$) and momentum (${\bf q}$) transferred from the medium to the photon. In the limits of either a weak core-hole potential or an ultrashort core-hole lifetime, this cross section is proportional to the dynamic structure factor $S({\bf q},\omega)$, which can be thought of as a $q$-resolved JDOS. RIXS can, in principle, provide direct access to the unoccupied states above $E_F$, which are otherwise difficult to probe spectroscopically, optical spectrum being restricted to $q=0$.  The connection with $S({\bf q},\omega)$ and hence with the spin and charge susceptibilities suggests that RIXS should be sensitive to spin-wave physics, a point to which we will return in the neutron scattering Subsection~\ref{S:Neu} below. 

When core-hole effects are strong, the RIXS intensity becomes $S({\bf q},\omega)$ to the leading order, modulated by the associated matrix element effects.\cite{MRB}  RIXS can thus be used to monitor the evolution of the pseudogap in cuprates.  We computed the RIXS spectrum for NCCO as a function of doping in the weak-core-hole limit, neglecting the GW self-energy but using the experimental dispersion, and obtained good agreement with the magnetic gap collapse determined by ARPES\cite{MBRIXS,RIXS}. However, the computed higher energy features lie at relatively low energies compared to the experimental results. This discrepancy between theory and experiment is removed when QP-GW self-energy is included in the RIXS computations.\cite{SusmitaRIXS} In this way, we have 
shown that signatures of the three key normal-state energy scales, the pseudogap, the charge transfer gap, and the Mott gap, can be identified in the RIXS spectra of cuprates.

Our RIXS modeling involves a three-band Hubbard Hamiltonian based on Cu
$d_{x^2-y^2}$ and O $p_x$,~$p_y$ orbitals with a self-energy correction applied to the antibonding band nearest to $E_F$ [Section~\ref{S:Ext_d}]. 
The Hamiltonian can be written as\cite{MBRIXS}:
\begin{eqnarray}\label{Hamil}
H&=&\sum_j (\Delta_{d0}d_j^{\dagger}d_j + Un_{j\uparrow}n_{j\downarrow}) + \sum_i
U_pn_{i\uparrow}n_{i\downarrow}
\nonumber\\
&+&\sum_{<i,j>} t_{CuO}(d_j^{\dagger}p_i+ (c.c)) + \sum_{<i,i^{\prime}>}
t_{OO}(p_i^{\dagger}p_i^{\prime}+ (c.c)),
\end{eqnarray}
where $\Delta_{d0}$ is the (bare) difference between the onsite energy levels of Cu $d_{x^2-y^2}$ and O $p-\sigma$, $t_{CuO}$ is the copper-d to oxygen-p hopping parameter, $t_{OO}$ the oxygen-oxygen hopping parameter, and $U$ ($U_p$) the
Hubbard interaction parameter on Cu (O). $n_{j} = d_j^{\dagger}d_j$ and $n_{i} =
p_i^{\dagger}p_i$ are the number operators for Cu-d and O-p electrons, respectively. The equations were solved at the Hartree-Fock (HF) level to obtain a self-consistent mean-field solution. Hartree corrections lead to a renormalized Cu-O splitting parameter $ \Delta = \Delta_{d0}+ Un_d/2- U_pn_p/2,$ where $n_d$ ($n_p$) is the average electron density on Cu(O) \cite{foot2}.
The AFM order splits the three bands into six bands as seen in  Fig.~\ref{3brixs}(a)\cite{MBRIXS}, while the QP-GW self-energy renormalizes the dispersion near $E_F$.  The hopping parameters are taken from LDA while interaction parameters $U$ and $\Delta_{d0}$ are adjusted to optimize agreement
between the antibonding band-splitting and the one-band results \cite{foot2,OKA}. When this is done, $\Delta_{d0}$ is found to be small and negative, while the three-band Hubbard $U$ is much larger than the effective one-band $U$, and it exhibits a much weaker doping dependence \cite{foot3a}.
[In this subsection, $U$ refers to the three-band value, unless noted otherwise.] The resulting small value of $\Delta_0$ is consistent with the common view of the undoped cuprates being charge-transfer insulators\cite{ZSA}, even though the true Mott gap between the Cu-$d_{x^2-y^2}$ orbitals is much larger. 

Note that the UHB and LHB of the Cu orbitals are intimately related to the antibonding and bonding bands of the three-band model. Remarkably, the Hubbard $U$ not only plays a role in splitting the hybridized Cu-O states into UHB and LHB separated by $\sim U/2$ from the nonbonding O states, but it is also involved in dehybridizing the Cu and O orbitals by opening an AFM gap in both UHB and LHB, the former being the pseudogap. [In analogy to the one-band model, we refer to the AFM-split subbands of the antibonding bands as the LMB and UMB.] While the AFM gap collapses rapidly with doping\cite{foot1b}, a residual charge-transfer gap persists at high energies, due to strong magnetic fluctuations.  A similar effect is seen in optical spectra, and the RIXS spectrum at $q=0$ closely resembles the optical spectrum. Our three-band calculation reproduces the experimental finding that the magnetization scales with the AF gap.\cite{Mang,MkII}


With the preceding considerations in mind, cuprate magnetism naturally separates into two regimes. At high energies, Mott physics produces localized spins on each copper site, splitting the Cu dispersion by an energy $\sim U$ into UHB and LHB. In the presence of hybridization with oxygens, the LHB [UHB] becomes identified with the bonding [antibonding] band of the three-band model. At lower energies, the spins on different sites interact, leading to magnetic gaps in both UHB and LHB of magnitude $\sim SU$ via the more conventional Slater physics associated with an AFM order, where $S$ is the magnetization on Cu. The Mott physics thus arises here as an emergent phenomenon. When the AFM gaps open at half-filling, hybridization between Cu and O is mostly lost. In particular, when the antibonding band develops magnetic order the electrons in the UMB develop mainly Cu character, while states near the top of the LMB are of nearly pure oxygen character \cite{MBRIXS}.  The opposite happens in the bonding band.  

Within the RPA framework, the K-edge RIXS cross section is given by: \cite{Igarashi,MBRIXS,RIXS}
\begin{eqnarray}\label{Wqw}
W({\bf q},\omega,\omega_i) &=& (2\pi)^3N|w(\omega,\omega_i)|^2
\nonumber\\
&\sum_{\mu}&{\rm Im}\left[Y^{+-}_{\mu,\mu} ({\bf q},\omega)\right]
|\alpha_{\mu}|^2 \cos{\big(2{\bf q}\cdot{\bf R}_{\mu}\big)},
\end{eqnarray}
where $\omega_i$ is the initial photon energy, and $\omega$ and $\textbf{q}$ are the energy and momentum transferred in the RIXS process. $w(\omega,\omega_i)$ contains the matrix-element of the initial and final state transition probabilities \cite{MBRIXS}, $N$ is the total number of Cu atoms and ${\bf R}_{\mu}$ is the position of the $\mu^{\rm th}$ orbital present in the intermediate state. The nearest neighbor (NN) O excitations and second NN
Cu excitations are included via $\alpha_1$ and $\alpha_2$, respectively. We assume small values $\alpha_1=0.1$ and $\alpha_2=0.05$, whereas $\alpha_0$ is taken to be equal to 1. $Y^{+-}_{\mu,\mu} ({\bf q},\omega)$ is the Fourier transform of the charge correlation function or the JDOS. 
$Y^{+ }_{\mu'\sigma',\mu\sigma}(q,t'-t) =
\langle \rho_{\bf{q},\mu' \sigma'}(t')\rho_{-\bf{q},\mu
\sigma}(t)\rangle$ with $\rho_{\bf{q},\mu \sigma}(t)$ 
the charge operator, is given by\cite{Igarashi,allen} 
\begin{eqnarray}\label{Y}
Y^{+-}_{\mu'\sigma',\mu\sigma}({\bf q},\omega) &=&
\sum^{\prime}_{k}\int d\omega_1\int d\omega_2
A_{\mu\mu'}(k+q,\omega_1)
\nonumber\\
&\times&A_{\mu'\mu}(k,\omega_2)\frac{f(\omega_2)-f(\omega_1)}{\omega+i\delta+\omega_2-\omega_1}
\end{eqnarray}
where $f(\omega)$ is Fermi function and $\sigma$ is the spin index. The prime in the $k-$summation means that the summation is restricted to the AFM zone.  In Ref.\onlinecite{SusmitaRIXS}, we computed the spectral weight $A$ of Eq.~\ref{Y} using the three-band Hubbard model, using the QP-GW self-energy for the antibonding band\cite{tanmoygoss}, and a constant broadening, $\Sigma^{''} = 0.5$ eV for $\omega > 4$ eV, consistent with the ARPES data \cite{lanzara2}.  

\begin{figure}[htp]
\centering
\hskip-0.7 cm
\rotatebox{270}{\scalebox{0.23}{\includegraphics{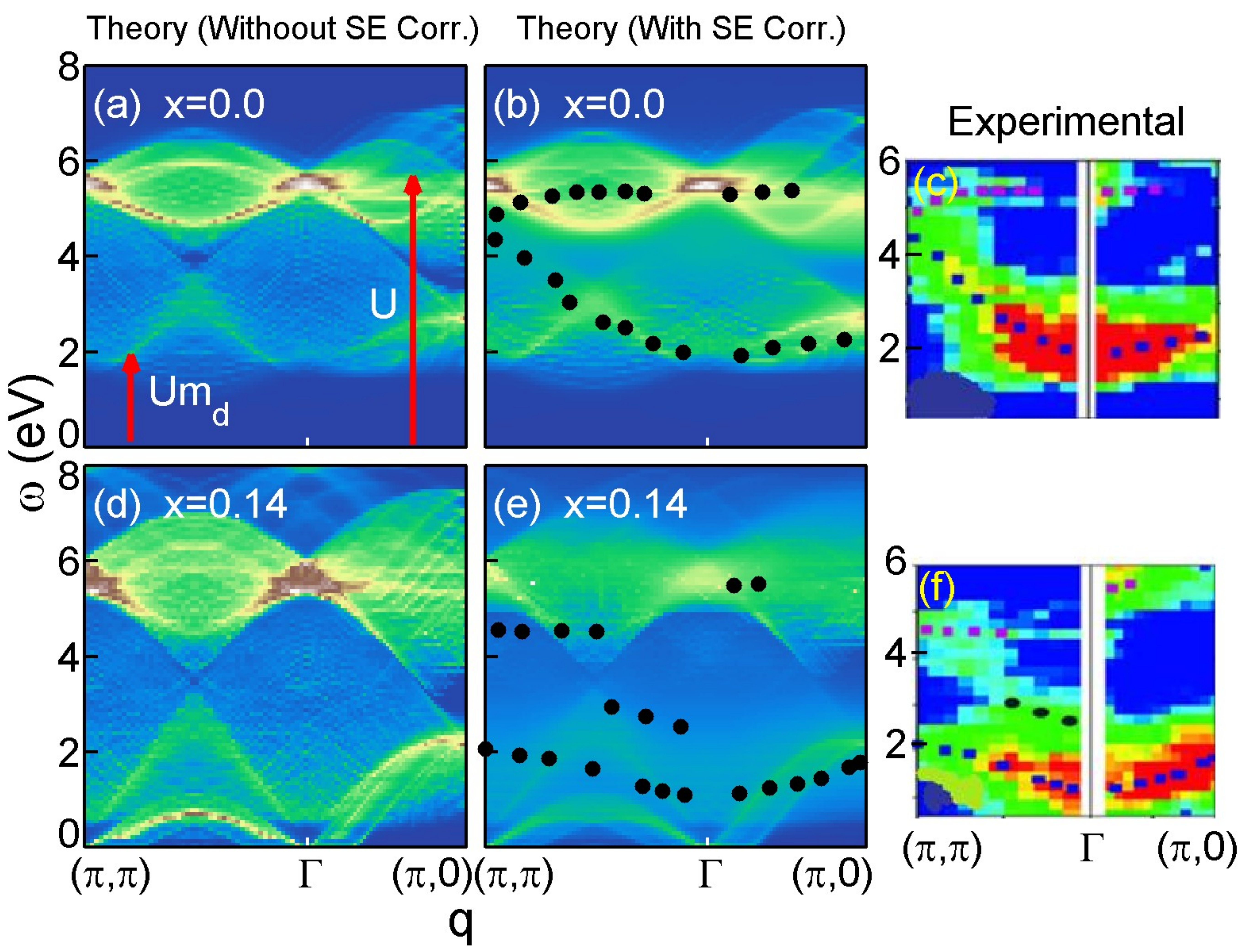}}}
\caption{ {RIXS spectra from NCCO} for $x=0$: (a) theory without and (b)
with self-energy corrections, and (c) experiment\cite{RIXS}. (d)-(f) Similar figures for $x=0.14$. Dotted lines in (b,e) give experimental peak positions.
[From Ref.~\protect\onlinecite{SusmitaRIXS}.]}
\label{figSE}
\end{figure}

Figure~\ref{figSE} shows the calculated RIXS spectra of NCCO for $x=0$ and
$x=0.14$. Frames (a) and (d) include AFM order but without self-energy corrections, whereas the calculations in frames (b) and (e) include the self-energy. The high intensities at energies around $5.6$ eV involve 
transitions from the lower magnetic band to the unoccupied states of the
antibonding band, reflecting the Mott gap feature. This `6~eV' feature is present for all dopings. At half-filling, in frames (a) and (b), the high
intensities around $2$ eV occur due to the transition within the antibonding 
Cu-O band across the AFM gap. This gap collapses with doping, and as a result, we find a smaller AFM gap at $14 \%$ electron doping in frames (d) and (e), close to the QCP, consistent with earlier results \cite{RIXS}.  A key result is that the self-energy produces a realistic broadening comparable to that observed experimentally.

The weak-core-hole RIXS calculations discussed so far cannot capture the distinction between well- and poorly-screened intermediate states. This is most apparent around 6~eV, where most experiments find an intense feature in the cuprates\cite{hill,Abbamonte,x0higherEcuts,Hasan,Ishii,Lu}, but there is an energy window where this feature is weak\cite{Hill2}, allowing more detailed analyses of the lower intensity  features, as in Figs.~\ref{figSE}(c,f)\cite{RIXS}.
The present RIXS calculations capture only the former, intense 6~eV feature.  Except for this feature, most features in the calculated RIXS intensities follow the experimental trends. 

As to the various energy scales reflected in the RIXS spectrum, we find numerically that the magnitude of the Mott gap is approximately equal to $U$ and the AFM gap to $SU$ as shown by the arrows in Fig.~\ref{figSE}(a).  Also, the charge transfer energy is the difference between the average oxygen energy and the upper Cu band\cite{ZSA}, which we find to be $\sim U/2$. Thus all these energy scales are controlled by $U$. In our calculations, the $6$ eV feature represents transitions across the true Mott gap, and the good agreement with experiment indicates that RIXS can be used to probe this important scale and how it is modified by hybridization with oxygens. In optical spectra at half-filling, the $\sim$2~eV charge transfer gap is indistinguishable from the AFM gap\cite{tanmoysw}.  At finite doping, these two features separate, with the AFM gap reflected as a mid-infrared peak which collapses rapidly with doping, while a residual charge transfer gap persists as a weak feature near $2$ eV even in the strongly doped regime.  A similar evolution is found in RIXS. The leading RIXS edge follows the doping dependence of the AFM gap \cite{MBRIXS,RIXS}, while a residual charge transfer gap feature can be seen in the RIXS spectra near the $\Gamma$ point, see Fig.~3 of Ref.\onlinecite{SusmitaRIXS}.  

%
%

\subsection{Neutron scattering}\label{S:Neu}

\begin{figure}[h]
\centering
\rotatebox{0}{\scalebox{0.53}{\includegraphics{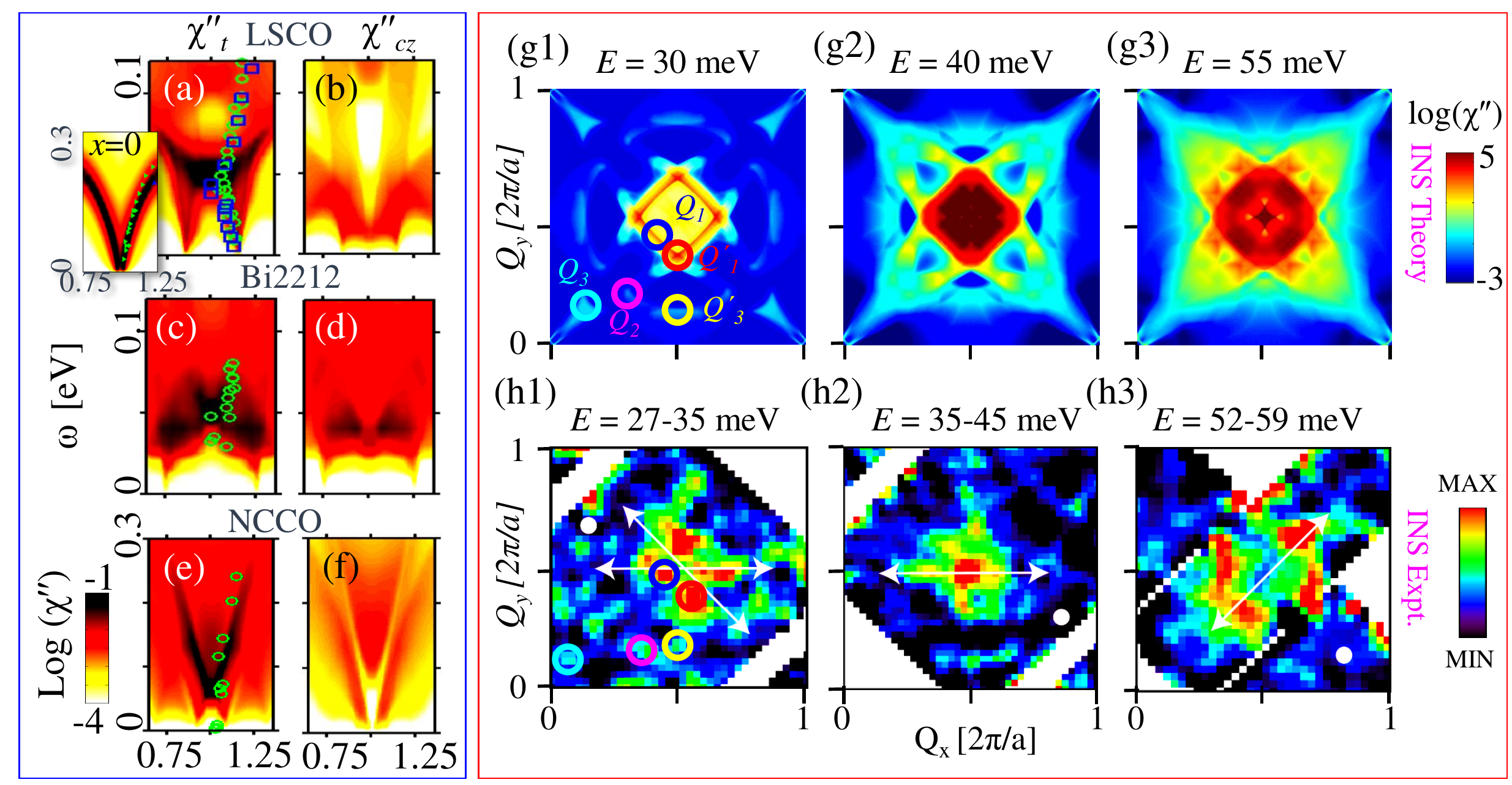}}}
\caption{ { Material dependent magnetic resonance modes}. (a-f): Blow-up of low-energy region of Fig.~\ref{chiop} in the vicinity of the magnetic-resonance mode in the SC state. Left column is the imaginary part of transverse susceptibility, while the right column gives the longitudinal part. Inset in (a) is the non-SC component at half-filling, exhibiting a gapless Goldstone mode at ${\bf Q}$. Various symbols are the same experimental data as in Fig.~\ref{chiop}. 
(g1-g3) Computed $\chi^{\prime\prime}(\omega)$ spectra (log scale) for three representative constant energy cuts: below, near, and above the resonance. 
(h1-h3) Corresponding experimental data on YBa$_2$Cu$_3$O$_{6+y}$ (YBCO) from Ref.~\cite{woo} for the acoustic channel, obtained after subtracting the corresponding normal-state data to enhance weak features away from ${\bf Q}$. In (g1) and (h1), circles of different colors depict various MQPS vectors, similar to the QPI vectors seen in STM studies. [After Ref.~\onlinecite{tanmoymagres}.]} 
\label{hour_glass}
\end{figure}

Many relevant results of neutron scattering studies of spin waves have already been discussed in Section~\ref{S:Boson} above, and therefore, here we focus on the modifications of the spin-wave spectrum in the SC state. The phenomenology of the low-energy neutron mode in cuprates is well-known, and a number of universal features have been identified experimentally \cite{wilson,vignolle,bourges,heyden,fong,woo,pailhes04,tranquada,daiybco} and interpreted theoretically.\cite{chubukov,norman,eremin,tanmoymagres} A distinct low energy magnetic mode is found near the AFM nesting vector ${\bf Q} = (\pi,\pi)$ in almost all cuprates. Intensity of this mode is enhanced in the SC state while its energy scales as, $\omega_{\rm res}({\bf Q}) \propto 2\Delta$,\cite{yu,pnictideResDas} suggesting a close connection with SC pairing. The dispersion of spectral weight away from this resonance peak also has a fairly universal character in that it forms an `hour-glass-like' pattern centered on the resonance mode, which undergoes a 45$^o$ rotation on passing through the resonance peak. Below the resonance energy, the spectral weight disperses along the Cu-O bond direction, while above the resonance, the dispersion peak lies along the diagonal direction. Despite these similarities, many details of the hour-glass dispersion are material specific.

The preceding phenomenology can be understood within the QP-GW model as follows. As superconductivity turns on, the normal state susceptibility in the particle-hole continuum, $\chi^1$ (Eq.~\ref{Eq:ch3_chi_0_sc}), becomes gapped by $\omega\le 2\Delta$, and within the SC gap, the particle-particle and hole-hole scatterings $\chi^{2,3}$ in Eq.~\ref{Eq:ch3_chi_0_scB} become active on the FS. However, their contribution to the total spin-resonance spectrum is controlled by the corresponding magnetic scattering form factor $C^{2,3}$ given in Eq.~\ref{t11}. On the normal-state FS [$E^{s\pm}({\bf k}) = 0$], the SC coherence factor reduces to $C^{2,3} = 1/2[1-{\rm sgn}(\Delta_{\bf k}){\rm sgn}(\Delta_{{\bf k}+{\bf q}})]$, which attains its maximum value of 1 whenever  $\Delta_{\bf k}$ and $\Delta_{{\bf k}+{\bf q}}$ have opposite signs.\cite{chubukov,norman,eremin,tanmoymagres} This scattering process may be called magnetic quasiparticle scattering (MQPS),\cite{tanmoymagres} in analogy with the very similar quasiparticle interference (QPI) pattern observed in STM. Furthermore, at small $|{\bf q}- {\bf Q}|$, the SDW coherence factor, Eq.~\ref{chiSDW}, simplifies\cite{SachdevSDWres} to $A\rightarrow 1-(\xi^-_{\bf k}+\xi^-_{{\bf k}+{\bf q}})^2(US)^2/4E_{0{\bf k}}^4\approx 1-\mathcal{O}([Q-q]^2)$, which attains its maximum value of 1 at $q = Q$. This is why the intensity in MQPS attains its maximum value near ${\bf Q}$ despite the presence of many other possible scattering vectors, which can accommodate sign-reversal of the $d$-wave SC pairing in cuprates. Taken together, two simple necessary conditions for the occurrence of the resonance are
\begin{eqnarray}
&&\Omega_{\rm res} = |\Delta_{{\bf k}_F}|+|\Delta_{{\bf k}_F+{\bf q}}| \\
\label{resonance1}
&&{\rm sgn}[\Delta_{{\bf k}_F}]\ne{\rm sgn}[\Delta_{{\bf k}_F+{\bf q}}].
\label{resonance2}
\end{eqnarray}
Figs.~\ref{hour_glass}(a-f) compare the dispersion of the spin-resonance spectrum in transverse (left column) and longitudinal (right column) susceptibilities around ${\bf Q}$ along the diagonal direction for two hole-doped and one electron-doped system with the corresponding INS data. Despite some material dependence in dispersion and intensity, a universal feature is seen for all materials in that an upward and a downward dispersion in the INS spectra meet at ${\bf Q}$ at finite energy. Our QP-GW model explains this hour-glass shape as originating from the intersection of the gapped residual spin-wave and the MQPS spectrum in the SC state. At this contact point, intensity attains its maximum value, resulting in the so-called spin-resonance.  Spin-waves are absent along the longitudinal direction, while pair-excitations, being scalar bosons, appear in both channels.

\begin{figure}[h]
\centering
\rotatebox{0}{\scalebox{0.65}{\includegraphics{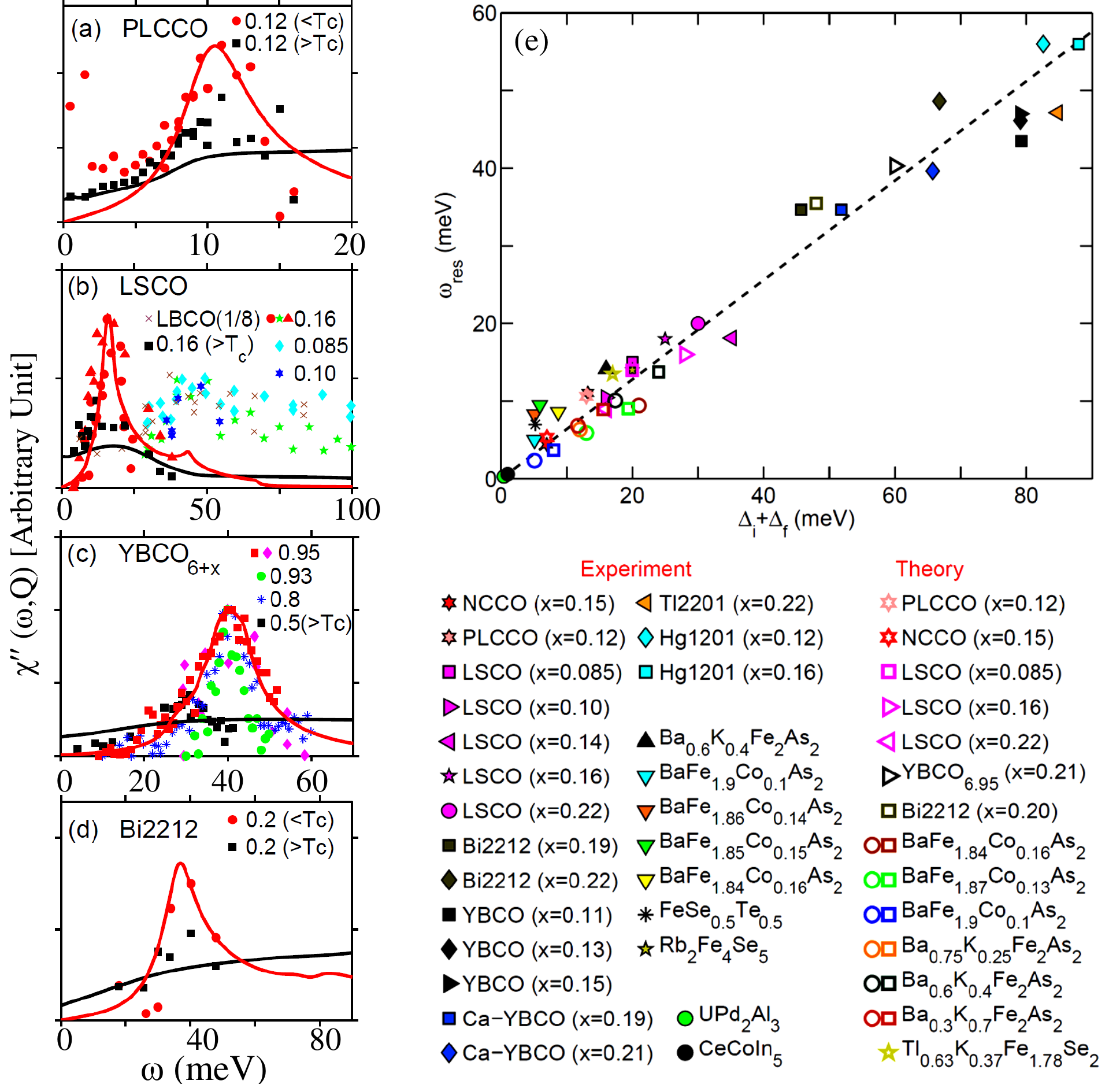}}}
\caption{ (a-d): { Magnetic resonance peak at $(\pi.\pi)$ in cuprates in the SC and normal states.} Solid red lines give theoretical $\chi^{\prime\prime}(Q,\omega)$ in the SC state. The corresponding normal state ($\Delta=0$ at $T>T_c$) spectrum near optimal doping is given by black lines. Symbols of various colors show experimental data as follows: PLCCO [Ref.~\cite{zaanen}]; LSCO [red circles and green stars from Ref.~\cite{vignolle}, red triangles and black squares from Ref.~\cite{christensen}, cyan diamond from Ref.~\cite{lipscombe}, blue stars from Ref.~\cite{kofu}, and brown cross for LBCO at 1/8 from Ref.~\cite{tranquada}]; YBCO- [red square, blue stars and green circles from Ref.~\cite{daiybco}, magenta diamond from Ref.~\cite{woo}, and black squares from Ref.~\cite{stock}];
and, Bi2212 [Ref.~\cite{fong,xu}]. Black and red symbols in all cases should be compared with the theoretical curves (lines with the same colors) as these experimental data are for similar doping and temperature values. All experimental datasets are normalized to their maximum values because some experimental results are not given as absolute values, while others are given only as differences between the SC- and normal-state results. Theoretical datasets are normalized to the maximum of the corresponding experimental values since the matrix element involved in neutron scattering is not included in the calculation\cite{walters}. Theoretical spectra have been broadened by 1-3 meV to account for impurity scattering. (e) Resonance energy vs superconducting gap for a various SC materials. Theoretical results are added to the redrawn experimental data from Ref.~\cite{yu} (see legend).  [From Ref.~\cite{Whiffnium}.]}
\label{resonance}
\end{figure}

The preceding scenario also explains the 45$^0$ rotation of the spectra as we move from the spin-wave energy scale to the MQPS as shown in Figs.~\ref{hour_glass}(g-h)\cite{Whiffnium}. Above the resonance, since spin-wave dispersion is aligned along the diagonal direction, the maximum intensity spot lies along this direction, see Fig.~\ref{hour_glass}(g3) and \ref{hour_glass}(h3) for theoretical and experimental results, respectively. In contrast, below the resonance, the bright spots in the INS spectra are dominated by FS-nesting, as given by conditions of Eqs.~\ref{resonance1}, and seen in Figs.~\ref{hour_glass}(g1) and \ref{hour_glass}(h1). Comparing the nodal-nesting vector ${\bf Q}_1=2(\pi\pm\delta,\pi\pm\delta)$, aligned along the commensurate direction, with its slightly incommensurate version along the bond direction,  ${\bf Q}^{\prime}_1=2(\pi\pm\delta,\pi)$ and $2(\pi,\pi\pm\delta)$, one finds that the latter connects twice as many Fermi momenta as any other ${\bf Q}_1$.\cite{tanmoymagres} Therefore, the INS spot along the bond direction is twice as intense as its counterpart along the diagonal direction, yielding a 45$^o$ rotation of the INS profile in going through the resonance. The MQPS dispersion is clearly different in the electron- and hole-doped cuprates due to differences in FS topologies. In particular, the downward MQPS dispersion is weak or absent because the nodal pocket in the electron-doped case is absent or small as discussed in Section~\ref{S:ElSpec}.1 above. The FS topology of both electron- and hole-doped samples become similar in the optimal to overdoped region, where the electron and hole-pockets coexist. For the hole doped sample, we have extended our INS calculation to such a FS topology and found that the presence of the electron-pocket gives an additional collective mode at an incommensurate wavevector above the hour-glass energy scale.\cite{Das_EP} This is because the electron-pocket lying around the $(\pi,0)$ momentum region possesses higher $d$-wave SC gap amplitude than the nodal pocket, and the corresponding collective mode appears almost around $\Omega_{\rm res}\approx 2\Delta$, according to Eq.~\ref{resonance1}. The second resonance mode is also observed in a YBCO sample in the computed energy and momentum scale.\cite{YBCO_secondmode}

Figure~\ref{resonance} shows that the energy scale of the resonance mode is directly related to the SC-gap\cite{yu,zaanen,vignolle,christensen,lipscombe,kofu,tranquada,daiybco,woo,stock,fong,xu,walters}. As pointed out already, the magnetic resonance mode in the theoretical spectra develops in the SC-state at the contact point of the spin-wave and MQPS, the spectrum being relatively featureless in the normal state. The computed lineshapes in Figs.~\ref{resonance}(a-d) are in good agreement with data in the SC-state. The experimentally observed broadening of the resonance peak is essentially accounted for by the theory in which only a small additional broadening ($\sim$ 1-3 meV) has been added to account for impurity scattering. Notably, the normal state experimental data are taken at temperatures just above $T_c$ where fluctuation and precursor SC effects might be present in the incoherent spectra as seen, for example, in the Nernst signal.\cite{NernstOng} Such incoherent SC gaps could sharpen the normal state resonance spectra compared to theory in which SC fluctuations are not included. 

As discussed earlier, the ${\bf Q}$ resonance, $\Omega_{\rm res}$, is related to the SC-gap by the universal relation, $\Omega_{\rm res}/2\Delta_{\rm max} = C_{res}g_{{\bf k}_o}$, where $g_{{\bf k}_o}$ is the d-wave structure factor at the hot-spot momentum ${\bf k}_o$. A fit to the whole dataset gives an average value of the constant $C_{res}\sim $0.64 in cuprates,\cite{yu}, and about 1 in the isotropic $s^{\pm}$-pairing pnictides.\cite{pnictideResDas} Our calculations are consistent with this value, Fig.~\ref{resonance}(e), with deviations from this average value coming from the material dependence of the hot-spot position.  As discussed in Sec.~\ref{S:LEK}, the coupling of the ${\bf Q}$ resonance mode to the electronic spectra governs the 50-70~meV energy LEK. 




\section{Doping dependent Gaps}\label{S:Gaps}
\begin{figure}
\centering
\rotatebox{0}{\scalebox{0.53}{\includegraphics{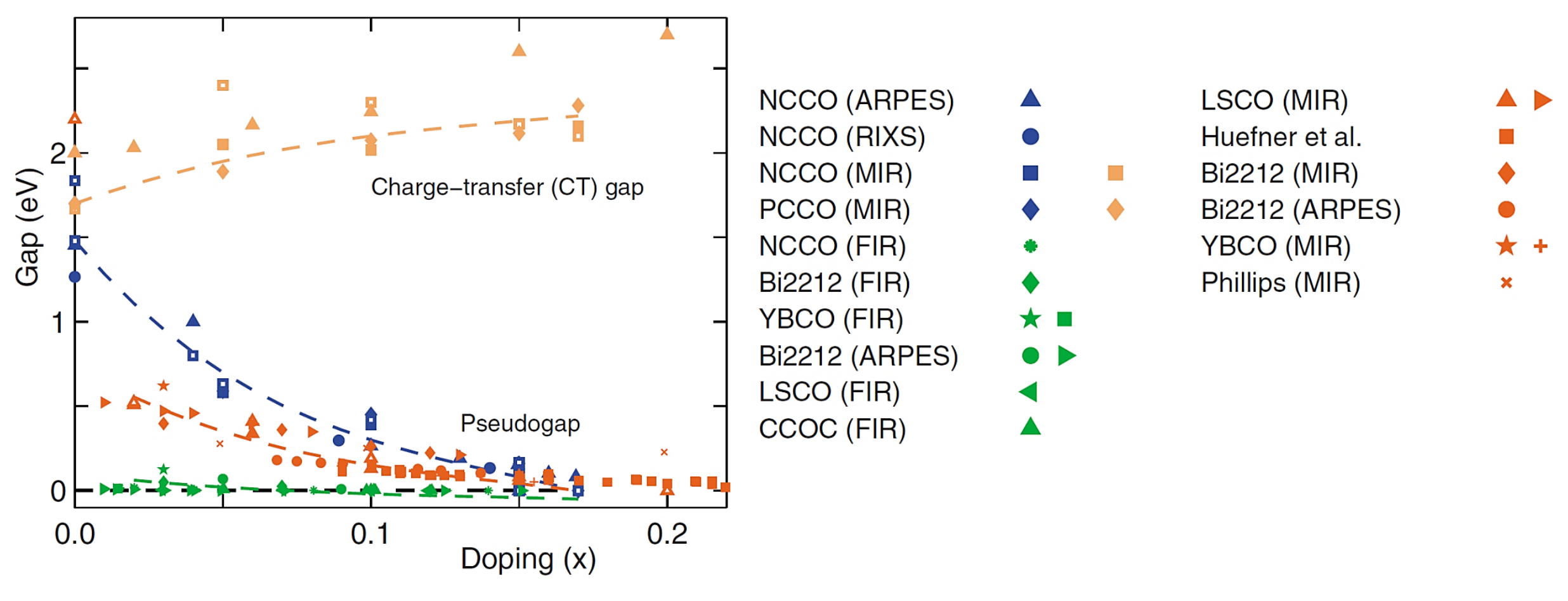}}}
\caption{ { Doping dependent magnetic
and CT gaps in the cuprates.} Theoretical results are
compared with optical MIR data\cite{arima,uchida,onoseprb} and pseudogap
data from ARPES\cite{armitage,matsui} and RIXS\cite{RIXS} for
electron-doped NCCO and various other experiments for hole-doped cuprates.
Red squares give pseudogap data for four different 
compounds [Bi2212, YBCO, TBCO and HBCO] from ARPES,
tunneling, Raman, Andreev reflection, and heat capacity
experiments (reproduced from the survey of
Ref.~\onlinecite{huefner}). Red triangles for the MIR data
of LSCO are taken from Lee {\it et al.}\cite{basov}. Red
circles and '+' symbols are the pseudogap measured via ARPES on
Bi2212\cite{Tanaka} and optical spectra on YBCO.\cite{liyu} 
Red 'x' symbols are theoretical results\cite{phillips} for the MIR
in a Mottness model ($t=0.4$ assumed in plotting this dataset).
Red diamond and star symbols represent independent MIR peak
positions, and the same symbols in green color give FIR gaps
measured simultaneosly in optical spectra of Bi2212 and
YBCO\cite{lupi}. Green circles give a Coulomb gap seen in
ARPES\cite{ding} [energy values obtained by assuming 
$2\Delta/k_BT=4$]. Remaining green symbols are reproduced
from Lupi {\it et al.}\cite{lupi} Open symbols of
same colors in all cases refer to our corresponding theoretical results, which are obtained self-consistently at each doping. Dashed lines are
guides to the eye for CT gaps (gold), pseudogaps for
electron- (blue) and hole-doping (red), and the FIR and Coulomb
gaps (green). [From Ref.~\onlinecite{tanmoyop}.]} 
 \label{gap}
\end{figure}

Figure~\ref{gap} summarizes the contrasting doping dependencies of the pseudogap and effective charge transfer (CT) gap extracted from various spectroscopies\cite{huefner,armitage,uchida,onoseprb,arima,matsui,RIXS,basov,Tanaka,liyu,phillips,lupi,ding}. While a strong case can be made that in electron-doped cuprates the pseudogap is associated with a coexisting $(\pi,\pi )$-SDW order, nature of the pseudogap in hole-doped cuprates remains controversial. There is a growing consensus, however, that it originates
from some form of density-wave-like competing order, which could
include coupling to phonons\cite{lee,muller,jarrell1}, and may
involve several competing modes driven by proximity to the VHS\cite{MGu2}; we will discuss the evidence for other, often incommensurate, competing phases in Section~\ref{S:DWs}.  Our analysis here has been based on invoking a 
$(\pi,\pi)-$-SDW order, but as noted already, the shape and doping dependence of the pseudogap is insensitive to the particular order chosen so long as this competing order vanishes in a QCP near optimal doping. This conclusion should hold even more strongly for the optical spectrum. Note that in the severely underdoped regime ($x\le 0.05$) the pseudogap must increase substantially to match the optical gap.  Our recent STM analysis of the doping dependent gaps, Fig.~\ref{phasediagN}(a), finds a larger gap at low doping, allowing an extrapolation to the undoped limit. In this doping range superconductivity is quenched, and there is a metal-insulator transition.  Fortunately, good spectroscopic data for deeply underdoped cuprates are starting to appear.\cite{ZhouUD} 

In NCCO, the MIR-gap corresponds to the true SDW gap, and its doping evolution is in good agreement with ARPES\cite{armitage,kusko,matsui} and RIXS\cite{RIXS} results, which are consistent with a QCP near $x=0.18$\cite{tanmoyprl,tanmoy2gap}. In contrast, the CT-gap shows an opposite doping dependence in that it increases (slowly) with doping. Although the CT gap does not show a real QCP, it rapidly loses intensity with doping, as discussed in more detail in Section~\ref{S:NFL}.1 below. 

Similar results are found in optical studies of  hole-doped LSCO\cite{uchida,basov},
Bi2212\cite{hwang}, YBCO\cite{cooper}, Tl$_2$Ba$_2$CuO$_{6+\delta}$ (TBCO), and
HgBa$_2$Ca$_2$Cu$_3$O$_{8+ \delta}$ (HBCO)\cite{mcguire} as well
as in x-ray absorption spectroscopy\cite{chen} and QMC
computations.\cite{jarrellb} Figure~\ref{gap} presents data on the MIR
gap and the pseudogap obtained from a wide range of experiments, including a recent comprehensive survey\cite{huefner}, as detailed in the caption to Fig.~\ref{gap}. In the intermediate and high doping regimes, there is very good agreement between the MIR and other measures of the pseudogap, providing evidence for the rapid growth of the pseudogap in the deeply underdoped regime, consistent with theory.  While all the hole-doped cuprates seem to have a very similar doping dependence of the pseudogap, they differ subtly from electron-doped systems with regard, for example, to the steepness of the rise
at low doping and the exact position of the QCP.\cite{huefner,tallon}  Some of these differences seem to be correlated with the doping dependence of the screened Hubbard $U$'s shown in Fig.~\ref{Ux}(a), and arise presumably from the strong screening associated with the VHS on the hole-doped side.\cite{markiecharge,kanamori} 

Although our calculations reproduce the gap structures of the cuprates, especially on energy scales of $\ge 100$~meV, additional effects would arise at lower energies involving coupling to phonons and impurities. In several cuprates, Lupi {\it et al.}\cite{lupi} find that in addition to the MIR feature there is an additional far-infrared peak which appears near 10\% doping, which shifts to higher energies at lower doping (green symbols in Fig.~\ref{gap}).  This seems to be a disorder effect, and displays a doping dependence similar to that of the so-called Coulomb gap seen in ARPES data from  Bi2212 (green circles and diamonds) in the low-temperature region.\cite{ding}

Our QP-GW model predicts that the MIR feature will extrapolate to the CT-gap, $\sim$1.8~eV in hole-doped cuprates, as $x\rightarrow 0$, whereas the experimental data at the lowest doping of $x\sim 0.02$ in Fig.~\ref{gap} have reached only about 1/3 of this value. We make several brief points on this thorny issue: (1) Looking at the experimental $\sigma (\omega )$ in Figs.~\ref{alp2F2}(a,b), there is a substantial transfer of spectral weight between the CT-peak at $x=0$ and the MIR feature at $x=0.02$, but it is not clear whether this spectral weight transfer arises via a peak-shift or by the growth of a new peak at intermediate energies.  This issue also underlies the ongoing theoretical debate as to whether the first doped holes go into the upper magnetic band or in midgap states; (2) We find that the screening of $U$ in hole-doped cuprates only begins to turn off when $x$ is reduced below 0.05, Fig.~\ref{Ux}(a), and in this underdoped regime results can be expected to be sensitive to details of the metal-insulator transition; (3) More generally, disorder is likely to play a significant role. In particular, in LSCO, holes get trapped by the dopant atoms at very low doping, and the screening of $U$ is turned off at very low temperatures \cite{Hirschf}. Notably, in this doping regime the measured Drude lifetimes are quite small, Fig.~\ref{fopsw}(b); and finally, (4) For hole-doped cuprates there is likely to be an issue of competing pseudogap phases\cite{MGu2}, a striped-phase being a candidate for introducing mid-gap states.

We emphasize that the dichotomy between the pseudogap and CT gap and the presence of the QCP are robust features of our QP-GW model of the cuprates, which does not involve any free parameters beyond essentially the one-band value of $U=1.7$~eV at half-filling. In particular, we have self-consistently computed the doping dependence of $U$ due to screening effects of charge fluctuations\cite{kanamori,markiecharge} to obtain the effective $U$ values shown Fig.~\ref{Ux}(a).

\section{Non-Fermi liquid physics}\label{S:NFL}

Understanding how non-Fermi-liquid (NFL) behavior\cite{senthil,sachdev2,Vidhyadhiraja,sakai,eder,zaanen2}
arises near the half-filled insulating state is one of the key
questions for unraveling the physics of not only the cuprates but
that of correlated electron systems more generally.  Features considered to be of a NFL form can arise from a number of different sources.  These include: (1) proximity to a Mott insulator, or a quantum phase transition, or a VHS; (2) some form of (nanoscale) phase separation; and, (3) strong impurity effects.  In narrow band systems, such as heavy-fermion materials, there can also be effects related to a coherence transition in the narrow band with the associated Kondo physics.  In this section we discuss two aspects of NFL physics in the cuprates.

\subsection{Anomalous Spectral Weight Transfer}

\begin{figure}
\centering\rotatebox{0}{\scalebox{0.53}{\includegraphics{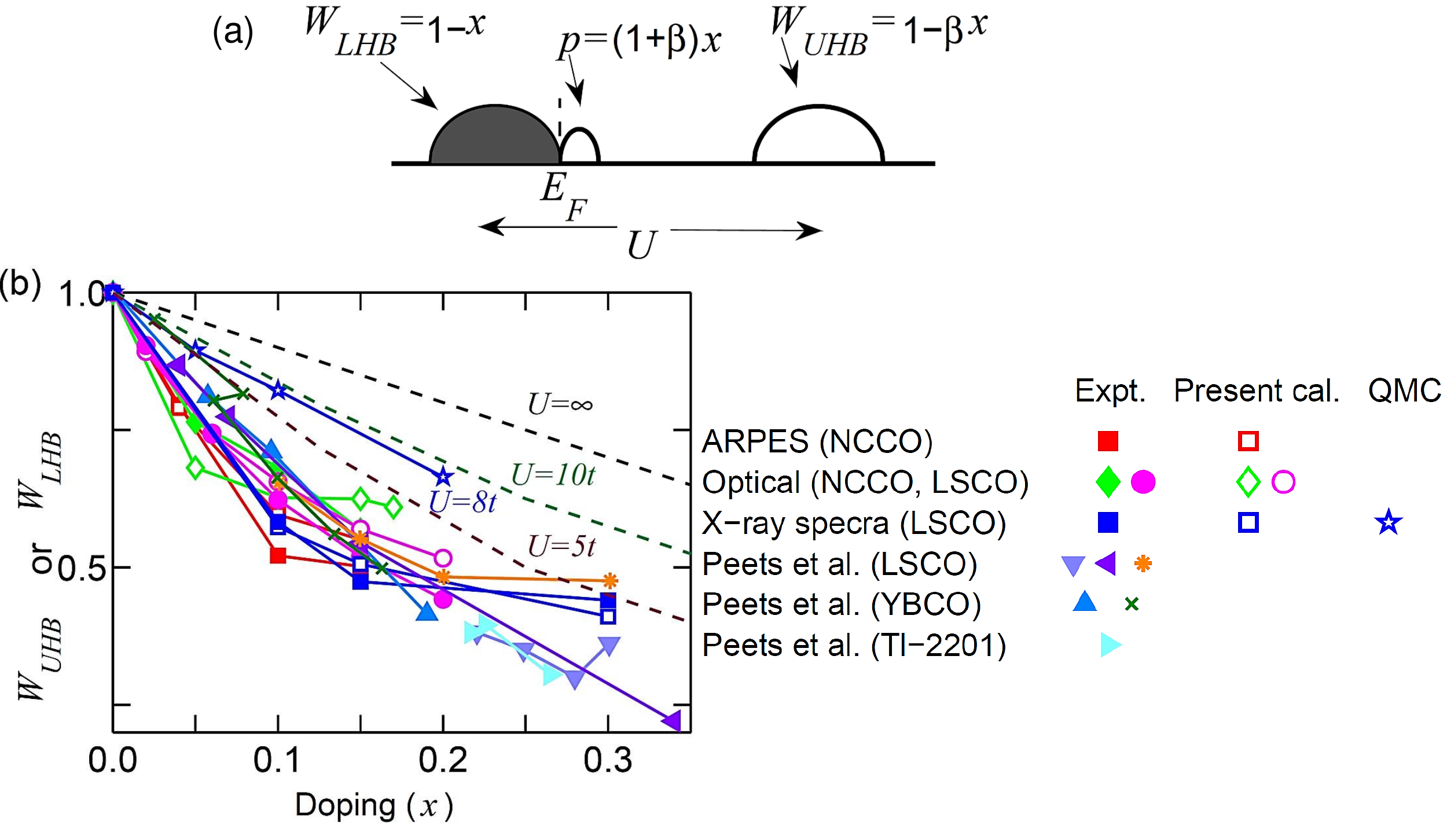}}}
\caption{ { ASWT in cuprates.} (a) Schematic diagram of ASWT for hole-doped cuprates. (b) Estimates of $W_{UHB}$ (for hole-doping) and $W_{LHB}$ (for electron-doping) from various experimental results (see legend)
\cite{armitage,Pellegrin,onose,onoseprb,uchida,XANESexp} are compared
with our theoretical results (open symbols of same color). Dashed
lines of various colors show exact diagonalization calculations
for different values of $U$ taken from Ref.~\onlinecite{EMS}. QMC
results\cite{jarrellb,maier} are plotted as blue stars.
All curves are normalized to $W_{UHB}\rightarrow1$ at half-filling. [From Ref.~\onlinecite{ASWT}.] }
 \label{aswt}
\end{figure}

In conventional insulators, each band holds a fixed number of states independent of doping. In this respect, Mott insulators are very anomalous: as holes are added, weight of the UHB decreases as spectral weight is transferred to the top of the LHB.  This anomalous spectral weight transfer (ASWT) with doping can be understood readily in the strong correlation limit, $U\rightarrow\infty$, Fig.~\ref{aswt}(a).\cite{EMS}  Since double-occupancy of an atom is forbidden in this case, when one electron is present, there is a large penalty $U$ for adding a second electron, i.e., there is an empty state in the UHB. If the electron is removed there is no more penalty, so that the hole is lost from the UHB and two holes appear at the top of the LHB.  For electron-doping the ASWT is associated with a loss of states in the LHB, and described by the mirror image of Fig~\ref{aswt}(a) with respect to $E_F$.

As $U$ decreases, an alternative source of ASWT arises.  At intermediate coupling, double-occupancy is reduced collectively via long-range magnetic order.  As magnetic order disappears at a QCP\cite{tanmoyprl,tanmoy2gap}, a much higher degree of double-occupancy is restored, and the UHB can completely vanish.  Indeed, the rate of ASWT is found to increase monotonically as $U$ decreases and, therefore, this rate can be used to quantify the degree of correlations in the cuprates. In a recent study\cite{ASWT}, we compared ASWT in XAS, ARPES, and optical measurements, finding similar doping dependencies in all these spectroscopies for both electron- and hole-doped cuprates. The results are consistent with intermediate coupling values of $U$, Fig.~\ref{aswt}(b), and also suggest that the effective $U$ varies with doping.

When $x$ electrons are removed from a half-filled insulator, Fig.~\ref{aswt}(a), $1+x$ holes are distributed between $p$ low-energy states, either at the top of the LHB or in the gap, and $W_{UHB}=1+x-p$ states in the UHB.  The ASWT can now be quantified in terms of the coefficient $\beta$, which is defined such that $W_{UHB}=1-\beta x$ and $p=(1+\beta)x$. For a conventional insulator $\beta =0$, while $\beta =1$ holds for a very strongly correlated Mott insulator ($t-J$ model or $U\rightarrow\infty$ Hubbard model). Exact diagonalization calculations on small clusters\cite{EMS,foot3b} (dashed lines in Fig.~\ref{aswt}(b)) find that reducing $U$ leads to larger values of $\beta$: $\beta\simeq1.5$ (at low doping) for $U=10t$ and $\beta\simeq2.0$ for $U=5t$. Shown also in Fig.~\ref{aswt}(b) are QMC results for $U=8t$ with $t'=0$ \cite{jarrellb,maier}, which are consistent with the exact diagonalization results.

In Ref.~\onlinecite{ASWT}, we extracted values of $\beta$ for cuprates from several spectroscopies, which are shown in Fig.~\ref{aswt}(b), along with XAS results on LSCO\cite{Pellegrin}, ARPES on NCCO\cite{armitage}, and optical absorption on both NCCO\cite{onose,onoseprb} and LSCO\cite{uchida}, including additional XAS data for LSCO, YBCO and TBCO from Ref.~\onlinecite{XANESexp}. All experimental measures of the high-energy spectral weight display a rapid falloff of the spectral weight with doping, and at low doping, this weight is seen to decrease almost linearly with doping with approximately the same slope of $\beta\simeq3.7$. These results are consistent with $U<5t$, indicating that the cuprates are far from being in the strong correlation limit. The observed falloff in Fig.~\ref{aswt}(b) supports a gap collapse around $x_{UHB}\sim 1/\beta =0.27$.  Notably, the value of $U\sim 5t$ is incompatible with the measured gap at half-filling. For example, optical spectra yield a gap consistent with $U\sim 8t$, but the value of $\beta$ calculated for fixed $U=8t$ is far from the experimental results in Fig.~\ref{aswt}(b). On the other hand, the experimental data can be explained by intermediate coupling model calculations\cite{markiewater,tanmoysw,waterfall} based on a {\it doping-dependent} effective $U$. The QP-GW results are plotted in Fig~\ref{aswt}(b)
as filled symbols of same color as the corresponding experimental data. A good agreement between QP-GW predictions and experiments is seen. 


Note that the ASWT plays out quite differently in different spectroscopies.  Due to strong electron-hole asymmetry in the cuprates, it is important to keep in mind the sensitivity of the probe to empty states for hole-doped cuprates, and to filled states for electron-doped cuprates. In particular, ARPES\cite{QPGW} or X-ray emission spectroscopies are well-suited for examining ASWT under electron doping, while XAS is appropriate for hole doping. Optical\cite{tanmoysw} and resonant inelastic x-ray scattering\cite{MBRIXS,RIXS} experiments would be applicable for both cases, as they measure the joint density of states. In principle, STM/STS\cite{Jouko} could follow either sign of charge, but would require a rather wide energy range of $\sim$2eV to see the full effect.

Within the QP-GW model, two main factors contribute to ASWT. Firstly, the
pseudogap collapses with doping, shifting the optical MIR peak to low energies while transferring weight to the Drude peak.  Secondly, the residual incoherent weight associated with the Hubbard bands decreases with doping\cite{tanmoysw} due to a decrease in magnon scattering.  Within our model, part of this decrease is due to the doping dependent value of the effective $U$. The separation of low and high energy spectral weight is made easier by the presence of a clear energy where the spectral weight is minimal in each spectroscopy.  In all cases, our model finds this minimum to be related to the HEK, which separates the coherent and incoherent parts of the spectra. In order to test our method for extracting high energy spectral weight, we considered optical spectra computed using the DCA with $U=8t$, which are in good agreement with the exact diagonalization results [blue stars in Fig~\ref{aswt}(b)].  Fig~\ref{aswt}(b) shows that neither exact diagonalization nor DCA results using a constant $U$ capture the ASWT, suggesting that the doping dependence of $U$ is an effect beyond the simplest $t-U$ one-band Hubbard model.  Recent DMFT calculations, which include next-nearest-neighbor hopping,  provide a better description of the doping evolution of the cuprates\cite{comanac,millis,DMFTb1}. Interestingly, the doping dependence of an effective $U$ can be explained by long-range Coulomb screening.\cite{tanmoysw}.  This effective doping dependence is reduced significantly in going to a three-band model\cite{MBRIXS,DMFTcuprate}. But multiple-band and long-range Coulomb effects are not included in the Hubbard model, and are therefore not captured in the QMC, exact diagonalization, and DMFT calculations.

\subsection{Non-Fermi-liquid Effects due to Broken Symmetry Phase}

Here we discuss how one property which is very difficult to understand within
the conventional Fermi liquid theory finds a natural explanation in a model of an AFM Fermi liquid.  In the Fermi liquid theory, dispersion and spectral weight would both have the same renormalization, unless the self-energy has a strong momentum dependence, which is not the case in cuprates\cite{markiewater,jarrellb}. In contrast, quasiparticles in cuprates are very fragile or gossamer-like\cite{goss}: the spectral weight of the quasiparticles fades away on approaching the insulator and renormalizes to zero at half-filling\cite{Yos06,yoshida}, even though the electronic dispersion remains finite or even appears to become unrenormalized with underdoping.\cite{sahrakorpi} In contrast, the electronic specific heat behaves in a more or less Fermi-liquid manner over the entire doping range from the overdoped metal to the insulator\cite{yoshida,Komiya,brugger,li}. These results clearly demonstrate that a non-Fermi-liquid or `strange metal' superconductor emerges from the Fermi-liquid background as doping is reduced. We have shown\cite{tanmoygoss} that this can be understood in a density-wave ground state, where the FS breaks up into pockets, and the self-energy develops a strong momentum dependence.  The model is simple enough that analytic expressions for the renormalization factors can be obtained. In addition to the SDW order, we have analyzed other candidates for the competing order including charge, flux and $d-$density waves\cite{tanmoy2gap}, and find that the results are insensitive to the particular nature of the competing order.  Our results are in good overall agreement with an intermediate coupling calculation by Paramekanti, et al.\cite{PRTRVB}, which approached the problem from the strong coupling (RVB) limit.

\begin{figure*}[top]
\centering
\rotatebox{0}{\scalebox{0.45}{\includegraphics{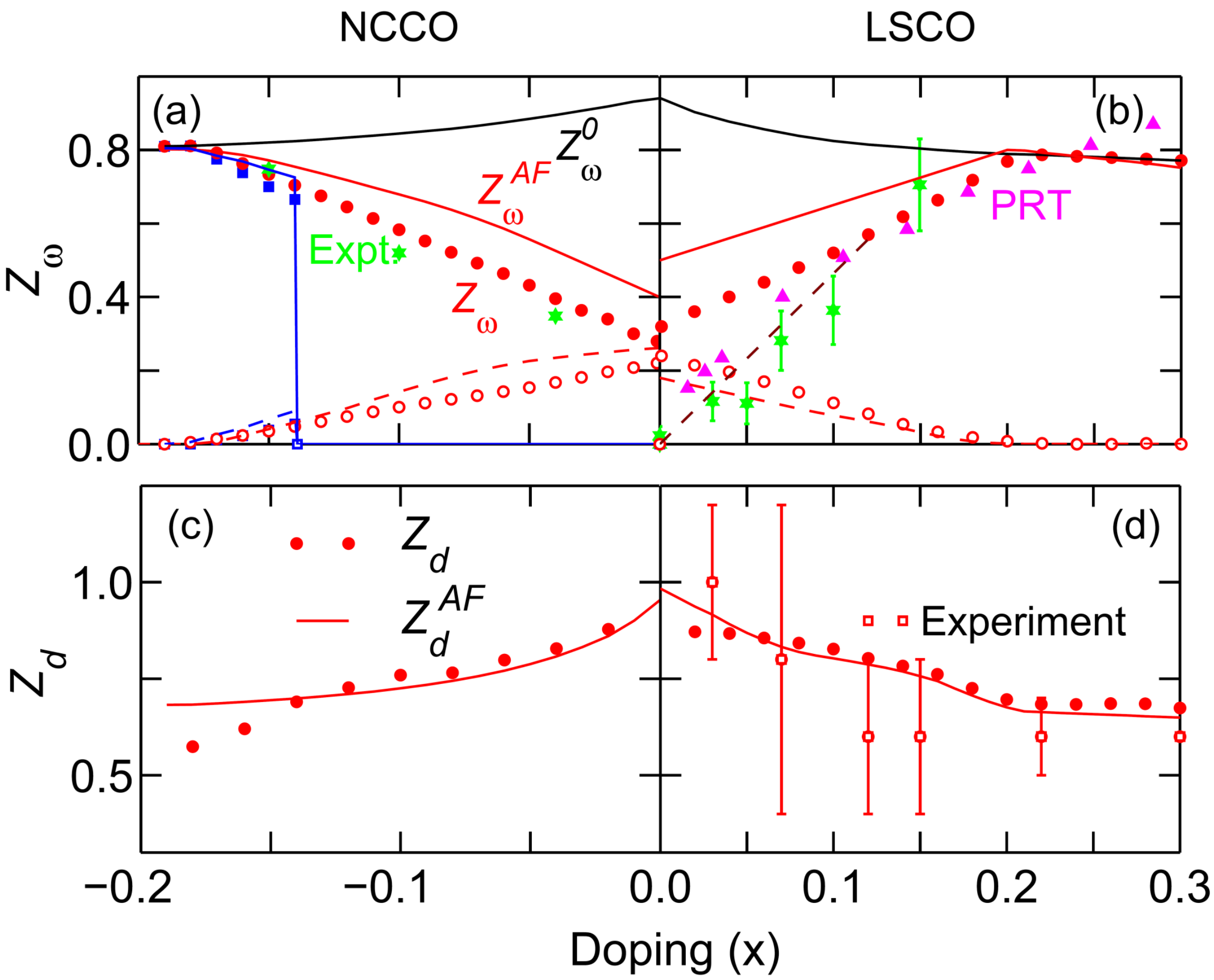}}}
\caption{ { Renormalization factors in cuprates.}
(a) $Z_{\omega}$ at $E_F$ along the antinodal (red) and nodal (blue) directions for NCCO. Filled (open) symbols give the main (shadow) bands, compared with the corresponding analytical approximation, $Z_{\omega}^{SDW}$, plotted as solid (dashed) line of same color. ARPES results (green symbols) are extracted from
Ref.~\onlinecite{armitage}. (b) Same as (a) but for LSCO along the nodal direction. These results are compared with calculations of Ref.~\onlinecite{PRTRVB} for hole doping, and ARPES results\cite{yoshida} for LSCO along the nodal direction. All experimental data and theory results of Ref.~\onlinecite{PRTRVB} in (a) and (b) are normalized to highlight their doping evolution. Brown dashed line shows that if there is nanoscale phase separation in LSCO then $Z_{\omega}$ would scale linearly with doping in the extreme underdoped region. 
(c) Dispersion renormalization $Z_d$ for NCCO along the antinodal direction, compared with the approximate analytical result $Z_d^{SDW}$ of Eq.~\ref{ZdAF} (solid line). (d) Same as (c) but for LSCO along the nodal direction, and the  related experimental data\cite{sahrakorpi}. [From Ref.~\onlinecite{tanmoygoss}.] 
} 
\label{md1d}
\end{figure*}
%

%

%
The spectral weight renormalization $Z_{\omega}$ is defined as the jump in the momentum density $n({\bf k})$ (see Section~\ref{S:ElSpec}.6) at the FS,
\begin{equation}\label{Zw}
Z_{\omega}=\Delta n(k_F),
\end{equation}
where $k_F$ is the Fermi momentum. $Z_{\omega}$ is plotted in Fig.~\ref{md1d}(a) for NCCO and in Fig.~\ref{md1d}(b) for LSCO as a function of doping. Shown also are other theoretical estimates for this renormalization factor.  The experiments are seen to be in good agreement with $Z_{\omega}$ calculated from Eq.~\ref{Zw}, but disagree strongly with the conventional Fermi-liquid form,
\begin{equation}\label{Zw0}
Z_{\omega}^0 = \left(1-\frac{\partial\Sigma^{\prime}(\omega)}
{\partial\omega}\right)_{\omega=0}^{-1}.
\end{equation}
It is interesting also to consider an approximate analytical form for $Z$ in an SDW metal,\cite{tanmoygoss}
\begin{equation}\label{ZwAF}
Z_{\omega}^{SDW}= \frac{Z_{\omega}^0}{2}\left[1\pm\left(1
+\left(\frac{2\Delta}{\xi_{k_F}-\xi_{k_F+Q}}\right)^2\right)^{-1/2} \right],
\end{equation}
which captures the correct doping dependence. Similar results are found for LSCO, Fig.~\ref{md1d}(b), except that in the underdoped region the ARPES data\cite{yoshida} seem to extrapolate smoothly to zero at half-filling (dashed line), which may be related to nanoscale phase separation in LSCO, as discussed below.

%
%

A similar analysis can be carried out for the Fermi velocity. Despite a substantial decrease of $Z_{\omega}$, the quasiparticle velocity $v_F$ does not diminish upon entering into the pseudogap phase. Although in the SDW-mean-field case $v_F$ decreases smoothly with underdoping as the gap grows, when a self-energy is introduced, the reduction of the coherent spectral weight ($Z_{\omega}$) with underdoping compensates this decrease, leading to a net enhancement of $v_F$. The corresponding dispersion renormalization factor,
\begin{eqnarray}\label{Zd}
Z_d=v_F/v_F^0,
\end{eqnarray}
is plotted in Figs.~\ref{md1d}(c,d), where $v_F^0$  is the bare (LDA) Fermi velocity, and compared with the usual Fermi liquid dispersion renormalization, 
\begin{eqnarray}\label{Zd0}
Z_d^0=Z_{\omega}^0Z_{k_F}^0=Z_{\omega}^0\left(1+\partial\Sigma'/v_F^0
\partial k\right).
\end{eqnarray}
Since the QP-GW self-energy is approximately $k$-independent, $Z_d^0\approx Z_{\omega}^0$. In sharp contrast, the calculated $Z_{d}$ in Fig.~\ref{md1d}(c) shows a striking opposite doping dependence to $Z_{\omega}$. This is because the SDW-gap introduces a new $k$-dependence in the dispersion renormalization given by \cite{tanmoygoss}
\begin{equation}\label{ZdAF}
Z_d^{SDW}=Z_{\omega}^0Z_{{\bf k}_F}^{SDW}=Z_{\omega}^0
\left(1+\frac{\Delta^2}{\xi_{k_F}\xi_{k_F+{\bf Q}}} \right).
\end{equation}
Figures~\ref{md1d}(c,d) compare $Z_{d}$ with $Z_d^{SDW}$. The doping dependence of $Z_{d}$ implies that as we go toward the Mott insulator, the dispersion tends towards the LDA-bands, consistent with LSCO results (blue open circles)\cite{sahrakorpi}. The opposite doping dependences of $Z_d$ and $Z_{\omega}$ can be understood from the analytical formulas, Eqs.~\ref{ZdAF} and~\ref{ZwAF}: $Z_d^{SDW}\sim\Delta_{SDW}$  increases with underdoping, while $Z_{\omega}^{SDW}\sim 1/\Delta_{SDW}$ decreases.

\begin{figure}[htop]
\centering
\rotatebox{0}{\scalebox{0.6}{\includegraphics{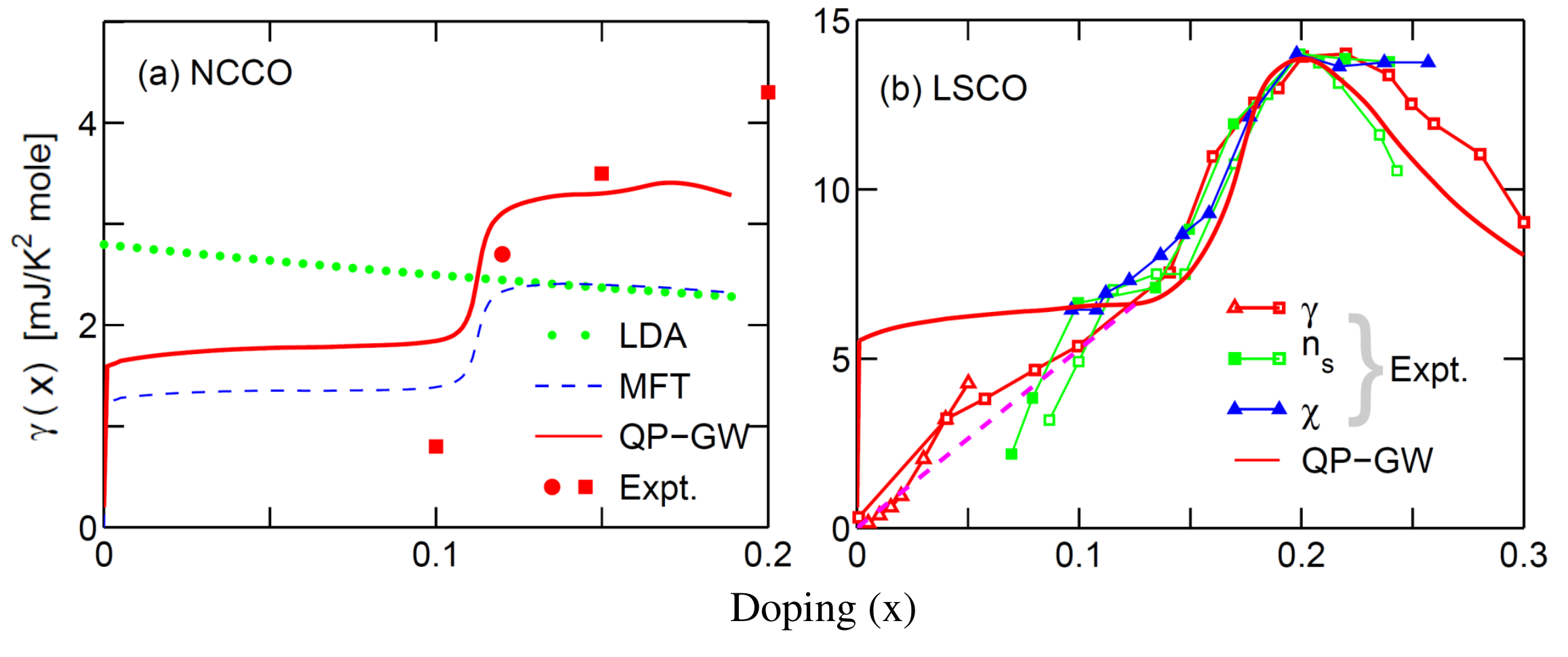}}}
\caption{ (a) Specific heat coefficient $\gamma(x)$ from various calculations, and experimental results on NCCO (red squares)\cite{brugger} and PLCCO
(red circles)\cite{li}. (b) Same as (a) but for LSCO (red squares\cite{yoshida} and triangles\cite{Komiya}). Results are compared with $n_s$ [filled green squares\cite{yoshida} and open green squares\cite{yang}] and $\chi$
(blue)\cite{LSCO4}), all normalized at the VHS. The red dashed line shows that in underdoped LSCO $\gamma$ scales linearly with doping. Theoretical $\gamma$ has been scaled by a factor of 1.1, consistent with a weak electron-phonon renormalization. [From Ref.~\onlinecite{tanmoygoss}.]}
\label{fspheat}
\end{figure}
%


The Sommerfeld specific heat coefficient $\gamma =c_V/T$ is given by\cite{abrikosov}
\begin{eqnarray}\label{gamma}
\gamma = c_V(T)/T \approx \frac{2\pi^2k_B^2}{3}N(0)/Z_{\omega}^0,
\end{eqnarray}
where $N_0(0)$ is the DOS in the SDW state without self-energy corrections.  Figure~\ref{fspheat}(a) compares experimental values of $\gamma$ in NCCO\cite{brugger} and Pr$_{1-x}$LaCe$_x$CuO$_4$ (PLCCO)\cite{li} with several calculations including bare LDA, mean-field theory (MFT) of SDW, and QP-GW model. The nearly flat $N(0)$ for $x<0.11$ reflects the quasi-two-dimensionality of the electron pocket (constant DOS) in cuprates. The step at $x\sim 0.11$ signals the appearance of the hole-pocket. Corresponding data\cite{yoshida,Komiya} for LSCO are plotted in Fig.~\ref{fspheat}(b), along with the related quantities $n_s$\cite{yoshida,yang} and $\chi$\cite{LSCO4}.
The large peak near $x=0.2$ is due to the presence of the VHS near $E_F$. These results are in striking contrast to the strong coupling limit where $\gamma$ would diverge with the effective mass as $x\rightarrow 0$.\cite{KOg} While agreement with experiment is quite good in LSCO for $x\ge 0.10$, at lower dopings $\gamma\rightarrow 0$ as $x \rightarrow 0$, an effect that is not captured by our model. This discrepancy could be due to a Coulomb gap\cite{Komiya} (green symbols in Fig.~\ref{gap}) and/or signal a nanoscale phase separation. The linear dashed line in Fig.~\ref{fspheat}(b) illustrates
the corrected form expected in the latter case.  Note that nanoscale phase separation would produce the dashed line seen in Fig.~\ref{md1d}(c) as well as explain the anomalous doping dependence of the chemical potential.\cite{FIMT}

%
%

\subsection{Role of VHS}
A number of studies show that the VHS plays a significant role in high-$T_c$ superconductivity and/or in driving competing phases in cuprates via the divergence in the DOS in an ideal 2D material\cite{VHS0,VHS00,VHS1}. It is puzzling, therefore, that many instabilities are found theoretically to optimize near the VHS, whereas the pseudogap order vanishes in a QCP near optimal doping where the VHS is expected.  Detailed susceptibility calculations reveal a factor that appears to have been overlooked in earlier models, namely, conventional FS nesting is also present in the cuprates, and could compete with VHS nesting\cite{MGu2}.  In particular, near the VHS, the antinodal parts of the FS are flat and nearly parallel, leading to a nearly 1D antinodal nesting (ANN).  The ANN has the important feature that when the VHS is approached, the nesting is lost as the FS pinches down to produce the VHS. ANN must thus have a QCP near the VHS, making it a prime candidate for the pseudogap, at least at higher dopings, Section~\ref{S:DWs}.2.2.  

Initially, it was thought that the superconducting $T_c$ would optimize at $x_{VHS}$, the point at which $E_F$ crosses the VHS, and that is approximately the case in LSCO\cite{VHS4}, but in other cuprates $x_{VHS}$ lies in the overdoped regime, presumably due to a competition between Mott physics and VHS physics.  There seems to be an {\it optimal VHS doping} away from half-filling, and the $T_c$ is suppressed if $x_{VHS}$ is too close (LSCO) or too far (Bi2201) from zero. Indeed, in Bi2201 there is tunneling evidence that the VHS crosses $E_F$ close to the termination of the superconducting dome as in Fig.~\ref{Bi2201}(b)\cite{VHS2,VHS3,nieminenPRB2}; see also Fig.~\ref{fig:2201domes}.  This is quite suggestive, since the linear-in-$T$ resistivity also seems to terminate at the same doping\cite{MGres}, and the VHS has long been known to be a source of linear-in-$T,\omega$ scattering rate.  Initially, there was some question whether this linearity in scattering rate translated into a linearity in resistivity,\cite{HluRi} but more recent work shows a qualitative agreement between the predicted doping- and temperature-dependence of the resistivity near a VHS with the corresponding experimental results for LSCO, at least in the overdoped regime.\cite{Rice2} In a number of correlated materials, a nearly pure linear-in-$T$ resistivity is found at the competing-order QCP, in a narrow doping range near optimal doping, but a linear-in-$T$ contribution to the resistivity persists over a wide doping range, finally vanishing near the end of the superconducting dome\cite{linT}.

Near a QCP, there are two relevant VHS's, the paramagnetic VHS and the VHS of the ordered phase.  Specifically, if we consider the $(\pi,\pi)$-magnetic order, the VHS of the lower magnetic band must cross $E_F$ close to the QCP, and if it crosses before the QCP, then three separate dopings will be dominated by linear scattering: (1) Doping where the VHS of the lower magnetic band crosses $E_F$;\cite{nieminenPRB2} (2) Doping of the QCP; and, (3) Doping where the VHS of the reconstructed large FS crosses $E_F$, which appears to be where the superconducting dome ends. In earlier work, it was assumed that all three crossings merged at a single doping\cite{VHS00}, but this does not have to be the case, as illustrated in Fig.~\ref{phasediagN}(a). 

\begin{figure}
\centering
\epsfxsize=0.9\textwidth\epsfbox{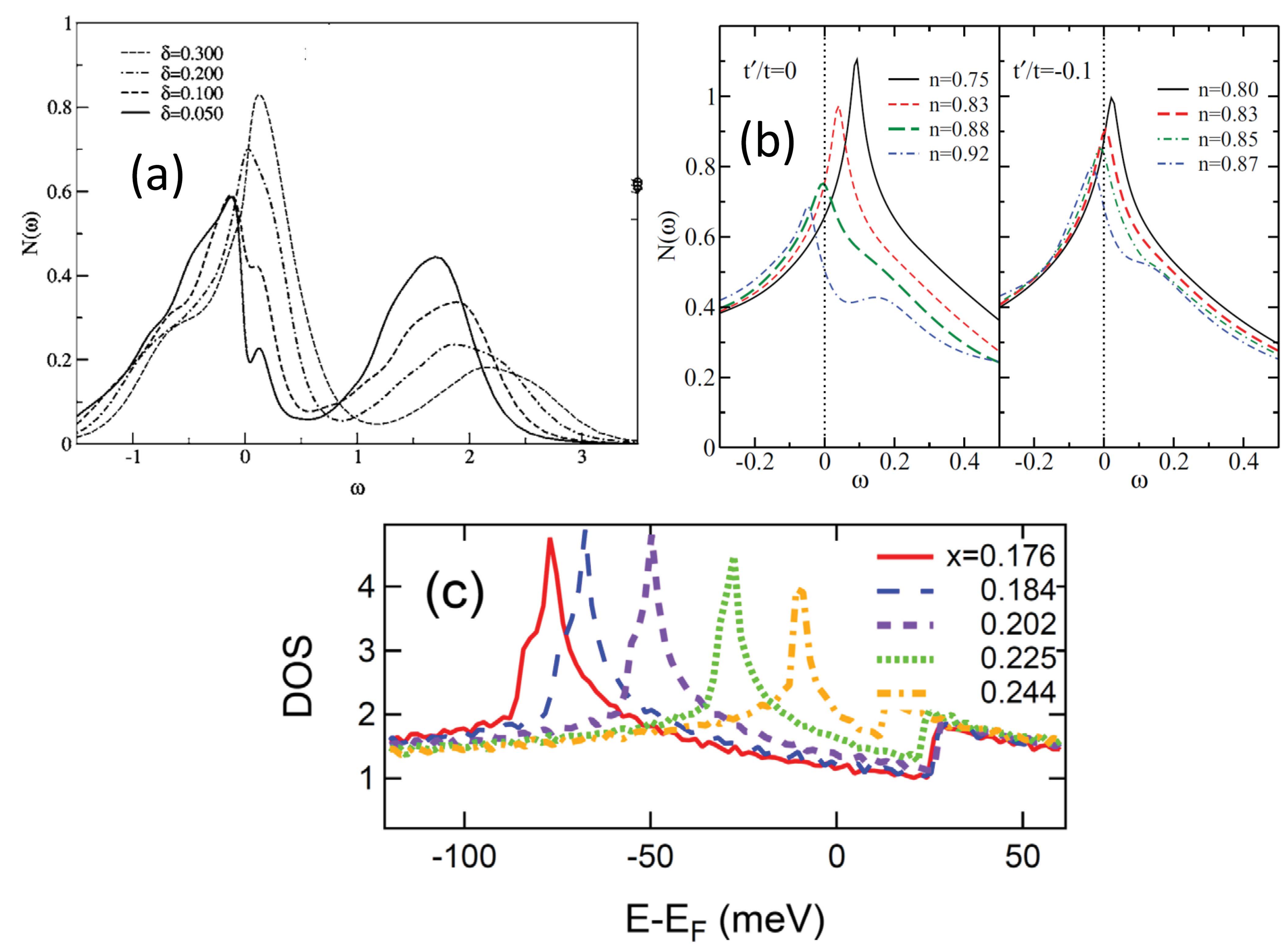}
\caption{ {Origin of the pseudogap.} (a) DOS for a series of dopings $\delta$ for the $t-U$ Hubbard model calculated within the DCA. [From Ref.~\onlinecite{jarrellb}.] (b) Similar calculations showing effect of $t'$. [From Ref.~\onlinecite{JarVHS2}.] In frames (a) and (b) the frequencies are given scaled by $t$.
(c) Calculated DOS for a series of hole-doped cuprates, assuming
$U_{eff}=2.3t$ for various dopings $x$: 0.176 (solid line), 0.184 (long dashed line),
0.202 (short dashed line), 0.225 (dotted line), and 0.244 (dot-dashed
line). [From Ref.~\onlinecite{MkII}.] 
}
\label{fig:n11}
\end{figure}

In this connection, recent DCA results were interpreted in terms of a VHS crossing $E_F$ near a quantum critical point.\cite{JarVHS,JarVHS2} However, Ref.~\cite{DCAVHAndy} pointed out that this cannot be a conventional VHS instability, since the gap opens after the Fermi level has crossed the VHS.  We suggest that these results could be understood if the QCP was SDW collapse, and the observed VHS was actually the VHS of the LMB. The computations of Ref.~\onlinecite{JarVHS} are for the $t-U$ Hubbard model ($t'=0$), for which the paramagnetic VHS falls at exactly half-filling, suggesting that what Ref.~\cite{JarVHS} may be seeing at finite $x=\delta$ (Fig.~\ref{fig:n11}(a)) is the VHS of the lower magnetic band. As doping $\delta$ increases, the VHS peak moves toward $E_F$, crossing it near $\delta =0.2$. Figure~\ref{fig:n11}(b) shows that a similar transition occurs for finite $t'$. Note that as the VHS approaches $E_F$, there is also a small pseudogap just at $E_F$ ($\omega =0$), which disappears after the VHS crosses $E_F$.

Insight into this pseudogap feature is obtained by considering the Hartree-Fock results of Fig.~\ref{fig:n11}(c), which show the gap-collapse as a function of doping of a form expected for a $(\pi,\pi)$-SDW.  The sharp dip near $\omega =0$ in Figure~\ref{fig:n11}(a) and~(b) is seen to have the same shape and doping dependence as the dips in Figure~\ref{fig:n11}(c) [see also features A and B in Fig.~\ref{phasediagN}(c)].  These features represent the VHS of the lower magnetic band and the bottom of the upper magnetic band, and as the magnetic gap collapses they merge and evolve into the VHS of the full band.  This is exactly what is seen in Fig.~\ref{fig:n11}(a), and the relation of this SDW phase to the pseudogap is discussed in Section~\ref{S:DWs}.1.2.
We note that the higher energy peak near $\omega =1.5$ in Fig.~\ref{fig:n11}(a) corresponds to the VHS of the upper magnetic band, feature D in Fig.~\ref{phasediagN}(c).  In a mean-field calculation, this peak vanishes when the gap collapses and features C and D merge, but when a self-energy is included, an incoherent remnant of this peak remains as seen in  Figs.~\ref{md2d4}(d) and (e) due to the kink at positive energies, and contributes to the residual Mott gap seen in the optical spectra of  Fig.~\ref{alp2F2}.  

%



Recent CDMFT calculations  using 2$\times$2 plaquettes have found a similar pseudogap phase diagram, with a transition at low doping between two metallic phases involving a change of FSs\cite{SoHauT,SoHauT2}. In these pure Hubbard model ($t'=0$) calculations, the small FS-metal at low doping plays the role of the doped $(\pi,\pi)$ SDW-metal found in DCA~\cite{jarrellb}, and the large FS-metal corresponds to the nonmagnetic state.  At low-$T$ the transition is first order, but at higher $T$, the two phase regime ends at a critical point, and emanating from that critical point is a `Widom line',\cite{SoSHT} as in conventional liquid-gas transitions\cite{HEStan}. 
The compressibility has a peak along this line, and diverges at its end point.  Similar results are found in mean-field studies of 
electron-doped cuprates.  The mean-field magnetic transition is mainly second order, but near the QCP it develops a tricritical point and ends in a first order transition between magnetic pockets and a large nonmagnetic FS\cite{Mkstripes}.  
Due to Mermin-Wagner physics\cite{MW}, the finite-$T$ crossover will have many properties of a Widom line.  
Thus, at the onset of a first order transition, the compressibility must diverge.  Most other properties will be spin-glass like due to the critical slowing down.  Different probes will detect the crossover at different temperatures due to the internal time-scale of the probe\cite{Andregas,MkCord}.  One caveat to the preceding picture of a Widom line should be noted.  The particular 2$\times$2 plaquette utilized in this study can produce anomalous results, as it does not allow a clear separation between the nodal and antinodal regions of the Fermi surface.\cite{DMFT4}  Hence, these calculations should be repeated on a larger cluster.

It is interesting to speculate whether the Widom line of the pure Hubbard model is also related to the trace of the VHS of the lower magnetic band, terminating in the critical end point. This interpretation is suggested by a number of factors: (1) Figure~9 of Ref.~\onlinecite{SoHauT2} shows the effective chemical potentials at three $k$-points vs doping and Hubbard energy $U$.  For $(\pi,0)$, corresponding to the VHS, $\mu_{eff}$ crosses $E_F$ very close to the doping of the first order transition for all $U$s for which there is a first order transition ($5.8<U<6.2$), but crosses at half-filling for smaller $U$'s, where there is no transition.  [Recall that when $t'=0$, the nonmagnetic VHS crosses $E_F$ at half-filling.]  For larger $U$'s there is no longer a first order transition, but  $\mu_{eff}$ crosses $E_F$ close to the point where the DOS at $(\pi,\pi)$ starts to decrease [Fig. 6 of Ref.~\onlinecite{SoHauT2}], marking the opening of a pseudogap; (2) As noted in Sections~\ref{S:QP-GW}.3 and~\ref{S:QP-GW}.5, upon entering a phase of short or long-range order, the self-energy develops a strong momentum dependence, Eqs.~\ref{SDWSE} and \ref{YRZ3}.  This in turn leads to a strong momentum dependence of $\mu_{eff,k}=\mu-\Sigma'_k(\omega\rightarrow 0)$, which represents the opening of a [pseudo]gap.  Such a strong momentum dependence of $\mu_{eff}$ is found in the small-FS phase, Fig.~9 of Ref.~\onlinecite{SoHauT2}; (3) The large compressibility\cite{SoSHT} and strong scattering rate along the Widom line [Fig.~10 of Ref.~\onlinecite{SoHauT2}], as well as the broad peak in spin susceptibility (Fig.~14 of Ref.~\onlinecite{SoHauT2}) are all consistent with the presence of a VHS. 


 In trying to construct a viable VHS model, it is important to keep in mind that the simple idea that $x_{VHS}$ can be found from a single LDA calculation, assuming rigid band filling, is of very limited applicability since $x_{VHS}$ varies with hole doping, and it is shifted by strong correlations\cite{JarVHS,JarVHS2} and ionic substitutions\cite{TallonThes}.

\section{Superconducting state}\label{S:SC}
\begin{figure}
\centering
\rotatebox{0}{\scalebox{0.55}{\includegraphics{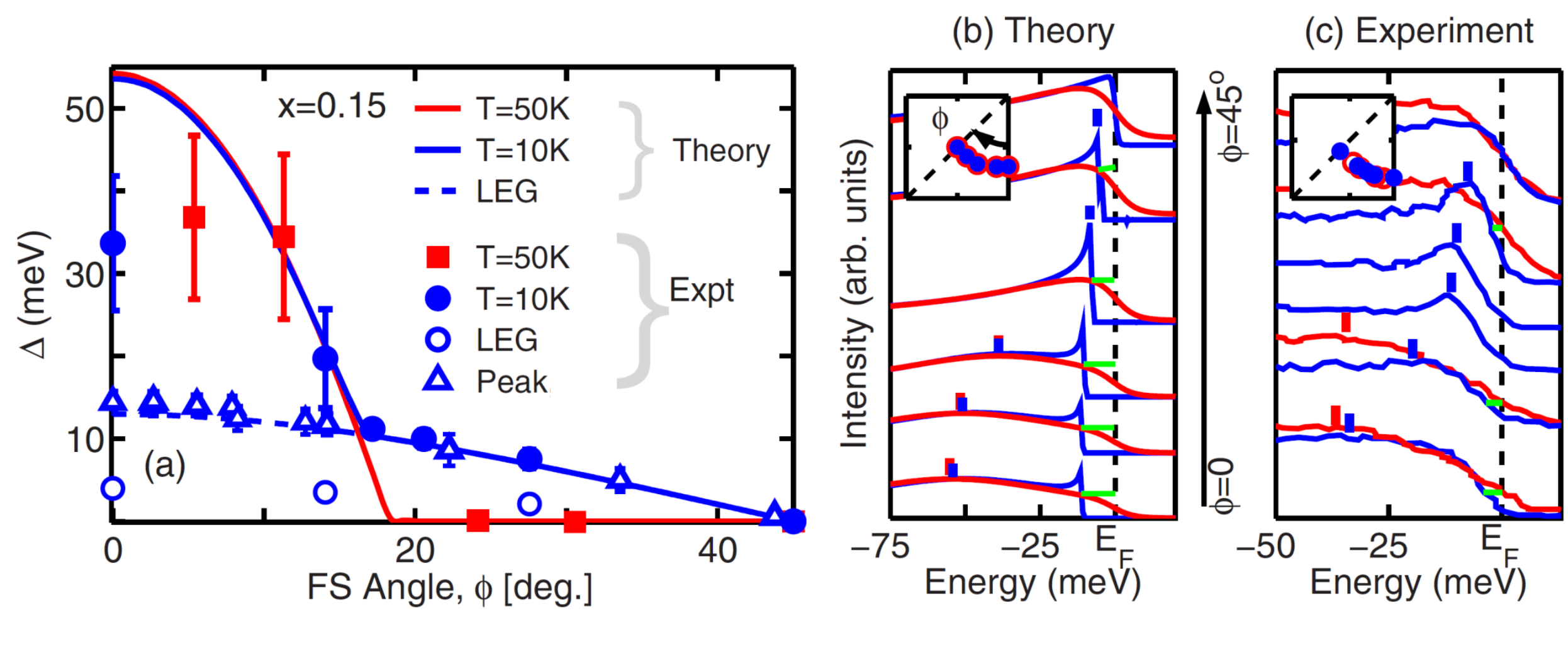}}}
\caption{ (a) Theoretical and experimental angle dependence of various gaps $\Delta$ in underdoped LSCO (x=0.15) along the FS, where $\phi$ =0 denotes the antinodal and $\phi=45^o$ the nodal direction. Solid lines give our results for the normal (T=50 K, red line) and the superconducting (T=10 K, blue line) states, and the corresponding experimental data are plotted with filled symbols of the same color (Ref.\onlinecite{Terashima}). Blue dashed line shows the computed coherence peak position at T=10 K, while blue open circles give
the corresponding experimental leading-edge-gap, green lines in frame (c). Blue
open triangles denote $\Delta$ based on peak positions from the data of
Ref.\onlinecite{Shi1}. (b) Computed spectral functions at different momentum points along the FS (see inset) for the normal (T
=50 K, blue lines) and the SC (T=10 K, red lines) states. Spectra at
the bottom of the figure refer to the antinodal direction ($\phi$ =0),
while those at the top to the nodal direction ($\phi =45^o$). Blue and red
tick marks on the spectra denote total gap values, while green lines
mark the leading-edge-gap. (c) Same as (b), except that this figure refers to the experimental EDCs taken from Figs. 2(a) and 2(b) of Ref.\onlinecite{Terashima}. In order to highlight spectral changes, normal state spectra (red) are plotted on top of those for the SC-state (blue) in several cases, even though these pairs of spectra are not taken at exactly the same angle $\phi$. [From Ref.~\onlinecite{tanmoy2gap}] }
\label{CohPeak}
\end{figure}

%
\begin{figure}[h]
\centering
\rotatebox{0}{\scalebox{0.53}{\includegraphics{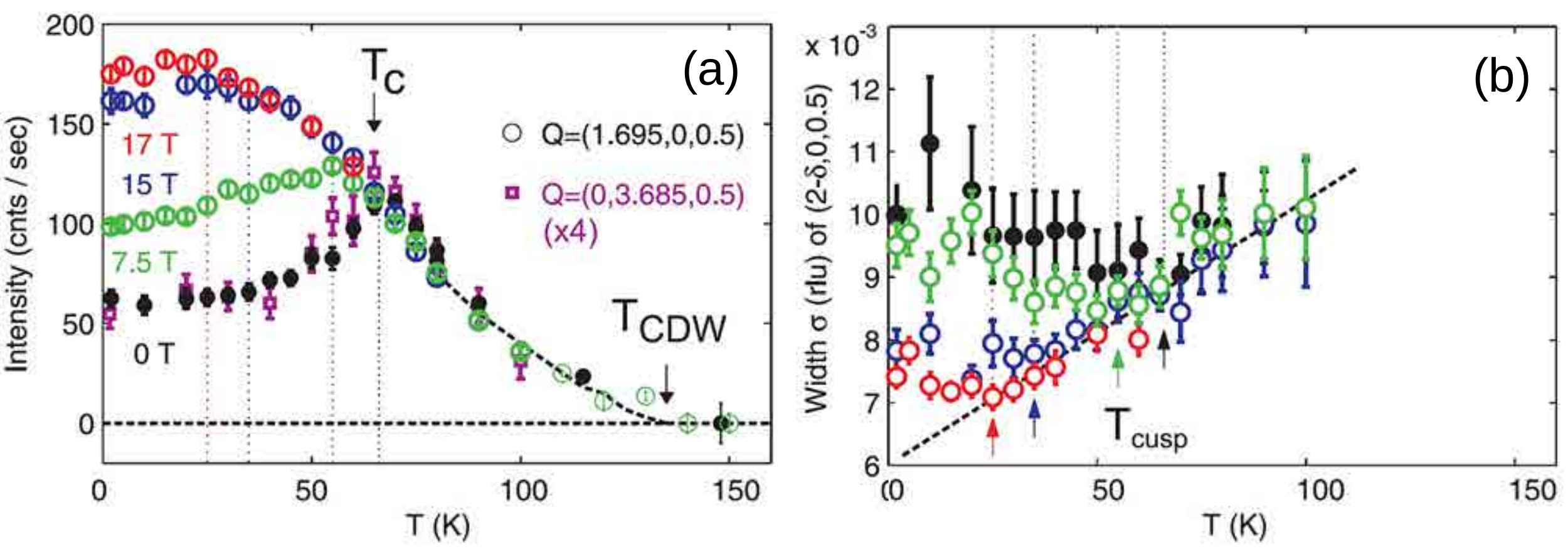}}}
\caption{ { CDW-SC competition in YBCO.} (a) Temperature dependence of the peak intensity at $Q$ = (1.695,0,0.5) (circles) and (0,3.691,0.5) (squares) for different applied magnetic fields. Square data points have been multiplied
by a factor of four. (b) Gaussian linewidth of the $Q$ = (1.695,0,0.5) CDW modulation as a function of temperature. [From Ref.~\onlinecite{YCDW4}.] } 
\label{CDWYBCO}
\end{figure}
\subsection{Separating the SC-Gap from Competing Orders}
Figure~\ref{CohPeak}(a) is a typical ARPES result demonstrating the two gap scenario\cite{Terashima,Shi1}.  At high temperatures, an antinodal gap is seen with a gapless Fermi arc or pocket in the nodal region.  At low temperatures, superconductivity turns on, and the arc becomes gapped with a $d$-wave gap, while the antinodal region remains nearly the same.  This is usually interpreted as superconductivity competing with some density wave order, which is responsible for the antinodal gap.  However, Bilbro and McMillan showed that when two gaps compete, the second gap grows at the expense of the first.\cite{bilbro}  This can be seen in Fig.~\ref{CDWYBCO}, where a CDW-gap develops in YBCO, growing with decreasing temperature until the SC-gap is reached, at which point superconductivity grows and the CDW-gap is suppressed.\cite{YCDW4} A magnetic field kills superconductivity and restores the CDW-gap. This suggests that, hidden under the antinodal gap in Fig.~\ref{CohPeak}(a) are contributions from both gaps.  Indeed, this can be seen in Figs.~\ref{CohPeak}(b,c): in the SC-state, non-pairbreaking scattering becomes ineffective, leading to a sharpening of the spectrum and the appearance of coherence peaks.  The antinodal coherence peak (bottom curves of Fig.~\ref{CohPeak}(b,c)) is consistent with data from Bi2212\cite{Kondo},  
developing at a $T_{pair}>T_c$.\cite{Kondo1}

\subsection{Nodeless d-wave gap due to competition with SDW}

The nodal d-wave SC-gap of the optimal and overdoped samples transforms into a nodeless or fully gapped SC state in the underdoped region of both electron-\cite{swave_edop93,swave_edop94,swave_edop94b,swave_edop98,swave_edop02,swave_edop02b} as well as many hole-doped cuprates.\cite{LSCO2000,ccoc2004,Tanaka,Inna,LSCO2013,bi22012013,YBCO2013} On the electron-doped side, we have demonstrated that strong SDW order removes the hole-pocket from the nodal region in the underdoped system, and the total band gap (=$\sqrt{\Delta_{SDW}^2+\Delta_{dSC}^2}$) appears to be nodeless even though the underlying pairing maintains $d$-wave symmetry.\cite{tanmoyprl,swaveTing} This explanation does not hold in the hole-doped case, where a hole-pocket is present at any doping. We have recently extended our QP-GW modeling to treat the transition from an SDW phase to d-wave superconductivity, accompanied by a robust triplet superconductivity, to obtain an odd parity Fulde-Ferrell-Larkin-Ovchinnikov state, consistent with the observed fully gapped SC-state on the hole-doped side.\cite{DasFFLO}

\subsection{Glue Functions}

In conventional superconductors, phonons responsible for superconductivity modify the electronic spectrum. This phononic imprint can be measured and used to extract a weighted electron-phonon coupling DOS, or Eliashberg function, $\alpha^2F$.\cite{Eliash,DougParks} For an electronic pairing mechanism, there is a similar function, $\sim U^2\chi_d^{\prime\prime}$, where $\chi_d$ is an appropriate $d$-wave susceptibility.\cite{flex2} Attempts have been made to extract this electronic `glue' function from photoemission, tunneling, and optical experiments.    

There is an ongoing debate concerning the dominant bosons responsible for pairing in cuprates.  Two sharply different scenarios have been proposed. One viewpoint holds that a `pairing glue' is not necessary as the pairs are bound by a superexchange interaction $J=4t^2/U$, and the dynamics of pairs involves virtual excitations above the Mott gap set by the energy scale $U$.\cite{PWA,glue,markiesc,Emery,CGP} In the opposing view, pairing is mediated by a bosonic glue which originates from SDW spin fluctuations\cite{resy1,resn,woo}. We may think of this as a debate between unconventional pairing with bosons of energies $>100$~meV ($J$ or $\sim$~2~eV $U$/CT scale) vs a more conventional pairing with lower energy bosons (around the scale of the magnetic resonance mode or lower energies).~\cite{glue} In fact, many studies adduce a finite high-energy contribution to pairing fluctuations\cite{markiesc,glue,Trem2,Civelli,Hankesc}, in addition to the low-energy AFM fluctuations, some exceptions notwithstanding\cite{Milless}. Notably, Ref.~\cite{Civelli} finds pairing fluctuations up to energies of $\sim 2t$, but the calculation was only carried out to energies of $\le 2.5t$, which is too low to see higher-energy fluctuations. Ref.~\cite{Hankesc} concludes that high-energy fluctuations are present in the one-band Hubbard model, but not in the three-band model. A recent analysis of the 2D Hubbard model based on the dynamical cluster approximation (DCA) indicates that superconductivity is not driven by resonating valence bond physics.\cite{nonRVB}

While many experimental probes show the presence of low-energy bosons in the cuprates\cite{SchaCar,DoHoTV,Casek,NoChu,Marel,GDM}, often with significant isotope effect\cite{lee,Lanz,DanDgl}, ARPES shows a HEK suggestive of significant coupling to electronic bosons in the 300-600~meV range.\cite{Ronning,graf,Non,Valla,Feng,Zhougl,Kordyuk,Ronning2}
These bosons are believed to be predominantly magnetic,\cite{markiewater,waterfall} with charge bosons lying at even higher energies extending to the charge-transfer energy scale of $\sim$2~eV.  We should keep in mind that many experimental studies concentrate on the coherent part of the spectrum, and are thus not suited for probing high energy fluctuations. 
Also, one usually extracts a susceptibility from experiments, which is not predominantly $d$-wave, and a model must be used to extract the appropriate $d$-wave component.

\subsubsection{Optical extraction techniques}

Ref.~\cite{Oglue} shows how optical measurements can be used to extract a $q$-averaged susceptibility 
\begin{equation}
\alpha^2F(\omega)= U^2[\bar\chi^{\prime\prime}_c(\omega)+3\bar\chi^{\prime\prime}_s(\omega )]/2,
\label{eq:00a}
\end{equation}
with $\chi_c$ [$\chi_s$] the charge [spin] susceptibility,
and $\bar\chi_i (\omega )=\int a^2d^2q\chi_i ({\bf q},\omega )/(2\pi)^2$, $i=s,c$.  Within the RPA (see Eq.~\ref{chirpat})
\begin{equation}
\chi_s =\chi_0/(1-U\chi_0),
\label{eq:00b}
\end{equation}
\begin{equation} 
\chi_c = \chi_{0}/\epsilon,
\label{eq:00c}
\end{equation}
where $\chi_{0}$ is the bare (LDA) susceptibility, and the dielectric constant within the Hubbard model can be written as
$\epsilon =\epsilon_0+U\chi_{0}$, with $\epsilon_0\sim 4.8$ being a background
dielectric constant.  Note that the Hubbard model does not properly describe plasmon physics, and longer range Coulomb interactions, not considered here, may need to be included.\cite{markiewater,Mkacplas}

We now discuss an approach for extracting the Eliashberg function based on analyzing the degree to which the experimental or theoretical optical conductivity $\sigma (\omega)$ of Eq.~\ref{sigma} can be represented by an extended Drude form\cite{hwangnature,hwangprl}
\begin{equation}
\sigma(\omega) =
\frac{i\omega_p^2}{4\pi}\frac{1}{\omega-2\Sigma_{op}(\omega)},
\label{eq:0}
\end{equation}
where $\omega_p=\sqrt{4\pi ne^2/m}$ is the plasma frequency, $n$ is the
carrier density, and $e$ and $m$ are the electronic charge and mass,
respectively. By assuming that the optical self-energy $\Sigma_{op}$ is related to the electronic self-energy $\Sigma$ by\cite{PBA2,Mars,Shulga,HEscale}
\begin{equation}
\Sigma_{op} =\frac{\int_0^{\omega}\Sigma (\omega')d\omega'}{\omega},
\label{eq:2g}
\end{equation}
at $T=0$, for $\omega >0$, the electronic self-energy becomes
\begin{equation}
\Sigma^{\prime\prime}(\omega )=-\int_0^{\omega}\alpha^2F(\Omega
)\tilde N(\omega-\Omega )d\Omega , \label{eq:4}
\end{equation}
where $\tilde N(\omega )=[N(\omega)+N(-\omega)]/2N_{av}$ is the average of the electron and hole DOS, and $N_{av}$ is the average of the DOS over the energy range of interest, chosen to make $\tilde N$ dimensionless. In this case, 
\begin{equation}
\Sigma_{op}^{\prime\prime}(\omega)=\int_0^{\omega}\frac{d\Omega}{\omega}\alpha^2F(\Omega
)n_{eh}(\omega-\Omega), 
\label{eq:3A}
\end{equation}
where
\begin{equation}
n_{eh}(\omega)=\int_0^{\omega}d\omega \tilde N(\omega). 
\label{eq:4A}
\end{equation}
Many groups have used a finite-temperature version of this result\cite{GDM,Maksimov}.

In many inversion schemes, DOS is approximated by a constant, and in this case Eq.~\ref{eq:4} becomes\cite{foot6gl}
\begin{equation}
\alpha^2F_0(\omega)=-{\partial\Sigma^{\prime\prime}(\omega )\over\partial\omega}.
\label{eq:2}
\end{equation}
The subscript `0' on $\alpha^2F$ here refers to the constant DOS case. The glue function given by Eq.~\ref{eq:2} can become negative unless $|\Sigma^{\prime\prime}|$ is a monotonically increasing function of $\omega$.\cite{foot2d} 
Similarly, from Eq.~\ref{eq:3A}, it follows that $-\partial\Sigma_{op}^{\prime\prime}/\partial\omega$ must be $>0$ (as for Eq.~\ref{eq:2}), and
\begin{equation}
\alpha^2F_1(\omega)=-\frac{\partial}{\partial\omega}\Bigl(\omega^2\alpha^2F_0(\omega)\Bigr),
\label{eq:2A}
\end{equation}
where $\alpha^2F_1$ is the glue function corresponding to Eq.~\ref{eq:3A}.  While this substitution is appropriate for phonon contributions to the self-energy, it is not appropriate over a 2-3~eV energy range. We emphasize that since $\Sigma^{\prime\prime}$ cannot be monotonic over the full bandwidth, one can expect to encounter negative $\alpha^2F$ values in analysis based on Eq.~\ref{eq:2}, unless the analysis is restricted to fairly low energies. In sharp contrast, these problems do not arise with Eqs.~\ref{eq:4} and ~\ref{eq:3A}.

\subsubsection{Application to Bi2212}

\begin{figure}[htop]
\hspace{-1cm}
\rotatebox{0}{\scalebox{.65}{\includegraphics{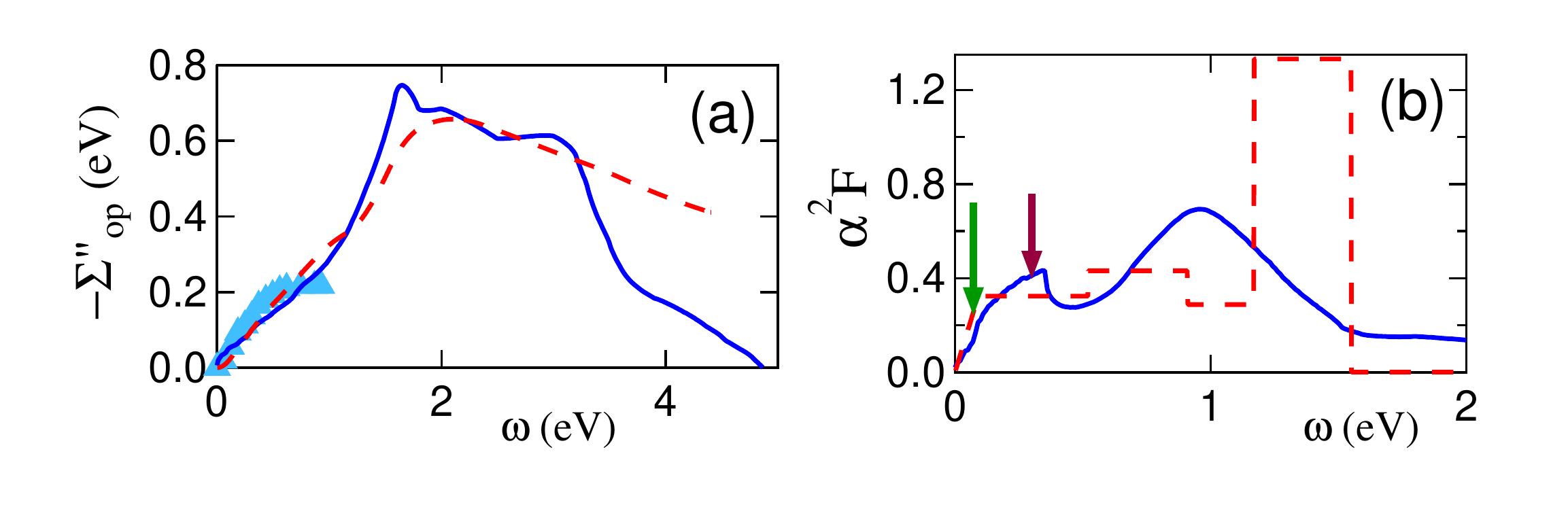}}}
\caption{ (a)
Self-energy in Bi2212 extracted from the optical data\cite{QT} (dark blue solid line), compared with the self-energy obtained via Eq.~\ref{eq:3A} (red dashed line), and the measured low-energy self-energy\cite{vdMMZ} (blue triangles). (b) The corresponding glue function (red dashed line) is compared to the sum of the calculated spin $\alpha^2F_s$ plus the charge glue function, Eq.~\ref{eq:00d}, for $U_{c,eff}=2~eV$ (blue solid line). 
[After Ref.~\onlinecite{Oglue}.]}
\label{fig:3b}
\end{figure}

Given the optical spectrum, we can extract the glue function over the full bandwidth. Figure~\ref{fig:3b}(a) shows the optical self-energy of Bi2212  (solid blue line) extracted from the data of Ref.~\onlinecite{QT} by using Eq.~\ref{eq:0}.  For simplicity, we calculated $\Sigma_{op}$ using the full optical spectrum, but features above $\sim$2.5~eV are probably due to interband transitions and should be disregarded.  In the low-energy limit, the self-energy is in reasonable agreement with earlier work\cite{vdMMZ} (blue triangles), which neglected the DOS factor (i.e., using Eq.~\ref{eq:2}).  Shown also is the model self-energy based on Eq.~\ref{eq:3A} (red dashed line), representing the glue function by a simple histogram shown in Fig.~\ref{fig:3b}(b) (red dashed line).
The blue solid line in Fig.~\ref{fig:3b}(b) represents the expected sum of spin and charge susceptibilities, Eq.~\ref{eq:00a}, which should describe the glue function.  Here, we use the calculated spin susceptibility, but since our calculated charge susceptibility does not capture plasmon physics, we used 
\begin{equation}
\alpha^2F_{c}(\omega)=-\frac{U_{c,eff}}{2}Im(\frac{1}{\epsilon (\omega)}),
\label{eq:00d}
\end{equation}
where $U_{c,eff}$ is a phenomenological charge vertex, and $\epsilon$ is the measured dielectric constant. Figure~\ref{fig:3b}(b) shows that the combination of spin and charge glue functions qualitatively reproduces the glue function extracted directly from the optical spectrum, including a significant contribution near 1~eV.  Differences above 1 eV may be due to limitations in extracting the true $\Sigma$ from $\Sigma_{op}$ as discussed in Ref,~\onlinecite{Oglue}. The contribution of the loss function to the optical glue has not been recognized previously.  
Note that Figure~\ref{fig:3b}(b) reproduces the three energy scales postulated by Anderson\cite{PWA}, namely, the spin-wave scale (green arrow), $J$-scale (red arrow), and $U$, or charge transfer scale (peak near $\sim 1$~eV).  These results have now been confirmed by Ref.~\cite{Carboglue}, who find that significant spectral weight of fluctuations extends up to 2.2~eV in Bi2212 and 1.2~eV in Bi2201, for all dopings.  Notably, a recent time-resolved optical study of near-optimally doped Bi2212\cite{GDM} argues that high-energy features are necessary to preserve the optical sum rule in the SC-state, including a 2.72~eV feature which they identify as the shifted remnant of the $\sim$2~eV optical gap of the undoped material. 

The present results are consistent with other recent quasi-first-principles calculations of the optical spectra, which find that the cuprate intraband optical spectrum extends up to $\sim$2.5~eV, with a residual charge transfer gap, associated with the incoherent part of the band, persisting well into the overdoped regime\cite{comanac,tanmoyop,DMFTb1,DMFTcuprate}. Moreover, several optical studies\cite{Little,GDM,Marel2} have found evidence that the onset of superconductivity affects spectral weight in an energy range extending beyond 1~eV.  
While optical experiments cannot separate out the relevant $d$-wave component of the glue, we find it has a similar energy dependence as seen by comparing Fig.~\ref{fig:3b}(b) with Fig.~\ref{fig:2SC}(a) below.

\begin{figure}[htop]
\centering
\rotatebox{0}{\scalebox{.65}{\includegraphics{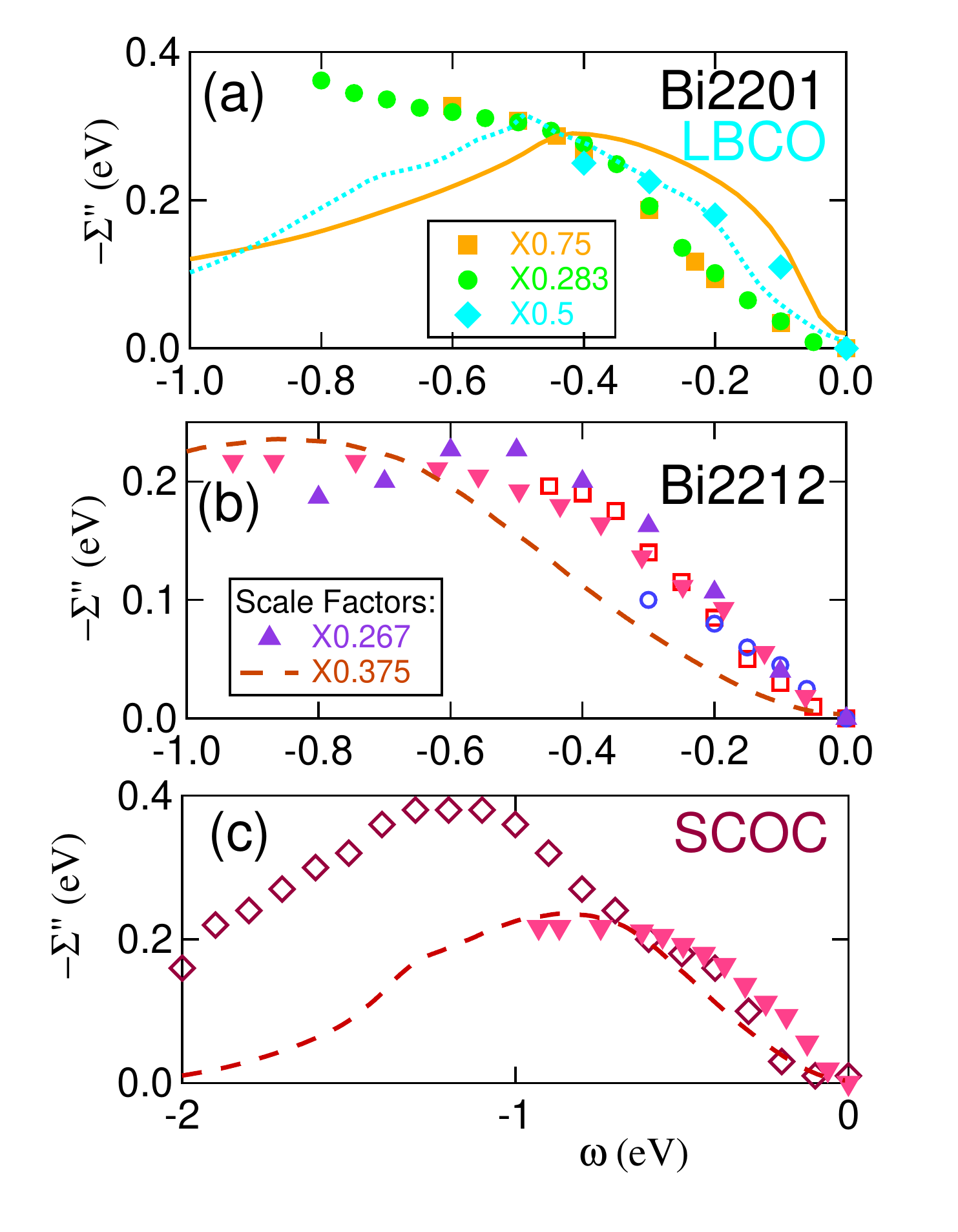}}}
\caption{ { Imaginary part of self-energy, $\Sigma^{\prime\prime}$},
at $T=0$ vs energy, comparing experimental and theoretical
results derived by several techniques. The impurity
contribution has been approximately removed by subtracting $\Sigma''(0)$ in all cases. Experimental points are from ARPES data for: (a) LBCO (blue diamonds,
Ref.~\protect\onlinecite{Valla}); Bi2201 (gold squares,
Ref.~\protect\onlinecite{Non} and green circles,
Ref.~\protect\onlinecite{Feng}); (b) Bi2212 (violet triangles,
Ref.~\protect\onlinecite{Valla}, open red squares,
Ref.~\protect\onlinecite{Zhougl}, and open blue circles,
Ref.~\protect\onlinecite{Kordyuk}); and, (c) CCOC (open
red-brown diamonds, Ref.~\protect\onlinecite{Ronning2}). 
Included in (b) and (c) are optical data from Bi2212 (inverted red
triangles, Ref.~\protect\onlinecite{vdMMZ}), taken at $T=130K>T_c$ to avoid
complications associated with superconductivity. Theoretical curves are from:
LSCO (light blue dotted line), Ref.~\protect\onlinecite{tanmoyop};
Bi2201 (gold line), Ref.~\protect\onlinecite{markiewater}; and, 
Bi2212 (red dashed line), Ref.~\protect\onlinecite{waterfall}.
Magnitudes of several data sets have been rescaled. [From Ref.~\onlinecite{Oglue}.]}
\label{fig:0}
\end{figure}

Overall, the glue extracted from optical experiments\cite{hwangoplscobi2212,hwang2,schachinger} displays 
two main peaks at energy scales in qualitative agreement with our QP-GW theory, as indicated by the arrows in Fig.~\ref{fig:3b}(b) at 300-400meV and $\sim 70$meV.  The high energy peak mainly stems from the strong intensities in the magnetic susceptibility near $(\pi,0)$, which are responsible for the HEK in ARPES, while the low-energy peak arises from the magnetic resonance mode, which dominates within the gap around $(\pi,\pi)$, giving rise to the LEK.  While our calculations reproduce both these energy scales, the 70~meV feature in the experimental spectra is more intense than that in our computations\cite{Marel}, even after superconductivity is included (see column 3 of Fig. 6 in Ref.~\onlinecite{tanmoymagres}). This is surprising, since our model reproduces both the magnetic resonance peak and the resulting LEK in ARPES. This discrepancy suggests a role for electron-phonon coupling, consistent with the time-resolved optical study of Ref. \cite{DCon}, which indicates that the 70~meV feature is more than 50\% of phononic origin. 



Ref.~\cite{Oglue} surveys a number of attempts to directly extract the self-energy from optical and photoemission experiments. While the energy dependence of the self-energy is readily extracted,  Fig.~\ref{fig:0}, there are substantial variations in the normalization of the spectra. Figure~\ref{fig:0} shows measured\cite{Valla,Non,Feng,Zhougl,Kordyuk,Ronning2} and calculated values\cite{foot8gl,foot9gl} of the imaginary part of the self-energy as a function of the excitation energy $\omega$.  Results for a number of different cuprates for various dopings are collected, and $\Sigma^{\prime\prime} 
(\omega)$ is seen to exhibit a similar shape in all cases, in good accord with our calculations. 

It should be noted that values of $\Sigma^{\prime\prime}$ extracted from experimental data differ by as much as a factor of four in magnitude for a given material, see Fig.~\ref{fig:0}(a). Recall that $\Sigma^{\prime\prime}$ is obtained as the product of the measured momentum-space width $\Delta k$ and the bare Fermi velocity, $v_{F0}$.  The variation in magnitude of  $\Sigma^{\prime\prime}$ arises from the uncertainty in estimating $v_{F0}$.
The best way to analyze the data may be to assume that the bare dispersion is given by the LDA.\cite{markietb}


\subsection{Calculation of $T_c$}

Here we discuss our calculations of $T_c$ based on solving the Eliashberg equation for spin-fluctuation mediated pairing in which both the low-energy magnon-like modes (near the LEK) as well as the high-energy paramagnon modes (near the HEK) are included. A good estimate of $T_c$ is obtained, highlighting the importance of high-energy fluctuations on the HEK energy scale in this connection. In order to assess effects of the pseudogap, we have also computed $T_c$ by a simpler BCS approach, self-consistently solving the mean-field gap equations in the presence of SC and SDW orders by assuming a momentum-independent pairing potential, see Table F2 for results. In particular, a dome-like pairing interaction must be assumed in order to reproduce the dome-like doping dependence of $T_c$. 




\subsubsection{Formalism}
 
Our Eliashberg calculations\cite{markiesc} are based on a one-band Hubbard Hamiltonian, extended to include pairing interaction. The singlet pairing potential is\cite{flex2}
\begin{equation}
V_s={U\over 1-U^2\chi_0^2(k'-k)}+{U^2\chi_0(k'+k)\over 1-U\chi_0(k'+k)},
\label{eq:1SC}
\end{equation}
in terms of the bare susceptibility $\chi_0$ and the Hubbard on-site repulsion $U$.
The mass renormalization potential (Eq.~[8] of Ref.~\onlinecite{flex2}(a)) is used: 
\begin{equation}
V_z={U^2\chi_0(k'-k)\over 1-U^2\chi_0^2(k'-k)}
+{U^3\chi_0^2(k'-k)\over 1-U\chi_0(k'-k)},
\label{eq:3SC} 
\end{equation}
where $p$ and $p'$ are the electron momenta, which are constrained to lie 
on the FS. $V_z$ [$V_s$] is the potential contributing to the normal [anomalous] part of the self-energy. These expressions give $T_c$s in good agreement with QMC results\cite{DCA2,DCA3}. The resulting coupling constants in various pairing channels $\alpha$ are
\begin{equation}
\bar\lambda_{\alpha}=-\int\int d^2kd^2k'\tilde g_{\alpha}(k)\tilde g_{\alpha}(k')
ReV(p,p',\omega =0)
\label{eq:4SC}
\end{equation}
where $V=V_s$ for the even parity channels. The normalized weighting 
function is $\tilde g_{\alpha}=g_{\alpha}(k)/(N_0|v_k|)$, where $v_k$ is the Fermi velocity and $N_0^2=(2\pi)^3\int{g_{\alpha}(k)^2d^2k/|v_k|}$.  The $g_{\alpha}$ are weighting functions of various symmetry\cite{flex2}, of which the most important are the lowest harmonics of the $s$-wave and $d_{x^2-y^2}$ symmetry, with $g_s=1$ and $g_d=\cos(k_xa)-\cos(k_ya)$.  We also define the coupling constant 
$\lambda_z$ via the $s$-wave version of Eq.~\ref{eq:4SC} with $V=V_z$.
Then the effective BCS coupling becomes $\lambda_{\alpha}=\bar\lambda_{\alpha}/
(1+\lambda_z)$, and the symmetrized Eliashberg functions are
\begin{equation}
\alpha^2F_{\alpha}(\omega)=-{1\over\pi}\int\int d^2kd^2k'\tilde g_{\alpha}(k)\tilde 
g_{\alpha}(k')V''(k,k',\omega),
\label{eq:5SC}
\end{equation}
where $V^{\prime\prime}$ is the imaginary part of the corresponding $V$.

\begin{figure}
\centering
\resizebox{13.5cm}{!}{\includegraphics{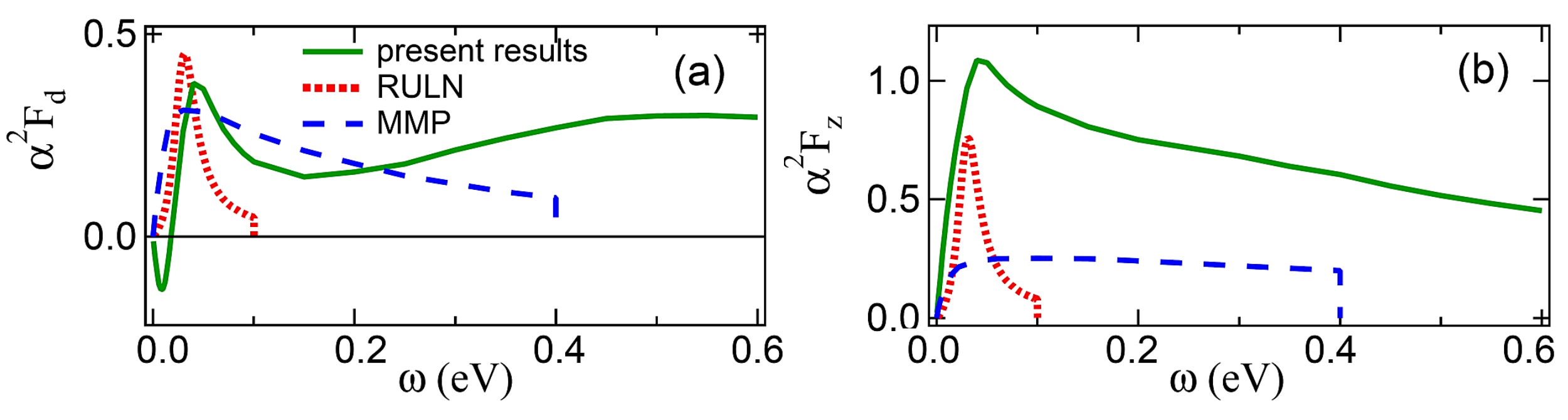}}
\caption{
{ Eliashberg functions $\alpha^2F_d$ and $\alpha^2F_z$} for hole-doping 
($x=0.30$) (green line) are compared with results of Refs.~\protect\onlinecite{RUIN} (red dotted line) and 
~\protect\onlinecite{MMP} (blue dashed line). [From Ref.~\onlinecite{markiesc}.]

}
\label{fig:1SC}
\end{figure}

Concerning technical details, we use a tight-binding parameterization of 
the dispersion of Bi2212 obtained by renormalizing the first-principles LDA results via $Z_0=1/Z=2$, which is appropriate for the overdoped regime\cite{hop, footz,foot2c}, with a reduced $U=3.2t$. Effects of bilayer splitting and the pseudogap are neglected. The latter approximation means that the $T_c$ is likely overestimated in the underdoped regime, and underestimated near optimal doping where critical fluctuations may be important. 
$\chi_0$, renormalized by $Z_0$, is first computed within the RPA throughout the BZ for frequencies up to 2.88 eV. $\alpha^2F$'s and $\lambda$'s are then computed from Eqs.~\ref{eq:1SC}-\ref{eq:5SC}. FS-restricted Eliashberg equations\cite{RUIN} are finally used to self-consistently obtain the $\Delta (\omega )$ and $Z(\omega )$, with $Z(0)\equiv Z=1+\lambda_z$.\cite{foot2d} Note that Migdal's theorem is not obeyed in the presence of the pseudogap\cite{Mont,Trem}, and our calculations do not account for modifications of Migdal's theorem in the 
SC-state. Nevertheless, our results provide a benchmark for the Eliashberg formulation in that we do not invoke empirical susceptibilities as has been the case in much of the existing literature. 

\subsubsection{Pure d-wave solution}

\begin{figure}
\centering
\resizebox{13.3cm}{!}{\includegraphics{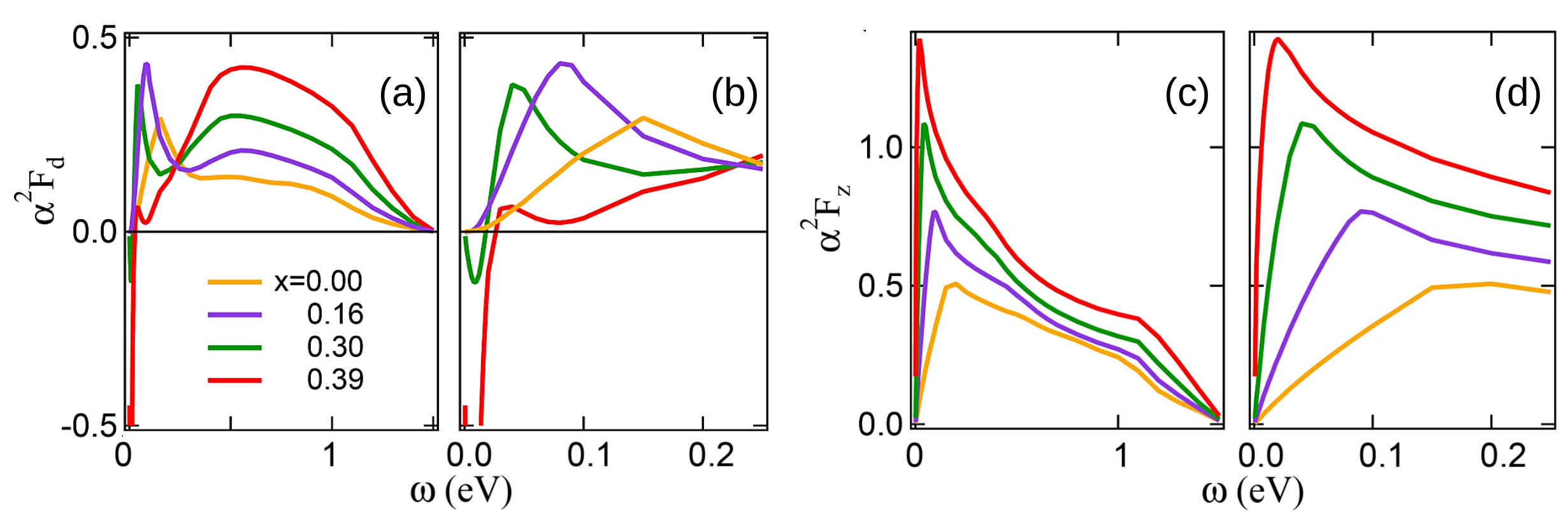}}
\caption{
{Eliashberg functions $\alpha^2F_d$ and $\alpha^2F_z$} over the doping range 
$x=0.0-0.4$. Lines of various colors refer to different dopings (see legend in (a)). Panels (a) and (c) give results over an extended frequency range of $0-1.5$ eV, while panels (b) and (d) highlight the low energy region of $0-250$ meV on an expanded energy scale. [From Ref.~\onlinecite{markiesc}.]
}
\label{fig:2SC}
\end{figure}

Fig.~\ref{fig:1SC} shows d-wave pairing weights $\alpha^2F_d$ and $\alpha^2F_z$, highlighting a key finding. The $\alpha^2F_d$ (green line) in (a) displays two clear features\cite{foot1}: A low energy peak (LEP) around 40 meV and a broad hump-like high-energy feature (HEF) extending from $\sim 0.5-1.0$ eV, see also Fig.~\ref{fig:2SC}(a) below. The LEP arises mainly from the magnetic resonance mode near $(\pi ,\pi )$, but the HEF is connected with the response from other parts of the BZ, particularly near $(\pi ,0)$ and $(\pi /2,\pi /2)$. [These fluctuations are also responsible for the HEK.] The negative dip in Fig.~\ref{fig:1SC}(a) at energies below 20 meV deserves comment. This dip reflects pair-breaking magnetic scattering (PBS) near $\Gamma$, which is related to earlier indications of ferromagnetic (FM) instabilities near the VHS\cite{SHluG,Chakra}. A similar scenario of competing d-wave pairing vs pair-breaking effects has been discussed in the context of electron-phonon pairing\cite{BuSca}.  Pair-breaking effects have also been reported in Ref.~\cite{Civelli}.

Figure~\ref{fig:1SC} also compares our results to early calculations of magnetic pairing in the cuprates, which employed parameterized models of susceptibility based on neutron scattering [Radtke, et al. (RULN)\cite{RUIN}] or NMR data
[Millis, et al. (MMP)\cite{MMP}].  Our LEP in panel (a) is similar to the weights assumed by RULN and MMP, and indeed provides a good description of the neutron scattering data near $(\pi,\pi)$, Section~\ref{S:ElSpec2}.3.  The MMP analysis, based on the NMR data, captures more of the weight, although it still misses the HEF and underestimates the total weight. Note that neutron scattering near $(\pi,\pi)$ accounts for only about 1/8th of the integrated spectral weight expected from a total scattering sum rule\cite{resn,LSC}. RULN and MMP models also strongly underestimate the renormalization weight $\alpha^2F_z$ as well as the PBS, both of which oppose the tendency for pairing.

Figure~\ref{fig:2SC} shows how $\alpha^2F$'s evolve with doping. In (a), the pairing weight in the high energy feature of $\alpha^2F_d$ is seen to increase 
monotonically with increasing doping, displaying an approximate isosbestic 
point at $\omega\sim 0.24$~eV. In the low energy region in (c), the peak in $\alpha^2F_d$ shifts to lower energies with increasing doping, while a negative pairbreaking peak grows dramatically, providing a plausible explanation for the termination of the superconducting dome. The nature of $\alpha^2F_d$ 
is seen to change quite substantially as $E_F$ approaches the 
VHS around $x=0.39$. Interestingly, by comparing (c) and (d), the low 
energy peak in $\alpha^2F_z$ is seen to follow that in $\alpha^2F_d$ to 
lower energies with doping.

\begin{figure}
\centering
         \resizebox{7.8cm}{!}{\includegraphics{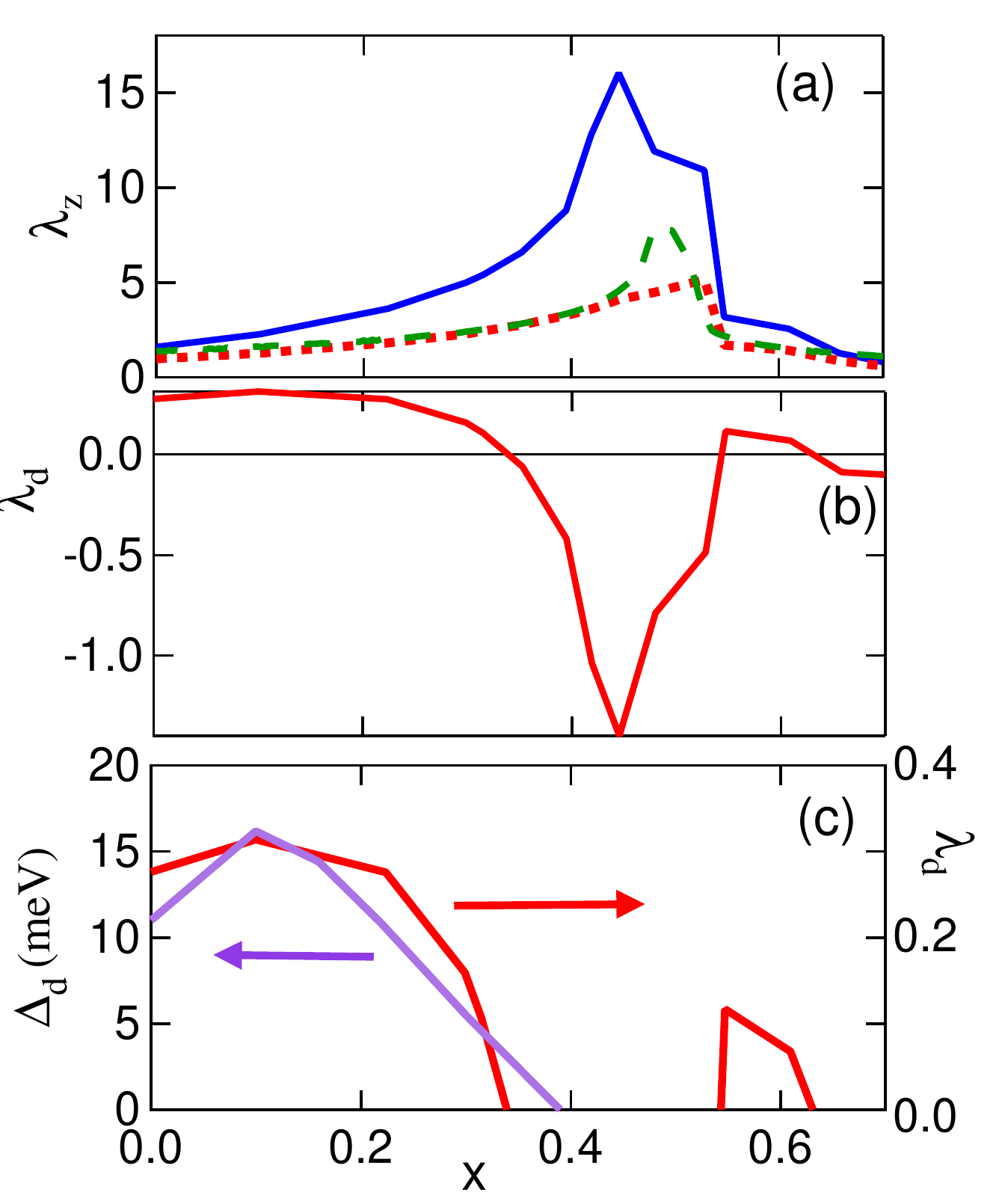}}
\caption{
Doping dependence of: (a) $\lambda_z$; (b) $\lambda_d$; and (c) $\Delta_d (T=0)$ (left scale), compared to that of $\lambda_d$ (right scale, note change in scale from (b)). In (a) three different computations of $\lambda_z$ are compared based on the full $V_z$ of Eq.~\ref{eq:3SC} (blue solid line), a simplified $V_{z0}=U^2\chi_0$ (red dotted line), and the estimate $N(0)U$ (green dashed line), where $N(0)$ is the DOS at $E_F$. [From Ref.~\onlinecite{markiesc}.]
}
\label{fig:3SC}
\end{figure}

Figure~\ref{fig:3SC} shows the doping dependence of $\lambda_z$, $\lambda_d$ 
and the low-temperature gap $\Delta_d (T=0)$. Three different estimates of 
$\lambda_z$ are compared in (a) for illustrative purposes. Values based on 
using the bare susceptibility, $V_{z0}=U^2\chi_0$ (red dashed line), are seen to be quite similar to the simple estimate $N(0)U$ (green dotted line), where 
$N(0)$ is the DOS at $E_F$. The full  $V_z$ (blue line), on the other hand, yields a significant enhancement of $\lambda_z$ over that obtained from $\chi_0$, especially near the region of the VHS peak, indicating that the system is close to a magnetic instability. Note that $\lambda_d$ is positive for dopings less than $\approx$ 0.4, but as $E_F$ enters the region of the VHS with increasing doping, $\lambda_d$ rapidly becomes large and negative due to FM fluctuations. Panel (c) shows that this doping 
dependence of $\lambda_d$ is well correlated with that of the pairing gap.
We stress that these results hold for a {\it pure $d_{x^2-y^2}$ order 
parameter.}  Harmonic content plays an important role, a point to which we return below.

\begin{figure}
\centering
            \resizebox{13.0cm}{!}{\includegraphics{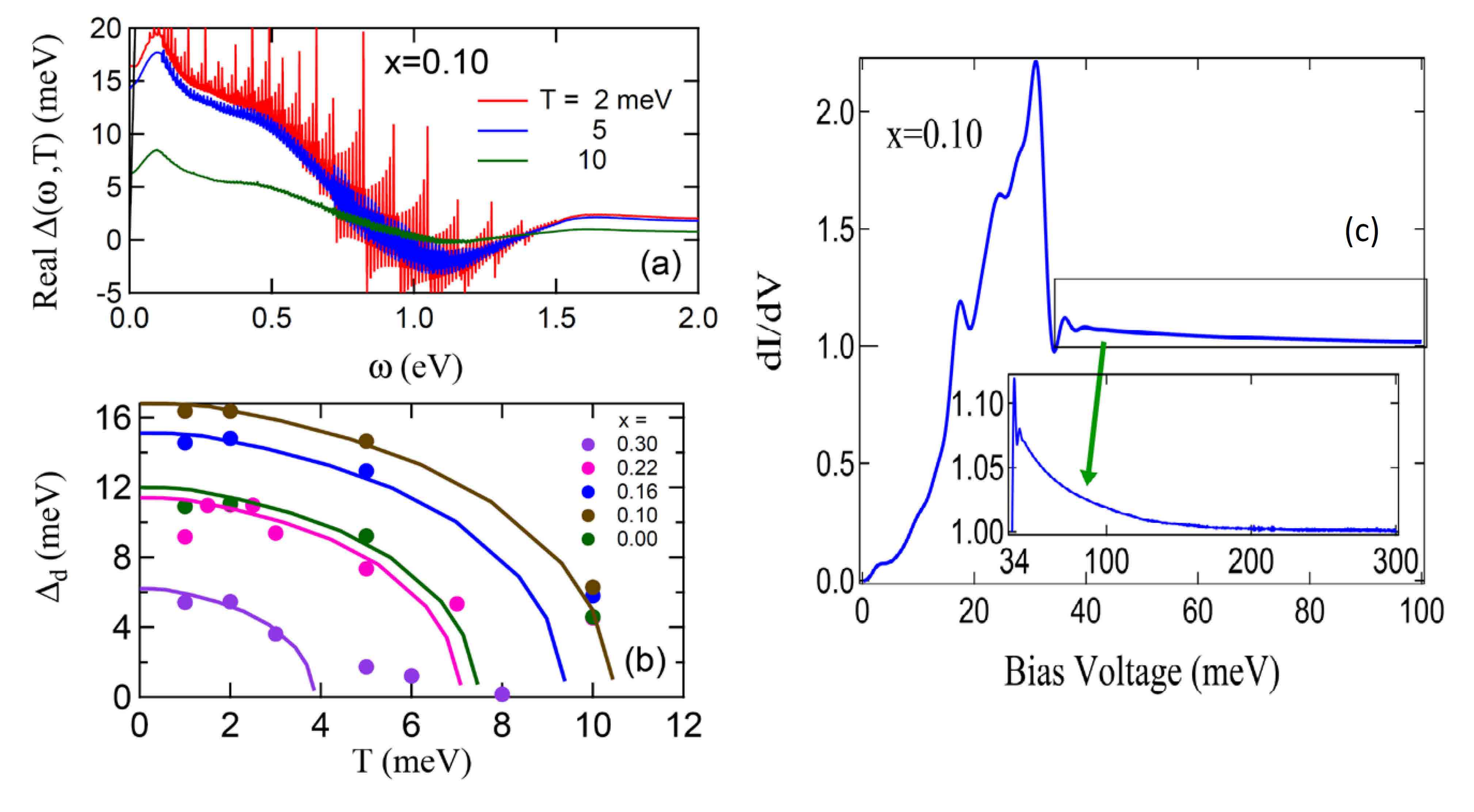}}
\caption{
(a) Real part of the gap function $\Delta (\omega,T)$ at $x$=0.10 as a 
function of frequency for a series of temperatures (see legend). Thin 
black line is the plot of $\Delta=\omega$ used to obtain the low 
energy gap as discussed in the text. (b) Computed temperature dependence of the low-energy gap $\Delta_d (T)$ at various dopings (see legend). (c) Typical computed superconductor-insulator-superconductor tunneling spectrum at $x=0.10$. Inset shows the high 
energy tail on an expanded scale. [From Ref.~\onlinecite{markiesc}.]

}
\label{fig:4SC}
\end{figure}

We turn now to discuss our solutions of the Eliashberg equations. Following common practice, we proceeded by discretizing the $\alpha^2F$'s on the real frequency axis.\cite{KMM} The results are sensitive to the number $N_m$ of points in the mesh. Our calculations are based on a 768-point non-uniform mesh over $0-2.88$ eV, and the resulting gap $\Delta (\omega )$ is approximately converged in the low-$\omega$ regime, allowing us to extract $\Delta_d(T)$. Figure~\ref{fig:4SC}(a) shows typical results for the real part of $\Delta (\omega )$ for a range of temperatures at $x=0.10$. The prominent oscillations in $\Delta (\omega )$ curves are the well-known consequence of discretizing 
$\alpha^2F$'s in solving the Eliashberg equations.\cite{MSC}  We define 
the gap by taking the intersection of the $\Delta(\omega)=\omega$ 
line (thin black line in Fig.~\ref{fig:4SC}(a)) with the $\Delta (\omega)$ curve. Fig.~\ref{fig:4SC}(b) shows how the computed low-energy gap $\Delta_d$ evolves with temperature at various dopings.  Due to the difficulty of finding 
well-converged solutions when $\Delta$ is small, we calculate $\Delta_d 
(T)$ at a few low temperatures, and use a fit to a $d$-wave BCS gap to 
estimate $T_c$.  We find $2\Delta_d(0)/k_BT_c\sim 3.2$ for different dopings. The resulting T$_c$'s are somewhat smaller than QMC values\cite{DCA2}, perhaps due to the effect of a finite $t'$. Note that there is a well-defined superconducting dome. The upper limit of the dome seems to be associated with the strong pair-breaking ferromagnetic scattering near the VHS, Figs.~\ref{fig:2SC}(a,c). Indeed, Storey, et al.\cite{TallStorey} find that in Bi-2212 the VHS induces strong pair-breaking, suppressing superconductivity, so that the optimum $T_c$ falls at a doping below the VHS.  This is consistent with the evidence for strong FM pair-breaking adduced by Kopp et al.\cite{Chakra}, and with the STM results of Fig.~\ref{Bi2201}(b).


It is striking that the gap features in Fig.~\ref{fig:4SC}(a) extend to very high energies, raising the obvious question as to how this high-energy tail 
would show up in the tunneling spectra.\cite{tunn1} Insight in this regard is 
provided by Fig.~\ref{fig:4SC}(c), where we show a typical tunneling spectrum 
computed\cite{tunn2} within our model. [Tunneling spectra computed at 
other dopings are similar, except that the features scale with $\Delta_d$.] The weight in Fig.~\ref{fig:4SC}(c) at energies above the peak-dip-hump feature is seen to be quite small with weak energy dependence (see inset) and would not be readily observable in the presence of an experimental background.


Our calculations will be seen to yield reasonable values of $T_c$'s. However, getting accurate gap values is more important as $T_c$ can be lowered by pair-breaking effects. For this reason, Ref.~\onlinecite{nieminenPRB2} compares our gap calculations to values derived from tunneling studies. There we found that the pseudogap splits the gap into separate gaps on two pockets, and the larger, antinodal gap is about twice as large as our estimate, but with a comparable doping dependence.  It is possible that this underestimate is due to neglect of the corresponding charge fluctuations, phonons, or critical fluctuations.

\subsubsection{Low vs High Energy Pairing Glue}

Within the present model, the LEP and HEF {\it both} play an essential role 
in generating large gaps. For example, at $x=0.3$, the HEF by itself produces 
a gap of only $\sim 0.4$~meV, while the LEP is virtually non-superconducting, 
even though the full $\alpha^2F_d$ yields a gap of 5.5 meV.  [Note, we separate $\alpha^2F$ into LEP and HEF at the minimum in $\alpha^2F$, $\omega_{min}=0.3$~eV.]  Similarly, for $x$=0.1, LEP [HEF] by itself has a  
$\sim$3 [0.4] meV gap, with a combined gap of $\sim$17~meV, with 
$\omega_{min}=0.16$~eV. This behavior can be readily understood from a 2-$\lambda$ model.\cite{Carb1} Since this is a purely electronic mechanism, we use a modified Allen-Dynes formula\cite{AlD,Carb2}
\begin{equation}
T_c={\omega_{ln}\over 
1.2}exp({-1.04(1+\lambda_z)\over\bar\lambda_d})
\nonumber \\
={\omega_{ln}\over 
1.2}exp({-1.04\over\lambda_d}),
\label{eq:1AD}
\end{equation}
$\Delta (0)=3.54T_c$, with
\begin{equation}
\bar\lambda_d=2\int_0^{\infty}{\alpha^2F(\omega )\over\omega}
\label{eq:2AD}
\end{equation}
and
\begin{equation}
ln(\omega_{ln})={2\over\bar\lambda_d}\int_0^{\infty}ln(\omega ){\alpha^2F(\omega 
)\over\omega}.
\label{eq:3AD}
\end{equation}
The Allen-Dynes equation has a well-known limitation\cite{Carb2} in that it 
predicts a maximum $T_c=\omega_{ln}/1.2$, whereas the Eliashberg equations 
have a solution that grows without limit $\sim\sqrt{\bar\lambda}$ as 
$\bar\lambda\rightarrow\infty$.  This leads to an underestimate 
of $\Delta_{LEP}$, while the model provides good estimates for the remaining 
gaps. For instance, at $x=0.1$, $\lambda_{LEP}=\lambda_{HEF}$=0.15, 
$\omega_{ln,LEP}=$~83~meV, $\omega_{ln,HEF}=$~530~meV, so that
$\Delta_{LEF}=$~0.26~meV and $\Delta_{HEF}=$~1.4~meV. When both features are combined, $\omega_{ln}=$~200~meV and $\lambda_d=0.3$, leading to $\Delta_d=19$~meV, in good agreement with the full calculation.  
While the Allen-Dynes model is highly simplified, it does capture the 
observed trend that both peaks contribute significantly.  Physically, the 
effective $\lambda$ is in the weak coupling regime, $\lambda\sim<<1$, so that high $T_c$ arises from the large $\omega_{ln}$, and the large boost from combining LEP and HEF arises since $e^{-1/2\lambda}>>2e^{-1/\lambda}$.
Clearly, an electron-phonon coupling could play a similar role in further enhancing $T_c$, and explaining the isotope effect.\cite{lee}

Our finding that both the low- and high-energy fluctuations are important is reminiscent of two-component $\alpha^2F$ models, with a strong peak at low frequencies and a weak [electronic] peak at very high
frequencies,\cite{Carb1,Carb2} and is consistent with the predictions of Refs.~\onlinecite{PWA,glue}, but contradicts the conclusion of Ref. \cite{Dahm}  that low energy fluctuations near the magnetic resonance alone can produce a 100K superconductor. This resonance mode is found to be far too weakly coupled to electron-hole pairs to explain high $T_c$s in cuprates \cite{resn,ACENS}.  Furthermore, experiments on overdoped LSCO find that strong kinks near 70~meV persist even at dopings where $T_c=0$, suggesting that pairing is mediated by a broad electronic spectrum.\cite{DanD2}  Ref. \cite{Dahm} may also have overlooked the role of ferromagnetic pair-breaking fluctuations, which are widely expected to be limiting $T_c$ on the overdoped side.\cite{TallStorey,Chakra} We should keep in mind that a key impediment to obtaining high $T_c$s is the emergence of competing phases\cite{CohAn}. 

\subsubsection{Competing SC gap symmetries}

\begin{figure}
\centering
            \resizebox{11.5cm}{!}{\includegraphics{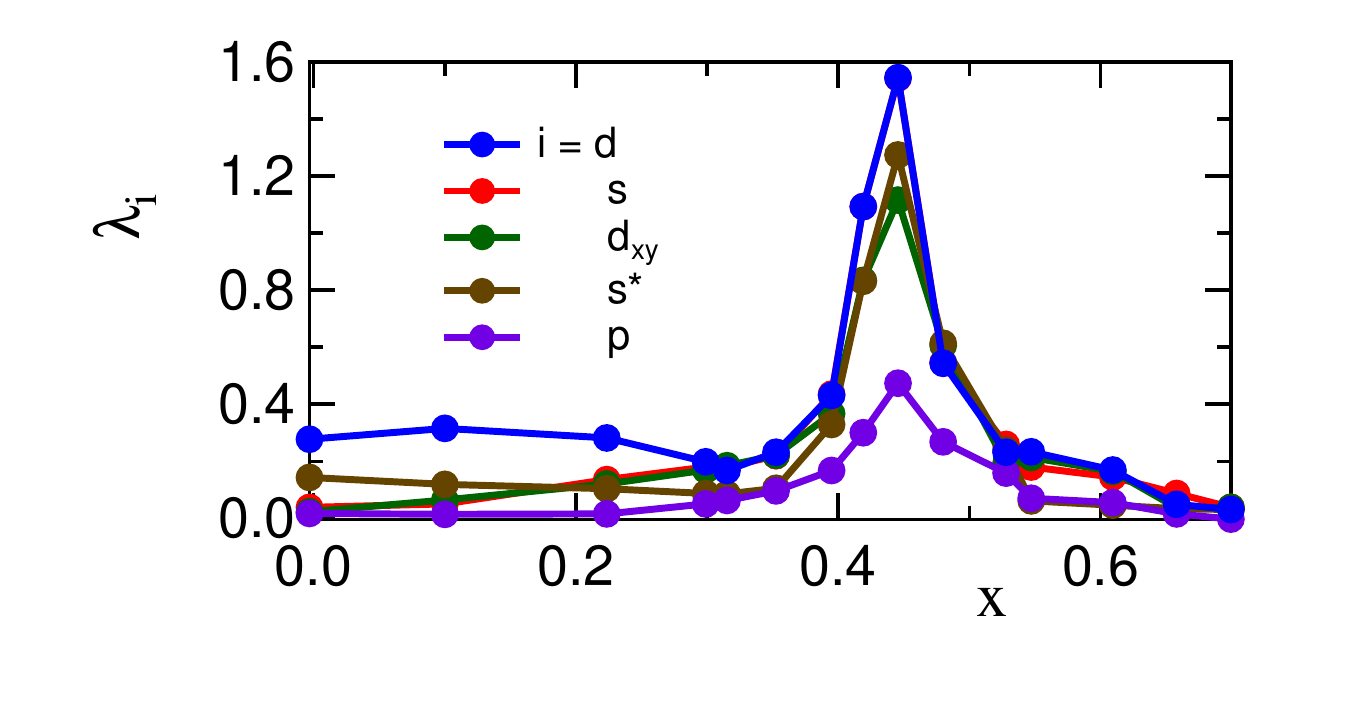}}
\caption{
Doping dependence of $\lambda_d$ (blue line), $\lambda_s$ (red line), 
$\lambda_{dxy}$ (green line), $\lambda_p$ (violet line), and $\lambda_{sp}$ 
(brown line), calculated from a 15$\times$15 harmonic matrix in each symmetry 
sector. [From Ref.~\onlinecite{markiesc}.]
}
\label{fig:6SC}
\end{figure}

The preceding analysis has been limited to a pure $d$-wave gap symmetry, 
without any harmonic content.  For tetragonal symmetry, there are five symmetry 
classes of the SC-gap, and each class can involve higher harmonics\cite{Hlub}. While we have not solved the tensor Eliashberg equations, it is straightforward to generalize our calculations to include harmonic structure to obtain the leading $\lambda$ eigenvalue for each symmetry class. 
The results are shown in Fig.~\ref{fig:6SC} following the analysis of Ref.~\onlinecite{GMV}. We see that: (1) The pure-$d$ analysis holds in the low-doping regime; (2) Near the VHS, harmonic content stabilizes d-wave symmetry, leading to the largest gaps; and, (3) in this regime, other symmetries can become comparable to d-wave.  In particular, there is a tendency toward $s$-wave pairing in the overdoped case.  

\subsubsection{Comparison with Other Calculations including DCA and CDMFT}

The SC transition is the most important property of the cuprates, and a variety of techniques have been applied to try to calculate the gaps and transition temperatures.  Here we briefly summarize some recent studies.

Baeriswyl {\it et al.}\cite{BEM} has reviewed a number of early variational\cite{PRTRVB,GiLh,EiBa,EiBa2}, quantum cluster\cite{DMFT13}, and Gaussian Monte Carlo\cite{AiIm} calculations of gaps and $T_c$s, finding an overall measure of agreement.  For the pure Hubbard or $t-J$ models ($t'=0$), most calculations indicate a $d$-wave superconducting dome, starting from half-filling and ending at possibly a $U$-dependent doping of $x\sim$ 0.2 [$U/t=8$]\cite{EiBa} or 0.36 [$U/t=10$, $t-J$ model]\cite{PRTRVB}.  Inclusion of AFM order yields an extended tail of pairing at higher doping\cite{GiLh}.  Notably, a Gaussian Monte Carlo calculation\cite{AiIm} failed to find superconductivity. However, since $U=6t$ was assumed, and only dopings larger that $x=0.18$ were probed, these results are, in fact, consistent with a recent DCA calculation which obtains superconductivity only for $x<0.13$ at $U=6t$.\cite{GuMi3}  For finite $t'=-0.3t$ and $U=8t$, variational\cite{EiBa2} and quantum cluster\cite{DMFT13} calculations yield results similar to the corresponding $t'=0$ results, but with a slightly larger upper limit of the superconducting dome (larger for hole- than electron-doping). All these results seem to be consistent with the notion of a universal superconducting dome, which terminates around $x=0.27$.


While QMC calculations are consistent with a magnetic fluctuation mechanism for cuprate superconductivity, the fermion sign problem restricts the calculations to small cluster sizes and high temperatures\cite{Scal1}.  The issue cannot be addressed by DMFT, since some momentum dependence is required to generate a $d$-wave gap. This has motivated a number of groups to apply cluster extensions of DMFT to the problem.\cite{DMFT3} It is important to recognize in this connection that since computations involve 2D-clusters, the Mermin-Wagner theorem would limit the transition temperature of conventional superconductors to $T_c=0$\cite{SuSu}.  Calculations on a finite cluster would see an effective $T_c>0$, which corresponds to the temperature at which the correlation length $\xi$ becomes comparable to the cluster size.  [For the analogous results on the Neel transition temperature, $T_N$, see Fig.~\ref{fig:64C}.]  Beyond the Mermin-Wagner physics, 2D-superconductors can also undergo a vortex-unbinding transition at a finite Kosterlitz-Thouless (KT) transition temperature\cite{KoTh}.

\begin{figure}
\centering
            \resizebox{10.5cm}{!}{\includegraphics{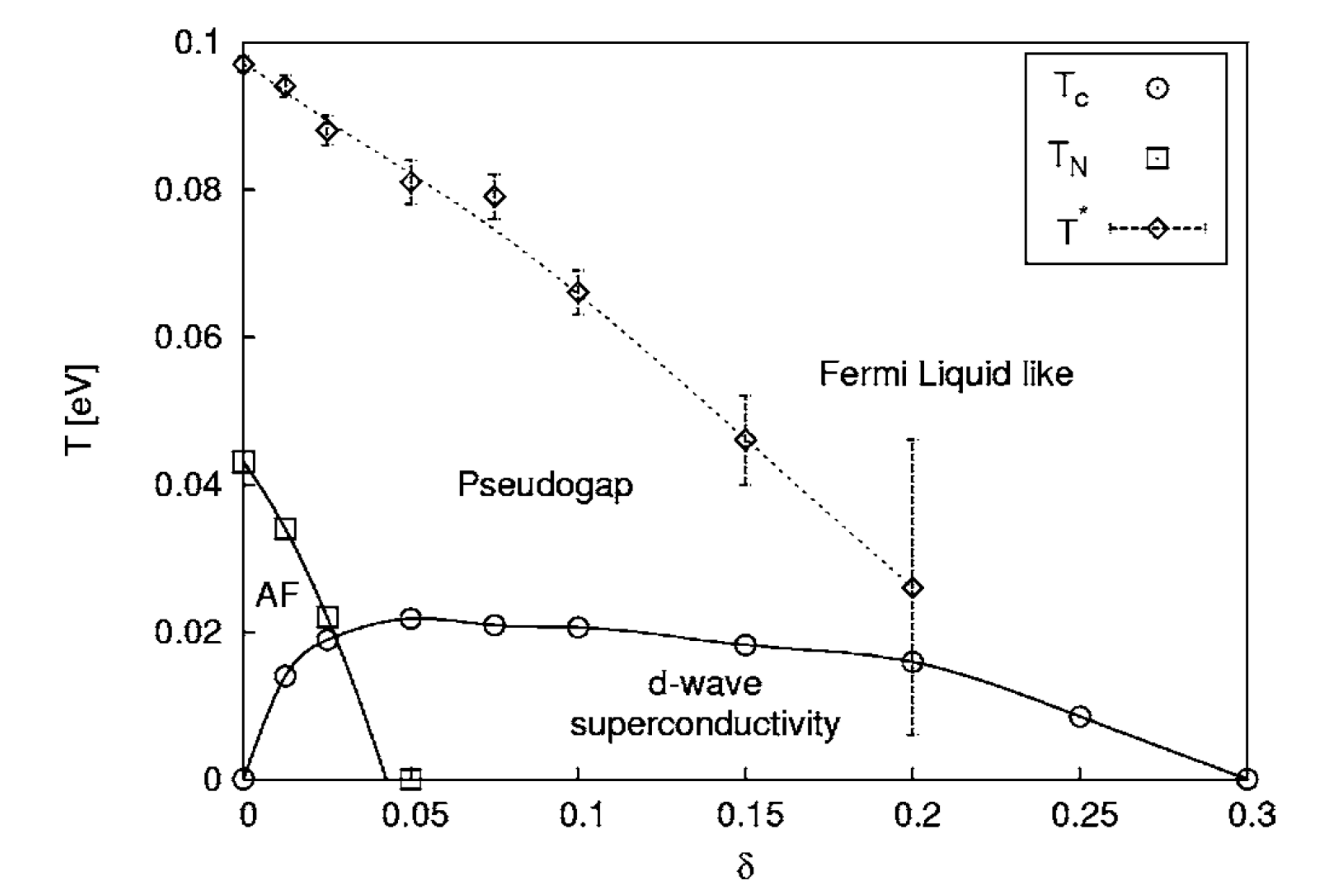}}
\caption{%
Phase diagram of the $t-U$ Hubbard model for $U=8t$ based on DCA calculations. [From Ref.~\onlinecite{DMFT3}.]
}
\label{fig:63C}
\end{figure}

With the preceding consideration in mind, and extending the discussion in Section~\ref{S:IntCII}.1 above, we comment on the effects of cluster size, $N_c$, on the results of cluster calculations. $N_c=4$, a $2\times 2$ cluster, is the smallest cluster that can support a $d$-wave pairing function.  Since pairs cannot fluctuate at this cluster size, a mean-field transition temperature is obtained, which nevertheless provides an estimate of the temperature $T'$ at which pairing fluctuations first begin in the system.  A DCA calculation\cite{DMFT3} finds a superconducting dome for a Hubbard model ($U=8t$, $t'=0$), Fig.~\ref{fig:63C}, much like the results discussed above, with  maximum $T'\simeq 230K$ at $x=0.05$ and a dome terminating near $x=0.30$.  
At low doping, Ref. \cite{DMFT3} sees an AFM order and a pseudogap driven by short-range AFM fluctuations, which start at $T^*=1100$K at $x=0$ and go to 0K somewhere around $x\sim$0.2-0.3 [with large error bars].  For $N_c>4$, the transition temperature drops, but seems to saturate for large $N_c$ to a KT transition ($T_c\sim$~55K) at $x=0.1$ (assuming $t=250$~meV).\cite{DMFT10}  However, the error bars are quite large so that a residual size dependence cannot be ruled out, and the computed $T_c$ is an upper bound. This suggests that interlayer coupling may be necessary to explain the larger experimental $T_c$s.

In contrast to the one-band results outlined above, computations based on three-band models derived from downfolded LDA bands\cite{OKA} are generally not in good agreement with experiments and mostly fail to reproduce $d$-wave superconductivity.  In one scheme that does find $d$-wave superconductivity, the order is reversed in that LSCO is predicted to have a larger $T_c$ than HgBa$_2$CuO$_4$.\cite{OKA}  These problems may be tied to the choice of a large $\Delta\equiv\epsilon_d-\epsilon_p$=3.25~eV, since such a large charge-transfer energy causes the cuprates to behave as Mott insulators rather than charge-transfer insulators, see Section~\ref{S:Ext}.4 below.

Concerning other properties, the gap is found to show a $d$-wave symmetry with a significant second harmonic component, and its eigenvector $\Phi_d({\bf K},i\omega_n)$ [with ${\bf K}=(\pi,0)$] varies with the Matsubara frequency $\omega_n$ in a very similar manner to the spin-susceptibility $\chi ({\bf Q},i\omega_n)$ [${\bf Q}=(\pi,\pi)$]\cite{maier}, suggesting that the pairing is driven by spin fluctuations.  While the authors of Ref.~\onlinecite{maier} claim that $\Phi_d$ cuts off at a scale of order $\sim J=4t^2/U$, it would appear from Fig. 7 of Ref.~\cite{maier} that it cuts off at an energy $\sim t\sim 250$~meV independent of $U/t$. Note that this scale is comparable to the HEK scale we found above.  At the temperature of the simulation, the lower ($\sim 55$~meV) scale could not be resolved.  While the $T_c$ is calculated from a Bethe-Salpeter equation involving the irreducible part of the four-point particle-particle vertex function, $\Gamma^{pp}$, a BCS-like pairing-energy $V_d$ can be defined\cite{maier}, which grows as $T$ is lowered and $x$ decreases toward half-filling.  It is further demonstrated in Ref.~\cite{maier} that the pairing interaction can be approximated by the RPA form 
\begin{equation}
V_{s,eff}=\frac{3}{2}\bar U^2\chi (K-K'),
\label{eq:2ADE}
\end{equation}
where $K=({\bf K},i\omega_n)$, $\bar U$ is a temperature and doping dependent effective Hubbard $U$ [see Fig.~\ref{Ux}], and $\chi$ is the (transverse) spin-susceptibility.\cite{DCA2} Eq.~\ref{eq:2ADE} should be compared with Eq.~\ref{eq:1SC} above.

The possible role of electron-phonon coupling in modifying $T_c$ has been explored via three popular models: the Holstein polaron, the buckling and the breathing modes.\cite{JM3}  All three models display qualitatively similar effects.  Anti-ferromagnetism and polaronic effects act to strengthen each other, and while all phonons enhance $d$-wave pairing strength, the polaronic localization dominates, and in all cases $T_c$ is suppressed. These results are not supportive of a combined electronic-phononic pairing mechanism.  



A CDMFT study of pairing in the $t-t'-t''-U$ model shows that a pairing glue description is appropriate\cite{Trem2}.  The principle contribution is associated with a low-frequency peak in the $(\pi,\pi)$ spin susceptibility $\chi''$ [for a 2$\times$2 plaquette, there are only three independent susceptibility components], but there is a significant contribution near $\omega =U/2$.  This is comparable to our results in Fig.~\ref{fig:3b}(b) in that, in both calculations, the low-energy peak seems to fall at energies larger than the magnetic resonance peak. For example, in optimally doped LSCO, Ref.~\onlinecite{Trem2} reports two peaks, which would lie at 70 and 126~meV for an LDA-like bare $t$ (420~meV), compared to the corresponding experimental values of 20 and 50~meV.\cite{Trem2R39}

\subsubsection{Universal Superconducting dome}

\begin{figure}
\leavevmode  
\centering
\rotatebox{0}{\scalebox{0.5}{\includegraphics{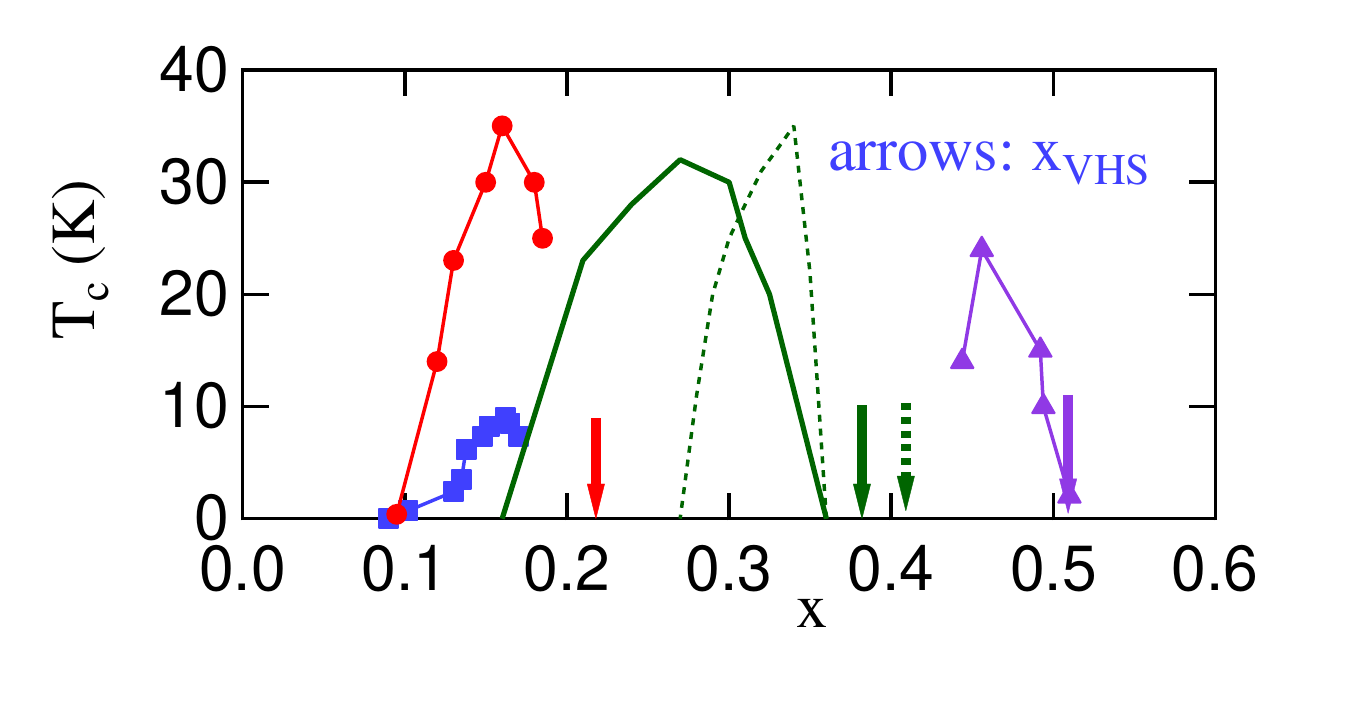}}}
\caption{
Superconducting dome in Bi2201. Results based on LDA-derived dispersion (green dotted line) are compared with various experimentally-derived dispersions based on: Ref. \cite{Ding201} (red circles and blue squares); 
Ref. \cite{Takeuchi201} (violet triangles); and, Ref. \cite{He201} (green solid line). Various colored arrows show the corresponding $x_{VHS}$, the doping where the VHS crosses $E_F$. [From Ref.~\cite{Whiffnium}.]
}
\label{fig:2201domes} 
\end{figure}

Models based on LDA dispersions can generally be expected to have difficulty in explaining a universal superconducting dome because the dispersion, and especially the position of the VHS, is quite material dependent.  In this connection, it is interesting to compare the superconducting domes ($T_c$ vs $x$) derived in Bi2201 on the basis of various experimentally determined dispersions, see Fig.~\ref{fig:2201domes}.\cite{Whiffnium,Ding201,Takeuchi201,He201} While some of the spread in the figure is associated with the different ways in which Bi2201 can be doped, much of the uncertainty reflects difficulty of obtaining dispersions from experimental spectra. 

 %
%
%


\section{Competing Phases}\label{S:DWs}

\subsection{$(\pi ,\pi )$-order}
The mean-field theory of the $(\pi,\pi)$ SDW order and the effects of self-energy corrections therein are discussed in Sections~\ref{S:QP-GW}.3 and \ref{S:QP-GW}.4 above. Our model provides a comprehensive picture of pseudogap physics in electron-doped cuprates and for hole-doped cuprates at very low doping as discussed in Sections~\ref{S:ElSpec}.1 and~\ref{S:ElSpec}.2 above.  However, at higher hole-doping we find evidence of competition from SDW and CDW orders with incommensurate $q$-vectors, see Section~\ref{S:ElSpec}.2. Nevertheless, we have shown that salient features of much spectroscopic data are reasonably well captured by assuming the $(\pi,\pi)$ SDW order for all hole-dopings.  The reason is that the self-energy is relatively insensitive to the specific nature of the competing order, as long as the ordering vector is $q=(\pi ,\pi )$ for SDW, CDW, or flux phase order.\cite{tanmoy2gap}  An incommensurate order would mainly manifest itself through a change in the small pockets of the residual FS, but such pockets are rarely seen clearly in experiments [see below]. In fact, in the following subsection, we will show that at high temperature ($T$) or energy ($\omega$), there is a commensurate-incommensurate crossover, and the high-$T$ and/or $\omega$ behavior is dominated by the $(\pi,\pi)$-plateau in susceptibility.  Details of possible incommensurate phases will also be discussed in this Section.

While this review is mainly concerned with competition between different phases, a more complicated intertwining of phases is also possible.  An example is of coexisting SDW and CDW phases, which are widely discussed in the stripe literature\cite{KT}, see also Ref.~\cite{MGu2} for some preliminary results.

\subsubsection{Commensurate-incommensurate crossover}

\begin{figure}
\leavevmode  
\centering
\rotatebox{0}{\scalebox{0.5}{\includegraphics{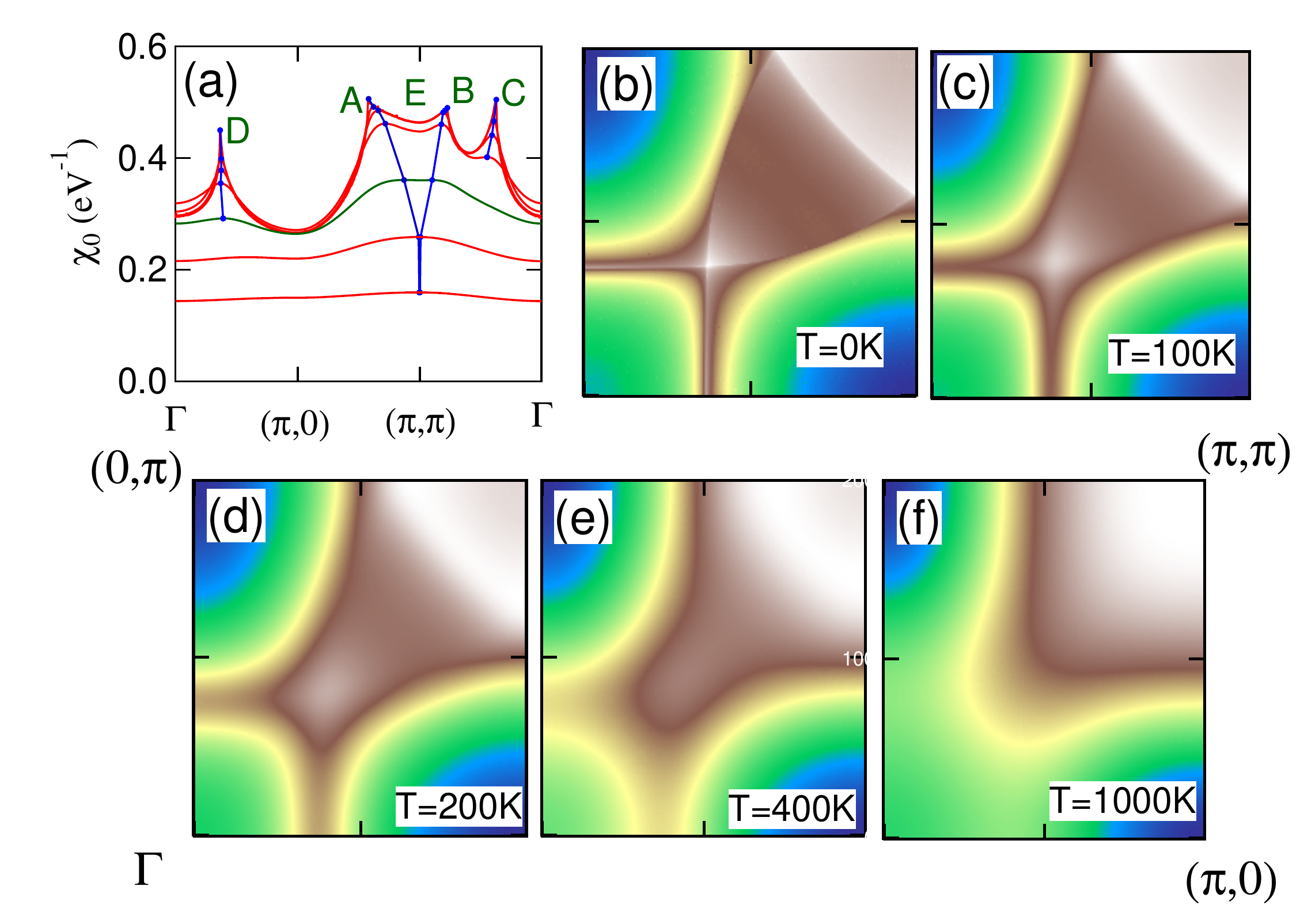}}}
\vskip0.5cm
\caption{{ $T$-dependence of susceptibility in Bi2212.}
(a) Susceptibility, $\chi_0$, along the high symmetry lines in the BZ at $\omega =0$ at a series of increasing temperatures: from top to bottom, $T$ = 0, 100, 200, 400, 1000, 2000, and 4000~K. (b-f) Maps of $\chi_0$ in the first BZ at various temperatures as marked on the figures.  [From Ref.~\cite{Whuffnium}.]
}
\label{fig:DW0a} 
\end{figure}

\begin{figure}
\leavevmode  
\centering
\rotatebox{0}{\scalebox{0.5}{\includegraphics{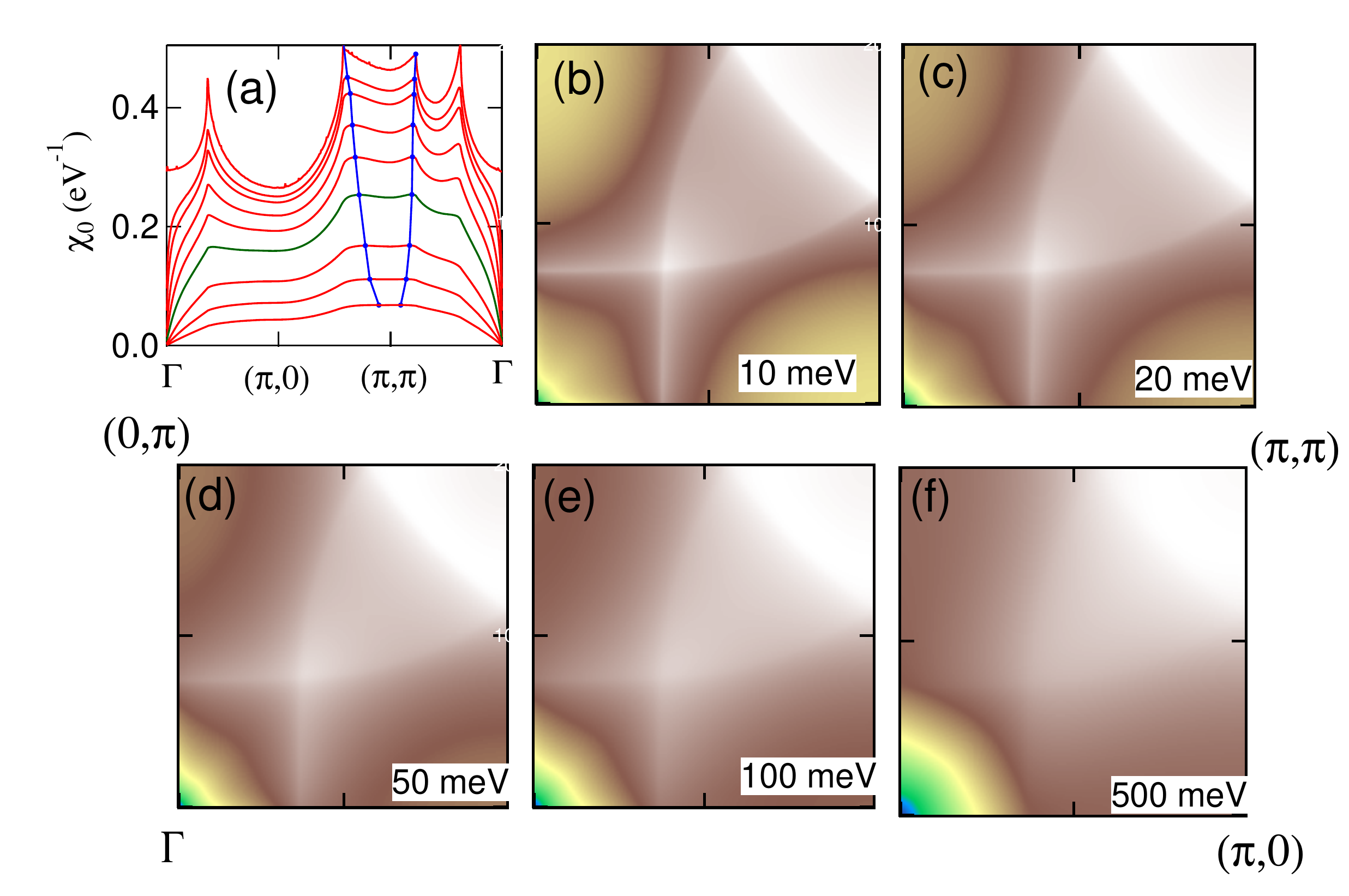}}}
\vskip0.5cm
\caption{{ $\omega$-dependence of the susceptibility in Bi2212.}
(a) Susceptibility, $\chi_0$, along the high symmetry lines in the BZ at $T=0$ for a series of increasing energies: from top to bottom, $\omega$ = 0, 10, 20, 50, 100, 200, 500, 1000, and 2000~meV. (b-f) Maps of $\chi_0$ in the first BZ at various energies $\omega$ as marked on the panels.   [From Ref.~\cite{Whuffnium}.]
} 
\label{fig:DW0b} 
\end{figure}

The broad, intense susceptibility plateau around $(\pi,\pi)$ leads to a commensurate-incommensurate crossover as we increase temperature (see  Fig.~\ref{fig:DW0a}) or energy $\omega$ (see Fig.~\ref{fig:DW0b}).\cite{Whuffnium}  We illustrate this generic effect with the example of Bi2212 for $x=0.134$.  With increasing temperature, features in $\chi_0$ wash away, leaving just the broadened $(\pi,\pi)$ plateau at high temperature.  Details of this evolution are interesting, and can be seen by following the labeled peaks in Figs.~\ref{fig:DW0a}(a).  Note that the four peaks A-D constitute four competing nesting vectors, with the most intense peak leading to the initial instability. At lowest $T$, there is no peak at the $(\pi,\pi)$ commensurate position (marked by E) at this doping. The peaks A and B represent near-$(\pi,\pi)$ or near-nodal nesting (NNN), while peaks C and D represent antinodal nesting (ANN).  Specifically, A drives nesting at $(\pi,\pi-\delta)$ and equivalent directions, B at $(\pi-\delta,\pi-\delta)$, C at $(\delta,\delta)$ and D at $(\delta,0)$, where $\delta/\pi$ is small.  D corresponds to nesting between the flat portions of the FS near $(\pi,0)$, and thus represents ANN. The results of Figs.~\ref{fig:DW0a} and~~\ref{fig:DW0b} neglect bilayer splitting in Bi2212, with the doping chosen so that at low $T$ peaks A and C have the same height.  For higher [lower] doping peak C [A] is stronger, leading to a transition in nesting vectors.

As to the temperature evolution, Fig.~\ref{fig:DW0a}\cite{Whuffnium}, as $T$ increases, peaks C and D, associated with ANN fade away rapidly, and at higher doping there can be a crossover from predominantly ANN nesting at C to NNN nesting at A.  With increasing temperature, the structure in susceptibility around the $(\pi,\pi)$-plateau in Fig.~\ref{fig:DW0a}(a) collapses in two stages.  At low temperatures, the ridge is anisotropic, with higher weight at A than at C.  As T increases, the ridge first becomes isotropic, then the weight along the ridge shifts toward $(\pi,\pi)$.  At this doping, coherent features disappear near $T_{coh}=$1000K or $\omega_{coh}=200$~meV $\sim 2T_{coh}$ [green lines in Figs.~\ref{fig:DW0a}(a),~\ref{fig:DW0b}(a)].
Similar effects are found on increasing the frequency $\omega$, Fig.~\ref{fig:DW0b}\cite{Whuffnium}, or the scattering rate (not shown).  

\subsubsection{Comparison with Other Calculations: DCA, DMFT, and effects of finite q-resolution}
\begin{figure}
\centering
\rotatebox{0}{\scalebox{0.5}{\includegraphics{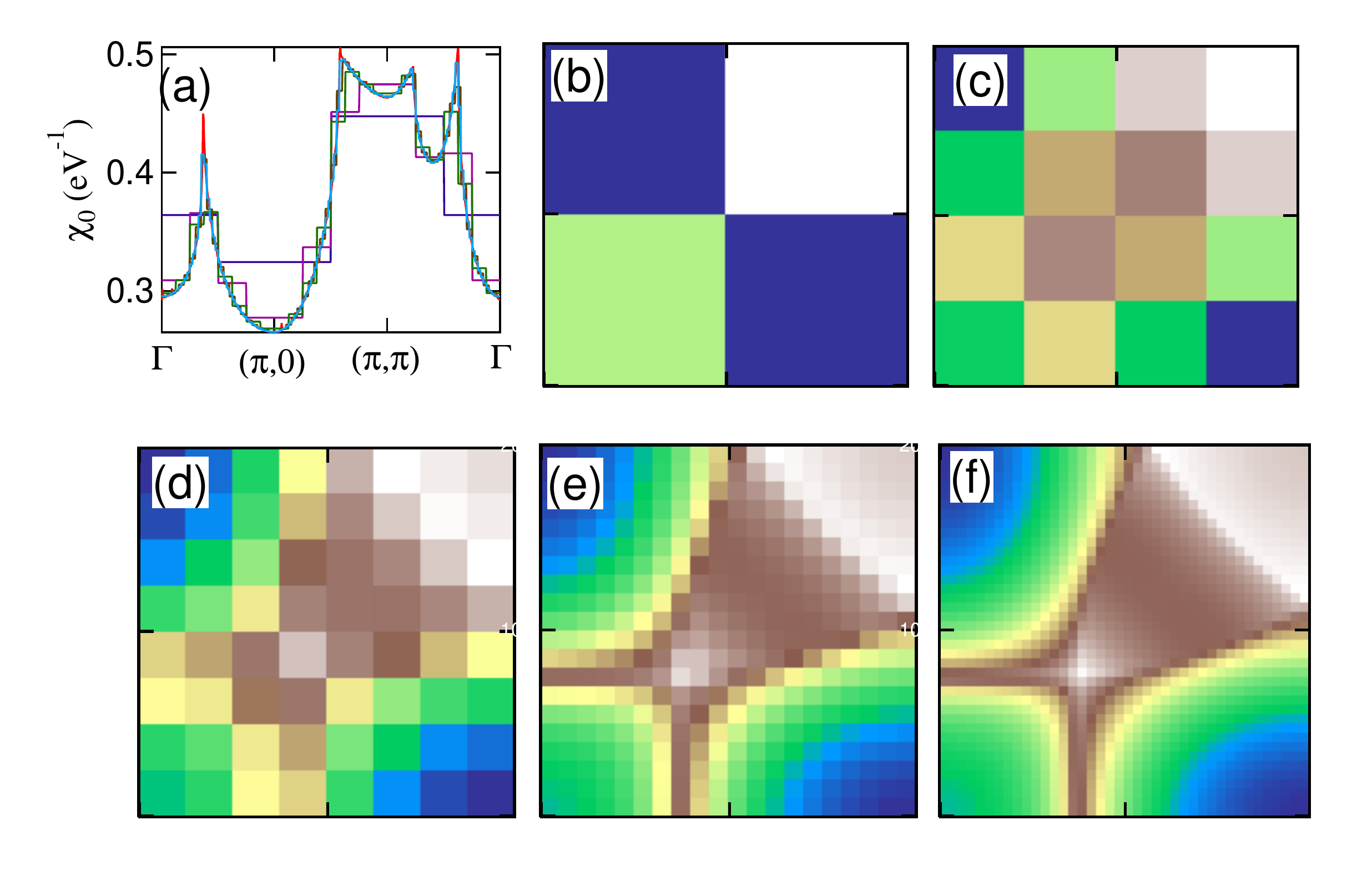}}}
\caption{
(a) {Patch susceptibility} ($\chi_0$) based on an increasing number of momentum points $N_c$ in the BZ using the Bi2212 data at $T=0$ and $\omega=0$ of Fig.~\ref{fig:DW0a}(b). Results for low to high resolution are shown: $N_c$ = 4 [dark blue line], 16 [violet], 64 [green], 400 [brown], 1600 [light blue], and 40000 [red]. (b-f) Maps of patch susceptitbility $\chi_0$ over the first BZ for $N_c$ = 4 (b), 16 (c), 64 (d), 400 (e), and 1600 (f). Figure~\ref{fig:DW0a}(b) corresponds to $N_c$ = 40000.  [From Ref.~\cite{Whuffnium}.]
} 
\label{fig:0c} 
\end{figure}

In the DCA computations, one typically obtains the susceptibility over a set of points in the BZ. Here we comment on how a limited momentum resolution of such a susceptibility impacts the determination of the dominant nesting vectors.  For this discussion, we define the `patch' susceptibility as the average of the susceptibility over all k-points in the patch, although the result would not be sensitive to a different choice, such as the susceptibility at a representative k-point in the patch.  Figure~\ref{fig:0c}\cite{Whuffnium} shows the susceptibility of Fig.~\ref{fig:DW0a}(b) when the averaging is carried out by using an increasing number of momentum points $N_c$ where a uniform ‘patch’ of intensity is associated with each momentum point. The susceptibility peak in Fig.~\ref {fig:0c} is seen to remain around $(\pi,\pi)$ until $N_c=16$ (frames (b) and (c)); at $N_c=64$ (frame (d)), one starts to see that the leading susceptibility is incommensurate, with its peak corresponding to the feature $A$ in Fig.~\ref{fig:DW0a}(a), while the corresponding feature $C$ is just starting to be resolved.  However, at this doping, features $A$ and $C$ are of equal intensity, whereas for the patches, feature $C$ remains weaker even for $N_c=1600$ (frame (f)).  These results suggest that the use of a susceptibility with limited momentum resolution would tend to favor the $(\pi,\pi)$-SDW order as the dominant competing order for essentially all dopings. This tendency is enhanced by restricting the calculations to models with only $t$ and $t'$ hopping, with small $t'$.

\begin{figure}
\centering
            \resizebox{9.3cm}{!}{\includegraphics{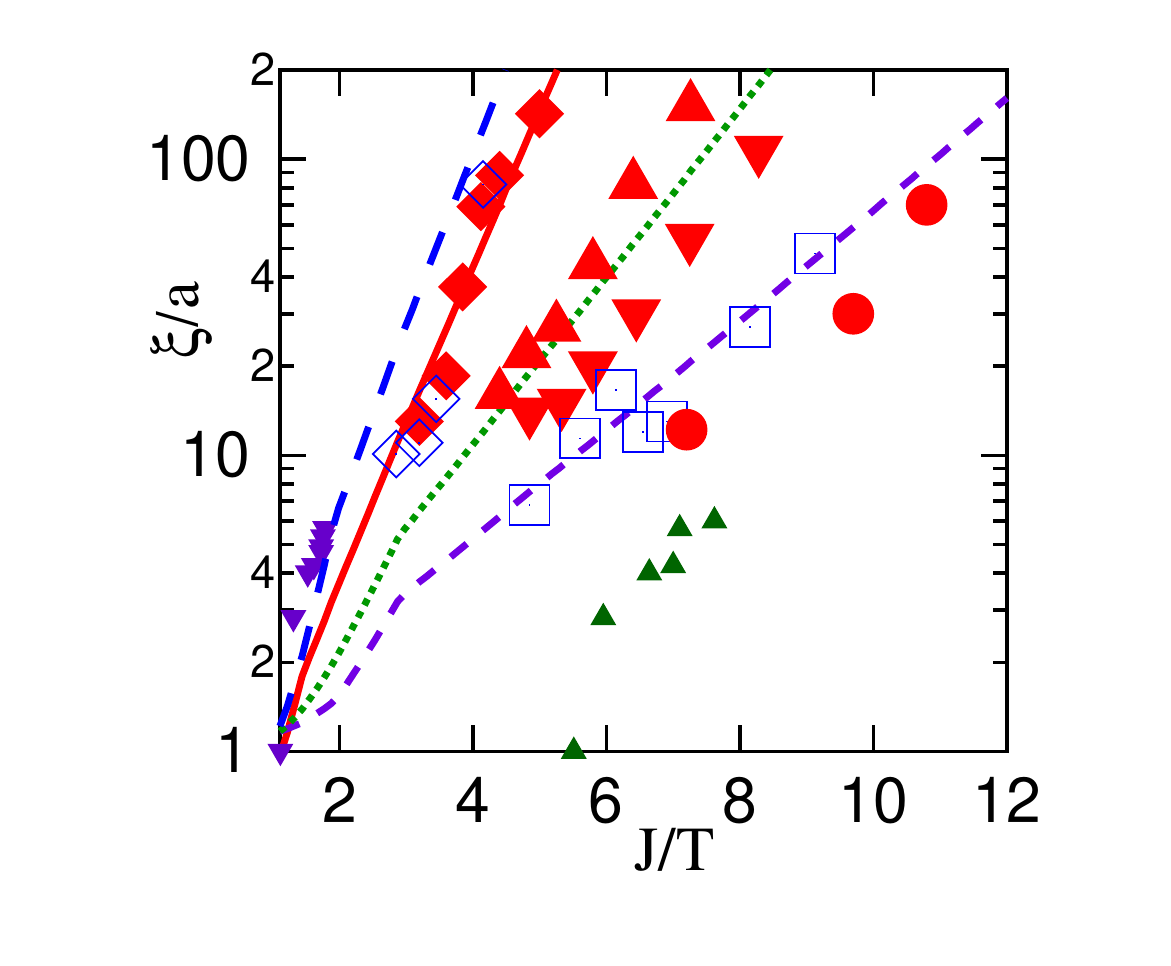}}
\caption{
{Calculated temperature dependence of the correlation length $\xi$} in NCCO\cite{MkII} for $x$=0 (red solid line), -0.04 (blue long-dashed line), -0.085 (green dotted line), and -0.10 (violet short-dashed line). Experimental data are from Ref.~\protect\onlinecite{Mat} for $x$ = 0 (blue open diamonds) and -0.15 (blue open squares); from Ref.~\protect\onlinecite{Mang} for $x$ = 0 (red solid diamonds), -0.10 (red solid up triangles), -0.14 (red solid down triangles), and -0.18 (red solid circles). Temperatures are shown in units of $J$ = 125meV. Also shown are DCA results of the $t-U$ Hubbard model with $U/t$ = 6 (green up triangles)\cite{JMHM} and 8 (violet down triangles)\cite{DMFT10}. [After Ref.~\onlinecite{MkII}.]
}
\label{fig:64C}
\end{figure}

As expected, the DCA pseudogap is associated with short-range $(\pi,\pi)$-SDW-order (Section~\ref{S:NFL}.3), which collapses in a QCP near optimal doping, Fig.~\ref{fig:63C}.  
These results are similar to a calculation incorporating Mermin-Wagner physics\cite{MkII}, which found that at $x=0$, a $(\pi,\pi)$ pseudogap turns on at $T^*\sim$1400~K, much larger than the 3D Neel temperature $T_{N3d}\sim$300~K associated with weak interlayer coupling, but much smaller than the mean-field transition temperature $T_{MF}= \Delta_{MF}/1.76k_B\sim$ 13,000~K for $\Delta_{MF}\sim$2~eV.  DCA finds $T^*\sim$1100~K.\cite{DMFT3}  We would expect that DCA captures Mermin-Wagner physics, so that the observed\cite{JMHM,DMFT10} patch size dependence of $T_N$ can be translated into a $T$-dependence of the SDW-correlation-length $\xi$, which is compared in Fig.~\ref{fig:64C} to the correlation length in NCCO at several dopings.\cite{MkII,Mat,Mang}  The agreement is seen to be surprisingly good.

\subsection{Incommensurate order}

The possibility of nanoscale phase separation or stripe physics is a large subfield in the study of cuprates and other correlated materials\cite{KT}.  We will concentrate our remarks on the nature of stripes and their relationship to CDWs and SDWs, especially on whether stripes are simply a form of a CDW with an associated SDW order, or are they better viewed as a form of SDW with an associated CDW order, or are stripes a new phenomenon altogether tied to frustrated phase separation.  In fact, we find that all three scenarios can  be realized in different materials and doping regimes.  As noted already in Section~\ref{S:QP-GW}.1, hole-doped cuprates are susceptible to SDW/CDW instabilities at FS-nesting $q$-vectors, including those associated with double-nesting \cite{MGu,MGu2, GML}. However, we also find that the energy vs doping curves can become concave, indicative of a tendency towards phase separation, especially at low doping\cite{GML}.  The resulting phase separation is confined by Coulomb repulsion at short length scales. It can thus appear as a 1D stripe, where the stripe periodicity is not related directly to nesting $q$-vectors. Hence, different types of order may be present in different doping ranges.  Moreover, there is competition between different DWs in certain doping ranges, which appears to be sensitive to band structure effects.

\begin{figure}[top]
\centering
\rotatebox{0}{\scalebox{0.5}{\includegraphics{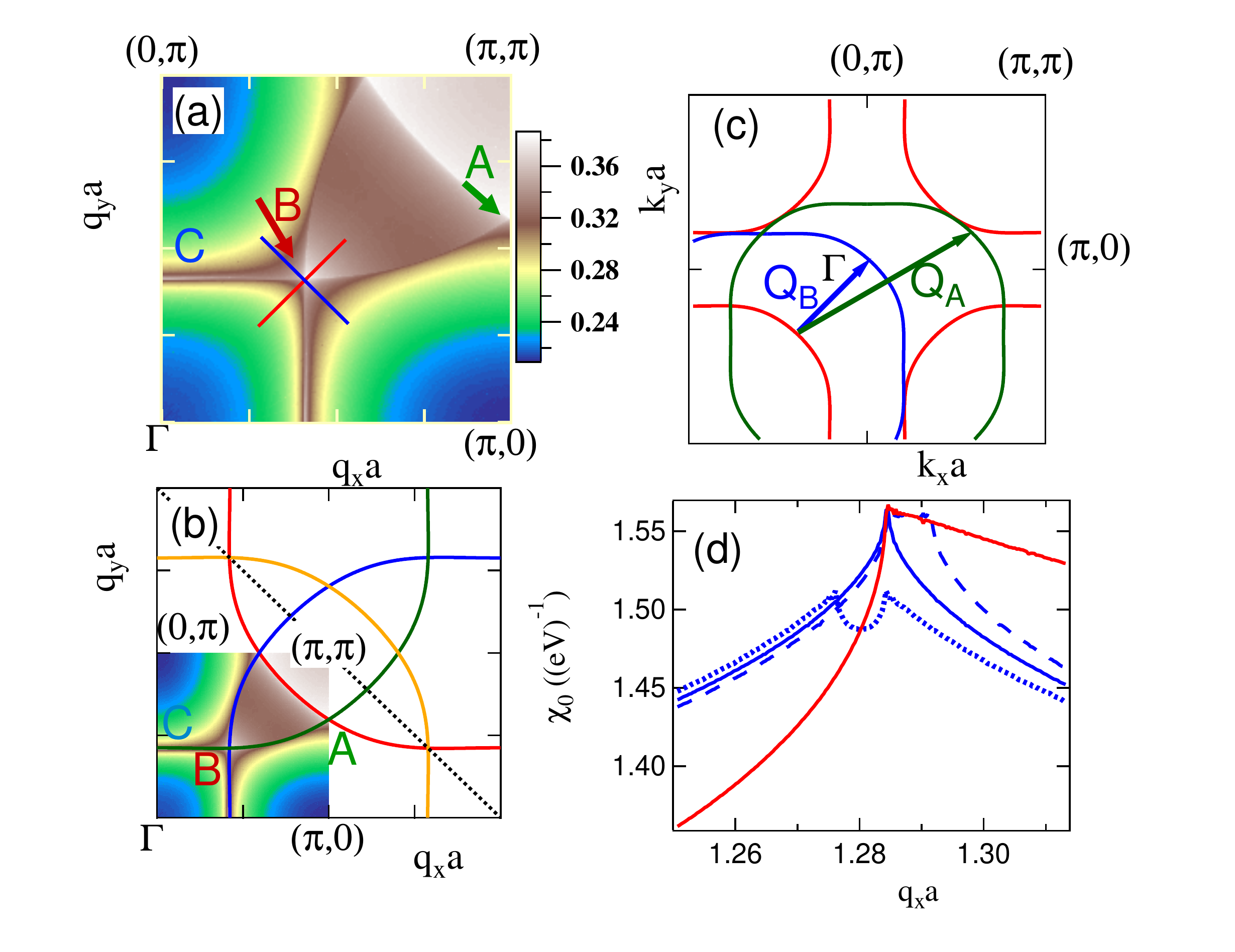}}}
\caption{ { Susceptibility and double-FS-nesting.}
(a) Real part of susceptibility, $\chi_0 '({\bf q})$, at doping
$x=0.12$. Arrows show dominant ANN (antinodal nesting) peak $B$ (red arrow)
and subdominant $(\pi ,\pi )$-plateau peak $A$ (green arrow). (b) 
$q=2k_F$ (red line) and $(\pi ,\pi )$-shifted images of the FS, 
illustrating the origin of the ridges and cusps $A$ and $B$ in (a). (c) Double- FS-nesting associated with cusps $A$ and $B$. (d) Cuts through the ANN-plateau- peak $B$ in panel (a). Solid red and blue lines give cuts along the solid lines of same color in (a), while the blue dashed and dotted lines are cuts along lines parallel to the blue line in (a), but displaced in opposite direction along the red line. [After Ref.~\onlinecite{MGu2}.]
}
\label{fig:2}
\end{figure}
Figure~\ref{fig:2}, which considers $\chi_0'({\bf q})$ (real part of bare magnetic susceptibility) at $\omega =0$, gives insight into how susceptibility peaks are related to FS-nesting.  Figures~\ref{fig:2}~(a) and~(b) show the intensity map of $\chi_0'$ over the BZ for NCCO\cite{MGu2} at $x =0.12$. Two different peaks are seen to dominate the susceptibility, leading to two competing SDW orders, as noted above in connection with Fig.~\ref{fig:DW0a}.  There is a clear plateau in $\chi_0'$ centered on $(\pi ,\pi )$ with peak at A (green arrow) and another peak of a competing order at B (red arrow).  These peaks and the connecting ridges are associated with $q=2k_F$ nesting as shown in Fig.~\ref{fig:2}(b).  Here, the red curve is the FS expressed in terms of the nesting variable ${\bf q}/2$, while the other curves are its $2\pi (n,m )$-shifted images with $n,m=0,1$.  The ridges in (a) match these FS images, indicating that these features provide calipers of the FS. For example, 
point $C$ is associated with antinodal nesting (ANN) across the flat sections of the FS-neck near $(\pi ,0)$, while peaks $A$ and $B$ originate from the crossing of two nesting curves or double-FS-nesting, see Fig.~\ref{fig:2}(c). All the aforementioned ridge features involve nonanalytic contributions to 
the susceptibility as seen, for example, from Figure~\ref{fig:2}(d), which shows various cuts through cusp B in Fig.~\ref{fig:2}(a). Similar non-analyticity is found for peaks A and B over the full doping range.

\subsubsection{SDWs}

\begin{figure}
\centering
\includegraphics[width=13.5cm,clip=true]{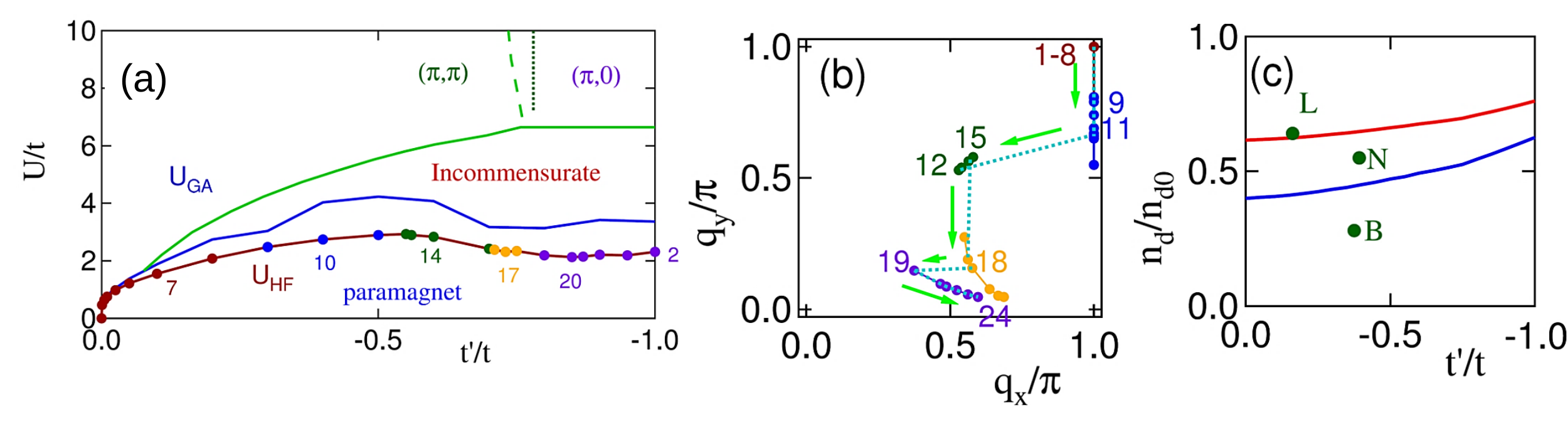}
\caption{
(a) {Phase diagram of the $t-t'-U$ Hubbard model at half-filling} including incommensurate phases. Results based on Hartree-Fock $U_{HF}$ (brown line) and Gutzwiller approximation $U_{GA}$ (blue line) are compared. Shown also is the line of first order transitions to commensurate $(\pi ,\pi )$ or $(\pi ,0)$ order (light green solid line), and the $(\pi ,\pi )$ to $(\pi ,0)$ crossover line (light green dashed line). For ease of comparison, the crossover line of Ref.~\onlinecite{TBPS} (green dotted line) from Fig.~\ref{fig:1d} is reproduced. 
The dominant $q$-vectors involved are marked on the HF+RPA curve and numbered consecutively. 
(b) Position of the dominant susceptibility peak in HF+RPA, color code and numbers match those shown in frame (a). Note that some metastable points are included here, which are not shown in (a). Blue dotted line traces evolution of stable points.
(c) $n_d/n_{d0}$ as a function of $t'$, assuming a constant $U=8t$ (red line) or
$10t$ (blue line). Here, $n_{d}$ denotes fractional double-occupancy, and $n_{d0}=0.25$ is the corresponding uncorrelated value. Symbols indicate experimental dispersion renormalization $Z_{disp}$ as a function of $t'$ (from Ref.~\onlinecite{markietb}), and give an experimental measure of $n_d/n_{d0}$. Letters refer to L = LSCO; N = NCCO; B = Bi2212. [From Ref.~\onlinecite{MGu}.]
}
\label{fig:1bb}
\end{figure}

While Fig.~\ref{fig:1d} showed the commensurate phases of the undoped $t-t'-U$ Hubbard model, it is generally found that, due to FS nesting, competing incommensurate phases become unstable at lower values of $U$. This is shown in Fig.~\ref{fig:1bb}(a), where $U_{GA}$ [$U_{HF}$] is the critical value of $U$ in the GA [HF] calculations above which the system is unstable to magnetic order.  While the HF+RPA calculations overestimate the stability of the magnetic phase, the overestimate is not very large.\cite{foot1e}  In all cases, the most unstable $q$-vector along the symmetry lines considered is the same in both the HF+RPA and GA+RPA 
results.  The main effect of the GA thus is to renormalize $U\rightarrow U_{GA}<U$ and reduce the range of ordered magnetic phases. In Fig.~\ref{fig:1bb}(a), the HF+RPA calculations are coded with variously colored circles which match the points in Fig.~\ref{fig:1bb}(b), and identify the ordering $q$-vectors involved.  These changes are associated with the evolution of the FS with $t'$.

\begin{figure}
\leavevmode  
\centering
\resizebox{13.2cm}{!}{\includegraphics{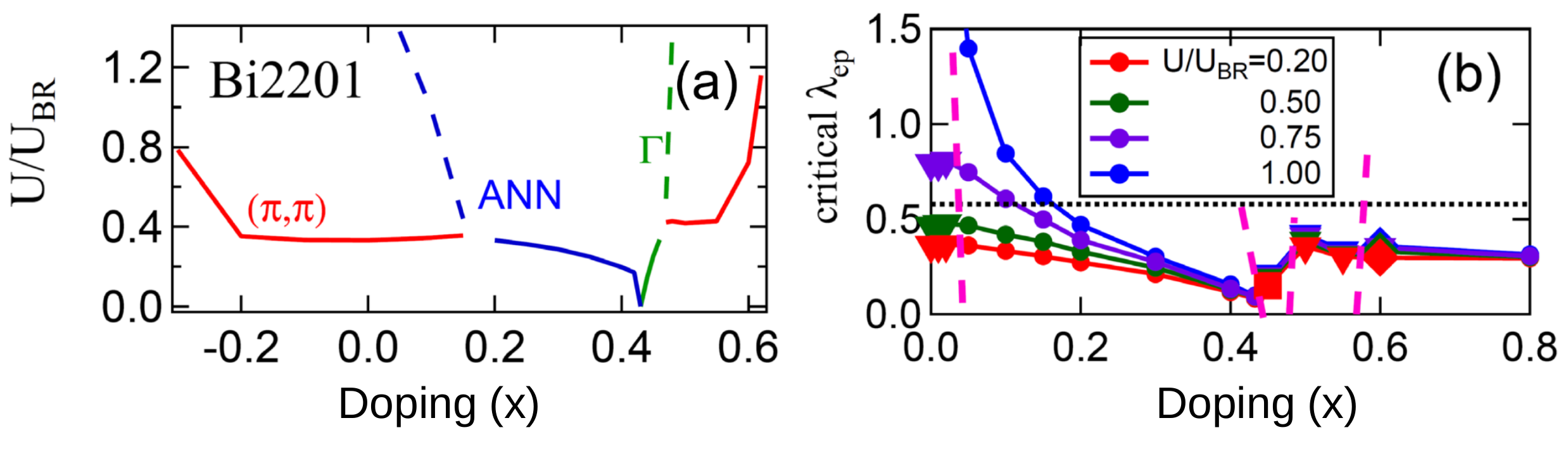}}
\vskip0.5cm
\caption{
{Gutzwiller-approximation based (a) magnetic and (b) charge phase diagrams} of Bi2201. (a) Dashed lines indicate metastable states, i.e. extensions of the condition $U_{Gutz}\chi = 1$ beyond the point where some phase becomes unstable. Red curves labeled $(\pi,\pi)$ refer to phases with near-$(\pi,\pi)$ or NNN ordering vectors. [From Ref.~\onlinecite{MGu2}.] (b) Threshold strengths of electron-phonon coupling $\lambda_{ep}$ as a function of $x$ for several values of $U$. Symbols represent symmetries of different CDW ordering vector: 
Vertical [diagonal] near-$(\pi ,\pi )$ (NNN) phase by triangles [diamonds], and   vertical [diagonal] ANN phase by squares [circles]. Dashed lines indicate transitions between different symmetries, while dotted line corresponds to the expected $\lambda_{ep}$ for Bi2201. [From Ref.~\onlinecite{zx8}.]}
\label{fig:8A}
\end{figure}  

Figure~\ref{fig:8A}(a) shows the magnetic phase diagram  for Bi2201 calculated in the Gutzwiller GA+RPA approach based on the Stoner criterion of Eq.~\ref{RPAgap2}. The phase diagram is fairly generic, and similar results are obtained for Bi2212 [neglecting bilayer splitting], SCOC, and even for electron-doped NCCO (for $x<0$). The phase diagram for LSCO [not shown], however, is different due to the smaller value of $t'$, and exhibits only the NNN order.  The phase diagram for electron doping is simpler, being dominated by the simple SDW order with ${\bf q}$ very close to $(\pi ,\pi)$, while for hole-doping there is a competition between the NNN and ANN orders, except in LSCO, where the NNN order dominates.  

\subsubsection{CDWs}

\begin{figure}
\centering
\includegraphics[width=13cm,clip=true]{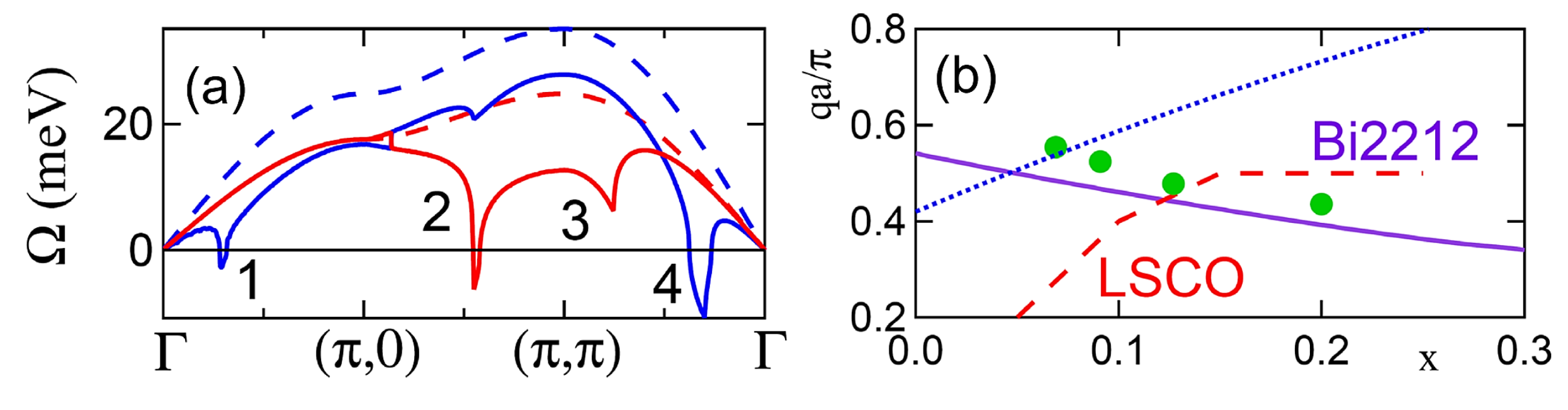}
\caption{ (a) {Phonon renormalization in Bi2201.}
Bare phonon dispersion (dashed lines) is compared to the dressed dispersion assuming $U/U_{BR}$= 0.60 (solid lines) at $x$ = 0.20. Longitudinal [transverse] phonons are plotted as blue [red] lines. For Bi2201, $U_{BR}=13.6t$. Only
modulation of the nearest-neighbor hopping $t$ is included.
Real $\Omega_{ph}$'s are shown as positive numbers, and imaginary $\Omega_{ph}$'s as negative numbers. 
(b) { Nesting vectors, $(q,q)$, in Bi2212} for the charge order state of the antibonding band (violet solid line) are compared with the experimental, $(0,q)$, charge-stripe vectors (green circles)\protect\cite{JCD}. Results for the NNN secondary charge-stripe order expected for LSCO (red dashed line) or Bi2212 (blue dotted line) are also shown.
[After Ref.~\onlinecite{zx8}.]
}
\label{fig:5}
\end{figure}

Although ANN order seems to be relevant to the `checkerboard' phase seen in STM studies of several cuprates,\cite{checkDavis} STM is sensitive to charge order rather than magnetic order.
While we have not found a charge ordered phase in a pure Hubbard model, the inclusion of electron-phonon coupling via a modulation of the hopping parameters leads to a phonon-softening CDW instability.\cite{zx8}  The phase diagram can be described in terms of a critical electron-phonon coupling $\lambda_{ep}$ vs $x$ for fixed $U$, as in Fig.~\ref{fig:8A}(b).  As doping changes, the threshold $q$-vector varies, and its symmetry can also change as indicated by various symbols in the figure. Despite some differences, the phase diagram of Fig.~\ref{fig:8A}(b) resembles the magnetic phase diagram of Fig.~\ref{fig:8A}(a), including the dominant $q$-vectors. 
This similarity arises because the CDW instability, driven by phonon softening, is also controlled by peaks in susceptibility, which lead to Kohn anomalies, Fig.~\ref{fig:5}(a).  Figure~\ref{fig:5}(b) shows that the dominant nesting vectors in STM studies of Bi2212\cite{JCD} are in reasonable agreement with the predictions for charge order. Note that if the density-wave order here was a magnetic NNN order, as in LSCO, then the {\it magnetic} $q$-vector would lie in the $(\pi,q)$-direction, while the secondary charge order would have an {\it opposite} doping dependence from experiment, see blue dotted line in Fig.~\ref{fig:5}(b).

Very recently, strong evidence for CDW order has been found in YBCO\cite{YCDW2,YCDW3,YCDW4,YCDW5,YCDW6}, Bi2201\cite{comin}, and Hg1201\cite{qGrev}, with nesting vectors consistent with Fig.~\ref{fig:5}(b).  A number of purely electronic models of the CDW have now appeared\cite{Metlitski,Pepin,LaPlaca,Hayward,Bulut,wang14,allais14}.  However, all proposed models are based on an assumed $(\pi,\pi)$-dominated spin susceptibility with a quantum critical point $x_c$ where $U\chi_{(\pi,\pi)}(T=0)\sim 1$, leaving strong, commensurate fluctuations for $x>x_c$, which induce a competition between CDW and superconductivity.  As we have seen in Fig.~\ref{fig:8A}(a), as doping increases, most cuprates crossover to a regime where the ANN susceptibility is the largest, and the near-$(\pi,\pi)$ fluctuations are cut off.  Although near-$(\pi,\pi)$ fluctuations dominate in LSCO at all dopings, the CDW seems to be absent, and the high-doping regime is consistent with a spin-density wave with an incommensurate $(\pi,\pi-\delta)$ nesting vector.\cite{MGu2} Notably, Ref.~\cite{YCDW4} estimates that x-ray diffraction intensities are enhanced by a factor of $\sim 600$ due to coupling of the CDW to a lattice  distortion, an effect which cannot be treated within a purely electronic CDW model. 


Quantum oscillations have been observed in YBCO\cite{QOsc,QOsc3,QOsc4,Seba1} and Hg1201\cite{qGrev}, indicating a strong FS renormalization, which leaves only a single small electron-like pocket. The area of this FS-pocket can be explained by a model of coherent crossed ANN CDWs\cite{Seba3}, similar to the model discussed in Ref.~\cite{MGu2} Also, ARPES studies indicate the existence of a trisected SC dome, suggestive of quantum critical points (QCPs) near both the lower and upper ends of the dome\cite{Inna}.  The dopings of the QCPs are consistent with the scenario developed here, with SDWs at low doping crossing over to the CDWs at higher doping, terminating in a Fermi liquid phase in the overdoped regime. The model proposal of the CDW induced electron pocket in the nodal region, however, has a serious issue. The electron pocket, by construction, has a band folding below the Fermi level, which means there should be a quasiparticle gap in the electronic structure along the nodal direction.\cite{DasCDWProb} No ARPES evidence of the nodal gap below the Fermi level is reported in any hole-hoped materials.

%




\section{Extensions of present model calculations}\label{S:Ext}

\subsection{Mermin-Wagner Physics and Quantum critical phenomena}\label{S:Ext_a}

In a 2D material, the MW theorem\cite{MW} states that fluctuations destroy density-wave order at finite temperatures, although a transition at $T=0$ is allowed.  
Our QP-GW calculations have been informed by earlier work on cuprates, which incorporated Mermin-Wagner (MW) physics\cite{MW} via mode coupling theory.\cite{MkII}  
This amounts to introducing a vertex correction which modifies the HF self-energy to
\begin{equation}\label{MWSE}
\Sigma_{MW}(k,\omega) = \frac{\bar\Delta^2(T)}{\omega -\epsilon_{k+Q}+i\Gamma_{MW}(T)},
\end{equation}
where $\bar\Delta (T)$ is a pseudogap and $\Gamma_{MW}(T)\rightarrow 0$ as $T\rightarrow 0$.  If there is a residual 3D coupling, the system can develop long range order at a finite N\'eel temperature $T_N$, with relatively little change in the gap if $T_N<<T_{MF}$. This justifies taking the MF gap and transition temperature as pseudogap $\Delta^*$ and pseudogap onset temperature $T^*$, respectively, as we generally do in our QP-GW modeling. In this connection, optical conductivity was calculated in the paramagnetic phase with thermally-disordered antiferromagnetism, and compared to experiments as well as other calculations using mean-field SDW order.\cite{Linlis} The paramagnetic results were found to approach the accuracy of mean-field calculations, but only when vertex corrections were included.  MW fluctuations have recently been introduced into diagrammatic extensions of DMFT.~\cite{DGA3})

Finally, we note that superconductivity is often found in the vicinity of a QCP, and the $T_c$ may well be enhanced by the associated quantum fluctuations.\cite{SachdevQPT} Our treatment at the mean-field level, however, does not account properly for the anomalous quantum critical phenomena expected in phase transitions near $T=0$, and would require MW-physics to be incorporated for addressing such issues. It would be interesting to check if a transition from a 3D long-range order to a 2D short-range order takes place as the QCP is approached.

\subsection{Disorder Effects}\label{S:Ext_b}

The QP-GW results discussed in this article neglect disorder effects. Their inclusion can be anticipated to give various corrections as follows. (1) Since long range order is quenched by charged impurities, which act as random fields.\cite{ImMa}, density wave correlations will be frozen at a finite size without a transition to long-range order.  As already noted, QP-GW captures dominant fluctuations at the mean-field level, so that our gap should generally be considered a pseudogap, or the onset of strong fluctuations, either dynamic (due to MW physics) or static (due to random fields). (2) In the deep underdoping limit, close to the metal-insulator transition, deviations from the QP-GW results can be seen in Section~\ref{S:NFL}.2 for LSCO, where both the renormalization factor $Z_{\omega}$, Fig.~\ref{md1d}(b), and the Sommerfeld coefficient $\gamma$, Fig.~\ref{fspheat}(b), approximately scale to zero as $x\rightarrow 0$. While this could be a stripe related effect, it is also possible that the first doped holes are localized on impurities.  There is evidence for Coulomb gap effects in this doping regime\cite{Komiya,PanDing}, which could explain these residual effects.\cite{Mist1} (3) Finally, we expect a density-wave order to break the FS into pockets, but ARPES shows instead the so-called FS-arcs where the spectral intensity from portions of the FS disappears. Disorder effects could contribute to this loss of intensity in addition to those of fluctuations.\cite{harrison} 

\subsection{Stronger Correlations}\label{S:Ext_c}

The transition between local moments and itinerant magnetism has proven one of the most intractable problems in condensed matter physics, and also lies at the heart of the debate on intermediate vs strong correlations in cuprates.  
One possibility\cite{AbSi} is that much of the evolution from weak to strong correlations takes place in the {\it incoherent part} of the spectrum.  Near half-filling there is a large gap associated with magnetic order, i.e., with significant local moments on the Cu sites.  As correlations increase, electrons will become more localized, destroying long range magnetic order, while local moments remain, leading to incoherent spectral weight resembling the magnetic gap.  This can be seen in CDMFT calculations, Fig.~\ref{md2d4}(j), where the incoherent bands have clearly become the UHB and LHB, while the coherent states are confined to a very narrow mid-gap feature.  The strong-coupling pseudogap\cite{Trem4} can then be seen as a small gap near $E_F$.  This is the remnant of the SDW gap in the coherent band as discussed in Section~\ref{S:DWs}.1.2, Fig.~\ref{fig:n11}. Such stronger coupling features however are not clearly seen  in the cuprates. Firstly, when the Mott gap is so large, it is hard to understand the rapidity of the anomalous spectral weight transfer with doping in Fig.~\ref{aswt} of Section~\ref{S:NFL}.1.  Secondly, cuprates are charge transfer insulators, not Mott insulators.  We will see in Section~\ref{S:Ext}.4 how this physics is captured in three-band models through a Cu-like Mott gap spanning mainly the O-like bands, with the low-energy physics driven by the SDW order in the UHB.  It is less clear, however, how one can understand both the Mott physics and the SDW order in the Cu bands above the oxygen bands. While our QP-GW model is designed to handle intermediate coupling, it also contains a number of features expected in the strongly correlated limit of the Hubbard model as follows.

\subsubsection{Suppression of double-occupancy}
Correlations in cuprates are expected to be most important at half-filling, where suppression of double-occupancy can lead to a Mott insulating phase. For example, when a Gutzwiller projection is applied to an assumed paramagnetic or superconducting ground state, double-occupancy is significantly reduced\cite{footB5}.  However, the mean-field SDW suppresses double-occupancy so well that Gutzwiller projection on an assumed SDW ground state actually {\it increases} double-occupancy as discussed in Ref.~\onlinecite{footB5}.

\subsubsection{Spin-wave dispersion}
Schrieffer, Wen, and Zhang\cite{SWZ} demonstrated that the mean-field SDW correctly reproduces the spin-wave spectrum of the Hubbard ($t-U$) model, and by including more distant hoppings ($t',t''$), the experimental spin-wave dispersions in the undoped cuprates can be described reasonably.  At half-filling, the predicted on-site magnetic moments are also of the correct order of magnitude. 

\subsubsection{Mott vs Slater Physics}
A Mott insulator is traditionally distinguished from a conventional Slater insulator or a band insulator in that the gap in the latter case is associated with symmetry breaking transitions and long-range order. In sharp contrast, the gap in a Mott insulator is driven by a reduction in double-occupancy, which opens a gap without an accompanying long-range order or a reduction in symmetry. The situation in 2D is, however, less clear cut. As discussed in Section~\ref{S:Ext}.1 above, the Mermin-Wagner theorem turns the density wave gap into a pseudogap at $T>0$.  On the other hand, in strong correlation models of cuprates it is often found that long-range order turns on at temperatures well below the temperature where the Mott gap first appears, indicating that the distinction between the density-wave gap and the Mott gap becomes less clear. For instance, the spin-correlation length in the doped Heisenberg model\cite{Manousakis} has an exponential divergence at low temperatures, which is of the same form as that in the SDW model with Mermin-Wagner corrections\cite{MkII}.

\subsubsection{Quasiparticle dispersion}
Even at the mean-field level, the large-$U$ limit antiferromagnet reproduces the one-hole [one-electron] dispersion of the lower [upper] Hubbard band in the $t-J$ model. Self-energy corrections add an incoherent part to the dispersion ($t$-scale). Thus, when $U$ is large, the magnetization saturates ($S=1/2$), so that $\Delta_{SDW}=U/2$, and the dispersion of the upper and lower magnetic bands approximately becomes
\begin{equation}\label{ULHB}
E_{k}^{\pm}=\epsilon_{k}^+\pm [\frac{U}{2}+\frac{(\epsilon_{k}^-)^2}{U}].
\end{equation}
This should be compared to the results of the $t-J$ model.  In the large $U$ limit, the hopping $t$ renormalizes to zero at half-filling, and the model reduces to the Heisenberg model.  Adding a hole to the Heisenberg model yields a finite dispersion, which can be calculated in the non-crossing approximation\cite{LiuMan}, leading to a dispersion of width $\sim 2J$.  Remarkably, the RPA solution of Eq.~\ref{ULHB} matches this dispersion of upper and lower Hubbard bands for arbitrarily large values of $U$.  Taking the simplest case of the pure Hubbard model with nearest neighbor hopping ($t$) only, with $Q=(\pi,\pi)$ and $\epsilon_{k+Q}=-\epsilon_k$, so that $\epsilon_k^+=0$ and $\epsilon_k^-=\epsilon_k=-2t[cos(k_xa)+cos(k_ya)]$, we obtain
\begin{eqnarray}\label{ULHB2}
E_{k}^{\pm}&=&\pm [\frac{U}{2}+J[\cos(k_xa)+\cos(k_ya)]^2]
\nonumber\\
&=&\pm [\frac{U}{2}+J(1+\frac{\cos(2k_xa)+\cos(2k_ya)}{2}+2\cos(k_xa)\cos(k_ya))],
\end{eqnarray}
where $J=4t^2/U$.  From the second form of Eq.~\ref{ULHB2}, it can be seen that all hopping is on the same SDW sublattice, and hence does not lead to any double-occupancy.  Equation~\ref{ULHB2} provides a good fit to the numerical results for the Heisenberg model\cite{LiuMan}.  Similarly, when more distant hoppings ($t'$ and $t''$) are included, Eq.~\ref{ULHB} provides the correct extension.  The underlying dispersion is thus given correctly by the HF result for arbitrarily large $U$.  The HF model has two important advantages over the $t-J$ model.  First, it includes both the upper and lower magnetic bands, which play the role of upper and lower Hubbard bands.  In contrast, the $t-J$ model describes only one, say the LHB, while sending the UHB infinitely far away as $U\rightarrow\infty$.  In this way, the HF model can describe optical interband transitions inaccessible to the $t-J$ model.  Moreover, as $U$ decreases into the intermediate coupling regime, the HF model would intrinsically provide a better approximation. 

\subsubsection{Transition temperature at strong coupling}  
A characteristic feature of Mott physics is that as $U$ increases the Mott gap grows but the magnetic transition temperature ultimately decreases.  The electrons become more localized, interacting weakly by the $t-J$ model, or Heisenberg model at half-filling. In this case the magnetic transition temperature should scale as $T_N\sim J \sim 1/U$.  

It is straightforward to see that a similar effect arises in the QP-GW formalism due to a breakdown of RPA associated with spin localization.  Even though the gap grows with increasing $U$, we expect the energy difference between any two magnetically ordered phases to scale as $J$.  We illustrate this with the example of a half-filled band with only nearest neighbor hopping.   Of all possible magnetic phases, the lowest in energy is the $(\pi,\pi)$-AFM, while the highest in energy is the FM.  Repeating the argument of Section~\ref{S:Ext}.3.4, the FM state has dispersion 
\begin{equation}\label{ULFB}
E_{k,FM}^{\pm}=\epsilon_{k}\pm \frac{U}{2}.
\end{equation}
For large $U$, one band is below $E_F$, and the other above, and the average energy per occupied electron is $\bar E_{FM}=-U/2$.  The average energy of the AFM state is lower than this by $\Delta\bar E_{AFM}=-J$.  By choosing a particular magnetic order, the energy per particle can be lowered by $\sim J$, but by allowing the spin direction to fluctuate the entropy is increased, lowering the free energy per particle by $\sim T$.  Thus when $T>J$, magnetic order will be lost, even though the moment per site or the average gap is large. 


\subsubsection{Anomalous spectral weight transfer (ASWT)}

As already noted in Section~\ref{S:NFL}.1, doping a Mott insulator leads to ASWT with doping: removing one electron from a Mott insulator leads to two empty states at low energies (in the lower Hubbard band), since there is no longer a penalty $U$ to be paid when the spin-reversed electron occupies the state\cite{EMS}.  The QP-GW
model captures most of this effect due to collapse of the
magnetic gap at a QCP, Section~\ref{S:NFL}.1.\cite{ASWT}, although it underestimates ASWT in the very low doping regime, where however this effect 
might become unobservable as a result of masking effects of strong disorder.

\subsubsection{Zhang-Rice Physics}

In a three-band model, the HF also reproduces much of the Zhang-Rice phenomenology\cite{ZR}, i.e., the first holes dope predominantly into oxygen states.  This point will be discussed further in Section~\ref{S:Ext}.4.

\subsubsection{One-dimensional Hubbard Model}

Finally, we recall the 1D Hubbard model to gain further insight.  Here one usually considers electrons fractionalizing into spinons and holons.  However, spinons and holons arise as topological defects in a 1D SDW background.  That is, the MF ground state would be SDW, but fluctuations destroy long range order. The holons and spinons thus describe fluctuations in a 1D SDW in the spirit of Mermin-Wagner physics.    

\subsection{3- and 4-band models of cuprates}\label{S:Ext_d}

\begin{figure}
\centering
\includegraphics[width=10cm,clip=true]{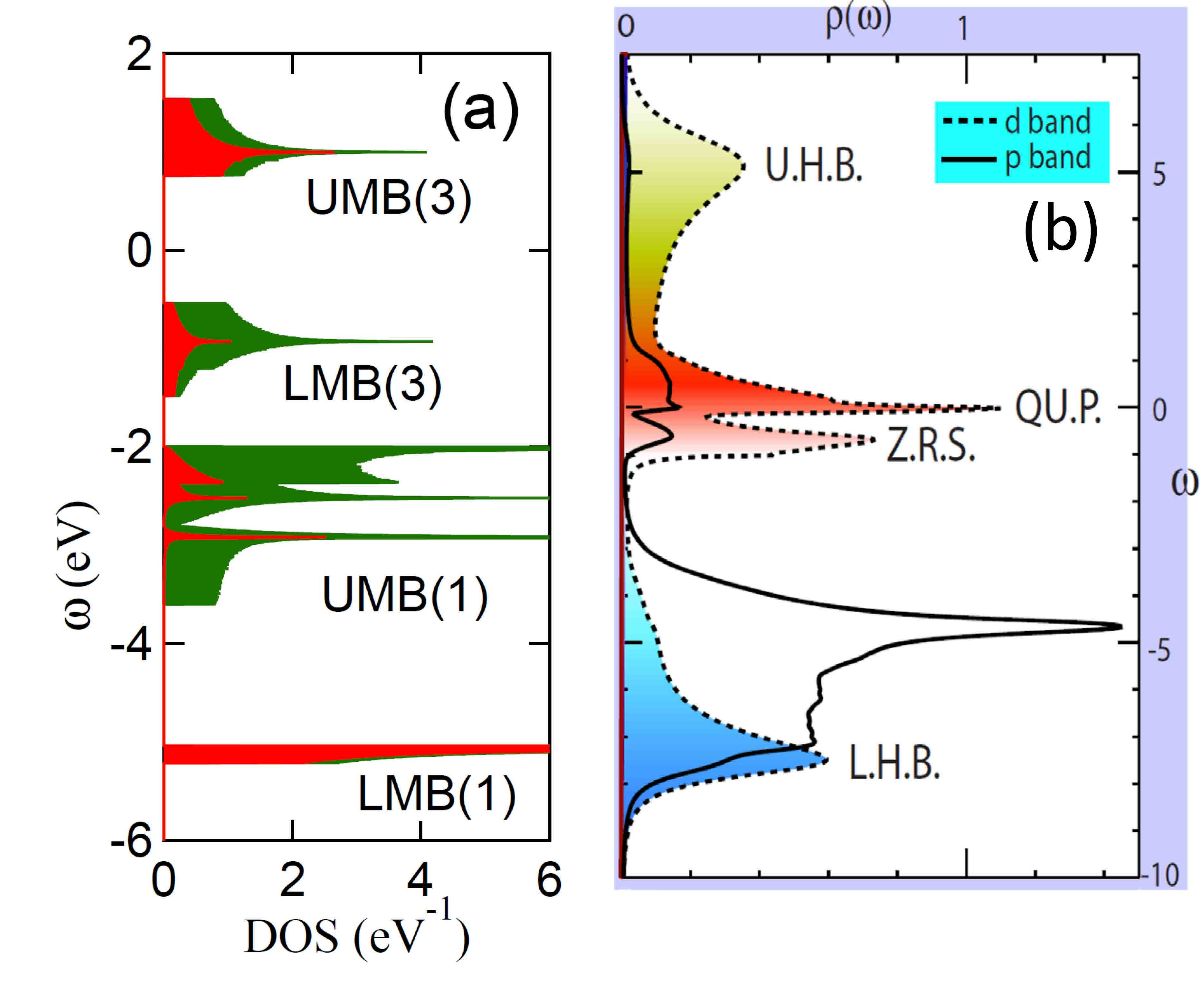}
\caption{ (a) { Three-band model} DOS for NCCO and its Cu (red) and O (green) components at half-filling using a bare three-band Cu-Cu interaction $U_{3b}$ = 7.5 eV on Cu-sites. [After Ref.~\onlinecite{MBRIXS}] (b) Corresponding DMFT results of DOS $\rho(\omega)$ for LSCO at $x=0.24$. [From Ref.~\onlinecite{WHK}] }
\label{3brixs}
\end{figure}

An important desideratum for a viable model of cuprates is that the key low-energy physics of the model be insensitive to the number of bands included in the model.  In particular, how can the Mott gap of a one band [copper only] model become the charge-transfer gap of a three-band model [Cu $d_{x^2-y^2}$, O$_1$ $p_x$, O$_2$ $p_y$]?  We address this issue with the example of our three-band model\cite{MBRIXS}.  In the absence of SDW order, the Cu-O hybridized antibonding band of the three-band model can be identified with the nonmagnetic band of the one-band model, with no charge transfer gap.  
However, the three bands are split into six bands when the SDW is introduced, accompanied by a {\it dehybridization} of Cu and O, so that the LMB of the bonding band (labeled LMB(1) in Fig.~\ref{3brixs}(a)) and the UMB of the antibonding band (labeled UMB(3)) are now nearly pure Cu.  This restores the picture of a charge transfer insulator, with the upper and lower Cu Hubbard bands straddling the O-bands, Fig.~\ref{3brixs}(a).\cite{ZSA}  The gap near $E_F$ is a magnetic gap between the antibonding LMB of mainly an O-character and the antibonding UMB of mainly Cu-character, Fig.~\ref{3brixs}(a).  Note that the true Mott gap is approximately $U_{3b}=7.5$~eV, but since $t_{CuO}=1.5$~eV, we obtain $U_{3b}/t_{CuO}\sim 5$. In this way, the model captures an important aspect of Zhang-Rice physics in that the LMB of the antibonding band is mainly oxygen-derived, and the first doped hole will have about 85\% oxygen character. In order to recover the charge transfer insulator model, we chose a very small value for the onsite energy difference $\Delta_{CuO}=\epsilon_{Cu}-\epsilon_O\sim 0$.  For large values of $\Delta_{CuO}$, most of the Cu-weight lies above the O-weight, and the system becomes a Mott insulator.  A small value for $\Delta_{CuO}$ has also been discussed in Refs.~\onlinecite{OKA,Varma}.


\begin{figure}
\centering
\includegraphics[width=14cm,clip=true]{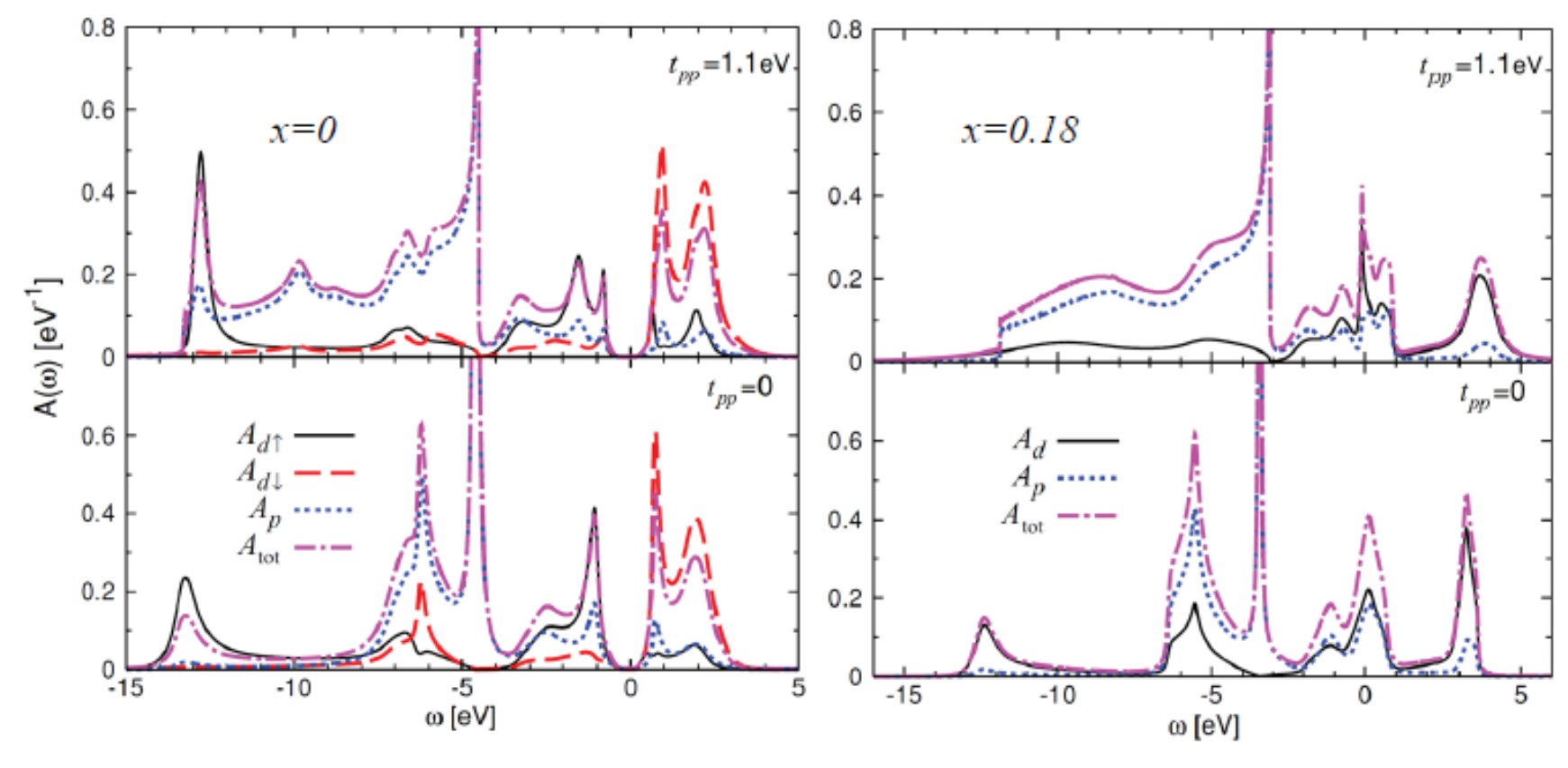}
\caption{ {DMFT results for DOS in a three-band model of the cuprates} for undoped (left column) and $x=0.18$ hole doping (right column), assuming the oxygen-oxygen hopping parameter $t_{pp}$ = 1.1~eV (top row) or 0 (bottom row). [From Ref.~\onlinecite{WM2}] }
\label{3brixs2}
\end{figure}

The results of Fig.~\ref{3brixs}(a) provide a conceptual basis for how Zhang-Rice physics arises within the intermediate coupling scenario, and show how a three-band charge transfer insulator is consistent with the model of a one-band Mott insulator.  The key is that both the Cu-like UMB(3) and the O-like LMB(3) in Fig.~\ref{3brixs}(a) are effectively half-bands, a feat that can be accomplished via Slater physics but is difficult in a pure Mott system.  Said somewhat differently, why should an ordinary charge transfer gap scale with SDW magnetization?\cite{MkII}  The reason is that the antibonding band of the three-band model is similar to the Cu-band of the one-band model, and one may view the spectrum as being derived from an SDW or a Mott-like framework.  These considerations provide a robust basis for invoking multi-band models for treating properties of the cuprates as we have done in modeling matrix element effects in a number of cases. 
For example, Figure~\ref{figSE}(a) shows how by using the dispersion of Fig.~\ref{3brixs}(a), the Hubbard $U$ generates both the large charge gap and a much smaller magnetic gap seen in RIXS experiments. 


Similar results have been obtained in DMFT calculations shown in  Figs.~\ref{3brixs}(b)\cite{WHK} and~\ref{3brixs2}\cite{WM2}, some differences notwithstanding. 
In comparing the DMFT and QP-GW results with reference to Figs.~\ref{3brixs}(b)\cite{WHK} and~\ref{3brixs2}\cite{WM2}, exact agreement is not expected, as the calculations assume different hopping and interaction parameters and Fig.~\ref{3brixs}(a) lacks self-energy corrections.  
Also, Fig.~\ref{3brixs}(b) is for $x=0.24$ hole-doping, which would move $E_F$ in frame (a) into LMB3 and open a pseudogap. 
Nevertheless, all calculations correctly capture, to varying degrees, the Cu character of the highest and lowest energy bands. 
Notably, we suggested in Section~\ref{S:ElSpec2}.2 that an intense feature seen in RIXS spectra of most cuprates represents the Mott gap, which corresponds to transitions between these two bands. 


\section{Other materials and multiband systems}\label{S:Other}

\begin{figure}
\centering
\includegraphics[width=13cm,clip=true]{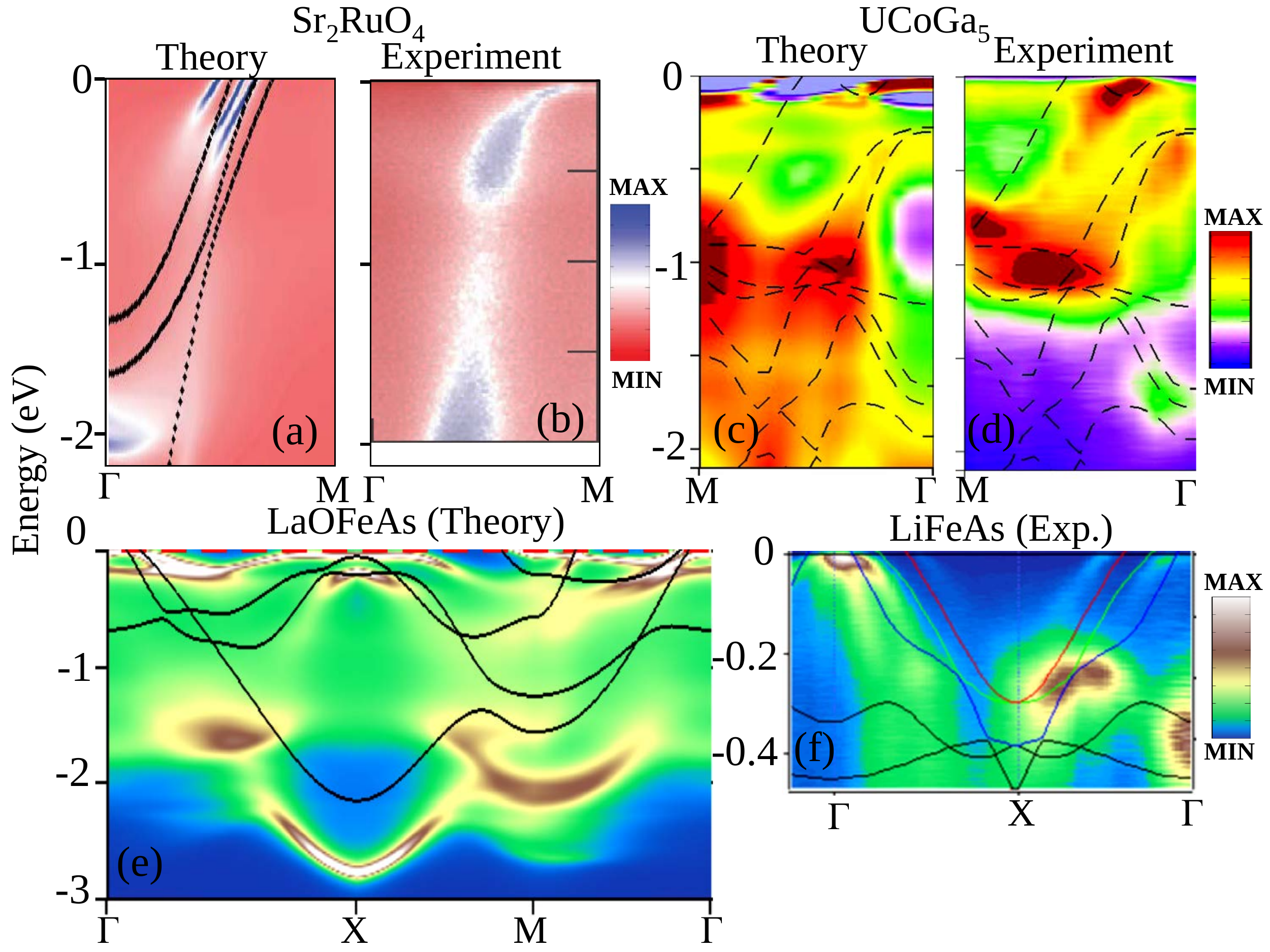}
\caption{Electronic spectrum of several different multiband systems including QP-GW self-energy corrections. Theory color plots give the one-particle spectral intensities without including effects of the ARPES matrix element. (a-b) Sr$_2$RuO$_4$.\cite{DasOtherMat,Sr2RuO4ARPES}, computed bare bands and spectral intensity (a) and experimental ARPES results. (c-d) Similar results for a representative actinide compound.\cite{DasActinidePRX} Related Pu-based compounds exhibit similar behavior.\cite{DasActinidePRL,Joyce} (e-f) A pnictide superconductor. }
\label{OtherMat}
\end{figure}

As discussed in connection with Fig.~\ref{Intermediate} above, the emergence of an intermediate coupling scenario involves strong dynamical fluctuations, which drive a sign reversal in $\Sigma^{\prime}$ and a related peak in $\Sigma^{\prime\prime}$, yielding a kink in dispersion and a peak-dip-hump feature in the DOS. In this connection, we have extended our QP-GW methodology to treat multiband systems\cite{DasActinidePRL,DasActinidePRX,DasActinideMRS}. Illustrative results for a number of materials are shown in Fig.~\ref{OtherMat} to highlight the generic nature of the QP-GW phenomenology in correlated materials more generally. Although a simple waterfall-like dispersion is more difficult to draw in multiband systems, nevertheless, a 
visual inspection of various spectra in Fig.~\ref{OtherMat} reveals the ubiquitous signature of the intermediate coupling scales.  A few material-specific comments are made in the following subsections.

\subsection{Sr$_2$RuO$_4$}
Figs.~\ref{OtherMat}(a) and (b) show that QP-GW calculations capture the HEK in Sr$_2$RuO$_4$. The low-energy spectrum of interest in Sr$_2$RuO$_4$ is dominated by $t_{2g}$-orbitals with moderate spin-orbit coupling. Theoretical results of Fig.\ref{OtherMat}(a) are based on tight-binding parameters obtained from Ref.~\cite{Sr2RuO4TB}.  Orbital- and momentum-dependent QP-GW self-energy is computed assuming a paramagnetic phase. The dressed single-particle spectrum is shown in Fig.~\ref{OtherMat}(a), and it is seen to be in reasonable accord with the corresponding ARPES data from Ref.~\cite{Sr2RuO4ARPES} reproduced in Fig.~\ref{OtherMat}(b). The strongest spin-fluctuations arise from particle-hole transitions between two flat regions of bands near $(\pi,0)$\cite{DasOtherMat}, and involve $d_{xz}/d_{yz}$ orbitals below $E_F$ and hybridized $d_{yz}/d_{xz}$ and $d_{xy}$ orbitals above $E_F$. Mass enhancement is estimated to be $Z\sim 0.41$ for all orbitals in good agreement with the value of 0.4 extracted from the ARPES data.\cite{Sr2RuO4Z} LDA+DMFT calculations predict a mass renormalization of 0.43$\pm$0.5.\cite{DMFT_Ruthenate}.

\subsection{UCoGa$_5$}
The 115 series of actinide compounds, superconducting PuCoGa$_5$, PuRhGa$_5$ and  PuCoIn$_5$, and the non-superconducting UCoGa$_5$, all appear to lie in the intermediate coupling regime with evidence of itinerant/local duality,\cite{DasActinidePRX} and a peak-dip-hump structure in their photoemission spectra\cite{Joyce}. In these materials, the low-energy 5$f$-electrons experience strong spin-orbit coupling of $\sim$1~eV (larger than Hund's coupling), which splits the low-energy spectrum into conduction and valence portions.\cite{JXZhu,Oppeneer} The resulting bands are quite flat near $(\pi,\pi)$ with substantial three-dimensionality in the particle and hole channels, and thus give rise to strong spin-fluctuations\cite{DasActinidePRL}, or more precisely, spin-orbit fluctuations.\cite{DasURu2Si2})  The computed QP-GW self-energy exhibits the characteristics of intermediate coupling, and splits the low-energy spectrum into itinerant and localized states as shown in Fig.~\ref{OtherMat}(c). These results were predicted in Ref.~\cite{DasActinidePRX} and observed subsequently in ARPES measurements on UCoGa$_5$, with a computed momentum averaged renormalization at $E_F$ of $Z\sim0.5$, comparable to the corresponding experimental estimate of 0.38.\cite{DasActinidePRX}

\subsection{Pnictides}
We comment on the pnictide superconductors with reference to Figs.~\ref{OtherMat}(e) and (f).  Theoretical results on LaOFeAs in Fig.~\ref{OtherMat}(e) are compared with  available data for LiFeAs over a large energy and momentum range\cite{BorisenkoLiFeAs} in Fig.~\ref{OtherMat}(f). In pnictides, the crystal-field splitting is relatively small while the Hund's coupling is large, which causes all five $d$-electrons to be present near $E_F$. This complicates the spin-fluctuation spectrum and the associated self-energy. \cite{DasOtherMat} QP-GW calculations in Fig.~\ref{OtherMat}(e) are based on tight-binding parameters taken from Ref.~\cite{pnictideTB}.  The computed renormalization of various orbitals is estimated as: $Z_{xz/yz}\sim$0.35; $Z_{xy}\sim$0.45; and, $d_{x^2-y^2}$ and $d_{z^2-r^2}$ around 0.48-0.53, in agreement with DMFT results\cite{DMFTpnictide}. The ARPES spectra in Figs.~\ref{OtherMat}(e-f) exhibit the itinerant/localized duality in low- and high-energy regions like other correlated materials.

\section{Conclusions and Outlook}\label{S:Disc}

We have shown that the intermediate coupling scenario provides a robust basis for gaining significant insight into wide-ranging properties of the electron- and hole-doped cuprates, including the interplay of their complicated doping and temperature dependencies and aspects of their phase diagrams. Our specific implementation of this scenario in the form of the comprehensive, beyond LDA, material-specific QP-GW scheme is shown to capture many key features of the ARPES, STM/STS, RIXS, optical, neutron and other spectroscopic measurements on the cuprates. Extensions of the intermediate coupling scheme to encompass multiple correlated bands, other orders than mainly the $(\pi, \pi)$-SDW-order considered in much of this review, and application to complex materials such as the pnictides, actinides, heavy-Fermion materials, and other oxides, as well as extensions to consider transport properties are warranted.

\section*{Acknowledgments}
It is a pleasure for us to acknowledge our many theoretical and experimental colleagues for collaborations and discussions over the years: 
T. Ahmed, W. Al-Sawai, Y. Ando, V. Arpiainen, M. C. Asensio, J. Avila, A. V. Balatsky, B. Barbiellini, S. Basak, S.V. Borisenko, M.C. Boyer, R. Cantelli, K. Chatterjee, Y.L. Chen, F. Cordero, T.P. Devereaux, H. Eisaki, J. Fink, A. Fujimori, M. Fujita, J.-M. Gillet, M. J. Graf, G. D. Gu, F. Guinea, M. Z. Hasan, R.H. He, Y. He, V. Hinkov, J.E. Hoffman, D. Hsieh, E.W. Hudson, Z. Hussain, H. Ikuta, M. Itou, Y. Kaga, T. Kakeshita, S. Kaprzyk, J.J. Kas, B. Keimer, S. Komiya, T. Kondo, A.A. Kordyuk, C. Kusko, A. Lanzara, W.S. Lee, Y. W. Li, H. Lin, M. Lindroos, J. Lorenzana, D.H. Lu, N. Mannella, W. Meevasana, P.E. Mijnarends, P. Mistark, N. Nagaosa, J. Nieminen, A. Paolone, D. Qian, J.J. Rehr, L. Roca, S. Sachdev,  S. Sahrakorpi, Y. Sakurai, T. Sasagawa, G. Seibold, Z.-X. Shen, A. Soumyanarayanan, I. Suominen, K. Tanaka, H. Takagi, T. Takeuchi, W.X. Ti, A. Tejeda, S. Uchida, F.D. Vila, M.A.H. Vozmediano, S. Wakimoto, Y. J. Wang, J. Wen, W.D. Wise, L. Wray, Y. Xia, J.W. Xiong, Z. Xu, K. Yamada, M.M. Yee, T. Yoshida, W.L. Yang, Y. Yin, J. Zaanen, M. Zech, I. Zeljkovic, Z.X. Zhao, X.J. Zhou, F. Zhou, J.-X. Zhu.
This work was supported by the US Department of Energy (DOE), Office of Science, Basic Energy Sciences contract number DE-FG02-07ER46352, and benefited from Northeastern University's Advanced Scientific Computation Center (ASCC), theory support at the Advanced Light Source, Berkeley and the allocation of time at the NERSC supercomputing center through DOE grant number DE-AC02-05CH11231.


\appendix
\section{Coexisting antiferromagnetic and superconducting orders}\label{S:AC}
\subsection{Phase Coexistence}

The Hamiltonian for a coexisting $(\pi ,\pi )$- AFM and $d$-wave superconductor is
\begin{eqnarray}\label{Hamafmsc}
H &=& \sum_{{\bf k},\sigma}\xi_{{\bf k}}c^{\dag}_{{\bf k},\sigma}c_{{\bf k},\sigma} + US^2 -
US\sum_{{\bf k},\sigma}\sigma c^{\dag}_{{\bf k}+{\bf Q},\sigma}c_{{\bf k},\sigma}\nonumber\\
&&-\sum_{i=1}^3\frac{\Delta_{i}^2}{V_{i}} +
\sum_{{\bf k}}\Delta_{{\bf k}}(c^{\dag}_{{\bf k},\uparrow}c^{\dag}_{-{\bf k},\downarrow}+
c_{-{\bf k},\downarrow}c_{{\bf k},\uparrow}).
\end{eqnarray}
It is convenient to rewrite the momentum sums into sums over the reduced
magnetic zone, in which case the Hamiltonian becomes
\begin{eqnarray}\label{Hamafmscred}
H&=&\sum_{{\bf k},\sigma}^{\prime}\Big[\xi_{{\bf k}}c^{\dag}_{{\bf k},\sigma}
c_{{\bf k},\sigma}+\xi_{{\bf k}+{\bf Q}}c^{\dag}_{{\bf k}+{\bf Q},\sigma}
c_{{\bf k}+{\bf Q},\sigma}\Big]-2US\sum_{{\bf k},\sigma}^{\prime}\sigma
c^{\dag}_{{\bf k}+{\bf Q},\sigma}c_{{\bf k},\sigma}\nonumber\\
&&+\sum_{{\bf k}}^{\prime}\Delta_{{\bf k}}\Big[c^{\dag}_{{\bf k},\uparrow}
c^{\dag}_{-{\bf k},\downarrow}+c_{-{\bf k},\downarrow}c_{{\bf k},\uparrow}-
c^{\dag}_{{\bf k}+{\bf Q},\uparrow}c^{\dag}_{-{\bf k}-{\bf Q},\downarrow}
+c_{-{\bf k}-{\bf Q},\downarrow}c_{{\bf k}+{\bf Q},\uparrow}\Big].
\end{eqnarray}

Defining a Nambu operator
\[\Psi_{{\bf k}} =
\left(\begin{array}{c}  c_{{\bf k},\sigma}\\
c_{{\bf k}+{\bf Q},\sigma} \\
c_{-{\bf k},\bar{\sigma}}^{\dag}\\
c_{-{\bf k}-{\bf Q},\bar{\sigma}}^{\dag}\\\end{array}\right)\] one
can obtain a $4\times 4$ Matsubara Green’s function matrix
$G({\bf k},\tau-\tau^{\prime})$ component-wise with components
$\Big[G({\bf k},\tau-\tau^{\prime})\Big]_{i,j}=-\Big\langle
T_{\tau}
\Psi_{{\bf k},i}(\tau)\Psi_{{\bf k},j}^{\dag}(\tau^{\prime})\Big\rangle$,
as follows

$G({\bf k},\tau-\tau^{\prime})=-$
\[\begin{centering}\left(\begin{array}{cccc}
\Big\langle T_{\tau}
c_{{\bf k},\sigma}(\tau)c_{{\bf k},\sigma}^{\dag}(\tau^{\prime})\Big\rangle
& \Big\langle T_{\tau}
c_{{\bf k},\sigma}(\tau)c_{{\bf k}+{\bf Q},\sigma}^{\dag}(\tau^{\prime})\Big\rangle
&
 \Big\langle T_{\tau}
c_{{\bf k},\sigma}(\tau)c_{-{\bf k},\bar{\sigma}}(\tau^{\prime})\Big\rangle
&
\Big\langle T_{\tau} c_{{\bf k},\sigma}(\tau)c_{-{\bf k}-{\bf Q},\bar{\sigma}}(\tau^{\prime})\\\\
 \Big\langle T_{\tau}
c_{{\bf k}+{\bf Q},\sigma}(\tau)c_{{\bf k},\sigma}^{\dag}(\tau^{\prime})\Big\rangle
& \Big\langle T_{\tau}
c_{{\bf k}+{\bf Q},\sigma}(\tau)c_{{\bf k}+{\bf Q},\sigma}^{\dag}(\tau^{\prime})\Big\rangle
&\Big\langle T_{\tau}
c_{{\bf k}+{\bf Q},\sigma}(\tau)c_{-{\bf k},\bar{\sigma}}(\tau^{\prime})\Big\rangle
& \Big\langle T_{\tau}
c_{{\bf k}+{\bf Q},\sigma}(\tau)c_{-{\bf k}-{\bf Q},\bar{\sigma}}(\tau^{\prime})\Big\rangle\\\\
\Big\langle T_{\tau}
c_{-{\bf k},\bar{\sigma}}^{\dag}(\tau)c_{{\bf k},\sigma}^{\dag}(\tau^{\prime})\Big\rangle
& \Big\langle T_{\tau}
c_{-{\bf k},\bar{\sigma}}^{\dag}(\tau)c_{{\bf k}+{\bf Q},\sigma}^{\dag}(\tau^{\prime})\Big\rangle
& \Big\langle T_{\tau}
c_{-{\bf k},\bar{\sigma}}^{\dag}(\tau)c_{-{\bf k},\bar{\sigma}}(\tau^{\prime})\Big\rangle
& \Big\langle T_{\tau}
c_{-{\bf k},\bar{\sigma}}^{\dag}(\tau)c_{-{\bf k}-{\bf Q},\bar{\sigma}}(\tau^{\prime})\Big\rangle\\\\
\Big\langle T_{\tau}
c_{-{\bf k}-{\bf Q},\bar{\sigma}}^{\dag}(\tau)c_{{\bf k},\sigma}^{\dag}(\tau^{\prime})\Big\rangle
& \Big\langle T_{\tau}
c_{-{\bf k}-{\bf Q},\bar{\sigma}}^{\dag}c_{{\bf k}+{\bf Q},\sigma}^{\dag}(\tau^{\prime})\Big\rangle
& \Big\langle T_{\tau}
c_{-{\bf k}-{\bf Q},\bar{\sigma}}^{\dag}c_{-{\bf k},\bar{\sigma}}(\tau^{\prime})\Big\rangle
&\Big\langle T_{\tau}
c_{-{\bf k}-{\bf Q},\bar{\sigma}}^{\dag}c_{-{\bf k}-{\bf Q},\bar{\sigma}}(\tau^{\prime})\Big\rangle \\
\end{array}\right).\end{centering}\]

With the Hamiltonian of Eq. \ref{Hamafmsc}, a set of
self-consistent equations can be derived for the Green's
function components $G_{11}$, $G_{12}$, $G_{13}$ and $G_{14}$. The
remaining components can then be derived from these four components
using the symmetries which will be discussed later. The unitary
matrix $\hat U$ that diagonalizes the Hamiltonian is given by

%
%
\[\left(\begin{array}{c} c_{{\bf k},\sigma} \\ c_{{\bf k}+{\bf Q},\sigma}\\ c_{-{\bf k},\bar{\sigma}}^{\dag} \\ c_{-{\bf k}-{\bf Q},\bar{\sigma}}^{\dag} \\\end{array} \right)=
\left(\begin{array}{cccc} \alpha_{{\bf k}}u^{+}_{{\bf k}}
&\sigma\beta_{{\bf k}}u_{{\bf k}}^{-} & -\alpha_{{\bf k}}
v_{{\bf k}}^{+}
& \sigma\beta_{{\bf k}}v^{-}_{{\bf k}}\\
-\sigma \beta_{{\bf k}}u^{+}_{{\bf k}}
&\alpha_{{\bf k}}u_{{\bf k}}^{-} &
\sigma\beta_{{\bf k}}v_{{\bf k}}^{+} &
\alpha_{{\bf k}}v^{-}_{{\bf k}}\\
\alpha_{{\bf k}}v^{+}_{{\bf k}} &
{\sigma}\beta_{{\bf k}}v_{{\bf k}}^{-} &
\alpha_{{\bf k}}u_{{\bf k}}^{+} &
-{\sigma}\beta_{{\bf k}} u^{-}_{{\bf k}}\\
{\sigma}\beta_{{\bf k}}v^{+}_{{\bf k}} &
-\alpha_{{\bf k}}v_{{\bf k}}^{-} &
{\sigma}\beta_{{\bf k}}u_{{\bf k}}^{+} &
\alpha_{{\bf k}}u^{-}_{{\bf k}}\\\end{array} \right)
\left(\begin{array}{c} B_{{\bf k},\sigma} \\
B_{{\bf k}+{\bf Q},\sigma}\\ B_{-{\bf k},\bar{\sigma}}^{\dag} \\
B_{-{\bf k}-{\bf Q},\bar{\sigma}}^{\dag} \\\end{array} \right),\]
where the coherence factors are defined in Eq.~\ref{eigenvec}. The diagonal Hamiltonian then is

\[H_{\rm diag}=\left(\begin{array}{cccc} E_{{\bf k}}^+ & 0& 0& 0\\
0&E_{{\bf k}}^- & 0& 0\\
0&0&-E_{{\bf k}}^+&0\\
0&0&0&-E_{{\bf k}}^-
\end{array}\right)\]
The resulting quasiparticle dispersion consists of upper ($\nu=+$) and lower ($\nu=-$) magnetic bands (U/LMB) further split by superconductivity:
\begin{eqnarray}\label{eigen2}
%
E_{\bf k}^{\nu}=\sqrt{\left(\xi_{\bf k}^++\nu
E_{0k}\right)^2+\Delta_{k}^2}.\nonumber
\end{eqnarray}
 Here,
\begin{eqnarray}\label{eigenv2}
%
E_{0{\bf k}}=\sqrt{\left(\xi_{\bf k}^-\right)^2+(US)^2}
\end{eqnarray}
 and
$\xi_{\bf k}^{\pm}=(\xi_{\bf k}\pm\xi_{{\bf k}+{\bf Q}})/2$.  The condensation energy then becomes:
\begin{eqnarray}\label{econ}
%
E_{con}=\sum_{{\bf k},\sigma}\left[2\xi_{\bf k}-E_{\bf k}^+-E_{\bf k}^-\right],
\end{eqnarray}
where the summation is over all $\bf k$. The equilibrium state is the state of minimum (most negative) condensation energy among the normal, AFM, SC, and AFM-SC states.  In these calculations, values of the order parameters and chemical potentials must be determined separately for each phase.

\subsection{Tensor Green's function}

Fourier transform of the Matsubara Green's function can be defined as
\begin{equation}\label{greenft}
\hat{G}({\bf k},{\bf Q},i\omega_n)=\hat{U}(i\omega_n-H_{\rm
diag})^{-1}\hat{U}^{\dag}
\end{equation}
This equation can be used to obtain components of the
Green's function as
{\allowdisplaybreaks\begin{eqnarray}\label{greencomponent}
G_{11}=-\Big\langle
T_{\tau}c_{{\bf k},\sigma}(\tau)c_{{\bf k},\sigma}^{\dag}(\tau^{\prime})\Big\rangle
&=&(\alpha_{{\bf k}}u^+_{{\bf k}})^2g(-E_{{\bf k}}^+)
+(\beta_{{\bf k}}u^-_{{\bf k}})^2g(-E_{{\bf k}}^-)\nonumber\\
&&+(\alpha_{{\bf k}}v^+_{{\bf k}})^2g(E_{{\bf k}}^+)
+(\beta_{{\bf k}}v^-_{{\bf k}})^2g(E_{{\bf k}}^-)\nonumber\\
G_{12}=-\Big\langle T_{\tau}
c_{{\bf k},\sigma}(\tau)c_{{\bf k}+{\bf Q},\sigma}^{\dag}(\tau^{\prime})\Big\rangle
&=&\sigma\alpha_{{\bf k}}\beta_{{\bf k}}[-(u^+_{{\bf k}})^2g(-E_{{\bf k}}^+)
+(u^-_{{\bf k}})^2g(-E_{{\bf k}}^-)\nonumber\\
&&-(v^+_{{\bf k}})^2g(E_{{\bf k}}^+)
+(v^-_{{\bf k}})^2g(E_{{\bf k}}^-)]\nonumber\\
G_{13}=- \Big\langle T_{\tau}
c_{{\bf k},\sigma}(\tau)c_{-{\bf k},\bar{\sigma}}(\tau^{\prime})\Big\rangle
&=&\alpha^2_{{\bf k}}u^+_{{\bf k}}v^+_{{\bf k}}g(-E_{{\bf k}}^+)
+\beta^2_{{\bf k}}u^-_{{\bf k}}v^-_{{\bf k}}g(-E_{{\bf k}}^-)\nonumber\\
&&-\alpha^2_{{\bf k}}u^+_{{\bf k}}v^+_{{\bf k}}g(E_{{\bf k}}^+)
-\beta^2_{{\bf k}}u^-_{{\bf k}}v^-_{{\bf k}}g(E_{{\bf k}}^-)\nonumber\\
G_{14}=- \Big\langle T_{\tau}
c_{{\bf k},\sigma}(\tau)c_{-{\bf k}-{\bf Q},\bar{\sigma}}(\tau^{\prime})
&=&\sigma\alpha_{{\bf k}}\beta_{{\bf k}}[u^+_{{\bf k}}v^+_{{\bf k}}g(-E_{{\bf k}}^+)
-u^-_{{\bf k}}v^-_{{\bf k}}g(-E_{{\bf k}}^-)\nonumber\\
&&-u^+_{{\bf k}}v^+_{{\bf k}}g(E_{{\bf k}}^+)
+u^-_{{\bf k}}v^-_{{\bf k}}g(E_{{\bf k}}^-)]\nonumber\\
G_{21}=-\Big\langle T_{\tau}
c_{{\bf k}+{\bf Q},\sigma}(\tau)c_{{\bf k},\sigma}^{\dag}(\tau^{\prime})\Big\rangle
&=&G_{12}\nonumber\\
G_{22}=-\Big\langle T_{\tau}
c_{{\bf k}+{\bf Q},\sigma}(\tau)c_{{\bf k}+{\bf Q},\sigma}^{\dag}(\tau^{\prime})\Big\rangle
&=&G_{11}(\alpha\rightarrow\beta;\beta\rightarrow\alpha)\nonumber\\
G_{23}=-\Big\langle T_{\tau}
c_{{\bf k}+{\bf Q},\sigma}(\tau)c_{-{\bf k},\bar{\sigma}}(\tau^{\prime})\Big\rangle
&=&-G_{14}\nonumber\\
G_{24}=-\Big\langle T_{\tau}
c_{{\bf k}+{\bf Q},\sigma}(\tau)c_{-{\bf k}-{\bf Q},\bar{\sigma}}(\tau^{\prime})\Big\rangle
&=&-G_{13}(\alpha\rightarrow\beta;\beta\rightarrow\alpha)\nonumber\\
G_{31}=-\Big\langle T_{\tau}
c_{-{\bf k},\bar{\sigma}}^{\dag}(\tau)c_{{\bf k},\sigma}^{\dag}(\tau^{\prime})\Big\rangle
&=&G_{13}\nonumber\\
G_{32}=-\Big\langle T_{\tau}
c_{-{\bf k},\bar{\sigma}}^{\dag}(\tau)c_{{\bf k}+{\bf Q},\sigma}^{\dag}(\tau^{\prime})\Big\rangle
&=&G_{23}\nonumber\\
G_{33}=-\Big\langle T_{\tau}
c_{-{\bf k},\bar{\sigma}}^{\dag}(\tau)c_{-{\bf k},\bar{\sigma}}(\tau^{\prime})\Big\rangle
&=&G_{11}(u\rightarrow v; v\rightarrow u)\nonumber\\
G_{34}=-\Big\langle T_{\tau}
c_{-{\bf k},\bar{\sigma}}^{\dag}(\tau)c_{-{\bf k}-{\bf Q},\bar{\sigma}}(\tau^{\prime})\Big\rangle
&=&-G_{12}(u\rightarrow v; v\rightarrow u)\nonumber\\
G_{41}=-\Big\langle T_{\tau}
c_{-{\bf k}-{\bf Q},\bar{\sigma}}^{\dag}(\tau)c_{{\bf k},\sigma}^{\dag}(\tau^{\prime})\Big\rangle
&=&G_{14}\nonumber\\
G_{42}=-\Big\langle T_{\tau}
c_{-{\bf k}-{\bf Q},\bar{\sigma}}^{\dag}c_{{\bf k}+{\bf Q},\sigma}^{\dag}(\tau^{\prime})\Big\rangle
&=&G_{24}\nonumber\\
G_{43}=-\Big\langle T_{\tau}
c_{-{\bf k}-{\bf Q},\bar{\sigma}}^{\dag}c_{-{\bf k},\bar{\sigma}}(\tau^{\prime})\Big\rangle
&=&G_{34}\nonumber\\
G_{44}=-\Big\langle T_{\tau}
c_{-{\bf k}-{\bf Q},\bar{\sigma}}^{\dag}c_{-{\bf k}-{\bf Q},\bar{\sigma}}(\tau^{\prime})\Big\rangle
&=&G_{11}(\alpha\rightarrow \beta; \beta\rightarrow
\alpha;u\rightarrow v; v\rightarrow u), 
\end{eqnarray}}
where $g(\nu E_{{\bf k}})=1/(ip_n-\nu E_{{\bf k}})$.  When $ip_n$ is analytically continued to $\omega +i\delta$, the Green's function $G_{11}$ is the full tensor extension of Eq.~\ref{SDW}.  As in Eq.~\ref{SDWSE}, we can write $G_{11}=1/(\omega -\epsilon_k-\tilde\Sigma_{11}(k,\omega))$, where 
\begin{eqnarray}\label{SigTens}
\tilde\Sigma_{11}(k,\omega)=\Sigma_{11}+\frac{\tilde g^2-B_0(\omega+\xi_{k+Q})}{\omega-\xi_{k+Q}-B_0},
\end{eqnarray}
\begin{eqnarray}\label{SigTens2}
B_0=\frac{2\xi_k^+\Delta_k^2}{(\omega+\xi_k)(\omega+\xi_{k+Q})-\tilde g^2},
\end{eqnarray}
\begin{eqnarray}\label{SigTens3}
\tilde g^2=g^2+\Delta_k^2.
\end{eqnarray}
In these equations, the QP-GW self-energy can be incorporated by the substitutions: $\xi_k\rightarrow\xi_k+\Sigma_{11}$; $\xi_{k+Q}\rightarrow\xi_{k+Q}+\Sigma_{22}$; $g\rightarrow g+\Sigma_{12}$; and, $\Delta_k\rightarrow \Delta_k+\Sigma_{13}$.  This is the origin of the explicit $\Sigma_{11}$ in Eq.~\ref{SigTens}.
%
%

\subsection{Transverse spin-susceptibility}

%
From the expression for the spin-susceptibility, we have the transverse component
\newline\\
{\allowdisplaybreaks\begin{eqnarray}\label{s3}
\chi_0^{+-}({\bf q},{\bf q}^{\prime},\tau)&=&\frac{1}{N}\sum_{{\bf k},{\bf k}^{\prime}}\Big\langle
T_{\tau}c_{{\bf k}+{\bf q},\uparrow}^{\dag}(\tau)c_{{\bf k},\downarrow}(\tau)
c_{{\bf k}^{\prime}-{\bf q}^{\prime},\downarrow}^{\dag}(0)c_{{\bf k}^{\prime},\uparrow}(0)
\Big\rangle\nonumber\\
&=&\frac{1}{N}\sum^{\prime}_{{\bf k},{\bf k}^{\prime}}\Big\langle
T_{\tau}\Big(c_{{\bf k}+{\bf q},\uparrow}^{\dag}(\tau)c_{{\bf k},\downarrow}(\tau)
+c_{{\bf k}+{\bf Q}+{\bf q},\uparrow}^{\dag}(\tau)c_{{\bf k}+{\bf Q},\downarrow}(\tau)\Big)\nonumber\\
&&~~~~~~~~~~~~~\times\Big(c_{{\bf k}^{\prime}-{\bf q}^{\prime},\downarrow}^{\dag}(0)c_{{\bf k}^{\prime},\uparrow}(0)
+c_{{\bf k}^{\prime}+{\bf Q}-{\bf q}^{\prime},\downarrow}^{\dag}(0)c_{{\bf k}^{\prime}+{\bf Q},\uparrow}(0)\Big)
\Big\rangle\nonumber\\
&=&\frac{1}{N}\sum^{\prime}_{{\bf k},{\bf k}^{\prime}}\left[\Big\langle
T_{\tau}c_{{\bf k}+{\bf q},\uparrow}^{\dag}(\tau)c_{{\bf k},\downarrow}(\tau)
c_{{\bf k}^{\prime}-{\bf q}^{\prime},\downarrow}^{\dag}(0)c_{{\bf k}^{\prime},\uparrow}(0)\Big\rangle\right.\nonumber\\
&&~~~~+\Big\langle
T_{\tau}c_{{\bf k}+{\bf q},\uparrow}^{\dag}(\tau)c_{{\bf k},\downarrow}(\tau)
c_{{\bf k}^{\prime}+{\bf Q}-{\bf q}^{\prime},\downarrow}^{\dag}(0)c_{{\bf k}^{\prime}+{\bf Q},\uparrow}(0)\Big\rangle\nonumber\\
&&~~~~+\Big\langle
T_{\tau}c_{{\bf k}+{\bf Q}+{\bf q},\uparrow}^{\dag}(\tau)c_{{\bf k}+{\bf Q},\downarrow}(\tau)
c_{{\bf k}^{\prime}-{\bf q}^{\prime},\downarrow}^{\dag}(0)c_{{\bf k}^{\prime},\uparrow}(0)\Big\rangle\nonumber\\
&&~~~~\left.+\Big\langle
T_{\tau}c_{{\bf k}+{\bf Q}+{\bf q},\uparrow}^{\dag}(\tau)c_{{\bf k}+{\bf Q},\downarrow}(\tau)
c_{{\bf k}^{\prime}+{\bf Q}-{\bf q}^{\prime},\downarrow}^{\dag}(0)c_{{\bf k}^{\prime}+{\bf Q},\uparrow}(0)\Big\rangle\right]\nonumber\\
&=&-\frac{1}{N}\sum^{\prime}_{{\bf k},{\bf k}^{\prime}}\left[\Big\langle
T_{\tau}c_{{\bf k},\downarrow}(\tau)c_{{\bf k}^{\prime}-{\bf q}^{\prime},\downarrow}^{\dag}(0)\Big\rangle
\Big\langle
T_{\tau}c_{{\bf k}^{\prime},\uparrow}(0)c_{{\bf k}+{\bf q},\uparrow}^{\dag}(\tau)\Big\rangle\right.\nonumber\\
&&~~~~~~~+\Big\langle
T_{\tau}c_{{\bf k},\downarrow}(\tau)c_{{\bf k}^{\prime},\uparrow}(0)\Big\rangle
\Big\langle  T_{\tau}c_{{\bf k}+{\bf q},\uparrow}^{\dag}(\tau)c_{{\bf k}^{\prime}-{\bf q}^{\prime},\downarrow}^{\dag}(0)\Big\rangle\nonumber\\
&&~~~~~~~+\Big\langle
T_{\tau}c_{{\bf k},\downarrow}(\tau)c_{{\bf k}^{\prime}+{\bf Q}-{\bf q}^{\prime},\downarrow}^{\dag}(0)\Big\rangle
\Big\langle
T_{\tau}c_{{\bf k}^{\prime}+{\bf Q},\uparrow}(0)c_{{\bf k}+{\bf q},\uparrow}^{\dag}(\tau)\Big\rangle\nonumber\\
&&~~~~~~~+\Big\langle T_{\tau}c_{{\bf k},\downarrow}(\tau)
c_{{\bf k}^{\prime}+{\bf Q},\uparrow}(0)\Big\rangle\Big\langle
T_{\tau}c_{{\bf k}+{\bf q},\uparrow}^{\dag}(\tau)c_{{\bf k}^{\prime}+{\bf Q}-{\bf q}^{\prime},\downarrow}^{\dag}(0)\Big\rangle\nonumber\\
&&~~~~~~~+\Big\langle
T_{\tau}c_{{\bf k}+{\bf Q},\downarrow}(\tau)c_{{\bf k}^{\prime}-{\bf q}^{\prime},\downarrow}^{\dag}(0)\Big\rangle
\Big\langle
T_{\tau}c_{{\bf k}^{\prime},\uparrow}(0)c_{{\bf k}+{\bf Q}+{\bf q},\uparrow}^{\dag}(\tau)\Big\rangle\nonumber\\
&&~~~~~~~+\Big\langle
T_{\tau}c_{{\bf k}+{\bf Q},\downarrow}(\tau)c_{{\bf k}^{\prime},\uparrow}(0)\Big\rangle
\Big\langle
T_{\tau}c_{{\bf k}+{\bf Q}+{\bf q},\uparrow}^{\dag}(\tau)c_{{\bf k}^{\prime}-{\bf q}^{\prime},\downarrow}^{\dag}(0)\Big\rangle\nonumber\\
&&~~~~~~~+\Big\langle
T_{\tau}c_{{\bf k}+{\bf Q},\downarrow}(\tau)c_{{\bf k}^{\prime}+{\bf Q}-{\bf q}^{\prime},\downarrow}^{\dag}(0)\Big\rangle
\Big\langle
T_{\tau}c_{{\bf k}^{\prime}+{\bf Q},\uparrow}(0)c_{{\bf k}+{\bf Q}+{\bf q},\uparrow}^{\dag}(\tau)\Big\rangle\nonumber\\
&&~~~~~~~\left.+\Big\langle
T_{\tau}c_{{\bf k}+{\bf Q},\downarrow}(\tau)c_{{\bf k}^{\prime}+{\bf Q},\uparrow}(0)\Big\rangle
\Big\langle
T_{\tau}c_{{\bf k}+{\bf Q}+{\bf q},\uparrow}^{\dag}(\tau)
c_{{\bf k}^{\prime}+{\bf Q}-{\bf q}^{\prime},\downarrow}^{\dag}(0)\Big\rangle\right]\nonumber\\
&=&-\frac{1}{N}\sum^{\prime}_{{\bf k},{\bf k}^{\prime}}\delta_{{\bf k}^{\prime},{{\bf k}+{\bf q}^{\prime}}}\left[\Big\langle
T_{\tau}c_{{\bf k},\downarrow}(\tau)c_{{\bf k},\downarrow}^{\dag}(0)\Big\rangle
\Big\langle
T_{\tau}c_{{\bf k}+{\bf q}^{\prime},\uparrow}(0)c_{{\bf k}+{\bf q},\uparrow}^{\dag}(\tau)\Big\rangle\right.\nonumber\\
&&~~~~~~~+\Big\langle
T_{\tau}c_{{\bf k},\downarrow}(\tau)c_{{\bf k}+{\bf Q},\downarrow}^{\dag}(0)\Big\rangle
\Big\langle
T_{\tau}c_{{\bf k}+{\bf Q}+{\bf q}^{\prime},\uparrow}(0)c_{{\bf k}+{\bf q},\uparrow}^{\dag}(\tau)\Big\rangle\nonumber\\
&&~~~~~~~+\Big\langle
T_{\tau}c_{{\bf k}+{\bf Q},\downarrow}(\tau)c_{{\bf k},\downarrow}^{\dag}(0)\Big\rangle
\Big\langle
T_{\tau}c_{{\bf k}+{\bf q}^{\prime},\uparrow}(0)c_{{\bf k}+{\bf Q}+{\bf q},\uparrow}^{\dag}(\tau)\Big\rangle\nonumber\\
&&~~~~~~~\left.+\Big\langle
T_{\tau}c_{{\bf k}+{\bf Q},\downarrow}(\tau)c_{{\bf k}+{\bf Q},\downarrow}^{\dag}(0)\Big\rangle
\Big\langle
T_{\tau}c_{{\bf k}+{\bf Q}+{\bf q}^{\prime},\uparrow}(0)c_{{\bf k}+{\bf Q}+{\bf q},\uparrow}^{\dag}(\tau)\Big\rangle\right]\nonumber\\
&&~~~+\delta_{{\bf k}^{\prime},{-{\bf k}}}\left[\Big\langle
T_{\tau}c_{{\bf k},\downarrow}(\tau)c_{{\bf k},\uparrow}(0)\Big\rangle
\Big\langle  T_{\tau}c_{{\bf k}+{\bf q},\uparrow}^{\dag}(\tau)c_{{\bf k}+{\bf q}^{\prime},\downarrow}^{\dag}(0)\Big\rangle\right.\nonumber\\
&&~~~~~~~+\Big\langle T_{\tau}c_{{\bf k},\downarrow}(\tau)
c_{{\bf k}+{\bf Q},\uparrow}(0)\Big\rangle\Big\langle
T_{\tau}c_{{\bf k}+{\bf q},\uparrow}^{\dag}(\tau)c_{{\bf k}+{\bf Q}+{\bf q}^{\prime},\downarrow}^{\dag}(0)\Big\rangle\nonumber\\
&&~~~~~~~+\Big\langle
T_{\tau}c_{{\bf k}+{\bf Q},\downarrow}(\tau)c_{{\bf k},\uparrow}(0)\Big\rangle
\Big\langle
T_{\tau}c_{{\bf k}+{\bf Q}+{\bf q},\uparrow}^{\dag}(\tau)c_{{\bf k}+{\bf q}^{\prime},\downarrow}^{\dag}(0)\Big\rangle\nonumber\\
&&~~~~~~~\left.+\Big\langle
T_{\tau}c_{{\bf k}+{\bf Q},\downarrow}(\tau)c_{{\bf k}+{\bf Q},\uparrow}(0)\Big\rangle
\Big\langle
T_{\tau}c_{{\bf k}+{\bf Q}+{\bf q},\uparrow}^{\dag}(\tau)c_{{\bf k}+{\bf Q}+{\bf q}^{\prime},\downarrow}^{\dag}(0)\Big\rangle\right]\nonumber\\
\end{eqnarray}}

As we did for the pure AFM state, we rewrite the susceptibility as a two-by-two matrix, with $\chi_{ij}^{+-}({\bf q},\omega )=\chi_0^{+-}({\bf q_i},{\bf q_j},\omega )$, with ${\bf q_1=q}$, ${\bf q_2=q+Q}$. For example, for ${\bf q}^{\prime}={\bf q}$, we get
%
{\allowdisplaybreaks\begin{eqnarray}\label{s4}
\chi_{11}^{+-}({\bf q},\tau)&=&-\frac{1}{N}\sum^{\prime}_{{\bf k}}
\Big[G_{11}({\bf k},\bar{\sigma},\tau)G_{11}({\bf k}+{\bf q},\sigma,-\tau)
+G_{12}({\bf k},\bar{\sigma},\tau)G_{21}({\bf k}+{\bf q},\sigma,-\tau)\nonumber\\
&&~~~~~~+G_{21}({\bf k},\bar{\sigma},\tau)G_{12}({\bf k}+{\bf q},\sigma,-\tau)
+G_{22}({\bf k},\bar{\sigma},\tau)G_{22}({\bf k}+{\bf q},\sigma,-\tau)\nonumber\\
&&~~~~~~+G_{13}({\bf k},\bar{\sigma},\tau)G_{31}({\bf k}+{\bf q},\bar{\sigma},-\tau)
+G_{14}({\bf k},\bar{\sigma},\tau)G_{41}({\bf k}+{\bf q},\bar{\sigma},-\tau)\nonumber\\
&&~~~~~~+G_{23}({\bf k},\bar{\sigma},\tau)G_{32}({\bf k}+{\bf q},\bar{\sigma},-\tau)\
+G_{24}({\bf k},\bar{\sigma},\tau)G_{42}({\bf k}+{\bf q},{\sigma},-\tau)\Big].\nonumber\\
\end{eqnarray}}
The ${\bf q}^{\prime}={\bf q}+{\bf Q}$ (Umklapp)-components are
{\allowdisplaybreaks\begin{eqnarray}\label{s5}
\chi_{12}^{+-}({\bf q},\tau)&=&-\frac{1}{N}\sum^{\prime}_{{\bf k}}
\Big[G_{11}({\bf k},\bar{\sigma},\tau)G_{21}({\bf k}+{\bf q},\sigma,-\tau)
+G_{12}({\bf k},\bar{\sigma},\tau)G_{11}({\bf k}+{\bf q},\sigma,-\tau)\nonumber\\
&&~~~~~~+G_{21}({\bf k},\bar{\sigma},\tau)G_{22}({\bf k}+{\bf q},\sigma,-\tau)
+G_{22}({\bf k},\bar{\sigma},\tau)G_{12}({\bf k}+{\bf q},\sigma,-\tau)\nonumber\\
&&~~~~~~+G_{13}({\bf k},\bar{\sigma},\tau)G_{41}({\bf k}+{\bf q},\sigma,-\tau)
+G_{14}({\bf k},\bar{\sigma},\tau)G_{31}({\bf k}+{\bf q},\bar{\sigma},-\tau)\nonumber\\
&&~~~~~~+G_{23}({\bf k},\bar{\sigma},\tau)G_{42}({\bf k}+{\bf q},\sigma,-\tau)\
+G_{24}({\bf k},\bar{\sigma},\tau)G_{32}({\bf k}+{\bf q},\sigma,-\tau)\Big]\nonumber\\
\end{eqnarray}}
A Fourier transformation yields
{\allowdisplaybreaks\begin{eqnarray}\label{afmscsft2}
&&\chi_{11}^{+-}({\bf q},i\omega_m)\nonumber\\
&=&-\frac{1}{N}\sum_{{\bf k}}\frac{1}{\beta}\sum_{p_n}
\Big[G_{11}({\bf k},\sigma,ip_n)G_{11}({\bf k}+{\bf q},\sigma,ip_n+i\omega_m)\nonumber\\
&&~~~~~~+G_{22}({\bf k},\sigma,ip_n)G_{22}({\bf k}+{\bf q},\sigma,ip_n+i\omega_m)\nonumber\\
&&~~~~~~+G_{13}({\bf k},\sigma,ip_n)G_{13}({\bf k}+{\bf q},\sigma,ip_n+i\omega_m)\nonumber\\
&&~~~~~~+G_{24}({\bf k},\sigma,i p_n)G_{24}({\bf k}+{\bf q},\sigma,ip_n+i\omega_m)\nonumber\\
&&~~~~~~-2\big\{G_{12}({\bf k},\sigma,ip_n)G_{12}({\bf k}+{\bf q},\sigma,ip_n+i\omega_m)\nonumber\\
&&~~~~~~+G_{14}({\bf k},\sigma,i p_n)G_{14}({\bf k}+{\bf q},\sigma,ip_n+i\omega_m)\big\}\Big]
\end{eqnarray}}
and
{\allowdisplaybreaks\begin{eqnarray}\label{afmscsft3}
&&\chi_{12}^{+-}({\bf q},i\omega_m)\nonumber\\
&=&-\frac{1}{N}\sum_{{\bf k}}\frac{1}{\beta}\sum_{p_n}
\Big[G_{11}({\bf k},\sigma,ip_n)G_{12}({\bf k}+{\bf q},\sigma,ip_n+i\omega_m)\nonumber\\
&&~~~~~~+G_{22}({\bf k},\sigma,ip_n)G_{12}({\bf k}+{\bf q},\sigma,ip_n+i\omega_m)\nonumber\\
&&~~~~~~+G_{13}({\bf k},\sigma,ip_n)G_{14}({\bf k}+{\bf q},\sigma,ip_n+i\omega_m)\nonumber\\
&&~~~~~~-G_{24}({\bf k},\sigma,ip_n)G_{14}({\bf k}+{\bf q},\sigma,ip_n+i\omega_m)\nonumber\\
&&~~~~~~-G_{12}({\bf k},\bar{\sigma},ip_n)G_{11}({\bf k}+{\bf q},\sigma,ip_n+i\omega_m)\nonumber\\
&&~~~~~~-G_{12}({\bf k},\bar{\sigma},ip_n)G_{22}({\bf k}+{\bf q},\sigma,ip_n+i\omega_m)\nonumber\\
&&~~~~~~-G_{14}({\bf k},\bar{\sigma},ip_n)G_{13}({\bf k}+{\bf q},\sigma,ip_n+i\omega_m)\nonumber\\
&&~~~~~~+G_{14}({\bf k},\bar{\sigma},ip_n)G_{24}({\bf k}+{\bf q},\sigma,ip_n+i\omega_m)\Big]
\end{eqnarray}}
Here we have used the symmetry of the Green's functions as described
earlier along with its spin symmetry.

\subsection{Writing a $4\times 4$ matrix for susceptibility}

We use Green's function notation to define components of susceptibilities. Some useful frequency summations are
{\allowdisplaybreaks\begin{eqnarray}\label{SCS1}
\chi_{01}^{\nu\nu^{\prime}}({\bf k},{\bf q},i\omega_m)&=&-\frac{1}{\beta}\sum_{ip_n}\left(\frac{1}{ip_n-E^{\nu}_{{\bf k}}}\right)\left(\frac{1}{ip_n+i\omega_m-E^{\nu^{\prime}}_{{\bf k}+{\bf q}}}\right)\nonumber\\
&=&-\frac{(f(E^{\nu}_{{\bf k}})-f(E^{\nu^{\prime}}_{{\bf k}+{\bf q}}))}{i\omega_m+(E^{\nu}_{{\bf k}}-E^{\nu^{\prime}}_{{\bf k}+{\bf q}})}\nonumber\\
\chi_{02}^{\nu\nu^{\prime}}({\bf k},{\bf q},i\omega_m)&=&-\frac{1}{\beta}\sum_{ip_n}\left(\frac{1}{ip_n+E^{\nu}_{{\bf k}}}\right)\left(\frac{1}{ip_n+i\omega_m+E^{\nu^{\prime}}_{{\bf k}+{\bf q}}}\right)\nonumber\\
&=&\frac{(f(E^{\nu^{\prime}}_{{\bf k}})-f(E^{\nu}_{{\bf k}+{\bf q}}))}{i\omega_m-(E^{\nu}_{{\bf k}}-E^{\nu^{\prime}}_{{\bf k}+{\bf q}})}\nonumber\\
\chi_{03}^{\nu\nu^{\prime}}({\bf k},{\bf q},i\omega_m)&=&-\frac{1}{\beta}\sum_{ip_n}\left(\frac{1}{ip_n-E^{\nu}_{{\bf k}}}\right)\left(\frac{1}{ip_n+i\omega_m+E^{\nu^{\prime}}_{{\bf k}+{\bf q}}}\right)\nonumber\\
&=&-\frac{1-(f(E^{\nu}_{{\bf k}})+f(E^{\nu^{\prime}}_{{\bf k}+{\bf q}}))}{i\omega_m-(E^{\nu}_{{\bf k}}+E^{\nu^{\prime}}_{{\bf k}+{\bf q}})}\nonumber\\
\chi_{04}^{\nu\nu^{\prime}}({\bf k},{\bf q},i\omega_m)&=&-\frac{1}{\beta}\sum_{ip_n}\left(\frac{1}{ip_n+E^{\nu}_{{\bf k}}}\right)\left(\frac{1}{ip_n+i\omega_m-E^{\nu^{\prime}}_{{\bf k}+{\bf q}}}\right)\nonumber\\
&=&-\frac{1-(f(E^{\nu}_{{\bf k}})+f(E^{\nu^{\prime}}_{{\bf k}+{\bf q}}))}{i\omega_m+(E^{\nu}_{{\bf k}}+E^{\nu^{\prime}}_{{\bf k}+{\bf q}})}\nonumber\\
\end{eqnarray}}
where $\nu,\nu^{\prime}=\pm$. When both ${\bf k}$ and ${\bf q}$ are summed over, we obtain,  
$\chi_{01}^{\nu,\nu^{\prime}}=\chi_{02}^{\nu,\nu^{\prime}}$.  The frequency sums of various terms of the
susceptibility thus are
{\allowdisplaybreaks\begin{eqnarray}\label{t11b}
T_{11}&=&\frac{1}{\beta}\sum_{p_n}
G_{11}({\bf k},\sigma,ip_n)G_{11}({\bf k}+{\bf q},\sigma,ip_n+i\omega_m)\nonumber\\
&=&\left[(\alpha_{{\bf k}}u^+_{{\bf k}})^2g(-E_{{\bf k}}^+)
+(\beta_{{\bf k}}u^-_{{\bf k}})^2g(-E_{{\bf k}}^-)
+(\alpha_{{\bf k}}v^+_{{\bf k}})^2g(E_{{\bf k}}^+)
+(\beta_{{\bf k}}v^-_{{\bf k}})^2g(E_{{\bf k}}^-)\right]\nonumber\\
&&\times\left[(\alpha_{{\bf k}+{\bf q}}u^+_{{\bf k}+{\bf q}})^2g(i\omega_n-E_{{\bf k}+{\bf q}}^+)
+(\beta_{{\bf k}+{\bf q}}u^-_{{\bf k}+{\bf q}})^2g(i\omega_n-E_{{\bf k}+{\bf q}}^-)\right.\nonumber\\
&&~~~~~~\left.+(\alpha_{{\bf k}+{\bf q}}v^+_{{\bf k}+{\bf q}})^2g(i\omega_n+E_{{\bf k}+{\bf q}}^+)
+(\beta_{{\bf k}+{\bf q}}v^-_{{\bf k}+{\bf q}})^2g(i\omega_n+E_{{\bf k}+{\bf q}}^-)\right]\nonumber\\
T_{22}&=&\frac{1}{\beta}\sum_{p_n}
G_{22}({\bf k},\sigma,ip_n)G_{22}({\bf k}+{\bf q},\sigma,ip_n+i\omega_m)\nonumber\\
&=&\left[(\beta_{{\bf k}}u^+_{{\bf k}})^2g(-E_{{\bf k}}^+)
+(\alpha_{{\bf k}}u^-_{{\bf k}})^2g(-E_{{\bf k}}^-)
+(\beta_{{\bf k}}v^+_{{\bf k}})^2g(E_{{\bf k}}^+)
+(\alpha_{{\bf k}}v^-_{{\bf k}})^2g(E_{{\bf k}}^-)\right]\nonumber\\
&&\times\left[(\beta_{{\bf k}+{\bf q}}u^+_{{\bf k}+{\bf q}})^2g(i\omega_n-E_{{\bf k}+{\bf q}}^+)
+(\alpha_{{\bf k}+{\bf q}}u^-_{{\bf k}+{\bf q}})^2g(i\omega_n-E_{{\bf k}+{\bf q}}^-)\right.\nonumber\\
&&~~~~~~\left.+(\beta_{{\bf k}+{\bf q}}v^+_{{\bf k}+{\bf q}})^2g(i\omega_n+E_{{\bf k}+{\bf q}}^+)
+(\alpha_{{\bf k}+{\bf q}}v^-_{{\bf k}+{\bf q}})^2g(i\omega_n+E_{{\bf k}+{\bf q}}^-)\right]\nonumber\\
T_{13}&=&\frac{1}{\beta}\sum_{p_n}
G_{13}({\bf k},\sigma,ip_n)G_{13}({\bf k}+{\bf q},\sigma,ip_n+i\omega_m)\nonumber\\
&=&\Big[\alpha^2_{{\bf k}}u^+_{{\bf k}}v^+_{{\bf k}}g(-E_{{\bf k}}^+)
+\beta^2_{{\bf k}}u^-_{{\bf k}}v^-_{{\bf k}}g(-E_{{\bf k}}^-)
-\alpha^2_{{\bf k}}u^+_{{\bf k}}v^+_{{\bf k}}g(E_{{\bf k}}^+)
-\beta^2_{{\bf k}}u^-_{{\bf k}}v^-_{{\bf k}}g(E_{{\bf k}}^-)\Big]\nonumber\\
&&\times\Big[\alpha^2_{{\bf k}+{\bf q}}u^+_{{\bf k}+{\bf q}}v^+_{{\bf k}+{\bf q}}g(i\omega_n-E_{{\bf k}+{\bf q}}^+)
+\beta^2_{{\bf k}+{\bf q}}u^-_{{\bf k}+{\bf q}}v^-_{{\bf k}+{\bf q}}g(i\omega_n-E_{{\bf k}+{\bf q}}^-)\nonumber\\
&&~~~~~~-\alpha^2_{{\bf k}+{\bf q}}u^+_{{\bf k}+{\bf q}}v^+_{{\bf k}+{\bf q}}g(i\omega_n+E_{{\bf k}+{\bf q}}^+)
-\beta^2_{{\bf k}+{\bf q}}u^-_{{\bf k}+{\bf q}}v^-_{{\bf k}+{\bf q}}g(i\omega_n+E_{{\bf k}+{\bf q}}^-)\Big]\nonumber\\
T_{24}&=&\frac{1}{\beta}\sum_{p_n}
G_{24}({\bf k},\sigma,ip_n)G_{24}({\bf k}+{\bf q},\sigma,ip_n+i\omega_m)\nonumber\\
&=&\Big[\beta^2_{{\bf k}}u^+_{{\bf k}}v^+_{{\bf k}}g(-E_{{\bf k}}^+)
+\alpha^2_{{\bf k}}u^-_{{\bf k}}v^-_{{\bf k}}g(-E_{{\bf k}}^-)
-\beta^2_{{\bf k}}u^+_{{\bf k}}v^+_{{\bf k}}g(E_{{\bf k}}^+)
-\alpha^2_{{\bf k}}u^-_{{\bf k}}v^-_{{\bf k}}g(E_{{\bf k}}^-)\Big]\nonumber\\
&&\times\Big[\beta^2_{{\bf k}+{\bf q}}u^+_{{\bf k}+{\bf q}}v^+_{{\bf k}+{\bf q}}g(i\omega_n-E_{{\bf k}+{\bf q}}^+)
+\alpha^2_{{\bf k}+{\bf q}}u^-_{{\bf k}+{\bf q}}v^-_{{\bf k}+{\bf q}}g(i\omega_n-E_{{\bf k}+{\bf q}}^-)\nonumber\\
&&~~~~~~-\beta^2_{{\bf k}+{\bf q}}u^+_{{\bf k}+{\bf q}}v^+_{{\bf k}+{\bf q}}g(i\omega_n+E_{{\bf k}+{\bf q}}^+)
-\alpha^2_{{\bf k}+{\bf q}}u^-_{{\bf k}+{\bf q}}v^-_{{\bf k}+{\bf q}}g(i\omega_n+E_{{\bf k}+{\bf q}}^-)\Big]\nonumber\\
-2T_{12}&=&-2\frac{1}{\beta}\sum_{p_n}
G_{12}({\bf k},\sigma,ip_n)G_{12}({\bf k}+{\bf q},\sigma,ip_n+i\omega_m)\nonumber\\
&=&-2\alpha_{{\bf k}}\beta_{{\bf k}}\alpha_{{\bf k}+{\bf q}}\beta_{{\bf k}+{\bf q}}
\Big[-(u^+_{{\bf k}})^2g(-E_{{\bf k}}^+)
+(u^-_{{\bf k}})^2g(-E_{{\bf k}}^-)\nonumber\\
&&~~~~~~~~~~~~~~~~~~~~~~~~~~-(v^+_{{\bf k}})^2g(E_{{\bf k}}^+)
+(v^-_{{\bf k}})^2g(E_{{\bf k}}^-)]\Big]\nonumber\\
&&~~~~~~\times\Big[-(u^+_{{\bf k}+{\bf q}})^2g(i\omega_n-E_{{\bf k}+{\bf q}}^+)
+(u^-_{{\bf k}+{\bf q}})^2g(i\omega_n-E_{{\bf k}+{\bf q}}^-)\nonumber\\
&&~~~~~~~~~~~~-(v^+_{{\bf k}+{\bf q}})^2g(i\omega_n+E_{{\bf k}+{\bf q}}^+)
+(v^-_{{\bf k}+{\bf q}})^2g(i\omega_n+E_{{\bf k}+{\bf q}}^-)]\Big]\nonumber\\
-2T_{14}&=&-2\frac{1}{\beta}\sum_{p_n}
G_{14}({\bf k},\sigma,ip_n)G_{14}({\bf k}+{\bf q},\sigma,ip_n+i\omega_m)\nonumber\\
&=&-2\alpha_{{\bf k}}\beta_{{\bf k}}\alpha_{{\bf k}+{\bf q}}\beta_{{\bf k}+{\bf q}}
\Big[u^+_{{\bf k}}v^+_{{\bf k}}g(-E_{{\bf k}}^+)
-u^-_{{\bf k}}v^-_{{\bf k}}g(-E_{{\bf k}}^-)\nonumber\\
&&~~~~~~~~~~~~~~~~~~~~~~~-u^+_{{\bf k}}v^+_{{\bf k}}g(E_{{\bf k}}^+)
+u^-_{{\bf k}}v^-_{{\bf k}}g(E_{{\bf k}}^-)\Big]\nonumber\\
&&~~~~~~\times\Big[u^+_{{\bf k}+{\bf q}}v^+_{{\bf k}+{\bf q}}g(-E_{{\bf k}+{\bf q}}^+)
-u^-_{{\bf k}+{\bf q}}v^-_{{\bf k}+{\bf q}}g(-E_{{\bf k}+{\bf q}}^-)\nonumber\\
&&~~~~~~~~~~~~-u^+_{{\bf k}+{\bf q}}v^+_{{\bf k}+{\bf q}}g(E_{{\bf k}+{\bf q}}^+)
+u^-_{{\bf k}+{\bf q}}v^-_{{\bf k}+{\bf q}}g(E_{{\bf k}+{\bf q}}^-)\Big]
\end{eqnarray}}
And for the cross-terms,
{\allowdisplaybreaks\begin{eqnarray}\label{l11}
L_{11}&=&\frac{1}{\beta}\sum_{p_n}
G_{11}({\bf k},\sigma,ip_n)G_{12}({\bf k}+{\bf q},\sigma,ip_n+i\omega_m)\nonumber\\
&=&\left[(\alpha_{{\bf k}}u^+_{{\bf k}})^2g(-E_{{\bf k}}^+)
+(\beta_{{\bf k}}u^-_{{\bf k}})^2g(-E_{{\bf k}}^-)
+(\alpha_{{\bf k}}v^+_{{\bf k}})^2g(E_{{\bf k}}^+)
+(\beta_{{\bf k}}v^-_{{\bf k}})^2g(E_{{\bf k}}^-)\right]\nonumber\\
&&\times\alpha_{{\bf k}+{\bf q}}\beta_{{\bf k}+{\bf q}}
\Big[-(u^+_{{\bf k}+{\bf q}})^2g(i\omega_n-E_{{\bf k}+{\bf q}}^+)
+(u^-_{{\bf k}+{\bf q}})^2g(i\omega_n-E_{{\bf k}+{\bf q}}^-)\nonumber\\
&&~~~~~~~~~~~~-(v^+_{{\bf k}+{\bf q}})^2g(i\omega_n+E_{{\bf k}+{\bf q}}^+)
+(v^-_{{\bf k}+{\bf q}})^2g(i\omega_n+E_{{\bf k}+{\bf q}}^-)]\Big]\nonumber\\
L_{22}&=&\frac{1}{\beta}\sum_{p_n}
G_{22}({\bf k},\sigma,ip_n)G_{12}({\bf k}+{\bf q},\sigma,ip_n+i\omega_m)\nonumber\\
&=&\left[(\beta_{{\bf k}}u^+_{{\bf k}})^2g(-E_{{\bf k}}^+)
+(\alpha_{{\bf k}}u^-_{{\bf k}})^2g(-E_{{\bf k}}^-)
+(\beta_{{\bf k}}v^+_{{\bf k}})^2g(E_{{\bf k}}^+)
+(\alpha_{{\bf k}}v^-_{{\bf k}})^2g(E_{{\bf k}}^-)\right]\nonumber\\
&&\times\alpha_{{\bf k}+{\bf q}}\beta_{{\bf k}+{\bf q}}
\Big[-(u^+_{{\bf k}+{\bf q}})^2g(i\omega_n-E_{{\bf k}+{\bf q}}^+)
+(u^-_{{\bf k}+{\bf q}})^2g(i\omega_n-E_{{\bf k}+{\bf q}}^-)\nonumber\\
&&~~~~~~~~~~~~-(v^+_{{\bf k}+{\bf q}})^2g(i\omega_n+E_{{\bf k}+{\bf q}}^+)
+(v^-_{{\bf k}+{\bf q}})^2g(i\omega_n+E_{{\bf k}+{\bf q}}^-)]\Big]\nonumber\\
%
L_{13}&=&\frac{1}{\beta}\sum_{p_n}
G_{13}({\bf k},\sigma,ip_n)G_{14}({\bf k}+{\bf q},\sigma,ip_n+i\omega_m)\nonumber\\
&=&\Big[\alpha^2_{{\bf k}}u^+_{{\bf k}}v^+_{{\bf k}}g(-E_{{\bf k}}^+)
+\beta^2_{{\bf k}}u^-_{{\bf k}}v^-_{{\bf k}}g(-E_{{\bf k}}^-)
-\alpha^2_{{\bf k}}u^+_{{\bf k}}v^+_{{\bf k}}g(E_{{\bf k}}^+)
-\beta^2_{{\bf k}}u^-_{{\bf k}}v^-_{{\bf k}}g(E_{{\bf k}}^-)\Big]\nonumber\\
&&\times\alpha_{{\bf k}+{\bf q}}\beta_{{\bf k}+{\bf q}}
\Big[u^+_{{\bf k}+{\bf q}}v^+_{{\bf k}+{\bf q}}g(-E_{{\bf k}+{\bf q}}^+)
-u^-_{{\bf k}+{\bf q}}v^-_{{\bf k}+{\bf q}}g(-E_{{\bf k}+{\bf q}}^-)\nonumber\\
&&~~~~~~~~~~~~-u^+_{{\bf k}+{\bf q}}v^+_{{\bf k}+{\bf q}}g(E_{{\bf k}+{\bf q}}^+)
+u^-_{{\bf k}+{\bf q}}v^-_{{\bf k}+{\bf q}}g(E_{{\bf k}+{\bf q}}^-)\Big]\nonumber\\
%
L_{24}&=&\frac{1}{\beta}\sum_{p_n}
G_{24}({\bf k},\sigma,ip_n)G_{14}({\bf k}+{\bf q},\sigma,ip_n+i\omega_m)\nonumber\\
&=&\Big[\beta^2_{{\bf k}}u^+_{{\bf k}}v^+_{{\bf k}}g(-E_{{\bf k}}^+)
+\alpha^2_{{\bf k}}u^-_{{\bf k}}v^-_{{\bf k}}g(-E_{{\bf k}}^-)
-\beta^2_{{\bf k}}u^+_{{\bf k}}v^+_{{\bf k}}g(E_{{\bf k}}^+)
-\alpha^2_{{\bf k}}u^-_{{\bf k}}v^-_{{\bf k}}g(E_{{\bf k}}^-)\Big]\nonumber\\
&&\times\alpha_{{\bf k}+{\bf q}}\beta_{{\bf k}+{\bf q}}
\Big[u^+_{{\bf k}+{\bf q}}v^+_{{\bf k}+{\bf q}}g(-E_{{\bf k}+{\bf q}}^+)
-u^-_{{\bf k}+{\bf q}}v^-_{{\bf k}+{\bf q}}g(-E_{{\bf k}+{\bf q}}^-)\nonumber\\
&&~~~~~~~~~~~~-u^+_{{\bf k}+{\bf q}}v^+_{{\bf k}+{\bf q}}g(E_{{\bf k}+{\bf q}}^+)
+u^-_{{\bf k}+{\bf q}}v^-_{{\bf k}+{\bf q}}g(E_{{\bf k}+{\bf q}}^-)\Big]\nonumber\\
-L_{12}&=&-2\frac{1}{\beta}\sum_{p_n}
G_{12}({\bf k},\sigma,ip_n)G_{11}({\bf k}+{\bf q},\sigma,ip_n+i\omega_m)\nonumber\\
&=&-\alpha_{{\bf k}}\beta_{{\bf k}}
\Big[-(u^+_{{\bf k}})^2g(-E_{{\bf k}}^+)
+(u^-_{{\bf k}})^2g(-E_{{\bf k}}^-)
-(v^+_{{\bf k}})^2g(E_{{\bf k}}^+)
+(v^-_{{\bf k}})^2g(E_{{\bf k}}^-)]\Big]\nonumber\\
&&\times\left[(\alpha_{{\bf k}+{\bf q}}u^+_{{\bf k}+{\bf q}})^2g(i\omega_n-E_{{\bf k}+{\bf q}}^+)
+(\beta_{{\bf k}+{\bf q}}u^-_{{\bf k}+{\bf q}})^2g(i\omega_n-E_{{\bf k}+{\bf q}}^-)\right.\nonumber\\
&&~~~~~~\left.+(\alpha_{{\bf k}+{\bf q}}v^+_{{\bf k}+{\bf q}})^2g(i\omega_n+E_{{\bf k}+{\bf q}}^+)
+(\beta_{{\bf k}+{\bf q}}v^-_{{\bf k}+{\bf q}})^2g(i\omega_n+E_{{\bf k}+{\bf q}}^-)\right]\nonumber\\
%
-L_{14}&=&-\frac{1}{\beta}\sum_{p_n}
G_{14}({\bf k},\sigma,ip_n)G_{13}({\bf k}+{\bf q},\sigma,ip_n+i\omega_m)\nonumber\\
&=&-\alpha_{{\bf k}}\beta_{{\bf k}}
\Big[u^+_{{\bf k}}v^+_{{\bf k}}g(-E_{{\bf k}}^+)
-u^-_{{\bf k}}v^-_{{\bf k}}g(-E_{{\bf k}}^-)
-u^+_{{\bf k}}v^+_{{\bf k}}g(E_{{\bf k}}^+)
+u^-_{{\bf k}}v^-_{{\bf k}}g(E_{{\bf k}}^-)\Big]\nonumber\\
&&\times\Big[\alpha^2_{{\bf k}+{\bf q}}u^+_{{\bf k}+{\bf q}}v^+_{{\bf k}+{\bf q}}g(i\omega_n-E_{{\bf k}+{\bf q}}^+)
+\beta^2_{{\bf k}+{\bf q}}u^-_{{\bf k}+{\bf q}}v^-_{{\bf k}+{\bf q}}g(i\omega_n-E_{{\bf k}+{\bf q}}^-)\nonumber\\
&&~~~~~~-\alpha^2_{{\bf k}+{\bf q}}u^+_{{\bf k}+{\bf q}}v^+_{{\bf k}+{\bf q}}g(i\omega_n+E_{{\bf k}+{\bf q}}^+)
-\beta^2_{{\bf k}+{\bf q}}u^-_{{\bf k}+{\bf q}}v^-_{{\bf k}+{\bf q}}g(i\omega_n+E_{{\bf k}+{\bf q}}^-)\Big]\nonumber\\
%
\end{eqnarray}}
The other terms can be found by using symmetry arguments. Finally, combining the preceding results, the transverse susceptibility can be written in the compact form:
\begin{eqnarray}\label{afmscsft20}
\chi_{ij}^{\sigma\sigma^{\prime}}({\bf q},i\omega_m)
&=&\frac{1}{N}\sum_{{\bf k}}^{\prime}\sum_{n,\nu\nu^{\prime}}
C_{n,ij}^{\nu\nu^{\prime},\sigma\sigma^{\prime}}\chi_{0n}^{\nu\nu^{\prime}}({\bf k},{\bf q},i\omega_m)
\end{eqnarray}
where $\sigma^{\prime}=\bar{\sigma}=-\sigma$ denotes transverse susceptibility, and $\sigma^{\prime}=\sigma$ is for the longitudinal and charge component, as introduced in Eq.~\ref{chi2}. The total coefficients are [subscript $n$ denotes $\chi$-components, and ($ij)$ denote SDW components]
\begin{eqnarray}\label{t11b2}
C_{n,ij}^{\nu\nu^{\prime},\sigma\sigma^{\prime}}&=&A_{n}^{\nu\nu^{\prime}}B_{ij}^{\nu\nu^{\prime},\sigma\sigma^{\prime}},
\end{eqnarray}
the $A_{n}^{\nu,\nu^{\prime}}$ are coefficients for superconductivity, and $B_{ij}^{\nu\nu^{\prime},\sigma\sigma^{\prime}}$ are for SDW. The momentum dependence in all coefficients is implicit. From symmetry, $\chi_{01}^{\nu\nu^{\prime}}=\chi_{02}^{\nu\nu^{\prime}}$, so that we
find
\begin{eqnarray}\label{afmscsft4}
\chi_{ij}^{+-}({\bf q},i\omega_m) &=&\frac{1}{N}\sum_{{\bf k}}^{\prime}\sum_{\nu,\nu^{\prime}}
 \Big[(A_{1}^{\nu\nu^{\prime}}+A_{2}^{\nu\nu^{\prime}})\chi_{01}^{\nu\nu^{\prime}}
+ A_{ 3}^{\nu\nu^{\prime}} \chi_{03}^{\nu\nu^{\prime}} 
+ A_{4}^{\nu\nu^{\prime}}\chi_{04}^{\nu\nu^{\prime}}\Big]
B_{ij}^{\nu\nu^{\prime},\sigma\sigma^{\prime}}\nonumber\\
\end{eqnarray}
Now the SC coefficients can be written as
{\allowdisplaybreaks\begin{eqnarray}\label{t11c}
A_{1}^{\nu\nu^{\prime}}+A_{2}^{\nu\nu^{\prime}}&=&
u^{\nu}_{{\bf k}}u^{\nu^{\prime}}_{{\bf k}+{\bf q}}(u^{\nu}_{{\bf k}}
u^{\nu^{\prime}}_{{\bf k}+{\bf q}}+v^{\nu}_{{\bf k}}v^{\nu^{\prime}}_{{\bf k}+{\bf q}})
+v^{\nu}_{{\bf k}}v^{\nu^{\prime}}_{{\bf k}+{\bf q}}(u^{\nu}_{{\bf k}}
u^{\nu^{\prime}}_{{\bf k}+{\bf q}}+v^{\nu}_{{\bf k}}v^{\nu^{\prime}}_{{\bf k}+{\bf q}})\nonumber\\
&=&\frac{1}{2}\left(1+\frac{(\xi^+_{{\bf k}}+\nu
E_{0{\bf k}})(\xi^+_{{\bf k}+{\bf q}}+\nu^{\prime}
E_{0{\bf k}+{\bf q}})+\Delta_{{\bf k}}\Delta_{{\bf k}+{\bf q}}}{E^{\nu}_{{\bf k}}E^{\nu^{\prime}}_{{\bf k}+{\bf q}}}\right)\nonumber\\
A_{3}^{\nu\nu^{\prime}}&=&
u^{\nu}_{{\bf k}}v^{\nu^{\prime}}_{{\bf k}+{\bf q}}(u^{\nu}_{{\bf k}}v^{\nu^{\prime}}_{{\bf k}+{\bf q}}
-v^{\nu}_{{\bf k}}u^{\nu^{\prime}}_{{\bf k}+{\bf q}})\nonumber\\
&=&\frac{1}{4}\left(1+\frac{\xi^+_{{\bf k}}+\nu
E_{0{\bf k}}}{E^{\nu}_{{\bf k}}}
-\frac{\xi^+_{{\bf k}+{\bf q}}+\nu^{\prime}
E_{0{\bf k}+{\bf q}}}{E^{\nu^{\prime}}_{{\bf k}+{\bf q}}}\right.\nonumber\\
&&~~~~~~~
\left.-\frac{(\xi^+_{{\bf k}}+\nu
E_{0{\bf k}})(\xi^+_{{\bf k}+{\bf q}}+\nu^{\prime}
E_{0{\bf k}+{\bf q}})+\Delta_{{\bf k}}\Delta_{{\bf k}+{\bf q}}}{E^{\nu}_{{\bf k}}E^{\nu^{\prime}}_{{\bf k}+{\bf q}}}\right)\nonumber\\
A_{4}^{\nu\nu^{\prime}}&=&
v^{\nu}_{{\bf k}}u^{\nu^{\prime}}_{{\bf k}+{\bf q}}(u^{\nu}_{{\bf k}}
v^{\nu^{\prime}}_{{\bf k}+{\bf q}}-v^{\nu}_{{\bf k}}u^{\nu^{\prime}}_{{\bf k}+{\bf q}})\nonumber\\
&=&\frac{1}{4}\left(1-\frac{\xi^+_{{\bf k}}+\nu
E_{0{\bf k}}}{E^{\nu}_{{\bf k}}}
+\frac{\xi^+_{{\bf k}+{\bf q}}+\nu^{\prime}
E_{0{\bf k}+{\bf q}}}{E^{\nu^{\prime}}_{{\bf k}+{\bf q}}}\right.\nonumber\\
&&~~~~~~~\left.-\frac{(\xi^+_{{\bf k}}+\nu
E_{0{\bf k}})(\xi^+_{{\bf k}+{\bf q}}+\nu^{\prime}
E_{0{\bf k}+{\bf q}})+\Delta_{{\bf k}}\Delta_{{\bf k}+{\bf q}}}{E^{\nu}_{{\bf k}}E^{\nu^{\prime}}_{{\bf k}+{\bf q}}}\right).
\end{eqnarray}}
The AFM coefficients are
{\allowdisplaybreaks\begin{eqnarray}\label{coeffafmsc2}
B_{11}^{\nu=\nu^{\prime},\sigma{\bar \sigma}}
&=&(\alpha_{{\bf k}}\alpha_{{\bf k}+{\bf q}}-\beta_{{\bf k}}\beta_{{\bf k}+{\bf q}})^2
=\frac{1}{2}\left(1+\frac{\xi_{{\bf k}}^{-}\xi_{{\bf k}+{\bf q}}^{-}-(US)^2}
{E_{0{\bf k}}E_{0{\bf k}+{\bf q}}}\right)\nonumber\\
B_{\rm 11}^{\nu \ne\nu^{\prime},\sigma{\bar \sigma}}
&=&(\alpha_{{\bf k}}\beta_{{\bf k}+{\bf q}}+\beta_{{\bf k}}\alpha_{{\bf k}+{\bf q}})^2
=\frac{1}{2}\left(1-\frac{\xi_{{\bf k}}^{-}\xi_{{\bf k}+{\bf q}}^{-}-(US)^2}
{E_{0{\bf k}}E_{0{\bf k}+{\bf q}}}\right)\nonumber\\
B_{22}^{\nu=\nu^{\prime},\sigma{\bar \sigma}}
&=&B_{11}^{\nu=\nu^{\prime},\sigma{\bar \sigma}}({\bf q}\rightarrow{\bf q}+{\bf Q})
=(\alpha_{{\bf k}}\beta_{{\bf k}+{\bf q}}-\beta_{{\bf k}}\alpha_{{\bf k}+{\bf q}})^2\nonumber\\
&=&\frac{1}{2}\left(1-\frac{\xi_{{\bf k}}^{-}\xi_{{\bf k}+{\bf q}}^{-}+(US)^2}
{E_{0{\bf k}}E_{0{\bf k}+{\bf q}}}\right)\nonumber\\
B_{22}^{\nu \ne \nu^{\prime},\sigma{\bar \sigma}}
&=&B_{11}^{\nu\ne\nu^{\prime},\sigma{\bar \sigma}}({\bf q}\rightarrow{\bf q}+{\bf Q})
=(\alpha_{{\bf k}}\alpha_{{\bf k}+{\bf q}}+\beta_{{\bf k}}\beta_{{\bf k}+{\bf q}})^2\nonumber\\
&=&\frac{1}{2}\left(1+\frac{\xi_{{\bf k}}^{-}\xi_{{\bf k}+{\bf q}}^{-}+(US)^2}
{E_{0{\bf k}}E_{0{\bf k}+{\bf q}}}\right)\nonumber\\
B_{12}^{\nu=\nu^{\prime},\sigma{\bar \sigma}}
&=&B_{21}^{\nu=\nu^{\prime},\sigma{\bar \sigma}}
=-\nu(\alpha_{{\bf k}}\beta_{{\bf k}}-\alpha_{{\bf k}+{\bf q}}\beta_{{\bf k}+{\bf q}})
=-\nu\frac{US}{2}\left(\frac{1}{E_{0{\bf k}}}-\frac{1}{E_{0{\bf k}+{\bf q}}}\right)\nonumber\\
B_{12}^{\nu \ne \nu^{\prime},\sigma{\bar \sigma}}
&=&B_{21}^{\nu\ne\nu^{\prime},\sigma{\bar \sigma}}
=-\nu(\alpha_{{\bf k}}\beta_{{\bf k}}+\alpha_{{\bf k}+{\bf q}}\beta_{{\bf k}+{\bf q}})
=-\nu\frac{US}{2}\left(\frac{1}{E_{0{\bf k}}}+\frac{1}{E_{0{\bf k}+{\bf q}}}\right).\nonumber\\
\end{eqnarray}}

\subsection{Charge and Longitudinal Spin Susceptibility:}

We now evaluate charge and longitudinal spin-susceptibilities by proceeding the same way as we did for
the SDW case above. The charge and longitudinal ($z$-component) susceptibilities are obtained from Eq.~\ref{chi2}:
\begin{eqnarray}\label{c5}
\chi^{\rho\rho}({\bf q},{\bf q}^{\prime},\tau)
&=&\frac{1}{N}\sum_{{\bf k},{\bf k}^{\prime},\sigma,\sigma^{\prime}}\Big\langle
T_{\tau}c_{{\bf k}+{\bf q},\sigma}^{\dag}(\tau)c_{{\bf k},\sigma}(\tau)
c_{{\bf k}^{\prime}-{\bf q}^{\prime},\sigma^{\prime}}^{\dag}(0)c_{{\bf k}^{\prime},\sigma^{\prime}}(0)
\Big\rangle
\end{eqnarray}
and
{\allowdisplaybreaks\begin{eqnarray}\label{c5b}
\chi^{zz}({\bf q},{\bf q}^{\prime},\tau)
&=&\frac{1}{N}\sum_{{\bf k},{\bf k}^{\prime},\sigma,\sigma^{\prime}}\sigma\sigma^{\prime}\Big\langle
T_{\tau}c_{{\bf k}+{\bf q},\sigma}^{\dag}(\tau)c_{{\bf k},\sigma}(\tau)
c_{{\bf k}^{\prime}-{\bf q}^{\prime},\sigma^{\prime}}^{\dag}(0)c_{{\bf k}^{\prime},\sigma^{\prime}}(0)
\Big\rangle\nonumber\\
&=&\frac{1}{N}\sum_{{\bf k},{\bf k}^{\prime},\sigma,\sigma^{\prime}}^{\prime}\sigma\sigma^{\prime}\Big\langle
T_{\tau}\left(c_{{\bf k}+{\bf q},\sigma}^{\dag}(\tau)c_{{\bf k},\sigma}(\tau)
+c_{{\bf k}+{\bf Q}+{\bf q},\sigma}^{\dag}(\tau)c_{{\bf k}+{\bf Q},\sigma}(\tau)\right)\nonumber\\
&&~~~\left(c_{{\bf k}^{\prime}-{\bf q}^{\prime},\sigma^{\prime}}^{\dag}(0)c_{{\bf k}^{\prime},\sigma^{\prime}}(0)
+c_{{\bf k}^{\prime}+{\bf Q}-{\bf q}^{\prime},\sigma^{\prime}}^{\dag}(0)c_{{\bf k}^{\prime}+{\bf Q},\sigma^{\prime}}(0)\right)
\Big\rangle\nonumber\\
&=&\frac{1}{N}\sum_{{\bf k},{\bf k}^{\prime},\sigma,\sigma^{\prime}}^{\prime}\sigma\sigma^{\prime}\left[\Big\langle
T_{\tau}c_{{\bf k}+{\bf q},\sigma}^{\dag}(\tau)c_{{\bf k},\sigma}(\tau)
c_{{\bf k}^{\prime}-{\bf q}^{\prime},\sigma^{\prime}}^{\dag}(0)c_{{\bf k}^{\prime},\sigma^{\prime}}(0)\Big\rangle\right.\nonumber\\
&&+\Big\langle
T_{\tau}c_{{\bf k}+{\bf q},\sigma}^{\dag}(\tau)c_{{\bf k},\sigma}(\tau)
c_{{\bf k}^{\prime}+{\bf Q}-{\bf q}^{\prime},\sigma^{\prime}}^{\dag}(0)c_{{\bf k}^{\prime}+{\bf Q},\sigma^{\prime}}(0)\Big\rangle\nonumber\\
&&+\Big\langle
T_{\tau}c_{{\bf k}+{\bf Q}+{\bf q},\sigma}^{\dag}(\tau)c_{{\bf k}+{\bf Q},\sigma}(\tau)
c_{{\bf k}^{\prime}-{\bf q}^{\prime},\sigma^{\prime}}^{\dag}(0)c_{{\bf k}^{\prime},\sigma^{\prime}}(0)\Big\rangle\nonumber\\
&&\left.+\Big\langle
T_{\tau}c_{{\bf k}+{\bf Q}+{\bf q},\sigma}^{\dag}(\tau)c_{{\bf k}+{\bf Q},\sigma}(\tau)
c_{{\bf k}^{\prime}+{\bf Q}-{\bf q}^{\prime},\sigma^{\prime}}^{\dag}(0)c_{{\bf k}^{\prime}+{\bf Q},\sigma^{\prime}}(0)\Big\rangle\right]\nonumber\\
&=&-\frac{1}{N}\sum_{{\bf k},{\bf k}^{\prime},\sigma,\sigma^{\prime}}^{\prime}\delta_{\sigma,\sigma^{\prime}}\left[
\Big\langle
T_{\tau}c_{{\bf k},\sigma}(\tau)c_{{\bf k}^{\prime}-{\bf q}^{\prime},\sigma^{\prime}}^{\dag}(0)\Big\rangle
\Big\langle T_{\tau}c_{{\bf k}^{\prime},\sigma^{\prime}}(0)c_{{\bf k}+{\bf q},\sigma}^{\dag}(\tau)\Big\rangle\right.\nonumber\\
&&+\Big\langle
T_{\tau}c_{{\bf k},\sigma}(\tau)c_{{\bf k}^{\prime}+{\bf Q}-{\bf q}^{\prime},\sigma^{\prime}}^{\dag}(0)\Big\rangle
\Big\langle T_{\tau} c_{{\bf k}^{\prime}+{\bf Q},\sigma^{\prime}}(0)c_{{\bf k}+{\bf q},\sigma}^{\dag}(\tau)\Big\rangle\nonumber\\
&&+\Big\langle
T_{\tau}c_{{\bf k}+{\bf Q},\sigma}(\tau)c_{{\bf k}^{\prime}-{\bf q}^{\prime},\sigma^{\prime}}^{\dag}(0)\Big\rangle
\Big\langle T_{\tau}c_{{\bf k}^{\prime},\sigma^{\prime}}(0)c_{{\bf k}+{\bf Q}+{\bf q},\sigma}^{\dag}(\tau)\Big\rangle\nonumber\\
&&\left.+\Big\langle T_{\tau}c_{{\bf k}+{\bf Q},\sigma}(\tau)c_{{\bf k}^{\prime}+{\bf Q}-{\bf q}^{\prime},\sigma^{\prime}}^{\dag}(0)
\Big\rangle \Big\langle T_{\tau}c_{{\bf k}^{\prime}+{\bf Q},\sigma^{\prime}}(0)
c_{{\bf k}+{\bf Q}+{\bf q},\sigma}^{\dag}(\tau)\Big\rangle\right]\nonumber\\
&&~~~+\delta_{\sigma,\bar{\sigma}^{\prime}}\left[ \Big\langle
T_{\tau}c_{{\bf k},\sigma}(\tau)c_{{\bf k}^{\prime},\sigma^{\prime}}(0)\Big\rangle
\Big\langle
T_{\tau}c_{{\bf k}^{\prime}-{\bf q}^{\prime},\sigma^{\prime}}^{\dag}(0)
c_{{\bf k}+{\bf q},\sigma}^{\dag}(\tau)\Big\rangle\right.\nonumber\\
&&+\Big\langle T_{\tau}c_{{\bf k},\sigma}(\tau)
c_{{\bf k}^{\prime}+{\bf Q},\sigma^{\prime}}(0)\Big\rangle
\Big\langle T_{\tau}c_{{\bf k}^{\prime}+{\bf Q}-{\bf q}^{\prime},\sigma^{\prime}}^{\dag}(0)
c_{{\bf k}+{\bf q},\sigma}^{\dag}(\tau)\Big\rangle\nonumber\\
&&+\Big\langle
T_{\tau}c_{{\bf k}+{\bf Q},\sigma}(\tau)c_{{\bf k}^{\prime},\sigma^{\prime}}(0)\Big\rangle
\Big\langle T_{\tau}c_{{\bf k}^{\prime}-{\bf q}^{\prime},\sigma^{\prime}}^{\dag}(0)
c_{{\bf k}+{\bf Q}+{\bf q},\sigma}^{\dag}(\tau)\Big\rangle\nonumber\\
&&\left.+\Big\langle T_{\tau}c_{{\bf k}+{\bf Q},\sigma}(\tau)
c_{{\bf k}^{\prime}+{\bf Q},\sigma^{\prime}}(0)\Big\rangle
\Big\langle
T_{\tau}c_{{\bf k}^{\prime}+{\bf Q}-{\bf q}^{\prime},\sigma^{\prime}}^{\dag}(0)
c_{{\bf k}+{\bf Q}+{\bf q},\sigma}^{\dag}(\tau)\Big\rangle\right]\nonumber\\
&=&-\frac{1}{N}\sum_{{\bf k},{\bf k}^{\prime},\sigma}^{\prime}\delta_{{\bf k}^{\prime},{\bf k}+{\bf q}^{\prime}}\left[
\Big\langle
T_{\tau}c_{{\bf k},\sigma}(\tau)c_{{\bf k},\sigma}^{\dag}(0)\Big\rangle
\Big\langle T_{\tau}c_{{\bf k}+{\bf q}^{\prime},\sigma}(0)c_{{\bf k}+{\bf q},\sigma}^{\dag}(\tau)\Big\rangle\right.\nonumber\\
&&+\Big\langle
T_{\tau}c_{{\bf k},\sigma}(\tau)c_{{\bf k}+{\bf Q},\sigma}^{\dag}(0)\Big\rangle
\Big\langle T_{\tau} c_{{\bf k}+{\bf q}^{\prime}+{\bf Q},\sigma}(0)c_{{\bf k}+{\bf q},\sigma}^{\dag}(\tau)\Big\rangle\nonumber\\
&&+\Big\langle
T_{\tau}c_{{\bf k}+{\bf Q},\sigma}(\tau)c_{{\bf k},\sigma}^{\dag}(0)\Big\rangle
\Big\langle T_{\tau}c_{{\bf k}+{\bf q}^{\prime},\sigma}(0)c_{{\bf k}+{\bf Q}+{\bf q},\sigma}^{\dag}(\tau)\Big\rangle\nonumber\\
&&\left.+\Big\langle
T_{\tau}c_{{\bf k}+{\bf Q},\sigma}(\tau)c_{{\bf k}+{\bf Q},\sigma}^{\dag}(0)\Big\rangle
\Big\langle T_{\tau}c_{{\bf k}+{\bf q}^{\prime}+{\bf Q},\sigma}(0)c_{{\bf k}+{\bf Q}+{\bf q},\sigma}^{\dag}(\tau)\Big\rangle\right]\nonumber\\
&&~~~+\delta_{{\bf k},-{\bf k}^{\prime}}\left[ \Big\langle
T_{\tau}c_{{\bf k},\sigma}(\tau)c_{{\bf k},\bar{\sigma}}(0)\Big\rangle
\Big\langle
T_{\tau}c_{{\bf k}+{\bf q}^{\prime},\bar{\sigma}}^{\dag}(0)
c_{{\bf k}+{\bf q},\sigma}^{\dag}(\tau)\Big\rangle\right.\nonumber\\
&&+\Big\langle T_{\tau}c_{{\bf k},\sigma}(\tau)
c_{{\bf k}+{\bf Q},\bar{\sigma}}(0)\Big\rangle \Big\langle
T_{\tau}c_{{\bf k}+{\bf Q}+{\bf q}^{\prime},\bar{\sigma}}^{\dag}(0)
c_{{\bf k}+{\bf q},\sigma}^{\dag}(\tau)\Big\rangle\nonumber\\
&&+\Big\langle
T_{\tau}c_{{\bf k}+{\bf Q},\sigma}(\tau)c_{{\bf k},\bar{\sigma}}(0)\Big\rangle
\Big\langle
T_{\tau}c_{{\bf k}+{\bf q}^{\prime},\bar{\sigma}}^{\dag}(0)
c_{{\bf k}+{\bf Q}+{\bf q},\sigma}^{\dag}(\tau)\Big\rangle\nonumber\\
&&\left.+\Big\langle T_{\tau}c_{{\bf k}+{\bf Q},\sigma}(\tau)
c_{{\bf k}+{\bf Q},\bar{\sigma}}(0)\Big\rangle \Big\langle
T_{\tau}c_{{\bf k}+{\bf Q}+{\bf q}^{\prime},\bar{\sigma}}^{\dag}(0)
c_{{\bf k}+{\bf Q}+{\bf q},\sigma}^{\dag}(\tau)\Big\rangle\right]
\end{eqnarray}}
The charge components are the same as the transverse ones with the only
difference that all the Green's function terms are
positive. This is also true in the pure SDW case, so that only the SDW
coefficients will be affected here, which we can write using
symmetries as 
{\allowdisplaybreaks\begin{eqnarray}\label{coeffafmsc3}
B_{11}^{\nu=\nu^{\prime},\sigma\sigma}
&=&(\alpha_{{\bf k}}\alpha_{{\bf k}+{\bf q}}+\beta_{{\bf k}}\beta_{{\bf k}+{\bf q}})^2
=\frac{1}{2}\left(1+\frac{\xi_{{\bf k}}^{-}\xi_{{\bf k}+{\bf q}}^{-}+(US)^2}
{E_{0{\bf k}}E_{0{\bf k}+{\bf q}}}\right)\nonumber\\
B_{11}^{\nu \ne \nu^{\prime},\sigma\sigma}
&=&(\alpha_{{\bf k}}\beta_{{\bf k}+{\bf q}}-\beta_{{\bf k}}\alpha_{{\bf k}+{\bf q}})^2
=\frac{1}{2}\left(1-\frac{\xi_{{\bf k}}^{-}\xi_{{\bf k}+{\bf q}}^{-}+(US)^2}
{E_{0{\bf k}}E_{0{\bf k}+{\bf q}}}\right)\nonumber\\
B_{22}^{\nu=\nu^{\prime},\sigma\sigma}
&=&B_{11}^{\nu=\nu^{\prime},\sigma\sigma}({\bf q}\rightarrow{\bf q}+{\bf Q})
=(\alpha_{{\bf k}}\beta_{{\bf k}+{\bf q}}+\beta_{{\bf k}}\alpha_{{\bf k}+{\bf q}})^2\nonumber\\
&=&\frac{1}{2}\left(1-\frac{\xi_{{\bf k}}^{-}\xi_{{\bf k}+{\bf q}}^{-}-(US)^2}
{E_{0{\bf k}}E_{0{\bf k}+{\bf q}}}\right)\nonumber\\
B_{22}^{\nu \ne \nu^{\prime},\sigma\sigma}
&=&B_{11}^{\nu\ne\nu^{\prime},\sigma\sigma}({\bf q}\rightarrow{\bf q}+{\bf Q})
=(\alpha_{{\bf k}}\alpha_{{\bf k}+{\bf q}}-\beta_{{\bf k}}\beta_{{\bf k}+{\bf q}})^2\nonumber\\
&=&\frac{1}{2}\left(1+\frac{\xi_{{\bf k}}^{-}\xi_{{\bf k}+{\bf q}}^{-}-(US)^2}
{E_{0{\bf k}}E_{0{\bf k}+{\bf q}}}\right)\nonumber\\
B_{12}^{\nu=\nu^{\prime},\sigma\sigma}
&=&B_{21}^{\nu=\nu^{\prime},\sigma\sigma}
=-\nu(\alpha_{{\bf k}}\beta_{{\bf k}}+\alpha_{{\bf k}+{\bf q}}\beta_{{\bf k}+{\bf q}})
=-\nu\frac{US}{2}\left(\frac{1}{E_{0{\bf k}}}+\frac{1}{E_{0{\bf k}+{\bf q}}}\right)\nonumber\\
B_{12}^{\nu \ne \nu^{\prime},\sigma\sigma}
&=&B_{21}^{\nu\ne\nu^{\prime},\sigma\sigma}
=-\nu(\alpha_{{\bf k}}\beta_{{\bf k}}-\alpha_{{\bf k}+{\bf q}}\beta_{{\bf k}+{\bf q}})
=-\nu\frac{US}{2}\left(\frac{1}{E_{0{\bf k}}}-\frac{1}{E_{0{\bf k}+{\bf q}}}\right),\nonumber\\
\end{eqnarray}}
where the subscript $L$ stands for longitudinal/charge.

\subsection{RPA Susceptibility:}

The RPA values of the susceptibilities above are the same as in the SDW case.
The $2\times2$ transverse susceptibility is obtained from the standard
formula\cite{SWZ}
\begin{eqnarray}\label{chirpat2}
\chi_{RPA}^{+-}({\bf q},i\omega_n)&=&\frac{\chi^{+-}({\bf q},i\omega_n)}{{\bf 1}-
U\chi^{+-}({\bf q},i\omega_n)},
\end{eqnarray}
where, away from half-filling, the charge and longitudinal parts get mixed, leading to the modified RPA formula \cite{chubukov}
\begin{eqnarray}\label{chirpaz}
\chi_{RPA}^{z\rho}({\bf q},i\omega_n)&=&\frac{\chi^{z\rho}({\bf q},i\omega_n)}{{\bf 1}-
\tilde U\chi^{z\rho}({\bf q},i\omega_n)}.
\end{eqnarray}
Here components of the mixed susceptibilties are $\chi^{z\rho}_{11}=\chi^{zz}({\bf q},\omega )$,  $\chi^{z\rho}_{22}=\chi^{\rho\rho}({\bf q+Q},\omega)$ and $\chi^{z\rho}_{12}=\chi^{z\rho}({\bf q,q+Q},\omega ),$ while $\tilde U$ was defined in Eq.~\ref{tildeu}.  Note that since $\chi$ is a $2\times2$ matrix, the denominators in the above equations are matrices as well.

\section{Antiferromagnetic order only}\label{S:AD}
From Eqs.~\ref{eigen} and~\ref{eigenvec}, when the superconducting order is absent, $u_{\bf k}^{\nu}=1$, $v_{\bf k}^{\nu}=0$, for $\nu =\pm 1$.  Hence from Eq.~\ref{t11b2}, $A_{1}^{\nu,\nu^{\prime}}+A_{2}^{\nu,\nu^{\prime}}=1$, $A_{3}^{\nu,\nu^{\prime}}=A_{4}^{\nu,\nu^{\prime}}=0$.  Thus, $\chi$ becomes
\begin{eqnarray}\label{afmscsft2AF}
\chi_{ij}^{\sigma\sigma^{\prime}}({\bf q},i\omega_m)
&=&\frac{1}{N}\sum_{{\bf k}}^{\prime}\sum_{\nu,\nu^{\prime}}
B_{ij}^{\nu\nu^{\prime},\sigma\sigma^{\prime}}\chi_{01}^{\nu,\nu^{\prime}}.
\end{eqnarray}
The general expression when both the lower and upper magnetic bands are partially filled is quite involved.  Results are simpler at half-filling and low temperatures when $\chi_{01}^{\nu,\nu}=0$.  In order to compare our results with those of Schrieffer, Wen, and Zhang (SWZ)\cite{SWZ}, we further assume that $\xi_{\bf k}^+=0$, so that $E_{\bf k}^{\nu}=\nu E_{0k}$. Then
{\allowdisplaybreaks\begin{eqnarray}\label{afmscsft2AF2}
\chi_{11}^{+-}({\bf q},\omega)
&=&\frac{1}{2N}\sum_{{\bf k}}^{\prime}\left[1-\frac{\xi_{{\bf k}}^{-}\xi_{{\bf k}+{\bf q}}^{-}-(US)^2}{E_{0{\bf k}}E_{0{\bf k}+{\bf q}}}\right]\nonumber\\
&&~~~~~~\times\left[{1\over\omega +E_{0k}+E_{0k+q}-i\delta}-{1\over\omega -E_{0k}-E_{0k+q}+i\delta}\right],
\end{eqnarray}}
{\allowdisplaybreaks\begin{eqnarray}\label{afmscsft2AF3}
\chi_{22}^{+-}({\bf q},\omega)
&=&\frac{1}{2N}\sum_{{\bf k}}^{\prime}\left[1+\frac{\xi_{{\bf k}}^{-}\xi_{{\bf k}+{\bf q}}^{-}+(US)^2}{E_{0{\bf k}}E_{0{\bf k}+{\bf q}}}\right]\nonumber\\
&&~~~~~\times\left[{1\over\omega +E_{0k}+E_{0k+q}-i\delta}-{1\over\omega -E_{0k}-E_{0k+q}+i\delta}\right],
\end{eqnarray}
{\allowdisplaybreaks\begin{eqnarray}\label{afmscsft2AF4}
\chi_{12}^{+-}({\bf q},\omega)
&=&-\frac{US}{2N}\sum_{{\bf k}}^{\prime}\left[{1\over E_{0{\bf k}}}+{1\over E_{0{\bf k}+{\bf q}}}\right]\nonumber\\
&&~~~~~~\times\left[{1\over\omega +E_{0k}+E_{0k+q}-i\delta}+{1\over\omega -E_{0k}-E_{0k+q}+i\delta}\right],
\end{eqnarray}}
which reproduces the SWZ result. For the longitudinal/charge modes, the only change is replacing Eq.~(\ref{coeffafmsc2}) by Eq.~(\ref{coeffafmsc3}), which gives
{\allowdisplaybreaks\begin{eqnarray}\label{afmscsft2AF5}
\chi^{z}({\bf q},\omega)&=&\chi_{11}^{z\rho}({\bf q},\omega)
=\frac{1}{2N}\sum_{{\bf k}}^{\prime}\left[1-\frac{\xi_{{\bf k}}^{-}\xi_{{\bf k}+{\bf q}}^{-}+(US)^2}{E_{0{\bf k}}E_{0{\bf k}+{\bf q}}}\right]\nonumber\\
&&~~~~~~\times\left[{1\over\omega +E_{0k}+E_{0k+q}-i\delta}-{1\over\omega -E_{0k}-E_{0k+q}+i\delta}\right],
\end{eqnarray}}
{\allowdisplaybreaks\begin{eqnarray}\label{afmscsft2AF6}
\chi^{c}({\bf q}+{\bf Q},\omega)&=&\chi_{22}^{z\rho}({\bf q},\omega)
=\frac{1}{2N}\sum_{{\bf k}}^{\prime}\left[1+\frac{\xi_{{\bf k}}^{-}\xi_{{\bf k}+{\bf q}}^{-}-(US)^2}{E_{0{\bf k}}E_{0{\bf k}+{\bf q}}}\right]\nonumber\\
&&~~~~~~\times\left[{1\over\omega +E_{0k}+E_{0k+q}-i\delta}-{1\over\omega -E_{0k}-E_{0k+q}+i\delta}\right],
\end{eqnarray}}
{\allowdisplaybreaks\begin{eqnarray}\label{afmscsft2AF7}
\chi_{12}^{z\rho}({\bf q},\omega)
&=&-\frac{US}{2N}\sum_{{\bf k}}^{\prime}\left[{1\over E_{0{\bf k}}}+{1\over E_{0{\bf k}-{\bf q}}}\right]\nonumber\\
&&~~~~\times\left[{1\over\omega +E_{0k}+E_{0k+q}-i\delta}+{1\over\omega -E_{0k}-E_{0k+q}+i\delta}\right]=0.
\end{eqnarray}}
Since the off-diagonal term vanishes, the longitudinal and charge susceptibilities decouple.  From Eq.~\ref{afmscsft2AF6}, shifted by $Q$, we see that $\chi^{c}({\bf q},\omega)=\chi^{z}({\bf q},\omega)$, Eq.~\ref{afmscsft2AF5}, the SWZ result.

\section{Superconducting order only}\label{S:AE}
Again, Eqs.~\ref{eigen} and~\ref{eigenvec} are used, but now assuming the SDW order is absent, $\alpha_{\bf k}\beta_{{\bf k}}=0$.  Unpacking the SC susceptibility is more complicated, and therefore we discuss it in
detail.   Firstly, $\alpha_k^2=(1+\xi_k/|\xi_k|)/2$ = 1 [0] if $\xi_k$ $>$0 [$<$0].  At the same
time, the sign of $\xi_k$ also influences the eigenenergies, such that $E^+({\bf k})$ =
$E_0({\bf k})=\sqrt{\xi_k^2+\Delta_k^2}$ if $\xi_k$ $>$0, and = $E_0({\bf k}+{\bf Q})$ if $\xi_k$
$<$0, with the opposite result for $E^-({\bf k})$ [recall that we are now assuming $US=0$.].  As $\xi_k$
and $\xi_{k+q}$ independently change sign, the factors $B$ switch from 1 to 0 in such a
way that we always remain on the correct branch, and the result is the same
as that we would get if we assume both $\xi_k$ and $\xi_{k+q}$ to be positive. Secondly, when there is no SDW order, $q$ and $q+Q$ are independent vectors, and we only need the formula for $\chi^{+-}_{11}$ [note $\chi^{+-}_{12}=B_{12}^{\nu\nu^{\prime},+-}=0$].  In this
case, $B_{11}^{\nu\nu^{\prime,+-}}=\delta_{\nu\nu'}$, and
{\allowdisplaybreaks\begin{eqnarray}
\chi_{11}^{+-}({\bf q},i\omega_m)
&=&\frac{1}{N}\sum_{{\bf k}}^{\prime}\sum_{n,\nu}
B_{n}^{\nu,\nu}(11)\chi_{0n}^{\nu,\nu},\\
\label{afmscsft2SC4}
&=&\frac{1}{2N}\sum_{{\bf k}}\left[
\left(1+\frac{\xi_{{\bf k}}\xi_{{\bf k}+{\bf q}}+\Delta_{{\bf k}}\Delta_{{\bf k}+{\bf q}}}{E^+_{{\bf k}}E^+_{{\bf k}+{\bf q}}}\right)\chi_{01}\right.\nonumber\\
&&+\frac{1}{2}\left(1+\frac{\xi_{{\bf k}}}{E^+_{{\bf k}}}
-\frac{\xi_{{\bf k}+{\bf q}}}{E^+_{{\bf k}+{\bf q}}}
-\frac{\xi_{{\bf k}}\xi_{{\bf k}+{\bf q}}+\Delta_{{\bf k}}\Delta_{{\bf k}+{\bf q}}}{E^+_{{\bf k}}E^+_{{\bf k}+{\bf q}}}\right)\chi_{03}\nonumber\\
&&\left. +\frac{1}{2}\left(1-\frac{\xi_{{\bf k}}}{E^+_{{\bf k}}}
+\frac{\xi_{{\bf k}+{\bf q}}}{E^+_{{\bf k}+{\bf q}}}
-\frac{\xi_{{\bf k}}\xi_{{\bf k}+{\bf q}}+\Delta_{{\bf k}}\Delta_{{\bf k}+{\bf q}}}{E^+_{{\bf k}}E^+_{{\bf k}+{\bf q}}}\right)\chi_{04}\right].
\end{eqnarray}}
$\chi_{0i}$ are given in Eq.~\ref{SCS1} above. This corrects the formula given in Ref.~\onlinecite{BScal}.

\section{Calculating the Self-Energy}\label{S:AF}

The bosonic modes cause an increase in the low-energy electronic mass and a shortening of the lifetime at
higher energies. This effect can be described in terms of a complex self-energy $\Sigma$ that the
electrons acquire as a consequence of electron-boson couplings.  Corresponding to each susceptibility component (transverse, longitudinal-plus-charge), there is a self-energy $\Sigma$ which is itself a 4$\times$4 tensor.  The transverse self-energy\cite{vignale,thesis} can be written as
\begin{eqnarray}\label{selfen1}
\Sigma^{+-}_{ij}({\bf q},\omega)&=&\sum_{k,{\bf k}}\int d\omega'\chi^{\prime\prime +-}_{ik}({\bf k},\omega')G_{kj}({\bf k-q},\omega,\omega')\Gamma({\bf k},{\bf q},\omega,\omega'),
\end{eqnarray}
where (Eq.~\ref{greenft})
\begin{equation}\label{greenftSE}
\hat{G}({\bf k},\omega ,\omega')=\hat{U}\hat g({\bf k},\omega ,\omega')\hat{U}^{\dag},
\end{equation}
where $\hat g$ is a diagonal matrix with $g_{ii}({\bf k},\omega ,\omega')=\tilde g(E_{{\bf k},i},\omega ,\omega')$, $E_{{\bf k},i}$ = $(E^+_{\bf k},E^-_{\bf k},-E^+_{\bf k},-E^-_{\bf k})$, $i=1,4$, and
\begin{equation}\label{greenftSE2}
\tilde g(E,\omega ,\omega')={f(E)\over \omega +\omega'+i\delta-E}
+{1-f(E)\over\omega-\omega' +i\delta -E}.
\end{equation}
In the previous section, $\chi^{+-}$ was a 2$\times$2 matrix, whereas Eq.~\ref{selfen1} involves a 4$\times$4 matrix, which is formed by
\begin{equation}\label{greenftSE3}
\hat\chi^{+-}=\begin{pmatrix} \chi^{+-}&0\cr
                    0&\chi^{+-}\cr\end{pmatrix}.
\end{equation}

The formalism for the longitudinal plus charge self-energy is similar, the only change being the substitution of the correct susceptibility in Eq.~\ref{selfen1}.  In our QP-GW formalism, we generally set the vertex function to $\Gamma\rightarrow 1/Z$.

The high-energy kink is associated with the peak in $\Sigma^{\prime\prime}$.  The strength of this peak may be described in terms of the area under the $\Sigma^{\prime\prime}$ curve, Fig.~\ref{sumrule}, which gives a direct measure of the tendency of the spectrum to split into coherent and incoherent parts, and hence a measure of the weight of
the Hubbard bands.  Fig.~\ref{sumrule} shows this quantity as a function of doping
above and below $E_F$ for both NCCO and LSCO. While high-energy kinks should be present both above and below $E_F$, for both materials we find that the occupied state kink, represented by $\int\Sigma^{\prime\prime}~d\omega$ below $E_F$, shows a much faster fall-off with doping.  Note that only this occupied kink can be seen in ARPES.  This fast fall-off seems to terminate around $x\sim 0.20-0.25$ close to the point where the high energy spectral weight extrapolates to
zero, $x_{UHB}=1/\beta\sim 0.25$ (Section~\ref{S:NFL}.1). This is also close to the doping where
AFM order ends in a critical point, suggesting an intimate connection
between the decrease of magnon scattering and the collapse of the AFM gap.
\begin{figure}[top]
\centering
\rotatebox{0}{\scalebox{0.1}{\includegraphics{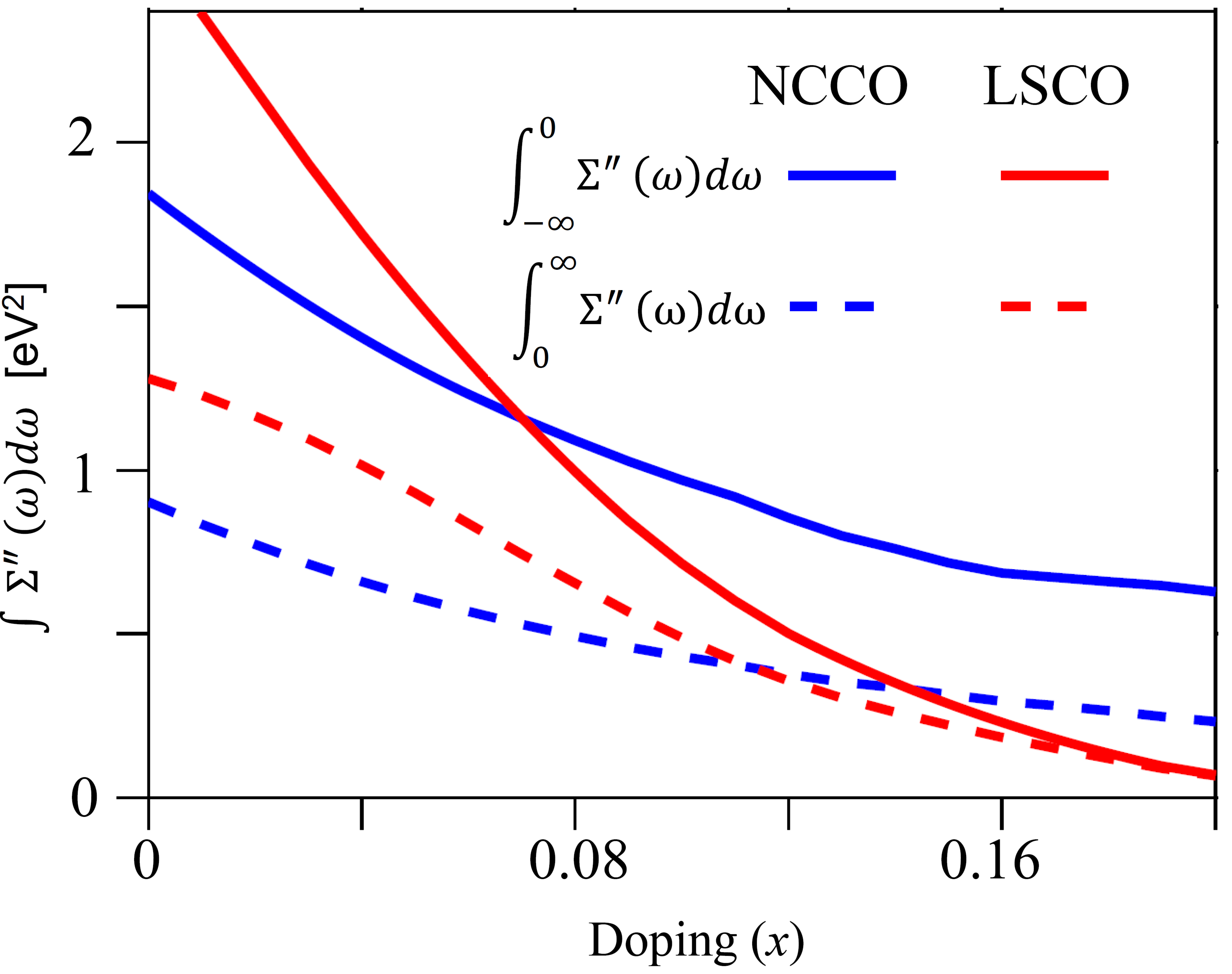}}}
\caption{ Integrated imaginary part of
the calculated self-energy as a function of doping broken up into contributions from positive and negative energies. [From Ref.~\onlinecite{ASWT}]} \label{sumrule}
\end{figure}

\section{Optical conductivity}\label{S:AG}

Within the framework of the linear response theory, optical conductivity is given by the Kubo formula in the limit ${\bf q}\rightarrow 0$
{\allowdisplaybreaks\begin{eqnarray}
&&\sigma_{ij}(i\omega_n)=\frac{ie^2}{\omega}\sum_{{\bf
k},\sigma}^{\prime} \Bigl\{ {\rm{Tr}}\left[\frac{n_{{\bf
k},\sigma}} {m_{{\bf k},ij}}\right]- \sum_{{\bf
k}^{\prime},\sigma^{\prime}}^{\prime}\int_{-\infty}^{\infty}d
\omega_1
\int_{-\infty}^{\infty}d\omega_2\nonumber\\
&&~~~{\rm{Tr}}\left[v_{{\bf k},i}A({\bf k},\sigma,\omega_1)
\Gamma^{op}_j({\bf k},{\bf k}^{\prime},\omega_1,\omega_2)
A({\bf k}^{\prime},\sigma^{\prime},\omega_2)\right]
\frac{f(\omega_2)-f(\omega_1)}{i\omega_n+\omega_2-\omega_1}
\Bigr\}
\label{opticon}
\end{eqnarray}}
where $(i,j) = x,y,z-$directions. Real frequency optical
conductivity is extracted from the Matsubara results by analytic
continuation $i\omega_n\rightarrow \omega +i/\tau$, where $\tau$
is the impurity scattering term. The first term in Eq.~\ref{opticon}
corresponds to the diamagnetic response kernel (Drude weight), 
which depends only on the FS topology, while the second
term (paramagnetic) gives the dynamical contribution to the
optical response\cite{allen}. The $v_{{\bf k},i}$ ($m_{{\bf
k},ij}$) are the band velocities (masses) of the coherent band, and
$n_{{\bf k},\sigma}$ is the momentum density of quasiparticles at
the FS. The spectral weights $A({\bf k},\omega)$ are
obtained from the imaginary part of the dressed
Green's function.  Finally, the optical vertex correction is
approximated by its lowest order term as $\Gamma^{op}_j({\bf
k},{{\bf k}^{\prime}},\omega_1,\omega_2)\rightarrow v_{{\bf
k},j}\delta_{{\bf k},{\bf k}^{\prime}}
\delta_{\sigma,\sigma^{\prime}}$.

\section{Parameters}\label{S:AH}

Our bare dispersions are all taken from self-consistent first-principles LDA calculations, which we have parameterized by tight-binding fits. These are collected in Table F1\cite{markietb,Jouko,markiesc,yoshida,huefner} for Bi$_2$Sr$_2$CaCu$_2$O$_{8}$ (Bi2212), La$_{2-x}$Sr$_x$CuO$_4$ (LSCO), Nd$_{2-x}$Ce$_x$CuO$_4$ (NCCO), and YBa$_2$Cu$_3$O$_{6+x}$ (YBCO).  More tight-binding model for YBCO with including the CuO chain state is given in Ref.~\cite{Das_EP_CuO}, and for HgBa$_2$Ca$_2$Cu$_2$O$_8$ (HBCCO) by including the low-lying Hg-O band can be obtained from Ref.~\cite{Das_HBCO}. The interaction parameters $V$ and $U$, discussed in Section~\ref{S:QP-GW}.2, are listed in Table F2, while the doping-dependent effective $U$ is plotted in Fig.~\ref{Ux}.

Note that Table F2 lists the experimental $\Delta_{pg}$ values, from which we obtained the $S$ values via Eq.~\ref{delselffinal}, shown in Fig.~\ref{Ux}(b). $U$ is then defined as $U=\Delta_{pg}/S$, and listed in Table F2 and plotted as symbols in Fig.~\ref{Ux}(a). We emphasize that these experimentally-derived values of $U$ are generally quite close to the parameter-free values obtained self-consistently to account for screening effects\cite{tanmoyop} as seen by comparing the symbols and solid lines in Fig.~\ref{Ux}(a). A similar procedure was used for obtaining the pairing potential $V$ from the experimental $\Delta_{sc}$, listed in Table F2.

%
%
%
%

%


\begin{table}[h]
\centering
\begin{tabular}{|c|c|c|c|c|c|c|c|}
\hline \hline
Material& $t$ & $t^{\prime}$ & $t^{\prime\prime}$& $t^{\prime\prime\prime}$&$t^{iv}$&$Z$\\
\hline
NCCO [Ref.~\onlinecite{markietb}]& 0.42 & -0.1 & 0.065 & 0.0075&0& 0.4 \\
LSCO [Ref.~\onlinecite{markietb}]& 0.4195 & -0.0375 & 0.018 & 0.034 &0& 0.3\\
Bi2212 [Ref.~\onlinecite{Jouko}]& 0.44 & -0.1 & 0.05 & 0.0 & 0&0.4\\
YBCO [Ref.~\onlinecite{tanmoymagres}]& 0.35 & -0.06 & 0.035 & -0.005 & -0.01& 0.55\\
\hline \hline
\end{tabular}
\vspace{0.5cm}
\caption{ Tight-binding parameters obtained by fitting to 
LDA band-structures and the associated band renormalization factors $Z$.}
\label{Tab:TB}
\end{table}

\begin{landscape}

\begin{table}[h]
\centering
\begin{tabular}{|c|c|c|c|c|c|c|}
\hline \hline
Material& Doping ($x$) &$\Delta_{pg}$ (meV)& $U/t$ & $\Delta_{sc}$ (meV)& Pairing Potential & $T_c$ (K) \\
& & (Exp./Theory) & (Theory)& (Exp./Theory) & $V$ (meV) (Theory)& Exp.(Theory)\\
 \hline
NCCO & 0.0&  950 (at hotspot) & 5.7 & -- & -- & -- \\
NCCO & 0.04&  628  & 4.4 & -- & -- & -- \\
NCCO & 0.10& 322 & 3.5 & 0.5 & -104 & $\sim$ 8 (7) \\
NCCO & 0.15& 160 & 3.1 & 5.5 & -113 & 24 (31) \\
NCCO & 0.18&  0 & 3.1 & 3.7 & -100 & 22 (20) \\
 LSCO & 0.0 & 770 & 5.5 & -- & -- & -- \\
 LSCO & 0.06 & 150 & 2.35 & 6 & -93 & 18 (27) \\
 LSCO &0.12& 120 & 2.27 & 11 & -63 & 30 (48)\\
 LSCO &0.16&63  & 2.25 & 15 & -51 & 40 (85)\\
 LSCO & 0.18 & 43 & 2.25 & 13 & -35 & 37 (75)  \\
LSCO &0.22&0 & 2.25 & 8  & -28 & 26 (48)\\
Bi2212 & 0.0 & 790 & 6.0 & -- & -- & -- \\
Bi2212 & 0.10 & 113 & 2.46 & 15 & -77 & 55 (85) \\
 Bi2212 & 0.12 & 95 & 2.42 & 17.5 & -75 & 65 (95) \\
 Bi2212 & 0.16 & 75 & 2.36 & 20 & -67 & 91 (115) \\
 Bi2212 & 0.19 & 50 & 2.36 & 17.5 & -58 & 70 (90)\\
 Bi2212 & 0.22 & 25 & 2.36 & 12.5 & -50 & 55 (75)\\
YBCO & 0.21 & 9 & 2.9 & 35 & -130 & 89 ( 90)\\ 
\hline \hline
\end{tabular}
\vspace{0.5cm}
\caption{Parameters for NCCO, LSCO and Bi2212 for various dopings $x$. $U/t$ values are chosen to reproduce the experimental pseudogaps
($\Delta_{pg}$) (third column) along the hot-spot
direction in NCCO, and the antinodal direction in LSCO and Bi2212. 
The pairing potential $V$ is similarly taken to reproduce the experimental superconducting
gap ($\Delta_{SC}$), whose maximum value along the antinodal direction
is given. Our mean-field calculations overestimate the
values of $T_c$, presumably due to the neglect of phase
fluctuations\cite{markiesc}. Experimental data are taken from: Ref.~\onlinecite{yoshida} for LSCO; Ref.~\onlinecite{tanmoyprl} for NCCO; Ref.~\onlinecite{huefner} for Bi2212; and Ref.~\onlinecite{tanmoymagres} for YBCO.}
\label{Tab:ch1_order}
\end{table}
\end{landscape}

%
\begin{figure}[top]
\centering
\rotatebox{0}{\scalebox{0.5}{\includegraphics{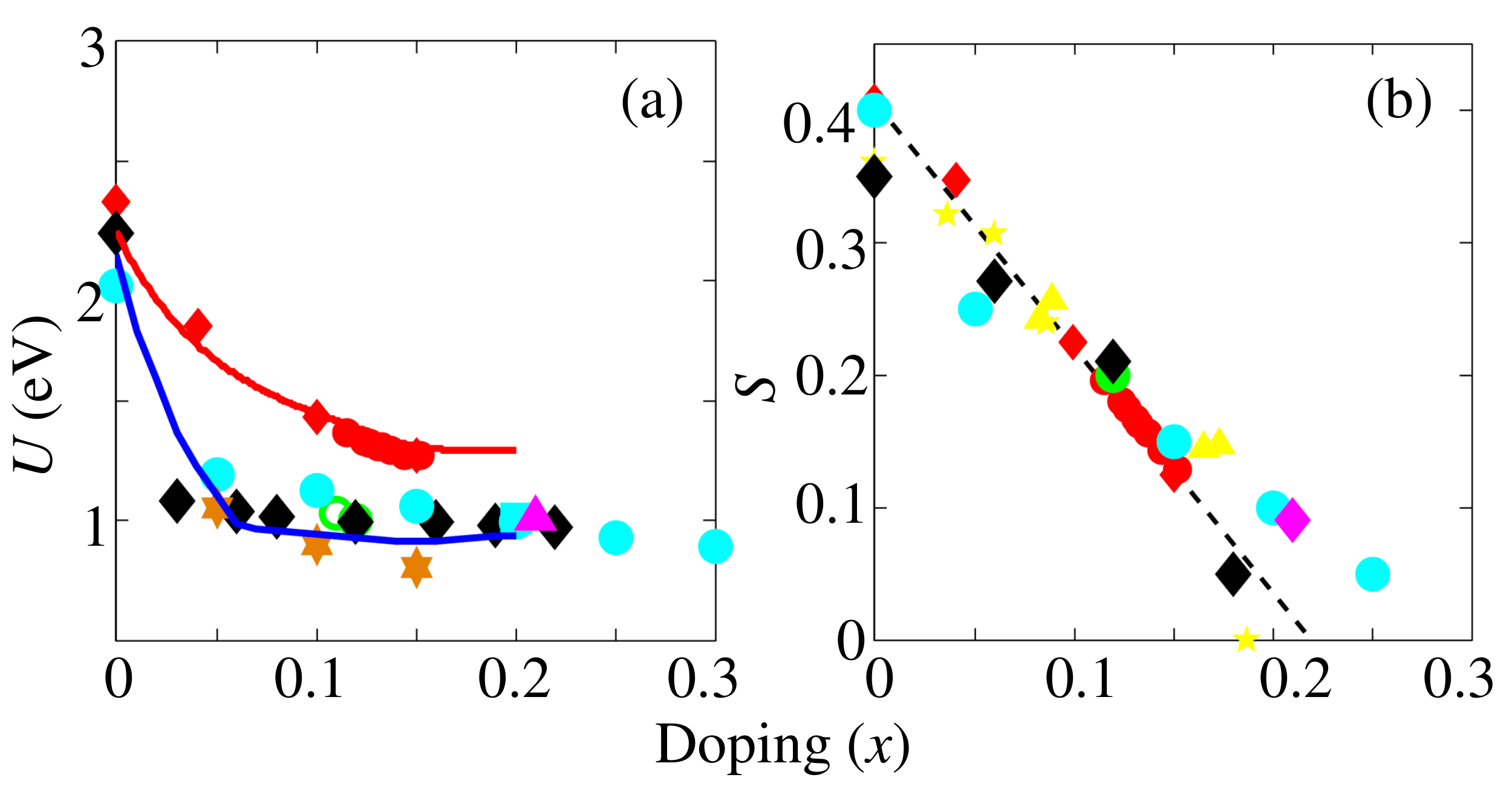}}}
\caption{ { Doping dependent values of effective $U$ and magnetization $S$.} (a) Solid lines give self-consistent, parameter-free values of $U$ as a function of doping in the QP-GW model\cite{tanmoyop} for electron- (red line) and hole-doped (blue line) cuprates. Various symbols give experimentally-derived values: NCCO: Red diamond, PCCO: red circles, PLCCO: green, LSCO: black, Bi212: cyan, YBCO: magenta.  Orange stars are an effective $\bar U(T)$ at $T/t=0.1$, taken as representative of the low-temperature limit assuming a bare $U=2$~eV, see Ref.~\onlinecite{DCA2}. (b) Corresponding values of the staggered (AFM) magnetization $S$. Yellow symbols are neutron data for NCCO from Ref.~\cite{Mang}. [After Ref.~\onlinecite{tanmoyLEK}.]}
\label{Ux}
\end{figure}
%






\section{Acronyms}\label{S:AI}

angle-resolved photoemission (ARPES) 

antinodal nesting (ANN)

anomalous spectral weight transfer (ASWT)

Brillouin zone (BZ)

Brinkman-Rice (BR)

cellular dynamical mean-field theory (CDMFT)

charge-density wave (CDW)

charge transfer (CT)

cluster perturbation theory (CPT)

coherent-potential-approximation (CPA)

density functional theory (DFT)

density-of-states (DOS)

dual fermion (DF)

dynamical cluster approximation (DCA)

dynamical vertex approximation (D$\Gamma$A)

dynamic mean-field theory (DMFT)

energy distribution curve (EDC)

the Fermi level ($E_F$)

Fermi surface (FS)

ferromagnetic (FM)

fluctuation-exchange (FLEX)

Gutzwiller approximation (GA)

Hartree-Fock (HF)

high-energy feature (HEF)

high-energy kink (HEK) 

inelastic neutron scattering (INS)

joint density-of-states (JDOS)

Kosterlitz-Thouless (KT) 

local-density approximation (LDA)

low energy peak (LEP)

lower Hubbard band (LHB)

lower magnetic band (LMB)

magnetic quasiparticle scattering (MQPS)

mean-field theory (MFT)

metal-insulator transition (MIT)

Mermin-Wagner (MW)

mid-infrared (MIR)


momentum distribution curve (MDC)

nearest neighbor (NN)

near-nodal nesting (NNN)

non-Fermi-liquid (NFL)

one-particle irreducible approach (1PI)

pair-breaking magnetic scattering (PBS) 

quantum critical point (QCP)

quantum Monte Carlo (QMC) 

quantum oscillations (QO) 

quasiparticle-GW (QP-GW)

quasiparticle interference (QPI)

quasiparticle self-consistent GW (QS-GW)


random-phase approximation (RPA)

resonant inelastic x-ray scattering (RIXS)

resonating valence bond (RVB)

scanning tunneling microscopy/spectroscopy (STM/STS)

spin-density wave (SDW) 

superconducting (SC)

topological transition (TT)

upper Hubbard band (UHB)

upper magnetic band (UMB)

Van-Hove singularity (VHS)

x-ray absorption near edge spectroscopy (XANES)

x-ray absorption spectroscopy (XAS)

Yang, Rice, and Zhang (YRZ)

{\it Materials}

Bi$_2$Sr$_2$CuO$_6$ (Bi2201)

Bi$_2$Sr$_2$CaCu$_2$O$_8$ (Bi2212)

Ca$_2$CuO$_2$Cl$_2$ (CCOC)

HgBa$_2$Ca$_2$Cu$_3$O$_{8+ \delta}$ (HBCO)

La$_{2-x}$Ba$_x$CuO$_4$  (LBCO)

La$_2$CuO$_4$ (LCO)

La$_{2-x}$Sr$_x$CuO$_4$ (LSCO)

Nd$_{2-x}$Ce$_x$CuO$_4$ (NCCO)

Nd$_2$CuO$_4$ (NCO)

Pr$_{2-x}$Ce$_x$CuO$_4$ (PCCO)

Pr$_{1-x}$LaCe$_x$CuO$_4$ (PLCCO)

Sr$_2$CuO$_2$Cl$_2$ (SCOC)

Tl$_2$Ba$_2$CuO$_{6+\delta}$ (TBCO)

YBa$_2$Cu$_3$O$_{6+y}$ (YBCO)


\begin{thebibliography}{99}
\bibitem{footSh}Shastry proposed a new scale for correlations.\cite{Shast}  Since his scale is in $U/t$, where $t$ is the nearest neighbor hopping energy, and $W\sim 8t$, the two scales are equivalent, but with different crossover values of $U$. Note however that in the new scheme Hartree-Fock transitions would only occur in the strongly correlated regime, while a system could be extremely strongly correlated and still have considerable double occupancy.
\bibitem{foottb}The TB parameters are taken to be doping
independent in the spirit of a rigid band picture. A more
realistic treatment of the doping dependence of the electronic
states (see, e.g., Refs.~\onlinecite{ft_sophi1,ft_sophi3,ft_sophi4}) was not
undertaken. However, we expect the rigid band model to be a good
approximation for doping away from the Cu-O planes.
\bibitem{foot6}The dimpling of the CuO$_2$ planes is taken into account by using two different bilayer distances: $d_1=2.9 \AA$ for O and $d_2=3.2 \AA$ for Cu \cite{dimpling}.
\bibitem{foot11}Spin-flip terms are neglected here.
\bibitem{footB1}The gap values are: $\Delta$ = $(25,35,35,45,45)$~meV. The magnetization values are: $m$ = $(0,0.04,0.08,0.12,0.16)$, for $x$ = (0.19,0.16,0.12,0.10,0.08).
\bibitem{footUD}At lower doping, we expect the competing order pseudogap to persist at higher temperatures.
\bibitem{AB4}Positron-annihilation spectroscopy can also probe the extended electron momentum density, although the interpretation is complicated by positron spatial distribution effects, see Ref.~\cite{AB5}.
\bibitem{foot7}In a one-band model, this insulating gap is often treated as an effective Mott gap\cite{ZR}, but we will not do so in this review.
\bibitem{foot2}Values of the three-band model parameters used are: $t_{CuO}=1.3$ eV, $t_{OO}=-0.65$ eV, $\Delta = -0.5,-0.38,-0.41$ eV and $m_d = 0.2, 0.12,0.06$, for $x = 0.0, 0.075, 0.14$, respectively.
\bibitem{foot3a}A smaller value of the effective Hubbard
parameter ($U=6.5$ eV instead of $7.2$ eV) leads to better agreement with experiments at
half-filling.  For 14\% doping, however, we use the same renormalized $U=5.7$~eV as Ref.  \cite{MBRIXS}.
\bibitem{foot1b}For finite hole doping, the AFM order can equally well be modeled by a competing magnetic, charge, or flux phase order.
\bibitem{foot3b}A single renormalization factor was used for both bands at all dopings.
\bibitem{foot6gl}This formula was generalized to finite temperatures in Ref.~\onlinecite{hwang2}.  Notably, however, Ref.~\onlinecite{hwang2} assumes a constant density of states, and hence does not account for the correction discussed here.
\bibitem{foot2d} A similar problem arises when the spectrum has a gap, see  Ref.~\onlinecite{DoHoTV}.
\bibitem{foot8gl}Only the self-energy due to magnetic fluctuations is computed since these dominate within $\sim$1~eV of $E_F$.
\bibitem{foot9gl}Since the momentum dependence of $\Sigma$ is weak, we take the result for $q=(\pi /2,\pi /2)$ as representative.
\bibitem{hop}The hopping parameters are taken from Ref.~\onlinecite{markiewater}: ($t,t',t'',t'''$) = (360,-100,35,10)~meV.
\bibitem{footz}We use $Z_0$ to distinguish the renormalization factor used  in calculating $\chi_0$; it is not identical to the doping dependent $Z$ 
found from the Eliashberg equations.
\bibitem{foot2c}This BCS renormalization factor is the inverse of that  normally used in Green's function theory.
\bibitem{foot1}Calculated $\alpha^2F$'s are mainly odd in $\omega$; a  small even component is subtracted out in the displayed results.
\bibitem{foot1e} In reality, the Mott physics lies at a higher energy,
comparable to the bonding-antibonding separation of the CuO$_2$-hybridized
bands.  The bands we refer to as U/LHB are the incoherent high
energy tails of the antibonding (LDA) band.
\end{thebibliography}
\end{document}